\documentclass[12pt,fleqn,a4paper]{article} 
\usepackage{epsfig,graphics,graphicx}
\usepackage{clshan-math,clshan-dm-dd}
\def \lsim {\:\raisebox{-0.7ex}{$\stackrel{\textstyle<}{\sim}$}\:}
\def \gsim {\:\raisebox{-0.7ex}{$\stackrel{\textstyle>}{\sim}$}\:}
\begin{document}
\thispagestyle{empty}
\begin{flushright}
 April 2011
\end{flushright}
\begin{center}
{\large\bf
 Effects of Residue Background Events
 in Direct Dark Matter \\ \vspace{-0.04cm} Detection Experiments on
 the Determinations of \\ \vspace{0.1cm} Ratios of WIMP--Nucleon Cross Sections} \\
\vspace{0.7cm}
 {\sc Chung-Lin Shan} \\
\vspace{0.5cm}
 {\it Department of Physics, National Cheng Kung University \\
      No.~1, University Road,
      Tainan City 70101, Taiwan, R.O.C.}                    \\~\\
 {\it Physics Division,
      National Center for Theoretical Sciences              \\
      No.~101, Sec.~2, Kuang-Fu Road,
      Hsinchu City 30013, Taiwan, R.O.C.}                   \\~\\
 {\it E-mail:} {\tt clshan@mail.ncku.edu.tw}                \\
\end{center}
\vspace{1cm}
\begin{abstract}
 In our work on the development of
 model--independent data analysis methods
 for determining ratios between different couplings/cross sections of
 Weakly Interacting Massive Particles (WIMPs)
 by using measured recoil energies
 from direct Dark Matter detection experiments directly,
 it was assumed that
 the analyzed data sets are background--free,
 i.e., all events are WIMP signals.
 In this article,
 as a more realistic study,
 we take into account
 a fraction of possible residue background events,
 which pass all discrimination criteria and
 then mix with other real WIMP--induced events
 in our data sets.

 Our simulations show that,
 assuming that
 the spin--dependent (SD) WIMP--nucleus interaction dominates
 over the spin--independent (SI) one,
 the maximal acceptable fraction of residue background events
 in the analyzed data sets
 for determining the ratio of the SD WIMP coupling
 on neutrons to that on protons
 is $\sim$ 20\% -- 40\%;
 whereas
 considering a general combination of
 the SI and SD WIMP interactions,
 the maximal acceptable background ratio
 for determining the ratio between
 two SD WIMP couplings
 as well as
 the ratios of the SD cross section
 on protons (neutrons) to the SI one
 is $\sim$ 10\% -- 20\%.
 Moreover,
 by considering different forms of background spectrum,
 we find that
 only background events in the {\em lowest} energy ranges
 could affect the reconstructions (significantly);
 those in high energy ranges
 would {\em almost not} change the reconstructed ratios
 or only {\em very slightly}.
\end{abstract}
\clearpage
\section{Introduction}
 Currently,
 direct Dark Matter detection experiments
 searching for Weakly Interacting Massive Particles (WIMPs)
 are one of the promising methods
 for understanding the nature of Dark Matter (DM)
 and identifying them among new particles produced at colliders
 as well as reconstructing the (sub)structure of our Galactic halo
 \cite{Smith90, Lewin96, SUSYDM96, Bertone05}.
 To this aim,
 model--independent methods for determining
 the WIMP mass
 \cite{DMDDmchi-SUSY07, DMDDmchi},
 the spin--independent (SI) WIMP coupling on nucleons
 \cite{DMDDfp2-IDM2008, DMDDfp2}
 as well as
 ratios between different WIMP couplings/cross sections
 \cite{DMDDidentification-DARK2009, DMDDranap}
 from direct detection experiments
 have been developed.

 These methods built basically on the work
 on the reconstruction of the (moments of the)
 one--dimensional velocity distribution function of halo WIMPs
 by using experimental data
 (measured recoil energies) directly
 \cite{DMDDf1v}.
 The spectrum of recoil energy
 is proportional to an integral over
 the one--dimensional WIMP velocity distribution,
 $f_1(v)$,
 where $v$ is the absolute value of the WIMP velocity
 in the laboratory frame.
 Since
 this integral is in fact just the minus--first moment of
 the velocity distribution function,
 which can be estimated from experimental data directly
 \cite{DMDDf1v, DMDDmchi},
 by assuming that
 the spin--dependent (SD) WIMP--nucleus interaction
 dominates over the spin--independent one,
 an expression for determining the ratio
 between the SD WIMP coupling on neutrons and that on protons
 has been derived
 \cite{DMDDidentification-DARK2009, DMDDranap}.
 Meanwhile,
 for a general combination of
 the SI and SD WIMP--nucleus interactions,
 by using detector materials with and without
 spin sensitivities on protons and/or on neutrons,
 a second expression for determining
 the ratio between two SD WIMP--nucleon couplings
 as well as two expressions for determining
 ratios of the SD WIMP--proton(neutron) cross section to the SI one
 have also been derived
 \cite{DMDDidentification-DARK2009, DMDDranap}.
 It was found that,
 by combining experimental data sets
 with different target nuclei,
 the ratios between different WIMP couplings/cross sections can be determined
 {\em without} making any assumption
 about the velocity distribution of halo WIMPs
 {\em nor} prior knowledge about their mass
 \cite{DMDDidentification-DARK2009, DMDDranap}.

 In the work on the development of
 these model--independent data analysis procedures
 for extracting information on WIMP couplings/cross sections
 from direct detection experiments,
 it was assumed that
 the analyzed data sets are background--free,
 i.e., all events are WIMP signals.
 Active background discrimination techniques
 should make this condition possible.
 For example,
 the ratio of the ionization to recoil energy,
 the so--called ``ionization yield'',
 used in the CDMS-II experiment
 provides an event--by--event rejection
 of electron recoil events
 to be better than $10^{-4}$ misidentification
 \cite{Ahmed09b}.
 By combining the ``phonon pulse timing parameter'',
 the rejection ability of
 the misidentified electron recoils
 (most of them are ``surface events''
  with sufficiently reduced ionization energies)
 can be improved to be $< 10^{-6}$ \cite{Ahmed09b}.
 Moreover,
 as demonstrated by the CRESST collaboration \cite{CRESST-bg},
 by means of inserting a scintillating foil,
 which causes some additional scintillation light
 for events induced by $\alpha$-decay of $\rmXA{Po}{210}$
 and thus shifts the pulse shapes of these events
 faster than pulses induced by WIMP interactions in the crystal,
 the pulse shape discrimination (PSD) technique
 can then easily distinguish WIMP--induced nuclear recoils
 from those induced by backgrounds%
\footnote{
 For more details
 about background discrimination techniques and status
 in currently running and projected direct detection experiments
 see e.g.,
 Refs.~\cite{Aprile09a, EDELWEISS-bg, Lang09b}.
}.

 However,
 as the most important issue in all
 underground experiments,
 the signal identification ability and
 possible residue background events
 which pass all discrimination criteria and
 then mix with other real WIMP--induced events in analyzed data sets
 should also be considered.
 Therefore,
 in this article,
 as a more realistic study,
 we follow our works
 on the effects of residue background events
 in direct Dark Matter detection experiments
 \cite{DMDDbg-mchi, DMDDbg-f1v, DMDDbg-fp2}
 and want to study
 how well we could determine
 the ratios of WIMP--nucleon couplings/cross sections model--independently
 by using ``impure'' data sets
 and how ``dirty'' these data sets could be
 to be still useful.

 The remainder of this article is organized as follows.
 In Sec.~2
 I review briefly the model--independent methods
 for determining ratios between different WIMP--nucleon couplings/cross sections
 by using experimental data sets directly.
 In Sec.~3
 the effects of residue background events
 in the analyzed data sets
 on the measured energy spectrum
 will be discussed.
 In Secs.~4 and 5
 I show numerical results of
 the reconstruction of ratios of WIMP--nucleon couplings/cross sections
 by using mixed data sets
 with different fractions of residue background events
 based on Monte Carlo simulations.
 I conclude in Sec.~6.
 Some technical details will be given in an appendix.
\section{Methods for determining ratios of
         WIMP--nucleon couplings/cross sections}
 In this section
 I review briefly
 the model--independent methods
 for determining the ratio of
 the SD WIMP coupling on neutrons
 to that on protons
 as well as the ratio
 between the SD and SI WIMP--proton cross sections%
\footnote{
 In this section
 I consider only the case with protons,
 but all formulae given in Section 2.3
 can be modified straightforwardly to the case with neutrons.
}.
 For more detailed illustrations and discussions
 about these precedures see \cite{DMDDranap}.
\subsection{Event rate for elastic WIMP--nucleus scattering}
 Considering the SI and SD
 WIMP--nucleus interactions together,
 the basic expression for the differential event rate
 for elastic WIMP--nucleus scattering can be given as
 \cite{SUSYDM96, DMDDranap}:
\beq
   \dRdQ
 = \frac{\rho_0}{2 \mchi \mrN^2}
   \bbigg{\sigmaSI \FSIQ + \sigmaSD \FSDQ}
   \int_{\vmin}^{\vmax} \bfrac{f_1(v)}{v} dv
\~.
\label{eqn:dRdQ_SISD}
\eeq
 Here $R$ is the direct detection event rate,
 i.e., the number of events
 per unit time and unit mass of detector material,
 $Q$ is the energy deposited in the detector,
 $\rho_0$ is the WIMP density near the Earth,
 $\sigma_0^{\rm (SI, SD)}$ are the SI/SD total cross sections
 ignoring the form factor suppression and
 $F_{\rm (SI, SD)}(Q)$ are the elastic nuclear form factors
 for the SI/SD WIMP interactions,
 respectively,
 $f_1(v)$ is the one--dimensional velocity distribution function
 of the WIMPs impinging on the detector,
 $v$ is the absolute value of the WIMP velocity
 in the laboratory frame.
 The reduced mass $\mrN$ is defined by
\beq
        \mrN
 \equiv \frac{\mchi \mN}{\mchi + \mN}
\~,
\label{eqn:mrN}
\eeq
 where $\mchi$ is the WIMP mass and
 $\mN$ that of the target nucleus.
 Finally,
 $\vmin$ is the minimal incoming velocity of incident WIMPs
 that can deposit the energy $Q$ in the detector:
\beq
   \vmin
 = \alpha \sqrt{Q}
\~,
\label{eqn:vmin}
\eeq
 with the transformation constant
\beq
        \alpha
 \equiv \sfrac{\mN}{2 \mrN^2}
\~,
\label{eqn:alpha}
\eeq
 and $\vmax$ is the maximal WIMP velocity
 in the Earth's reference frame,
 which is related to
 the escape velocity from our Galaxy
 at the position of the Solar system,
 $\vesc~\gsim~600$ km/s.

 Through e.g., squark and Higgs exchanges with quarks,
 WIMPs could have a ``scalar'' interaction with nuclei%
\footnote{
 Besides of the scalar interaction,
 WIMPs could also have a ``vector'' interaction
 with nuclei
 \cite{SUSYDM96, Bertone05}.
 However,
 for Majorana WIMPs ($\chi = \bar{\chi}$),
 e.g., the lightest neutralino in supersymmetric models,
 there is no such vector interaction.
}.
 The SI scalar WIMP--nucleus cross section
 can be expressed as \cite{SUSYDM96, Bertone05}
\beqn
           \sigmaSI
 \=        \afrac{4}{\pi} \mrN^2
           \bBig{Z f_{\rm p} + (A - Z) f_{\rm n}}^2
           \non\\
 \eqnsimeq \afrac{4}{\pi} \mrN^2 A^2 |f_{\rm p}|^2
           \non\\
 \=        A^2 \afrac{\mrN}{\mrp}^2 \sigmapSI
\~.
\label{eqn:sigma0SI}
\eeqn
 Here
\beq
   \sigmapSI
 = \afrac{4}{\pi} \mrp^2 |f_{\rm p}|^2
\label{eqn:sigmapSI}
\eeq
 is the SI WIMP cross section on protons,
 $f_{\rm p(n)}$ are the effective
 $\chi \chi {\rm p p (nn)}$ four--point couplings,
 $A$ is the atomic mass number of the target nucleus,
 and $\mrp$ is the reduced mass of
 the WIMP mass $\mchi$ and the proton mass $m_{\rm p}$.
 Note that
 I have used here
 the theoretical prediction
 for the lightest supersymmetric neutralino
 (and for all WIMPs which interact primarily through Higgs exchange)
 that
 the scalar couplings are approximately the same
 on protons and on neutrons:
\( 
        f_{\rm n}
 \simeq f_{\rm p}
\);
 the tiny mass difference between a proton and a neutron
 has also been neglected.

\begin{table}[t!]
\small
\begin{center}
\renewcommand{\arraystretch}{1.35}
\begin{tabular}{|| c   c   c   c   c   c   c   c ||}
\hline
\hline
 \makebox[1.3cm][c]{Isotope}        &
 \makebox[0.9cm][c]{$Z$}            & \makebox[0.9cm][c]{$J$}     &
 \makebox[1.5cm][c]{$\Srmp$}        & \makebox[1.5cm][c]{$\Srmn$} &
 \makebox[1.8cm][c]{$-\Srmp/\Srmn$} & \makebox[1.8cm][c]{$\Srmn/\Srmp$} &
 \makebox[3.7cm][c]{Natural abundance (\%)} \\
\hline
\hline
 $\rmXA{F}{19}$   &  9 & 1/2 &                   0.441  & \hspace{-1.8ex}$-$0.109 &
      4.05  &  $-$0.25   &       100   \\
\hline
 $\rmXA{Na}{23}$  & 11 & 3/2 &                   0.248  &                   0.020 &
  $-$12.40  &     0.08   &       100   \\
\hline
 $\rmXA{Cl}{35}$  & 17 & 3/2 & \hspace{-1.8ex}$-$0.059  & \hspace{-1.8ex}$-$0.011 &
   $-$5.36  &     0.19   &        76   \\
\hline
 $\rmXA{Cl}{37}$  & 17 & 3/2 & \hspace{-1.8ex}$-$0.058  &                   0.050 &
      1.16  &  $-$0.86   &        24   \\
\hline
 $\rmXA{Ge}{73}$  & 32 & 9/2 &                   0.030  &                   0.378 &
   $-$0.08  &    12.6    &  7.8 / 86 (HDMS) \cite{Bednyakov08a}\\
\hline
 $\rmXA{I}{127}$  & 53 & 5/2 &                   0.309  &                   0.075 &
   $-$4.12  &     0.24   &       100   \\
\hline
 $\rmXA{Xe}{129}$ & 54 &  1/2 &                  0.028  &                   0.359 &
   $-$0.08  &    12.8    &        26   \\
\hline
 $\rmXA{Xe}{131}$ & 54 &  3/2 & \hspace{-1.8ex}$-$0.009 & \hspace{-1.8ex}$-$0.227 &
   $-$0.04  &    25.2    &         21   \\
\hline
\hline
\end{tabular}
\end{center}
\caption{
 List of the relevant spin values of the nuclei
 used for simulations presented in this paper.
 More details can be found in
 e.g., Refs.~\cite{SUSYDM96, Tovey00, Giuliani05, Girard05}.
}
\end{table}

 On the other hand,
 through e.g., squark and Z boson exchanges with quarks,
 WIMPs could also couple to the spin of target nuclei,
 an ``axial--vector'' (spin--spin) interaction.
 The SD WIMP--nucleus cross section
 can be expressed as \cite{SUSYDM96, Bertone05}:
\beq
   \sigmaSD
 = \afrac{32}{\pi} G_F^2 \~ \mrN^2
   \afrac{J + 1}{J} \bBig{\Srmp \armp + \Srmn \armn}^2
\~.
\label{eqn:sigma0SD}
\eeq
 Here $G_F$ is the Fermi constant,
 $J$ is the total spin of the target nucleus,
 $\expv{S_{\rm (p, n)}}$ are the expectation values of
 the proton and neutron group spins%
\footnote{
 Note that
 detailed nuclear spin structure calculations show that
 not only unpaired nucleons contribute
 to the total cross section,
 the even group of nucleons has sometimes
 also a non--negligible spin
 (see Table 1 and
  e.g., data given in Refs.~\cite{SUSYDM96, Tovey00, Giuliani05}).
},
 and $a_{\rm (p, n)}$ are the effective SD WIMP couplings
 on protons and on neutrons.
 Since
 for a proton or a neutron
 $J = \frac{1}{2}$ and $\Srmp$ or $\Srmn = \frac{1}{2}$,
 the SD WIMP cross section on protons or on neutrons
 can be given as
\beq
   \sigma_{\chi {\rm (p, n)}}^{\rm SD}
 = \afrac{24}{\pi} G_F^2 \~ m_{\rm r, (p, n)}^2 |a_{\rm (p, n)}|^2
\~.
\label{eqn:sigmap/nSD}
\eeq
 As shown in Eq.~(\ref{eqn:sigma0SI}),
 due to the coherence effect with the entire nucleus,
 the cross section for SI scalar WIMP--nucleus interaction
 scales approximately as the square of
 the atomic mass of the target nucleus.
 Hence,
 in most supersymmetric models,
 the SI cross section for nuclei with $A~\gsim~30$ dominates
 over the SD one \cite{SUSYDM96, Bertone05}.
 However,
 as discussed in Refs.~\cite{Bertone07, Barger08, Belanger08},
 in Universal Extra Dimension (UED) models,
 the SD WIMP interaction with nucleus is less suppressed
 and could be compatible or even larger than the SI one.
\subsection{Only a dominant SD WIMP--nucleus cross section}
 Consider at first the case that
 the SD WIMP--nucleus interaction strongly dominates over the SI one
 and thus neglect
 the first SI term, $\sigmaSI \FSIQ$,
 in the bracket on the right--hand side of Eq.~(\ref{eqn:dRdQ_SISD}).
 By using a time--averaged recoil spectrum,
 and assuming that no directional information exists,
 the normalized one--dimensional
 velocity distribution function of halo WIMPs, $f_1(v)$,
 has been solved analytically \cite{DMDDf1v}
 and,
 consequently,
 its generalized moments can be estimated by
 \cite{DMDDf1v, DMDDmchi}%
\footnote{
 Here we have implicitly assumed that
 $\Qmax$ is so large that
 a term $2 \Qmax^{(n+1)/2} r(\Qmax) / F^2(\Qmax)$
 is negligible.
}
\beqn
    \expv{v^n}(v(\Qmin), v(\Qmax))
 \= \int_{v(\Qmin)}^{v(\Qmax)} v^n f_1(v) \~ dv
    \non\\
 \= \alpha^n
    \bfrac{2 \Qmin^{(n+1)/2} r(\Qmin) / \FQmin + (n+1) I_n(\Qmin, \Qmax)}
          {2 \Qmin^{   1 /2} r(\Qmin) / \FQmin +       I_0(\Qmin, \Qmax)}
\~.
\label{eqn:moments}
\eeqn
 Here $v(Q) = \alpha \sqrt{Q}$,
 $Q_{\rm (min, max)}$ are
 the experimental minimal and maximal
 cut--off energies of the data set,
 respectively,
\beq
        r(\Qmin)
 \equiv \adRdQ_{{\rm expt}, \~ Q = \Qmin}
\label{eqn:rmin}
\eeq
 is an estimated value of
 the {\em measured} recoil spectrum $(dR/dQ)_{\rm expt}$
 ({\em before} normalized by
  an experimental exposure, $\cal E$)
 at $Q = \Qmin$,
 and $I_n(\Qmin, \Qmax)$ can be estimated through the sum:
\beq
   I_n(\Qmin, \Qmax)
 = \sum_{a = 1}^{N_{\rm tot}} \frac{Q_a^{(n-1)/2}}{F^2(Q_a)}
\~,
\label{eqn:In_sum}
\eeq
 where the sum runs over all events in the data set
 that satisfy $Q_a \in [\Qmin, \Qmax]$
 and $N_{\rm tot}$ is the number of such events.
 Then,
 since the integral on the right--hand side of Eq.~(\ref{eqn:dRdQ_SISD})
 is just the minus--first generalized moment of
 the velocity distribution function, $\expv{v^{-1}}$,
 which can be estimated by Eq.~(\ref{eqn:moments}),
 by setting $Q = \Qmin$ and
 using the definition (\ref{eqn:alpha}) of $\alpha$,
 one can obtain straightforwardly that
\beq
   \rho_0 \sigmaSD
 = \afrac{1}{\calE}
   \mchi \mrN \sfrac{\mN}{2}
   \bbrac{\frac{2 \Qmin^{1/2} r(\Qmin)}{F_{\rm SD}^2(\Qmin)} + I_0}
\~.
\label{eqn:rho_sigma}
\eeq
 Now,
 in order to eliminate $\rho_0$ here,
 we combine two experimental data sets
 with different target nuclei, $X$ and $Y$.
 By substituting the expression (\ref{eqn:sigma0SD})
 for $\sigmaSD$ into Eq.~(\ref{eqn:rho_sigma})
 and using again the definition (\ref{eqn:alpha}) of $\alpha$
 for both target nuclei,
 the ratio between two SD WIMP--nucleon couplings
 has been solved analytically as
 \cite{DMDDidentification-DARK2009, DMDDranap}%
\footnote{
 Note that,
 although the constraints on two SD WIMP--nucleon couplings
 have conventionally been shown in the $\armp - \armn$ plane,
 I will always use the $\armn / \armp$ ratio
 in this article.
}
\beq
   \afrac{\armn}{\armp}_{\pm, n}^{\rm SD}
 =-\frac{\SpX \pm \SpY \abrac{\calR_{J, n, X} / \calR_{J, n, Y}} }
        {\SnX \pm \SnY \abrac{\calR_{J, n, X} / \calR_{J, n, Y}} }
\~,
    ~~~~~~~~~~~~~~~~ 
    n \ne 0
.
\label{eqn:ranapSD}
\eeq
 Here I have used the following relation \cite{DMDDmchi}:
\beq
   \frac{\alpha_X}{\alpha_Y}
 = \frac{\calR_{n, Y}}{\calR_{n, X}}
\~,
\label{eqn:ralphaXY}
\eeq
 and defined
\beq
        \calR_{J, n, X}
 \equiv \bbrac{\Afrac{J_X}{J_X + 1}
               \frac{\calR_{\sigma, X}}{\calR_{n, X}}}^{1/2}
\~,
\label{eqn:RJnX}
\eeq
 with%
\footnote{
 Note that
 $\calR_{\sigma, (X, Y)}$ and $\calR_{n, (X, Y)}$
 defined here as well as
 the estimator for $I_n$ given in Eq.~(\ref{eqn:In_sum})
 can be used for either the SI or the SD case
 with a corresponding form factor.
 However,
 since we consider here only the SD interaction,
 $F^2(Q)$ needed for using Eqs.~(\ref{eqn:rho_sigma}),
 (\ref{eqn:RsigmaX_min})
 and (\ref{eqn:RnX_min}) should be substituted by
 form factors for the SD cross section.
}
\beq
        \calR_{\sigma, X}
 \equiv \frac{1}{\calE_X}
        \bbrac{\frac{2 \QminX^{1/2} r_X(\QminX)}{\FQminX} + \IzX}
\~,
\label{eqn:RsigmaX_min}
\eeq
 and
\beq
        \calR_{n, X}
 \equiv \bfrac{2 \QminX^{(n+1)/2} r_X(\QminX) / \FQminX + (n+1) \InX}
              {2 \QminX^{   1 /2} r_X(\QminX) / \FQminX +       \IzX}^{1/n}
\~;
\label{eqn:RnX_min}
\eeq
 $\calR_{J, n, Y}$, $\calR_{\sigma, Y}$,
 and $\calR_{n, Y}$
 can be defined analogously%
\footnote{
 Hereafter,
 without special remark
 all notations defined for the target $X$
 can be defined analogously for the target $Y$
 and occasionally for the target $Z$.
};
 $F_{(X, Y)}(Q)$
 are the form factors of the nucleus $X$ and $Y$,
 $r_{(X, Y)}(Q_{{\rm min}, (X, Y)})$
 refer to the counting rates for the target $X$ and $Y$
 at the respective lowest recoil energies included in the analysis,
 and $\calE_{(X, Y)}$ are the experimental exposures
 with the target $X$ and $Y$.
 Note that,
 firstly,
 Eq.~(\ref{eqn:ranapSD}) can be used
 once {\em positive} signals are observed
 in two (or more) experiments;
 information on the local WIMP density $\rho_0$,
 on the velocity distribution function
 of incident WIMPs, $f_1(v)$,
 as well as on the WIMP mass $\mchi$
 are {\em not} necessary.
 Secondly,
 because the couplings in Eq.~(\ref{eqn:sigma0SD}) are squared,
 we have two solutions for $\armn / \armp$ here;
 if exact ``theory'' values for ${\cal R}_{J, n , (X, Y)}$ are taken,
 these solutions coincide for
\beq
   \afrac{\armn}{\armp}_{+, n}^{\rm SD}
 = \afrac{\armn}{\armp}_{-, n}^{\rm SD}
 = \cleft{\renewcommand{\arraystretch}{1}
          \begin{array}{l l l}
           \D -\frac{\SpX}{\SnX}         \~, & ~~~~~~~~ &
           {\rm for}~\calR_{J, n, X} = 0 \~, \\ \\ 
           \D -\frac{\SpY}{\SnY}         \~, &          &
           {\rm for}~\calR_{J, n, Y} = 0 \~,
          \end{array}}
\label{eqn:ranapSD_coin}
\eeq
 which depends only on properties of two used target nuclei
 (see Table 1).
 Moreover,
 it can be found from Eq.~(\ref{eqn:ranapSD}) that
 one of these two solutions has a pole
 at the middle of two intersections,
 which depends simply on the signs of $\SnX$ and $\SnY$:
 since $\calR_{J, n, X}$ and $\calR_{J, n, Y}$ are always positive,
 if both $\SnX$ and $\SnY$ are positive or negative,
 the ``$-$ (minus)'' solution $(\armn / \armp)^{\rm SD}_{-, n}$
 will diverge and
 the ``$+$ (plus)'' solution $(\armn / \armp)^{\rm SD}_{+, n}$
 will be the ``inner'' solution;
 in contrast,
 if the signs of $\SnX$ and $\SnY$ are opposite,
 the ``$-$ (minus)'' solution
 will be the ``inner'' solution.

 On the other hand,
 it has been found \cite{DMDDranap} that,
 in order to reduce the statistical uncertainty
 on $(\armn / \armp)^{\rm SD}_{\pm, n}$
 estimated by Eq.~(\ref{eqn:ranapSD}),%
\footnote{
 It is true with {\em non--negligible}
 experimental threshold energies \cite{DMDDranap}.
 Later we will see that,
 with {\em negligible} threshold energies,
 $(\armn / \armp)^{\rm SD}_{\pm, n}$ estimated
 with $r_{(X, Y)}(Q_{s, 1, (X, Y)})$
 could be a little bit larger.
}.
 one can practically
 use the estimate of the counting rate,
 instead of at the experimental minimal cut--off energy,
 at the shifted point $Q_{s, 1}$
 (from the central point of the first bin, $Q_1$)
 defined by
\beq
   Q_{s, 1}
 = Q_1 + \frac{1}{k_1} \ln\bfrac{\sinh (k_1 b_1 / 2)}{k_1 b_1 / 2}
\~,
\label{eqn:Qs1}
\eeq
 where $k_1$ is the logarithmic slope of
 the reconstructed recoil spectrum
 in the first $Q-$bin and $b_1$ is the bin width.
 Then,
 according to Eq.~(\ref{eqn:rmin_eq}),
 the {\em measured} recoil spectrum
 at $Q = Q_{s, 1}$ can be estimated by
\beq
   r(Q_{s, 1})
 = \afrac{dR}{dQ}_{{\rm expt}, \~ 1, \~ Q = Q_{s, 1}}
 = r_1
 = \frac{N_1}{b_1}
\~,
\label{eqn:rmin_Qs1}
\eeq
 where $N_1$ is the event number
 in the first bin.

 As shown in Ref.~\cite{DMDDranap},
 the statistical uncertainties
 on $(\armn / \armp)^{\rm SD}_{\pm, n}$
 estimated with different $n$
 (namely with different moments of
  the WIMP velocity distribution function)
 with $r_{(X, Y)}(Q_{s, 1, (X, Y)})$
 are clearly reduced and,
 interestingly,
 almost equal.
 Therefore,
 since
\beq
   \calR_{J, -1, X}
 = \bbrac{\afrac{J_X}{J_X + 1}
          \frac{2 \~ r_X(Q_{s, 1, X})}{\calE_X F_X^2(Q_{s, 1, X})}}^{1/2}
\~,
\label{eqn:JmaX}
\eeq
 one would need practically {\em only} events
 in the {\em lowest} energy ranges
 for estimating $\armn / \armp$.
 And,
 consequently,
 one has to estimate the values of form factors
 only at $Q = Q_{s, 1}$,
 and the zero momentum transfer approximation
 $F^2(Q \simeq 0)) \simeq 1$ can then be used.

\subsection{Combination of the SI and SD cross sections}
 Now I consider the case with
 a {\em non--negligible} SI WIMP--nucleus cross section.
 At first,
 by combining Eqs.~(\ref{eqn:sigma0SI}), (\ref{eqn:sigma0SD}),
 and (\ref{eqn:sigmap/nSD}),
 we can find
\beq
   \frac{\sigmaSD}{\sigmaSI}
 = \afrac{32}{\pi} G_F^2 \~ \mrp^2 \Afrac{J + 1}{J}
   \bfrac{\Srmp + \Srmn (\armn / \armp)}{A}^2 \frac{|\armp|^2}{\sigmapSI}
 = \calCp \afrac{\sigmapSD}{\sigmapSI}
\~,
\label{eqn:rsigmaSDSI}
\eeq
 where I have defined
\beq
        \calCp
 \equiv \frac{4}{3} \afrac{J + 1}{J}
        \bfrac{\Srmp + \Srmn (\armn/\armp)}{A}^2
\~.
\label{eqn:Cp}
\eeq
 Then the expression (\ref{eqn:dRdQ_SISD})
 for the differential event rate
 can be rewritten as
\beq
   \adRdQ_{\rm expt}
 = \calE
   A^2 \! \afrac{\rho_0 \sigmapSI}{2 \mchi \mrp^2} \!\!
   \bbrac{\FSIQ + \afrac{\sigmapSD}{\sigmapSI} \calCp \FSDQ}
   \int_{\vmin}^{\vmax} \bfrac{f_1(v)}{v} dv
\~.
\label{eqn:dRdQ_SISD_expt}
\eeq
 Set $Q = \Qmin$.
 One can find straightforwardly that,
 for this general case,
 Eq.~(\ref{eqn:rho_sigma}) becomes to
\beq
    \rho_0
    \bbrac{A^2 \afrac{\mrN}{\mrp}^2 \sigmapSI}
 =  \afrac{1}{\calE}
    \mchi \mrN \sfrac{\mN}{2}
    \bbrac{\frac{2 \Qmin^{1/2} r(\Qmin)}{F'^2(\Qmin)} + I_0}
\~,
\label{eqn:rho_sigma_SISD}
\eeq
 where $I_n$ should be estimated by Eq.~(\ref{eqn:In_sum})
 with the replacement of $\FQ$ by
\(
         F'^2(Q)
 \equiv  \FSIQ
       + \abrac{\sigmapSD / \sigmapSI} \calCp \FSDQ
\).

 By combining two targets $X$ and $Y$
 and using
 the relation (\ref{eqn:ralphaXY}) for $\alpha_X / \alpha_Y$
 with $n = -1$,
 the ratio of the SD WIMP--proton cross section
 to the SI one has been solved analytically as
 \cite{DMDDranap}
\beq
   \frac{\sigmapSD}{\sigmapSI}
 = \frac{\FSIQminY (\calR_{m, X}/\calR_{m, Y}) - \FSIQminX}
        {\calCpX \FSDQminX - \calCpY \FSDQminY (\calR_{m, X} / \calR_{m, Y})}
\~,
\label{eqn:rsigmaSDpSI}
\eeq
 where $m_{(X, Y)} \propto A_{(X, Y)}$ has been assumed,
 ${\cal C}_{{\rm p}, (X, Y)}$ have been defined in Eq.~(\ref{eqn:Cp}),
 and
\beq
        \calR_{m, X}
 \equiv \frac{r_X(\QminX)}{\calE_X \mX^2}
\~.
\label{eqn:RmX}
\eeq
 As the estimator (\ref{eqn:ranapSD}) for $\armn / \armp$,
 one can use Eq.~(\ref{eqn:rsigmaSDpSI})
 to estimate $\sigma_{\chi {\rm p}}^{\rm SD} / \sigmapSI$
 {\em without} a prior knowledge of the WIMP mass $\mchi$.
 Moreover,
 since ${\cal C}_{{\rm p}, (X, Y)}$
 depend only on the nature of the detector materials,
 $\sigma_{\chi {\rm p}}^{\rm SD} / \sigmapSI$
 is practically only a function of $\calR_{m, (X, Y)}$,
 i.e., the counting rate
 at the experimental minimal cut--off energies,
 which can be estimated by using events
 in the lowest available energy ranges.

 Meanwhile,
 for the general combination of
 the SI and SD WIMP--nucleus cross sections,
 the $\armn / \armp$ ratio
 appearing in Eq.~(\ref{eqn:Cp})
 has been solved analytically
 by introducing a {\em third} nucleus
 with {\em only} an SI sensitivity:
\(
   \Srmp_Z
 = \Srmn_Z
 = 0
\),
 i.e.,
\(
   {\cal C}_{{\rm p}, Z}
 = 0
\)
 as \cite{DMDDranap}
\beqn
    \afrac{\armn}{\armp}_{\pm}^{\rm SI + SD}
 \= \frac{-\abrac{\cpX \snpX - \cpY \snpY}
          \pm \sqrt{\cpX \cpY} \vbrac{\snpX - \snpY}}
         {\cpX \snpX^2 - \cpY \snpY^2}
    \non\\
 \= \cleft{\renewcommand{\arraystretch}{0.5}
           \begin{array}{l l l}
            \\
            \D -\frac{\sqrt{\cpX} \mp \sqrt{\cpY}}{\sqrt{\cpX} \snpX \mp \sqrt{\cpY} \snpY}\~, &
            ~~~~~~~~ & ({\rm for}~\snpX > \snpY), \\~\\~\\ 
            \D -\frac{\sqrt{\cpX} \pm \sqrt{\cpY}}{\sqrt{\cpX} \snpX \pm \sqrt{\cpY} \snpY}\~, &
                     & ({\rm for}~\snpX < \snpY). \\~\\
           \end{array}}
\label{eqn:ranapSISD}
\eeqn
 Here I have defined
\cheqna
\beq
        \cpX
 \equiv \frac{4}{3} \Afrac{J_X + 1}{J_X} \bfrac{\SpX}{A_X}^2
        \bbrac{  \FSIQminZ \afrac{\calR_{m, Y}}{\calR_{m, Z}} \!
               - \FSIQminY} \!
        \FSDQminX
\~,
\label{eqn:cpX}
\eeq
\cheqnb
\beq
        \cpY
 \equiv \frac{4}{3} \Afrac{J_Y + 1}{J_Y} \bfrac{\SpY}{A_Y}^2
        \bbrac{  \FSIQminZ \afrac{\calR_{m, X}}{\calR_{m, Z}} \!
               - \FSIQminX} \!
        \FSDQminY
\~,
\label{eqn:cpY}
\eeq
\cheqn
 and
\beq
        \snpX
 \equiv \frac{\SnX}{\SpX}
\~.
\label{eqn:snpX}
\eeq
 Note that,
 firstly,
 $(\armn / \armp)_{\pm}^{\rm SI + SD}$ and $c_{{\rm p}, (X, Y)}$
 given in Eqs.~(\ref{eqn:ranapSISD}), (\ref{eqn:cpX}), and (\ref{eqn:cpY})
 are functions of only
 $r_{(X, Y, Z)}(Q_{{\rm min}, (X, Y, Z)})$
 (or $r_{(X, Y, Z)}(Q_{s, 1, (X, Y, Z)})$),
 which can be estimated with events
 in the lowest energy ranges.
 Secondly,
 while the decision of the inner solution of
 $(\armn / \armp)_{\pm, n}^{\rm SD}$
 depends on the signs of $\SnX$ and $\SnY$,
 the decision with $(\armn / \armp)_{\pm}^{\rm SI + SD}$
 depends {\em not only} on the signs of
 \mbox{$\snpX = \SnX / \SpX$} and \mbox{$\snpY = \SnY / \SpY$},
 {\em but also} on the {\em order} of the two targets.
 For e.g., a Ge + Cl combination,
 since \mbox{$ s_{{\rm n/p}, \rmXA{Ge}{73}} = 12.6
             > s_{{\rm n/p}, \rmXA{Cl}{37}} = -0.86$},
 one should use the {\em upper} expression
 in the second line of Eq.~(\ref{eqn:ranapSISD}),
 and since $s_{{\rm n/p}, \rmXA{Ge}{73}}$
 and $s_{{\rm n/p}, \rmXA{Cl}{37}}$
 have the opposite signs,
 the ``$-$ (minus)'' solution of this expression
 (or the ``$+$ (plus)'' solution of the expression in the first line)
 is the inner solution.
 In contrast,
 for the F + I combination used in our simulations,
 since \mbox{$ s_{{\rm n/p}, \rmXA{F}{19}}  = -0.247
             < s_{{\rm n/p}, \rmXA{I}{127}} = 0.243$}
 and since $s_{{\rm n/p}, \rmXA{F}{19}}$
 and $s_{{\rm n/p}, \rmXA{I}{127}}$
 have the opposite signs,
 the ``$-$ (minus)'' solution of the {\em lower} expression
 in the second line of Eq.~(\ref{eqn:ranapSISD})
 (or the ``$-$ (minus)'' solution of the expression in the first line)
 is then the inner solution.

 Furthermore,
 in order to reduce the statistical uncertainty,
 one can choose at first a nucleus
 with {\em only} an SI sensitivity
 as the second target:
\(
   \SpY
 = \SnY
 = 0
\),
 i.e.,
\(
   {\cal C}_{{\rm p}, Y}
 = 0
\).
 The expression in Eq.~(\ref{eqn:rsigmaSDpSI})
 can thus be reduced to
\beq
   \frac{\sigmapSD}{\sigmapSI}
 = \frac{\FSIQminY (\calR_{m, X} / \calR_{m, Y}) - \FSIQminX}
        {\calCpX \FSDQminX}
\~.
\label{eqn:rsigmaSDpSI_even}
\eeq
 Then we choose a nucleus with (much) larger
 proton (or neutron) group spin
 as the first target:
\(
        \SpX
 \gg    \SnX
 \simeq 0
\),
 in order to eliminate the $\armn/\armp$ dependence of
 $\calCpX$ given in Eq.~(\ref{eqn:Cp}):%
\footnote{
 Analogously,
 we can define
\beq
        \calCn
 \equiv \frac{4}{3} \Afrac{J + 1}{J}
        \bfrac{\Srmp (\armp/\armn) + \Srmn}{A}^2
\~,
\label{eqn:Cn}
\eeq
 and choose $\SnX \gg \SpX \simeq 0$
 to eliminate its $\armn/\armp$ dependence.
}
\beq
        \calCpX
 \simeq \frac{4}{3} \Afrac{J_X + 1}{J_X} \bfrac{\SpX}{A_X}^2
\~.
\label{eqn:CpX_p}
\eeq
\section{Effects of residue background events}
 In this section
 I first show some numerical results of
 the energy spectrum of WIMP recoil signals
 mixed with a few background events.
 For generating WIMP--induced signals,
 we use the shifted Maxwellian velocity distribution
 \cite{Lewin96, SUSYDM96, DMDDf1v}:
\beq
   f_{1, \sh}(v)
 = \frac{1}{\sqrt{\pi}} \afrac{v}{\ve v_0}
   \bbigg{ e^{-(v - \ve)^2 / v_0^2} - e^{-(v + \ve)^2 / v_0^2} }
\~,
\label{eqn:f1v_sh}
\eeq
 with $v_0 \simeq 220~{\rm km/s}$
 and $\ve = 1.05 \~ v_0$,
 which are the Sun's orbital velocity
 and the Earth's velocity in the Galactic frame%
\footnote{
 The time dependence of the Earth's velocity
 will be ignored in our simulations.
},
 respectively;
 the maximal cut--off
 of the velocity distribution function
 has been set as $\vmax = 700$ km/s.
 The commonly used elastic nuclear form factor
 for the SI WIMP--nucleus cross section
 \cite{Engel91, SUSYDM96, Bertone05}:
\beq
   F_{\rm SI}^2(Q)
 = \bfrac{3 j_1(q R_1)}{q R_1}^2 e^{-(q s)^2}
\label{eqn:FQ_WS}
\eeq
 as well as
 the thin--shell form factor
 for the SD WIMP cross section
 \cite{Lewin96, Klapdor05, DMDDranap}:
\beqn
    F_{\rm SD}^2(Q)
 \= \cleft{\renewcommand{\arraystretch}{1.5}
           \begin{array}{l l l}
            j_0^2(q R_1)                      \~, & ~~~~~~~~ &
            {\rm for}~q R_1 \le 2.55~{\rm or}~q R_1 \ge 4.5 \~, \\ 
            {\rm const.} \simeq 0.047         \~, &          &
            {\rm for}~2.5 5 \le q R_1 \le 4.5 
           \end{array}}
\label{eqn:FQ_TS}
\eeqn
 will also be used%
\footnote{
 Other commonly used analytic forms
 for the one--dimensional WIMP velocity distribution
 as well as for the elastic nuclear form factor
 for the SI WIMP--nucleus cross section
 can be found in Refs.~\cite{DMDDf1v, DMDDmchi-NJP}.
}.
 Meanwhile,
 in order to check
 the need of a prior knowledge about
 an (exact) form of the residue background spectrum,
 two forms for the background spectrum
 have been considered.
 The simplest choice is a constant spectrum:
\beq
   \adRdQ_{\rm bg, const}
 = 1
\~.
\label{eqn:dRdQ_bg_const}
\eeq
 More realistically,
 we use the target--dependent exponential form
 introduced in Ref.~\cite{DMDDbg-mchi}
 for the residue background spectrum:
\beq
   \adRdQ_{\rm bg, ex}
 = \exp\abrac{-\frac{Q /{\rm keV}}{A^{0.6}}}
\~.
\label{eqn:dRdQ_bg_ex}
\eeq
 Here $Q$ is the recoil energy,
 $A$ is the atomic mass number of the target nucleus.
 The power index of $A$, 0.6, is an empirical constant,
 which has been chosen so that
 the exponential background spectrum is
 somehow {\em similar to},
 but still {\em different from}
 the expected recoil spectrum of the target nucleus;
 otherwise,
 there is in practice no difference between
 the WIMP scattering and background spectra.
 Note that,
 among different possible choices,
 we use in our simulations the atomic mass number $A$
 as the simplest, unique characteristic parameter
 in the general analytic form (\ref{eqn:dRdQ_bg_ex})
 for defining the residue background spectrum
 for {\em different} target nuclei.
 However,
 it does {\em not} mean that
 the (superposition of the real) background spectra
 would depend simply/primarily on $A$ or
 on the mass of the target nucleus, $\mN$.
 In other words,
 it is practically equivalent to
 use expression (\ref{eqn:dRdQ_bg_ex})
 or $(dR / dQ)_{\rm bg, ex} = e^{-Q / 13.5~{\rm keV}}$ directly
 for a $\rmXA{Ge}{76}$ target.

 Note also that,
 firstly,
 as argued in Ref.~\cite{DMDDbg-mchi},
 two forms of background spectrum given above
 are rather naive;
 however,
 since we consider here
 only {\em a few residue} background events
 induced by perhaps {\em two or more} different sources,
 which pass all discrimination criteria,
 and then mix with other WIMP--induced events
 in our data sets of ${\cal O}(50)$ {\em total} events,
 exact forms of different background spectra
 are actually not very important and
 these two spectra,
 in particular,
 the exponential one,
 should practically not be unrealistic%
\footnote{
 Other (more realistic) forms for background spectrum
 (perhaps also for some specified targets/experiments)
 can be tested on the \amidas\ website
 \cite{AMIDAS-web, AMIDAS-eprints}.
}.
 Secondly,
 as demonstrated in Ref.~\cite{DMDDranap}
 and reviewed in the previous section,
 the model--independent data analysis procedures
 for determining ratios between different WIMP--nucleon couplings/cross sections
 require only measured recoil energies
 (induced mostly by WIMPs and
  occasionally by background sources)
 from direct detection experiments.
 Therefore,
 for applying these methods to future real data,
 a prior knowledge about (different) background source(s)
 is {\em not required at all}.

 Moreover,
 for our numerical simulations
 presented here as well as in the next two sections,
 the actual numbers of signal and background events
 in each simulated experiment
 are Poisson--distributed around their expectation values
 {\em independently};
 and the total event number recorded in one experiment
 is then the sum of these two numbers.
 Additionally,
 we assumed that
 all experimental systematic uncertainties
 as well as the uncertainty on
 the measurement of the recoil energy
 could be ignored.
 The energy resolution of most existing detectors
 is so good that its error can be neglected
 compared to the statistical uncertainty
 for the foreseeable future
 with pretty few events.

\begin{figure}[p!]
\begin{center}
\vspace{-0.75cm}
\hspace*{-1.6cm}
\includegraphics[width=9.8cm]{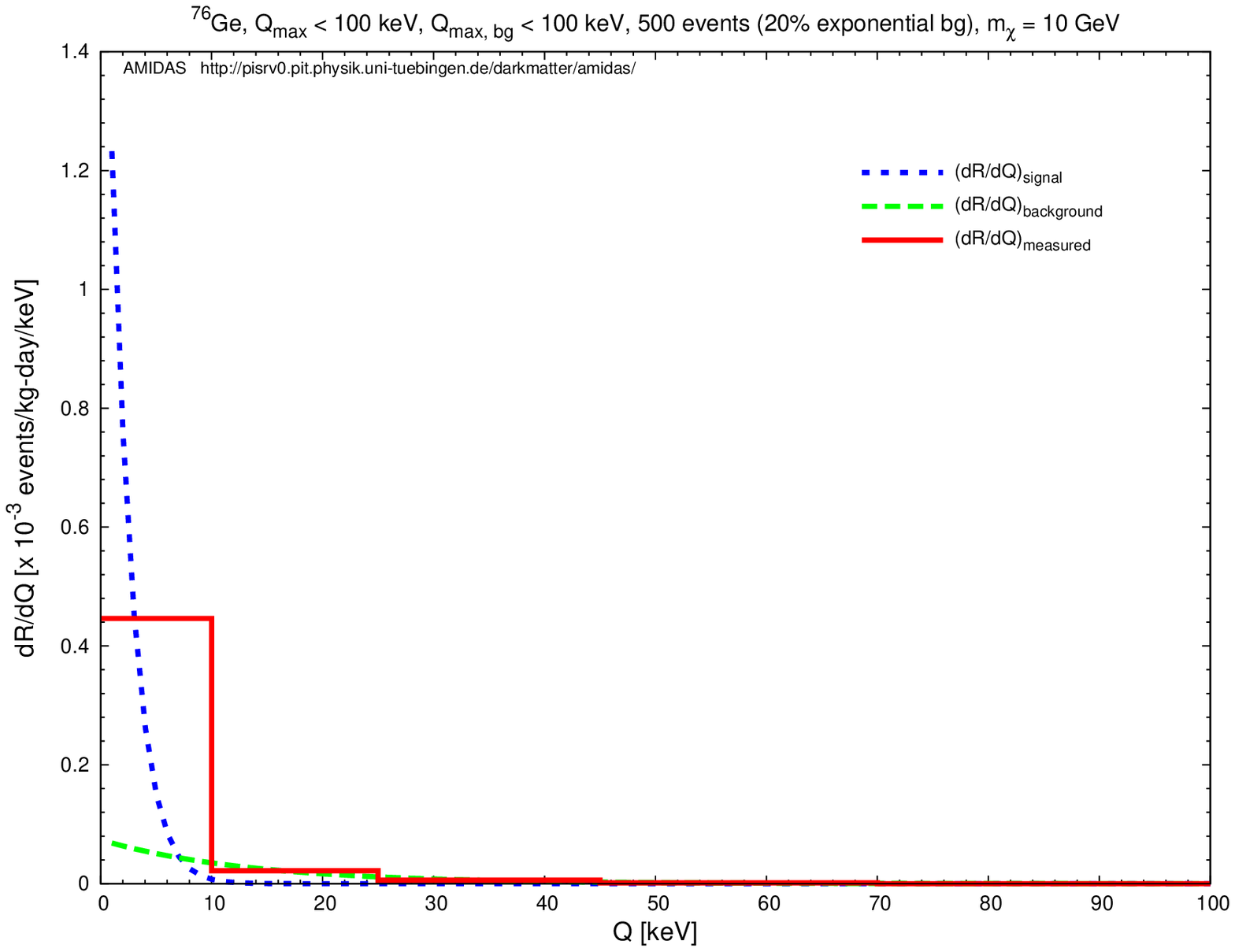} \hspace{-1.1cm}
\includegraphics[width=9.8cm]{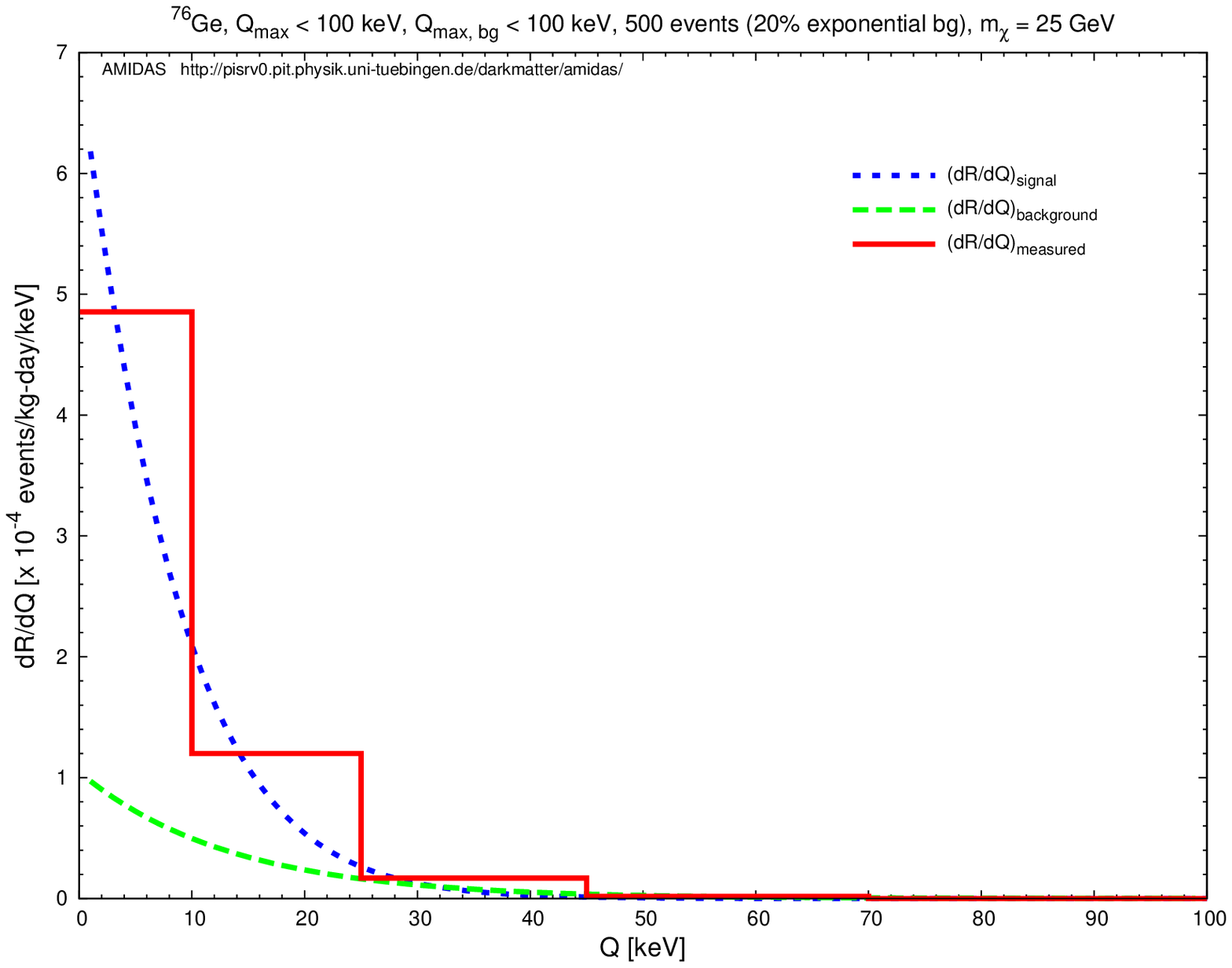} \hspace*{-1.6cm} \\
\vspace{0.5cm}
\hspace*{-1.6cm}
\includegraphics[width=9.8cm]{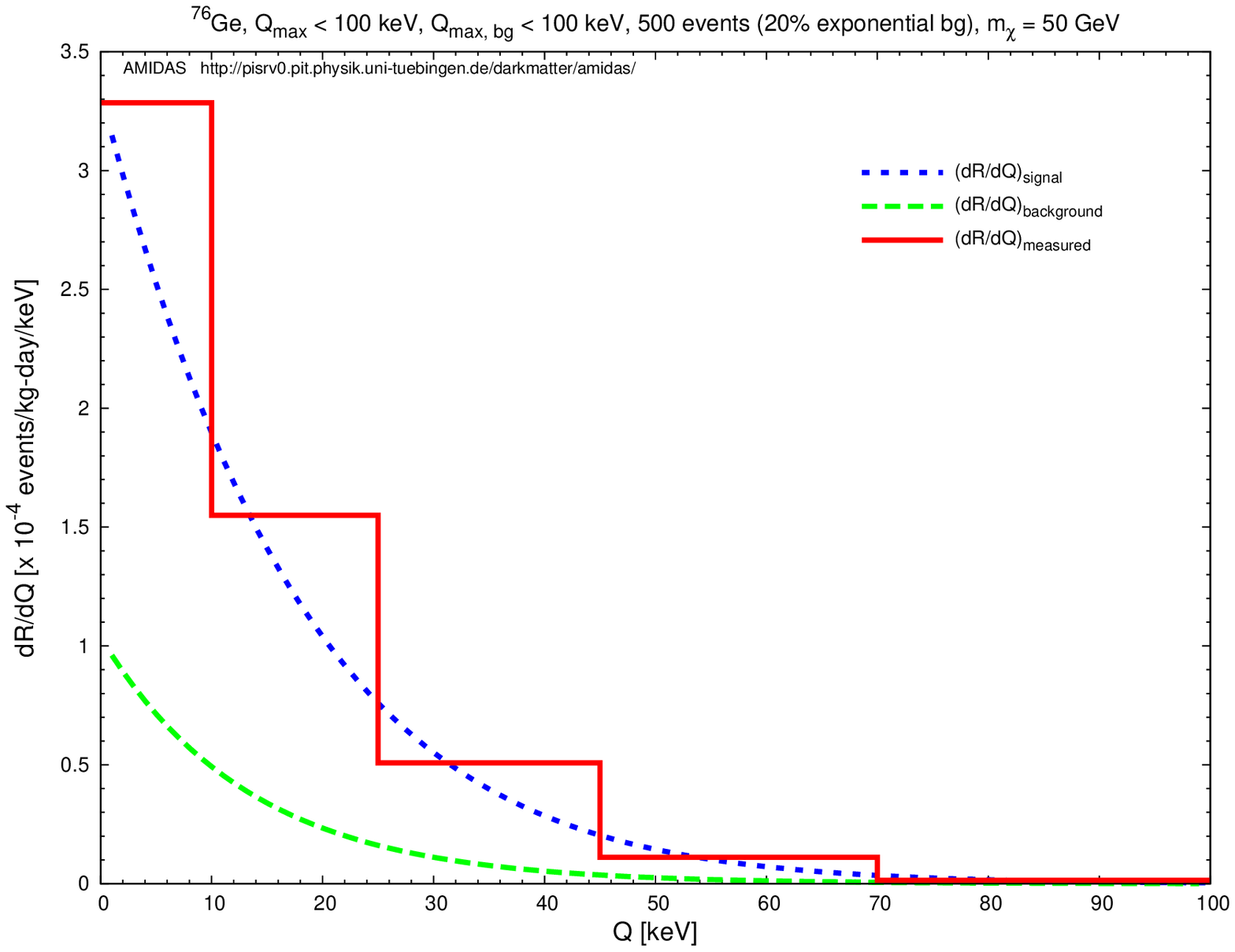} \hspace{-1.1cm}
\includegraphics[width=9.8cm]{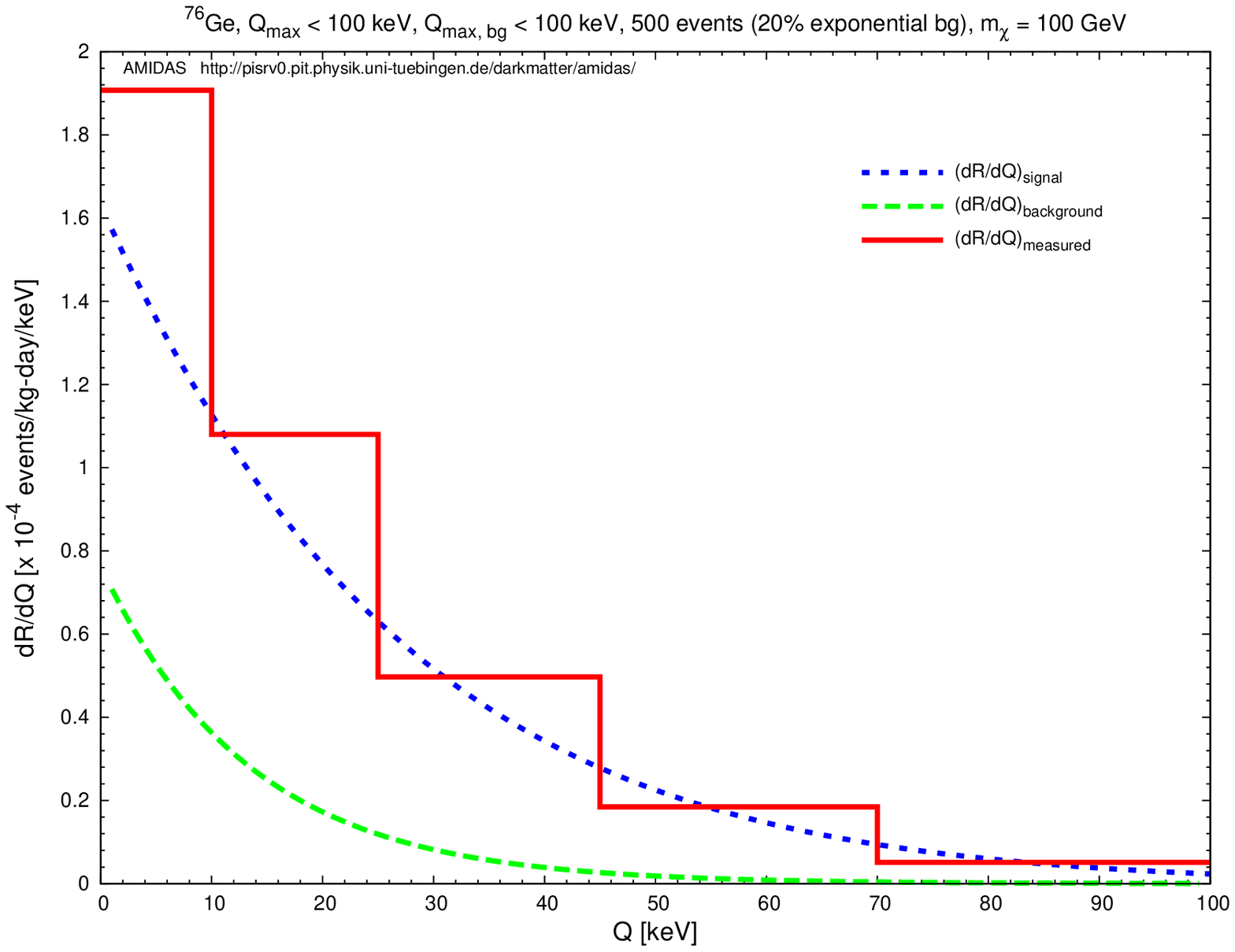} \hspace*{-1.6cm} \\
\vspace{0.5cm}
\hspace*{-1.6cm}
\includegraphics[width=9.8cm]{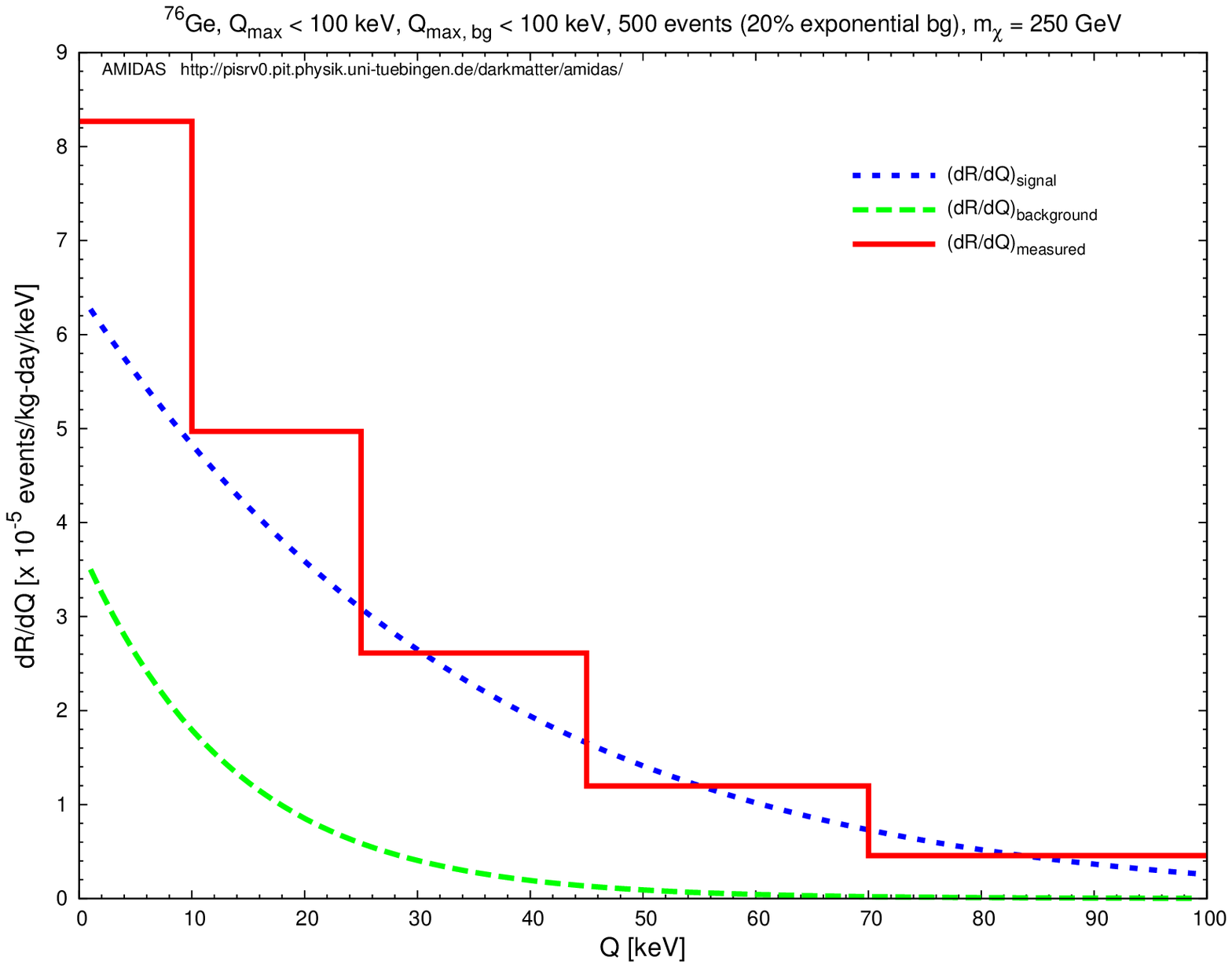} \hspace{-1.1cm}
\includegraphics[width=9.8cm]{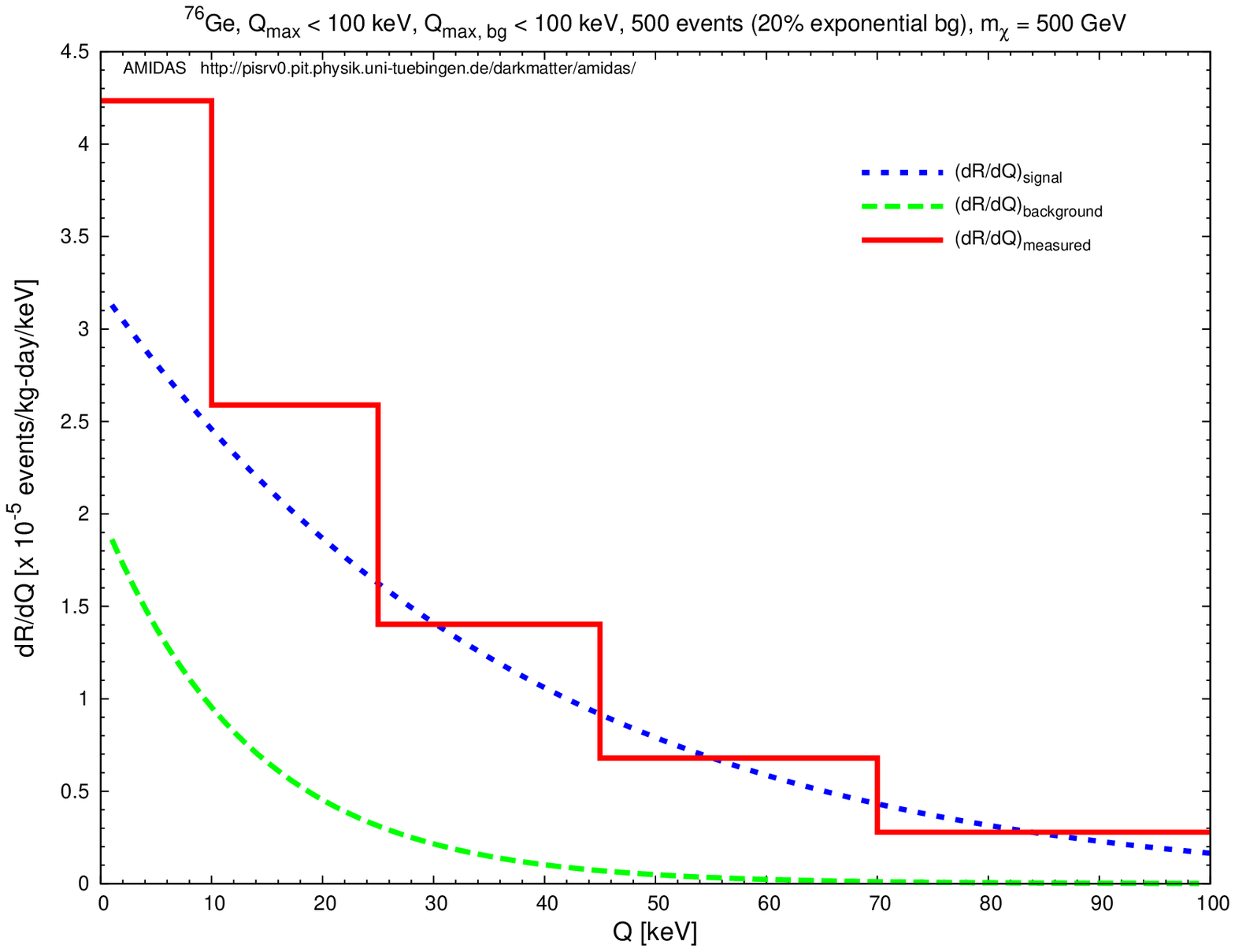} \hspace*{-1.6cm} \\
\vspace{-0.75cm}
\end{center}
\caption{
 Measured energy spectra (solid red histograms)
 for a $\rmXA{Ge}{76}$ target
 with six different WIMP masses:
 10, 25, 50, 100, 250, and 500 GeV.
 The dotted blue curves are
 the elastic WIMP--nucleus scattering spectra,
 whereas
 the dashed green curves are
 the exponential background spectra
 normalized to fit to the chosen background ratio,
 which has been set as 20\% here.
 The experimental threshold energies
 have been assumed to be negligible
 and the maximal cut--off energies
 are set as 100 keV.
 The background windows
 have been assumed to be the same as
 the experimental possible energy ranges.
 5,000 experiments with 500 total events on average
 in each experiment have been simulated.
 See the text for further details
 (plots from Ref.~\cite{DMDDbg-mchi}).
}
\label{fig:dRdQ-bg-ex-Ge-000-100-20}
\end{figure}
\begin{figure}[t!]
\begin{center}
\hspace*{-1.6cm}
\includegraphics[width=9.8cm]{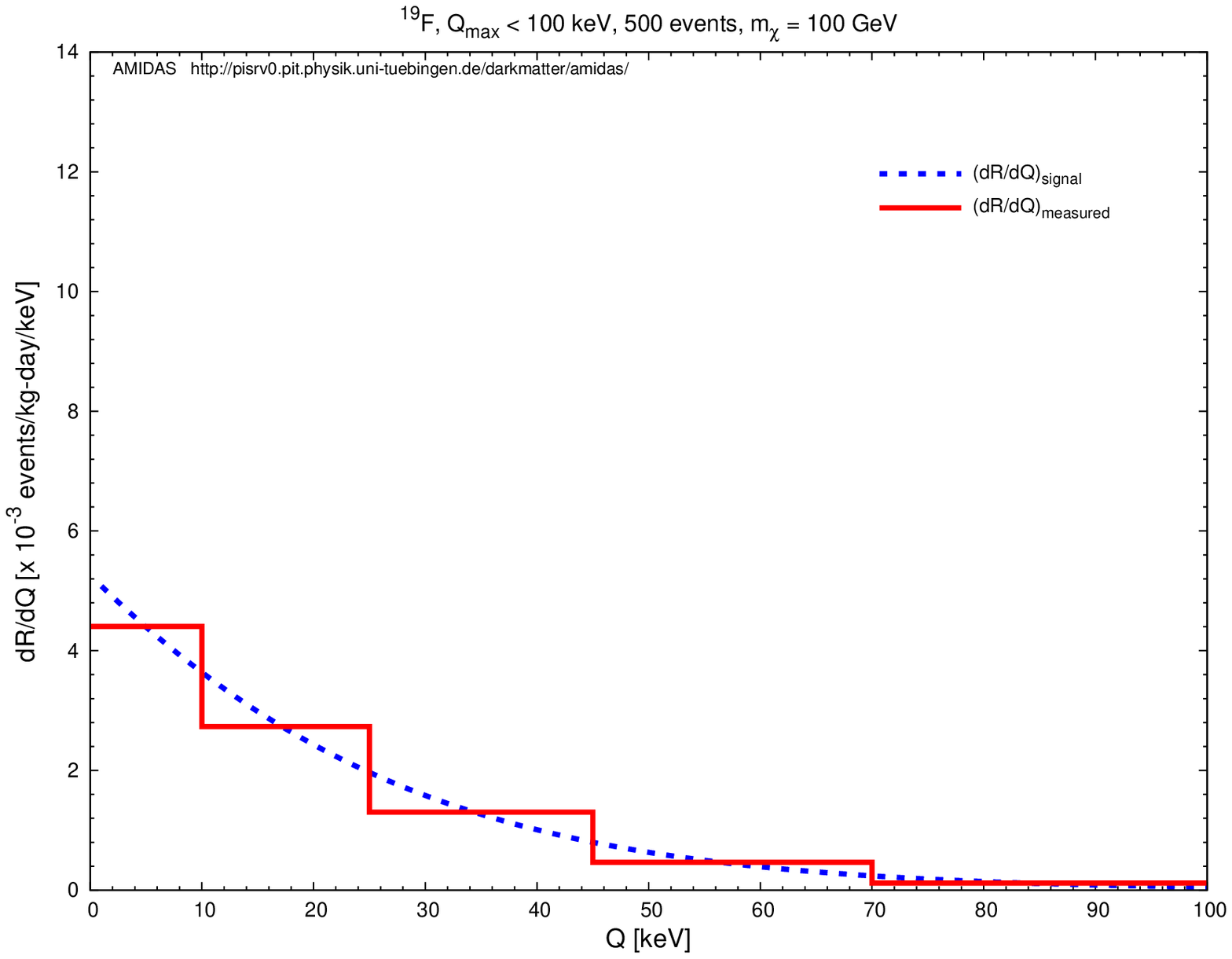} \hspace{-1.1cm}
\includegraphics[width=9.8cm]{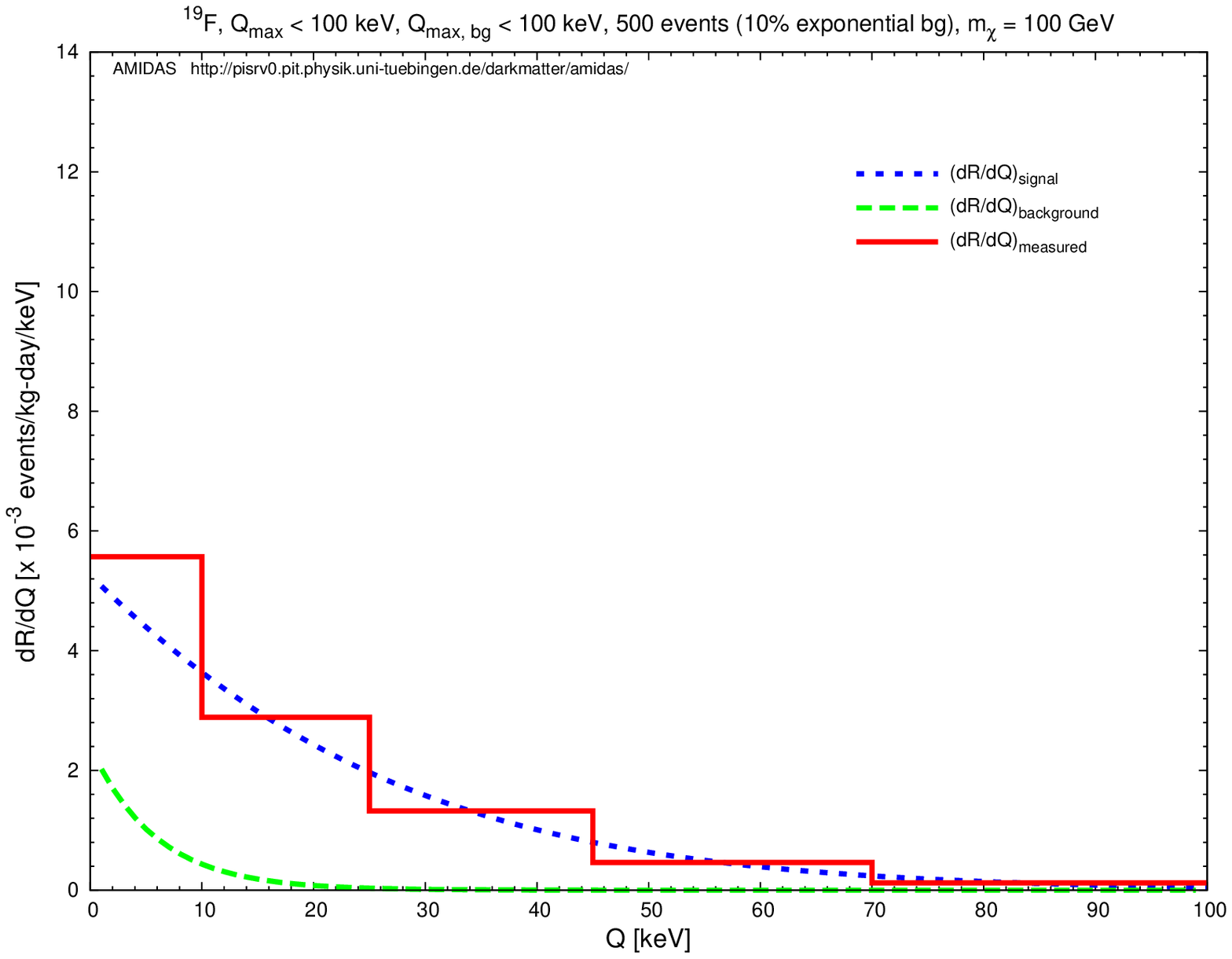} \hspace*{-1.6cm} \\
\vspace{0.5cm}
\hspace*{-1.6cm}
\includegraphics[width=9.8cm]{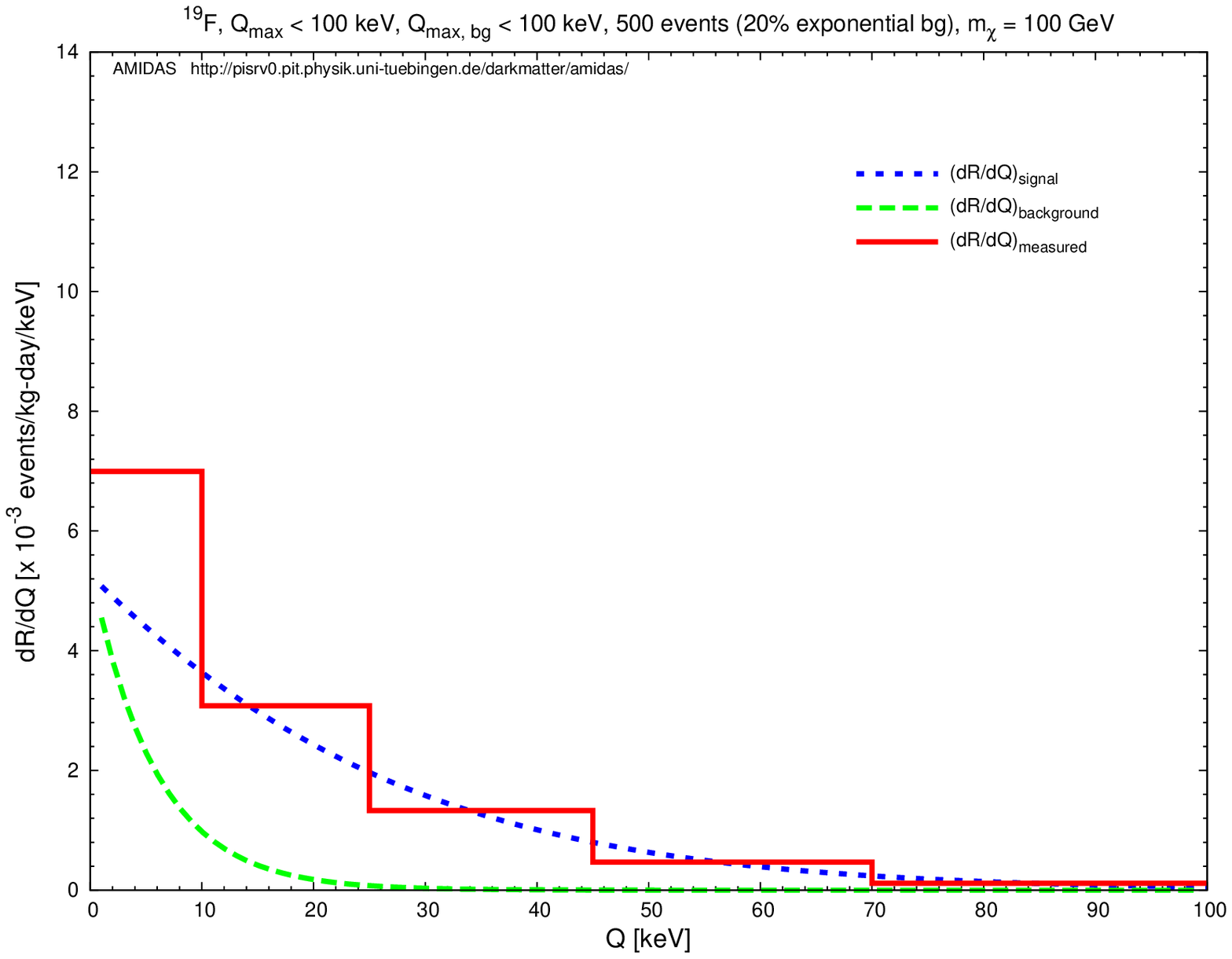} \hspace{-1.1cm}
\includegraphics[width=9.8cm]{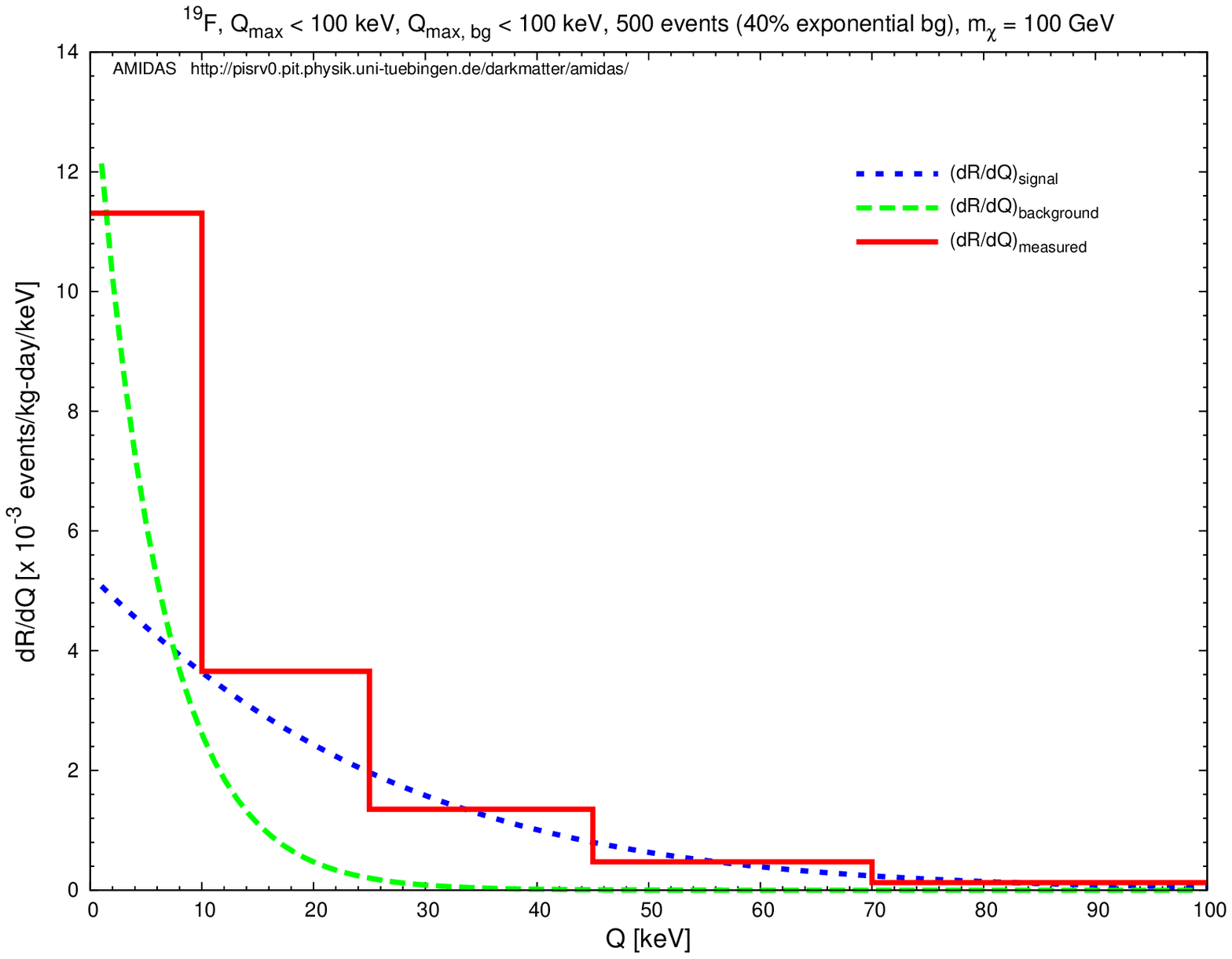} \hspace*{-1.6cm} \\
\vspace{-0.25cm}
\end{center}
\caption{
 As in Figs.~\ref{fig:dRdQ-bg-ex-Ge-000-100-20},
 except that
 a $\rmXA{F}{19}$ target has been used,
 the input WIMP mass has been fixed as 100 GeV,
 and four different background ratios:
 no background (top left),
 10\% (top right),
 20\% (bottom left),
 and 40\% (bottom right)
 are shown here.
}
\label{fig:dRdQ-bg-ex-F-000-100-100}
\end{figure}
\begin{figure}[t!]
\begin{center}
\hspace*{-1.6cm}
\includegraphics[width=9.8cm]{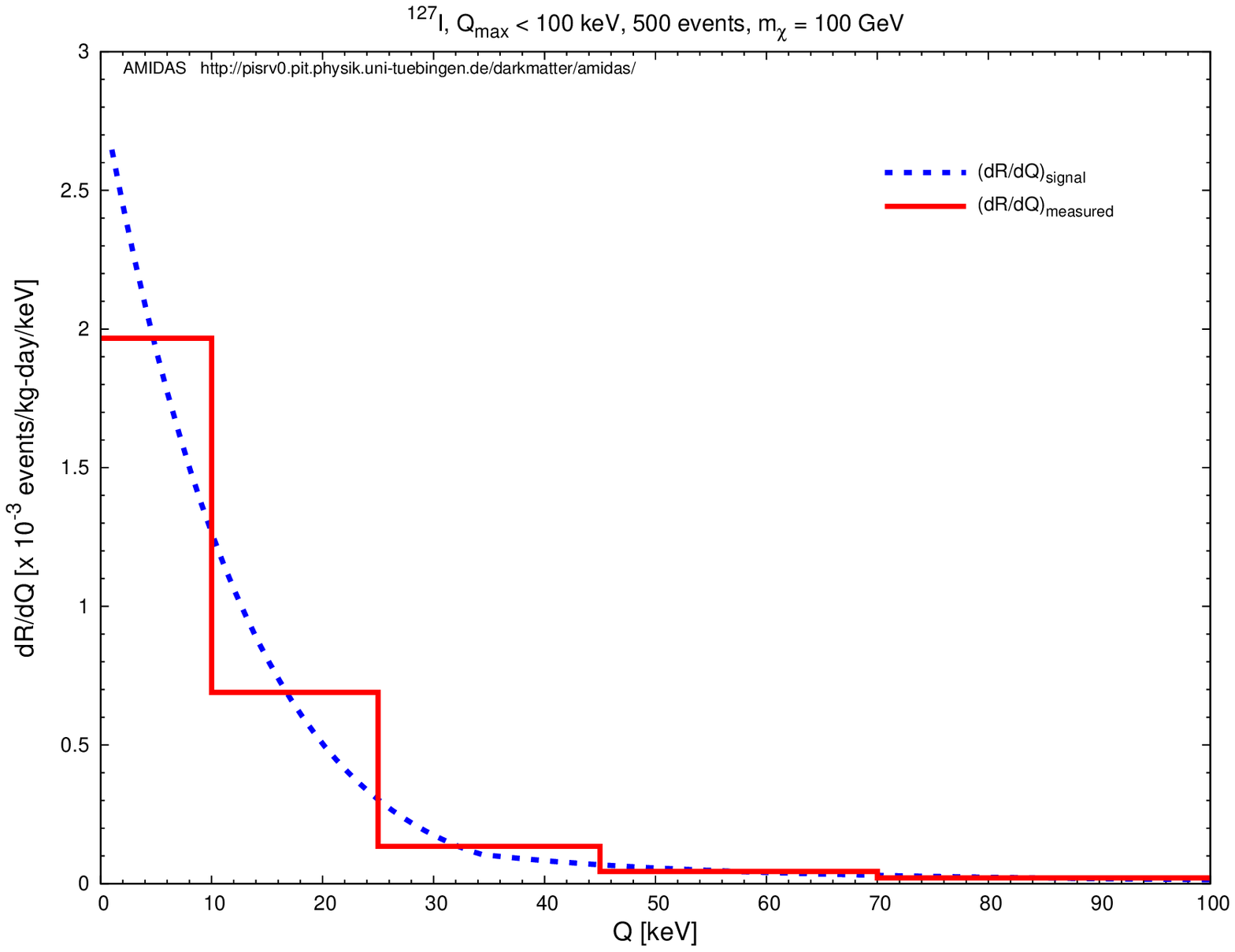} \hspace{-1.1cm}
\includegraphics[width=9.8cm]{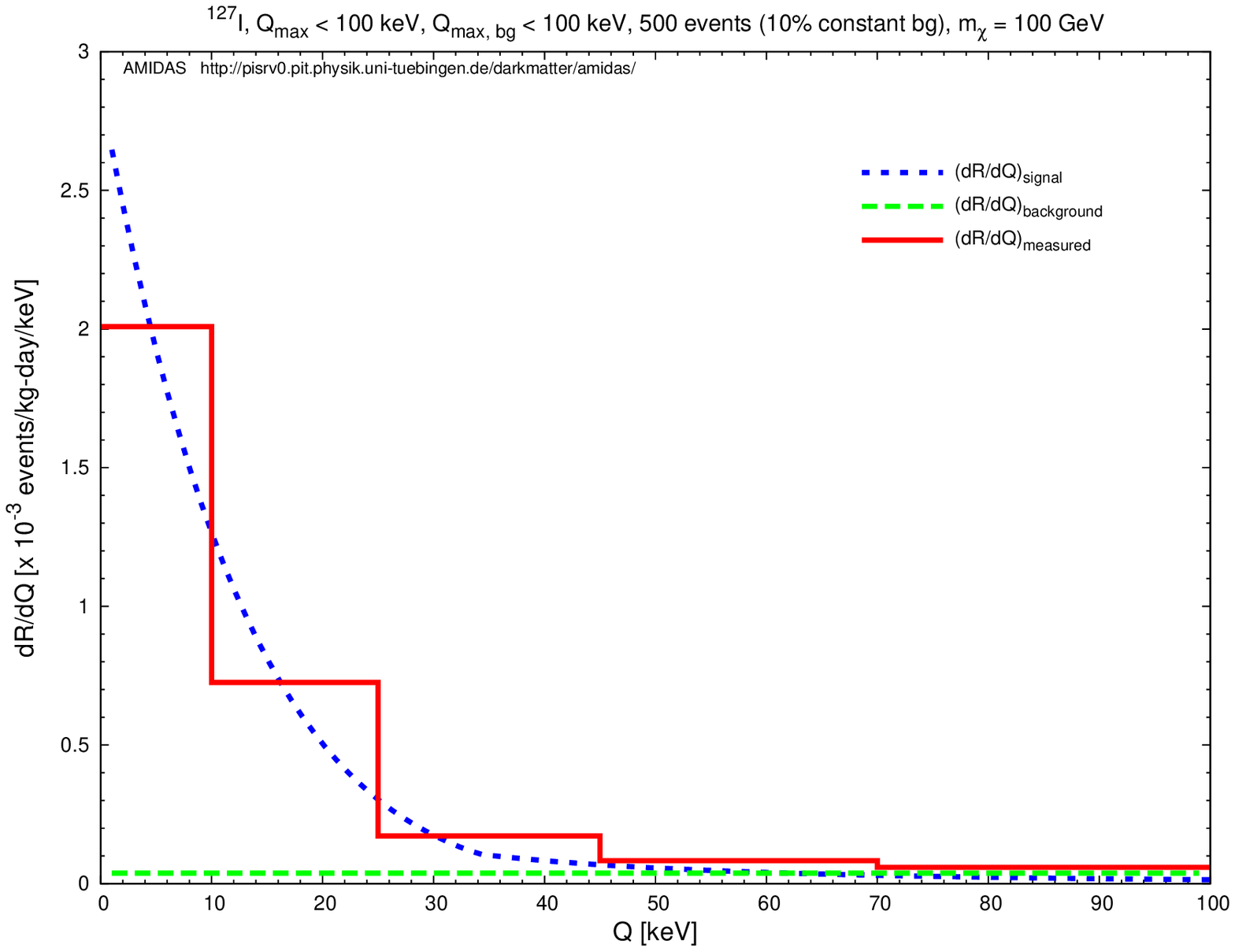} \hspace*{-1.6cm} \\
\vspace{0.5cm}
\hspace*{-1.6cm}
\includegraphics[width=9.8cm]{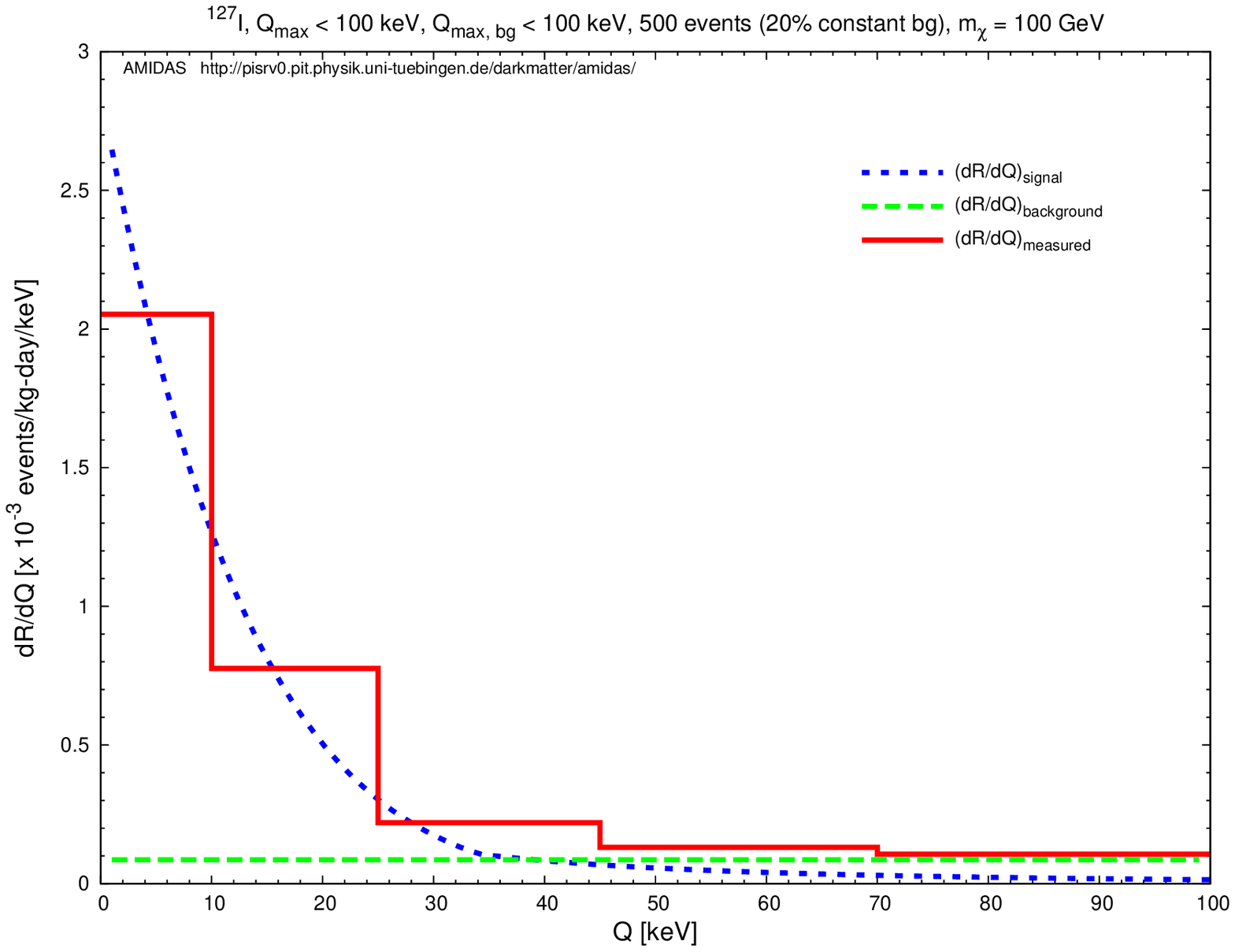} \hspace{-1.1cm}
\includegraphics[width=9.8cm]{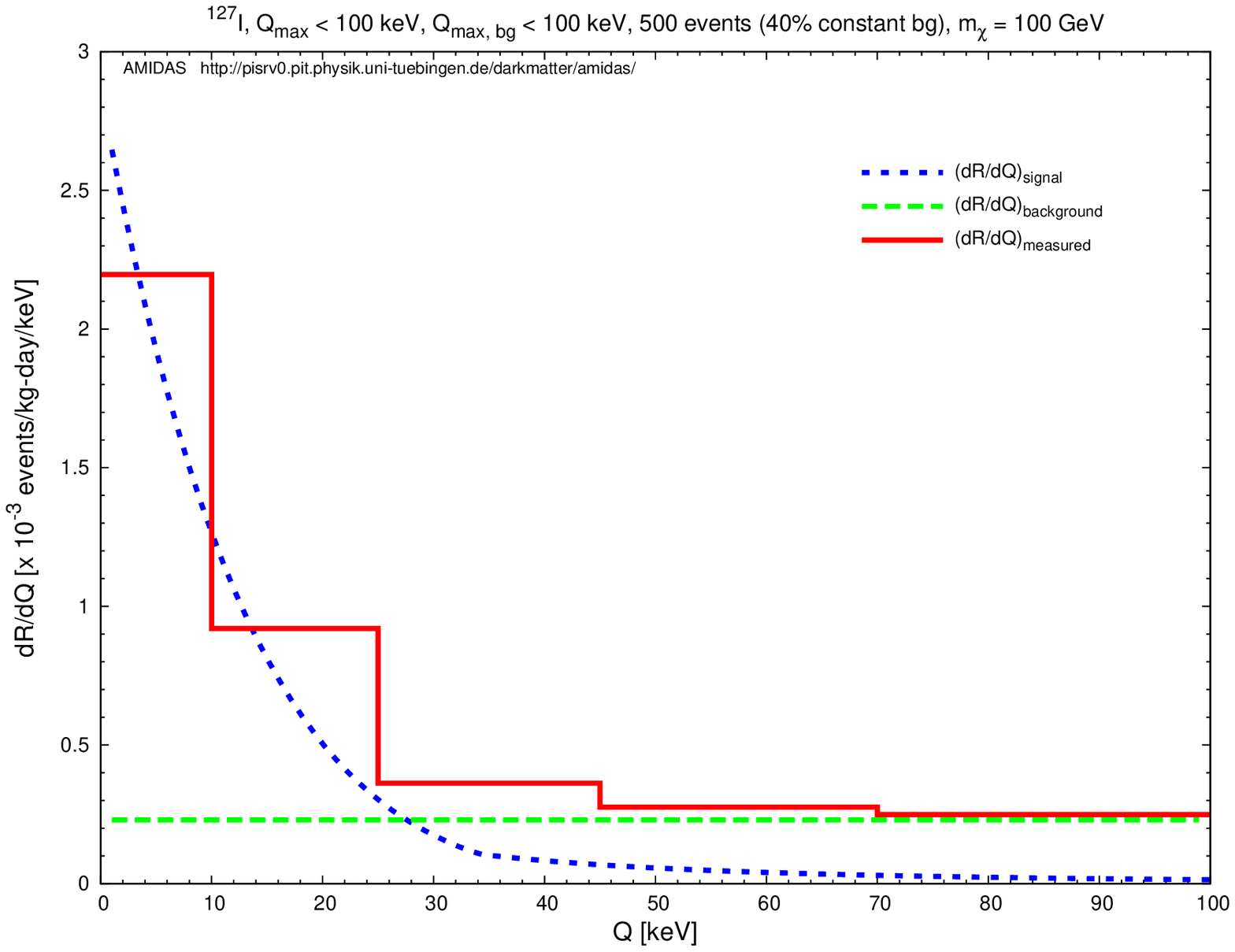} \hspace*{-1.6cm} \\
\vspace{-0.25cm}
\end{center}
\caption{
 As in Figs.~\ref{fig:dRdQ-bg-ex-F-000-100-100},
 except that
 an $\rmXA{I}{127}$ target
 and the constant background spectrum
 have been used here.
}
\label{fig:dRdQ-bg-const-I-000-100-100}
\end{figure}

 In Figs.~\ref{fig:dRdQ-bg-ex-Ge-000-100-20}
 I show measured energy spectra (solid red histograms)
 for a $\rmXA{Ge}{76}$ target
 with six different WIMP masses:
 10, 25, 50, 100, 250, and 500 GeV
 based on Monte Carlo simulations.
 The dotted blue curves are
 the elastic WIMP--nucleus scattering spectra,
 whereas
 the dashed green curves are
 the exponential background spectra
 given in Eq.~(\ref{eqn:dRdQ_bg_ex}),
 which have been normalized so that
 the ratios of the areas under these background spectra
 to those under the (dotted blue) WIMP scattering spectra
 are equal to the background--signal ratio
 in the whole data sets
 (e.g.,~20\% backgrounds to 80\% signals
  shown in Figs.~\ref{fig:dRdQ-bg-ex-Ge-000-100-20}).
 The experimental threshold energies
 have been assumed to be negligible
 and the maximal cut--off energies
 are set as 100 keV.
 The background windows
 (the possible energy ranges
  in which residue background events exist)
 have been assumed to be the same as
 the experimental possible energy ranges.
 5,000 experiments with 500 total events on average
 in each experiment have been simulated.

 Remind that
 the measured energy spectra shown here
 are averaged over the simulated experiments.
 Five bins with linear increased bin widths
 have been used for binning
 generated signal and background events.
 As argued in Ref.~\cite{DMDDf1v},
 for reconstructing the one--dimensional
 WIMP velocity distribution function,
 this unusual, particular binning has been chosen
 in order to accumulate more events
 in high energy ranges
 and thus to reduce the statistical uncertainties
 in high velocity ranges.
 However,
 as shown in Sec.~2,
 for the determinations of ratios between different WIMP couplings/cross sections,
 one needs either events in the {\em first} energy bins
 or {\em all} events in the whole data sets.
 Hence,
 there is in practice no difference
 between using an equal bin width for all bins
 or a (linear) increased bin widths.

 It can be found
 in Figs.~\ref{fig:dRdQ-bg-ex-Ge-000-100-20} that
 the shape of the WIMP scattering spectrum
 depends highly on the WIMP mass:
 for light WIMPs ($\mchi~\lsim~50$ GeV),
 the recoil spectra drop sharply with increasing recoil energies,
 while for heavy WIMPs ($\mchi~\gsim~100$ GeV),
 the spectra become flatter.
 In contrast,
 the exponential background spectra shown here
 depend only on the target mass
 and are rather flatter (sharper)
 for light (heavy) WIMP masses
 compared to the WIMP scattering spectra.
 This means that,
 once input WIMPs are light (heavy),
 background events would contribute relatively more to
 high (low) energy ranges,
 and, consequently,
 the measured energy spectra
 would mimic scattering spectra
 induced by heavier (lighter) WIMPs.
 Moreover,
 for heavy WIMP masses,
 since background events would contribute relatively more to
 low energy ranges,
 the estimated value of the measured recoil spectrum
 at the lowest experimental cut--off energy, $r(\Qmin)$,
 could thus be (strongly) overestimated.

 Furthermore,
 in Figs.~\ref{fig:dRdQ-bg-ex-F-000-100-100}
 we use a $\rmXA{F}{19}$ nucleus as detector material
 and fix the input WIMP mass as 100 GeV.
 Four different background ratios have been considered:
 no background (top left),
 10\% (top right),
 20\% (bottom left),
 and 40\% (bottom right)
 background events in the analyzed data sets.
 It can be seen clearly that,
 for lighter nuclei e.g., F or Si,
 the WIMP scattering spectra are flatter than
 that of a Ge target,
 and more importantly,
 the exponential background spectra
 contribute (almost) {\em only} to {\em low} energy ranges
 (mostly into the first $Q-$bin
  and a bit into the second one)
 and does {\em not} affect the measured spectra
 (solid red histograms) in high energy ranges.
 In contrast,
 Figs.~\ref{fig:dRdQ-bg-const-I-000-100-100}
 show that,
 for heavier nuclei e.g., I or Xe,
 the WIMP scattering spectra are shaper than
 that of a Ge target
 and,
 since the constant background spectra
 contribute relatively {\em mainly} to {\em high} energy ranges
 (mostly into the last two $Q-$bin
  and a bit into that in the middle),
 the measured spectra in low energy ranges
 changes only very slightly.
 Consequently,
 the estimated value of $r(\Qmin)$
 would only be (very) slightly overestimated.

 More detailed illustrations and discussions
 about the effects of residue background events
 with different spectrum forms
 on the measured energy spectrum
 and on the determination of the WIMP mass
 can be found in Ref.~\cite{DMDDbg-mchi}.
\section{Results of the reconstructed ratios of
         WIMP--nucleon couplings/cross sections I:
         with exponential background spectra}
 In this and the next sections
 I present simulation results
 of the reconstructed ratios between
 different WIMP--nucleon couplings/cross sections
 with mixed data sets
 from WIMP--induced and background events
 by means of the model--independent procedures
 described in Sec.~2.%
\footnote{
 Note that,
 rather than the mean values,
 the (bounds on the) reconstructed ratios
 are always the median values
 of the simulated results.
}
 Considering the natural abundances of
 spin--sensitive detector materials (see Table 1),
 a $\rmXA{F}{19}$ and an $\rmXA{I}{127}$ nuclei
 have been chosen as our targets.
 The threshold energies of all experiments
 have been assumed to be negligible%
\footnote{
 Different from our setup used
 in Refs.~\cite{DMDDidentification-DARK2009, DMDDranap},
 where the threshold energies have been set as 5 keV
 for all targets.
}
 and the maximal cut--off energies
 are set the same as \mbox{100 keV}.
 The exponential and constant background spectra
 given in Eqs.~(\ref{eqn:dRdQ_bg_ex}) and (\ref{eqn:dRdQ_bg_const})
 have been used for generating background events
 in windows of the entire experimental possible ranges
 in this and the next sections,
 respectively.
 2 (3) $\times$ 5,000 experiments have been simulated.
 Each experiment contains 50 {\em total} events
 on average.
 Note that
 {\em all} events recorded in our data sets
 are treated as WIMP signals in the analyses,
 although statistically we know that
 a fraction of these events could be backgrounds.
\subsection{Reconstructed \boldmath$(\armn / \armp)_{\pm, n}^{\rm SD}$}
\begin{figure}[t!]
\begin{center}
\includegraphics[width=8.5cm]{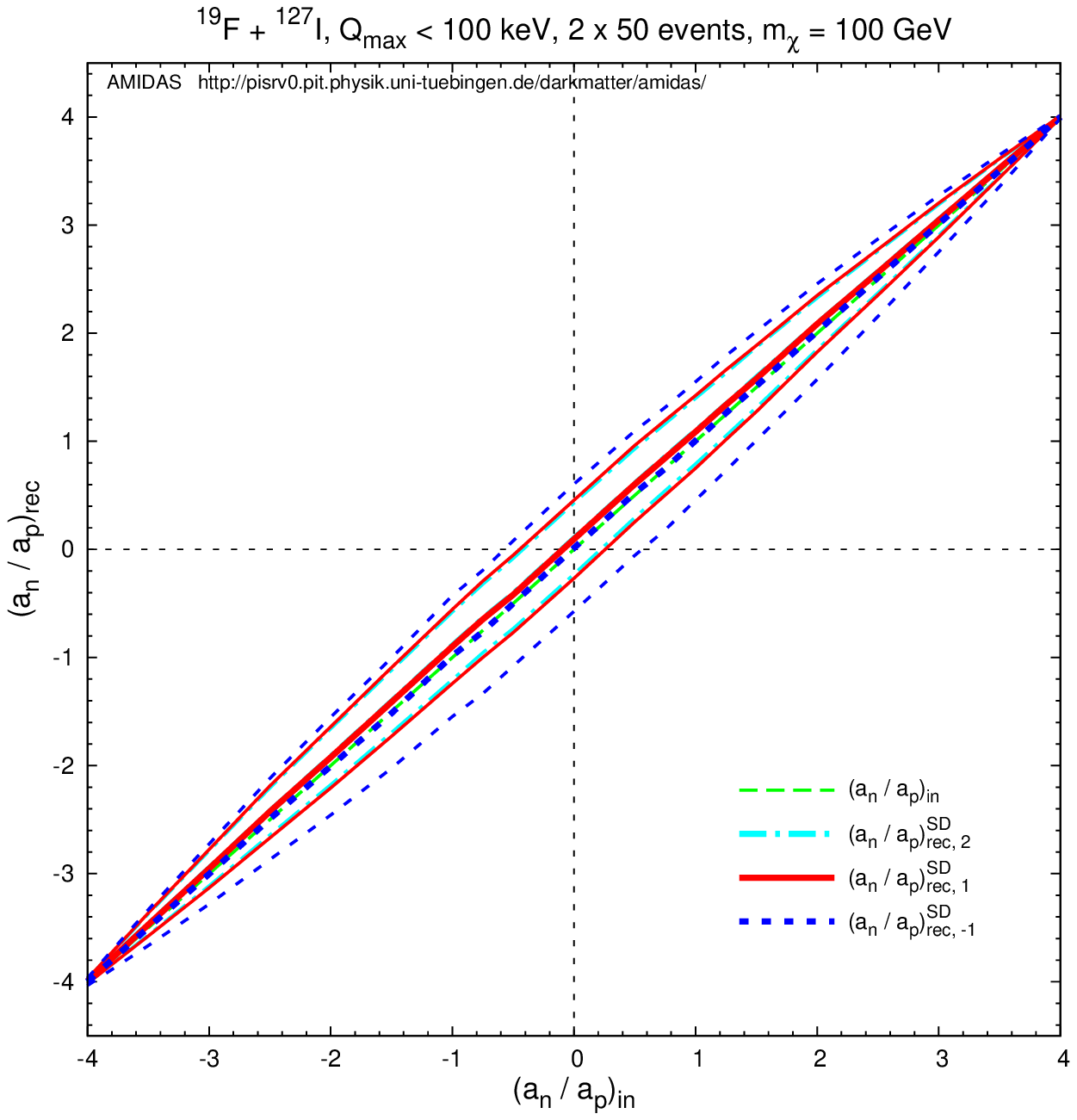}
\includegraphics[width=8.5cm]{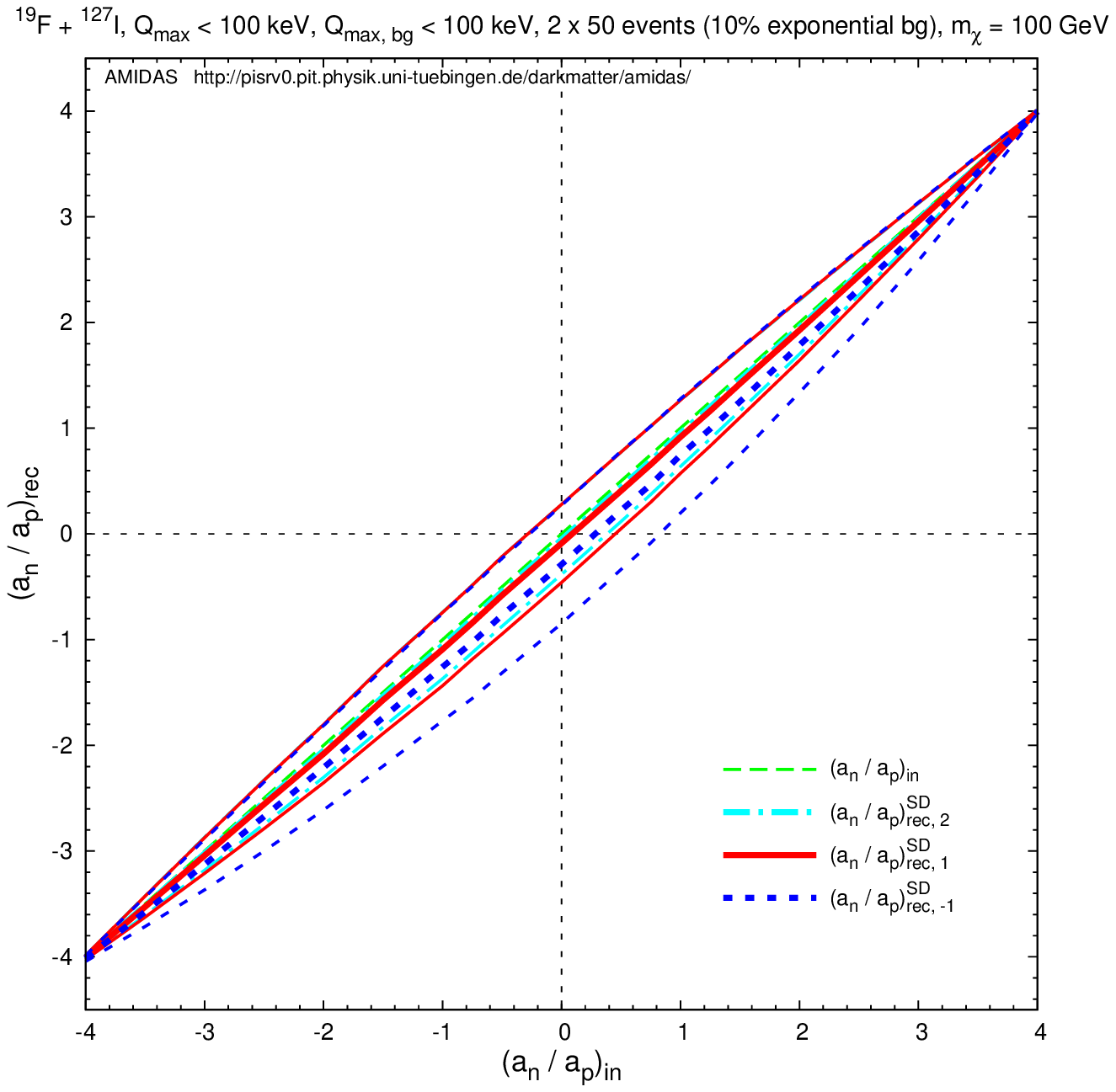} \\
\vspace{0.5cm}
\includegraphics[width=8.5cm]{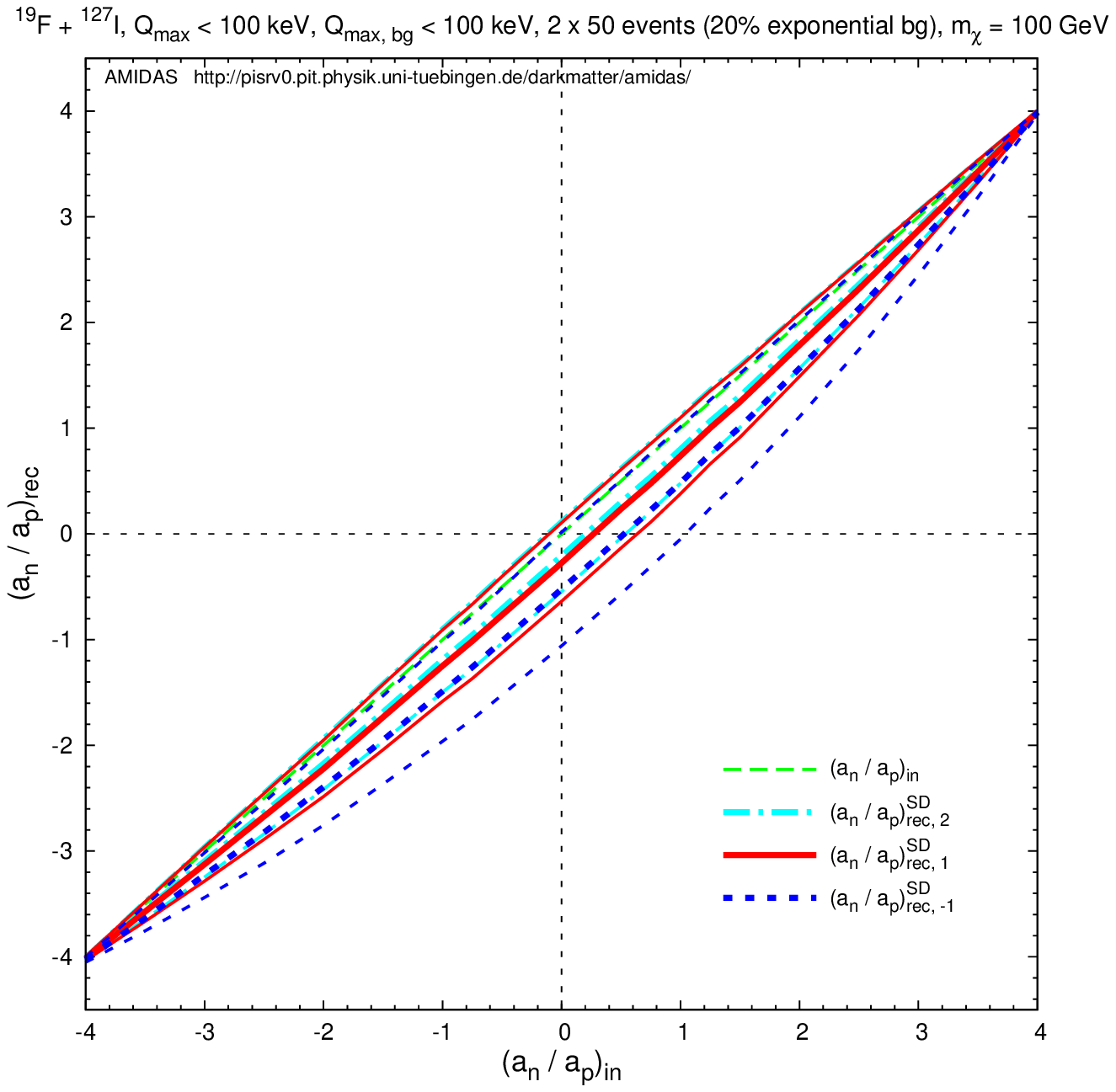}
\includegraphics[width=8.5cm]{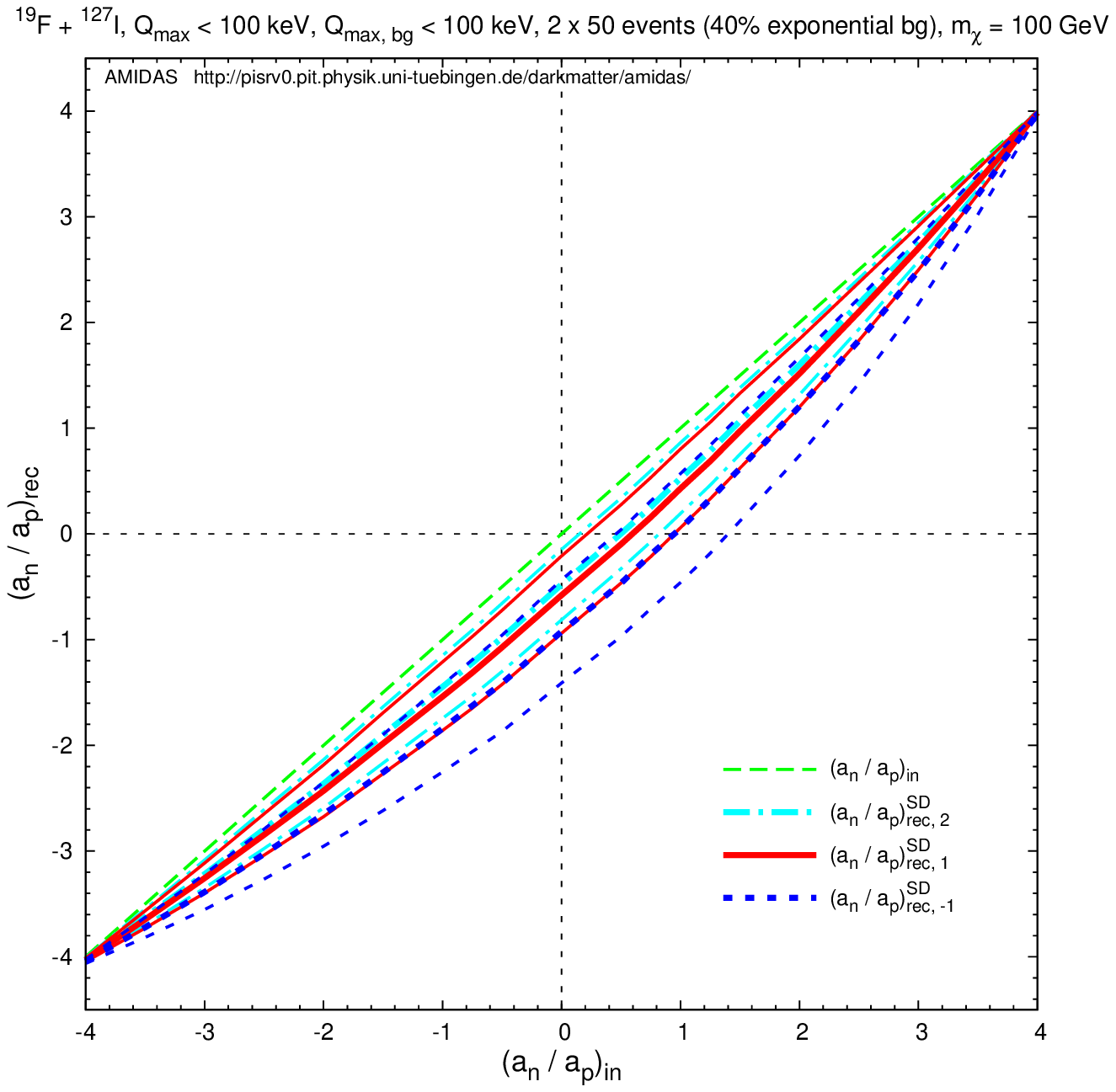} \\
\vspace{-0.25cm}
\end{center}
\caption{
 The reconstructed $\armn / \armp$ ratios
 estimated by Eq.~(\ref{eqn:ranapSD})
 and the lower and upper bounds of
 their 1$\sigma$ statistical uncertainties
 estimated by Eq.~(\ref{eqn:sigma_ranapSD})
 with $n = -1$ (dashed blue), 1 (solid red),
 and 2 (dash--dotted cyan)
 as functions of the input $\armn / \armp$ ratio.
 Here the ``$-$ (minus)'' solution has been used
 (as the ``inner'' solution) \cite{DMDDranap}.
 The mass of incident WIMPs
 has been set as 100 GeV and
 each experiment contains 50 {\em total} events
 on average. 
 The other parameters are as
 in Figs.~\ref{fig:dRdQ-bg-ex-F-000-100-100}.
 and \ref{fig:dRdQ-bg-const-I-000-100-100}.
}
\label{fig:ranapSD-ranap-rec-ex}
\end{figure}

 Consider at first
 the case of a dominant SD WIMP--nucleus interaction.
 Figs.~\ref{fig:ranapSD-ranap-rec-ex}
 show the reconstructed $\armn / \armp$ ratios
 and the lower and upper bounds of
 their 1$\sigma$ statistical uncertainties
 estimated by Eqs.~(\ref{eqn:ranapSD})
 and (\ref{eqn:sigma_ranapSD})
 with $n = -1$ (dashed blue), 1 (solid red),
 and 2 (dash--dotted cyan)
 as functions of the input $\armn / \armp$ ratio.
 The background ratios shown here
 are no background (top left),
 10\% (top right),
 20\% (bottom left),
 and 40\% (bottom right)
 background events in the analyzed data sets.
 Here the ``$-$ (minus)'' solution has been used
 (as the ``inner'' solution) \cite{DMDDranap}.
 The mass of incident WIMPs
 has been set as 100 GeV.

 It can be found here that,
 firstly,
 the statistical uncertainty on
 the reconstructed $\armn / \armp$ ratio with $n = 2$
 is a little bit smaller than
 the uncertainty on the ratio reconstructed with $n = 1$,
 and both of them are much smaller than that reconstructed with $n = -1$.
 Secondly,
 due to the non--negligible background ratio
 in the analyzed data sets,
 the reconstructed $\armn / \armp$ ratios
 become {\em underestimated};
 the larger the background ratio
 the larger this systematic deviation of
 the reconstructed $\armn / \armp$.
 However,
 for the same data sets,
 the larger the $n$ value
 (or, equivalently, the larger the moment of $f_1(v)$ used),
 the smaller this systematic deviation.
 This implies that
 the larger the background ratio,
 the more incompatibile
 between the $\armn / \armp$ ratios
 reconstructed with different $n$.
 This (in)compatibility could thus offer us a simple check
 for the purity/availability of our data sets.

\begin{figure}[t!]
\begin{center}
\includegraphics[width=8.5cm]{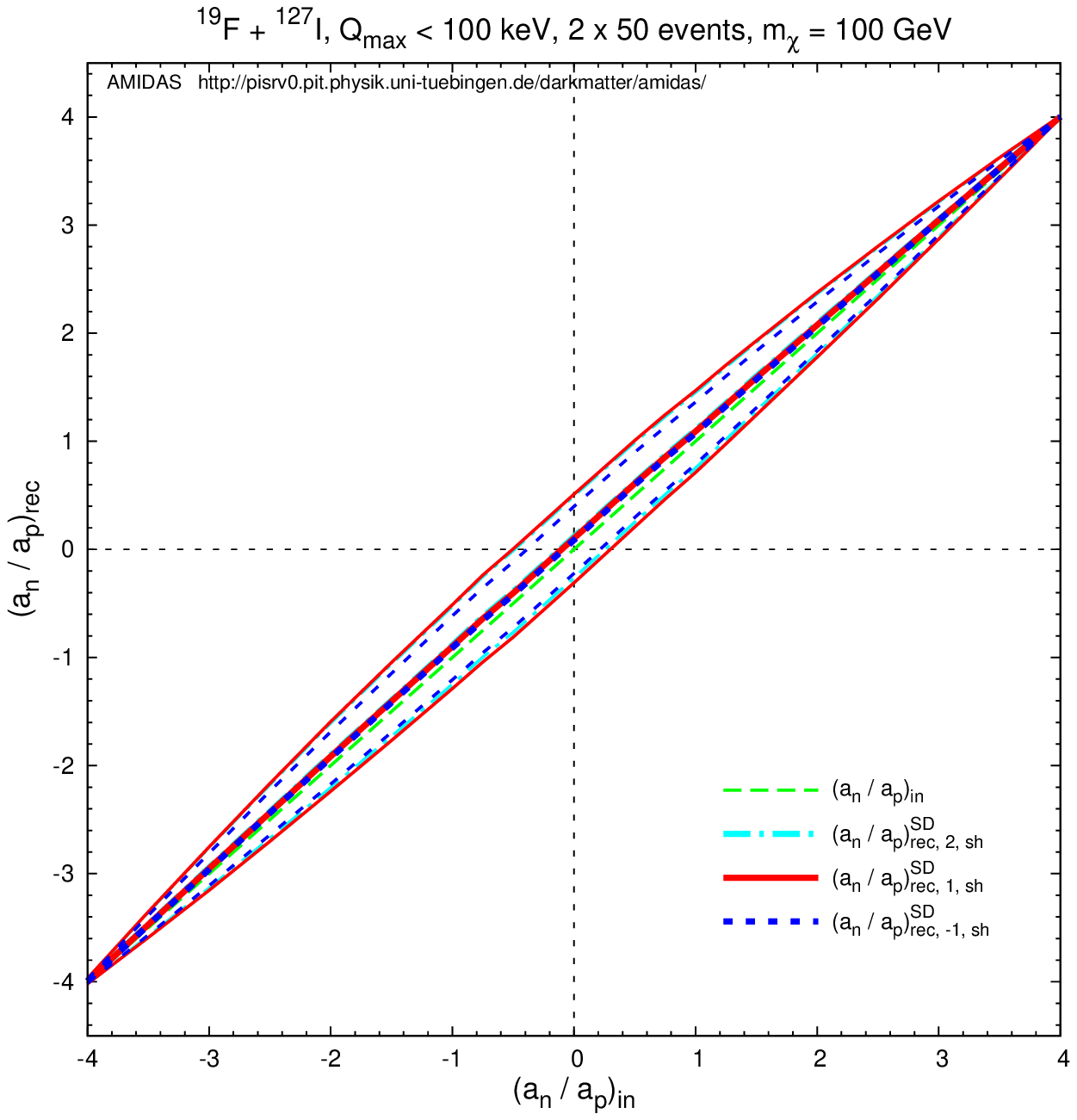}
\includegraphics[width=8.5cm]{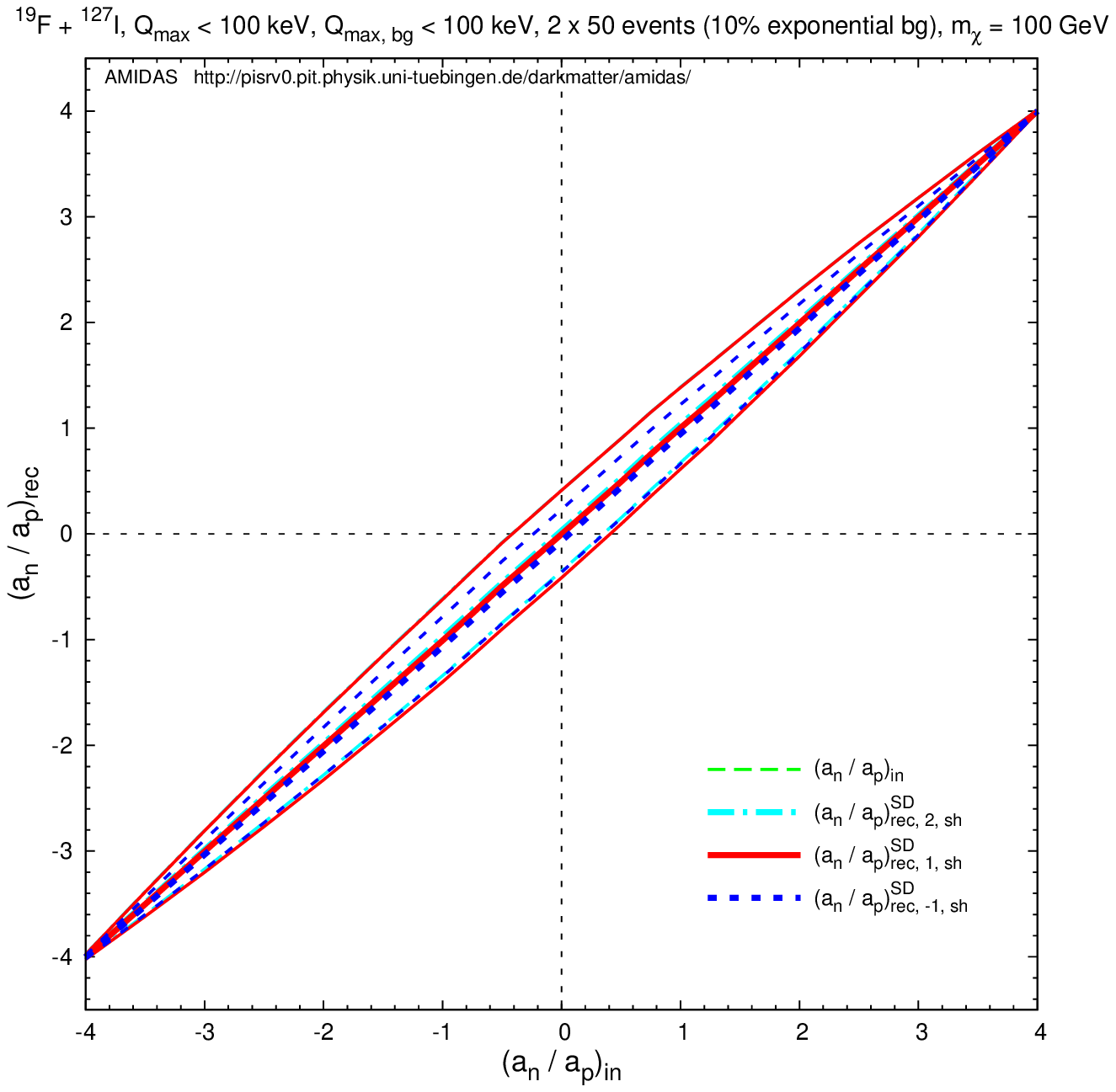} \\
\vspace{0.5cm}
\includegraphics[width=8.5cm]{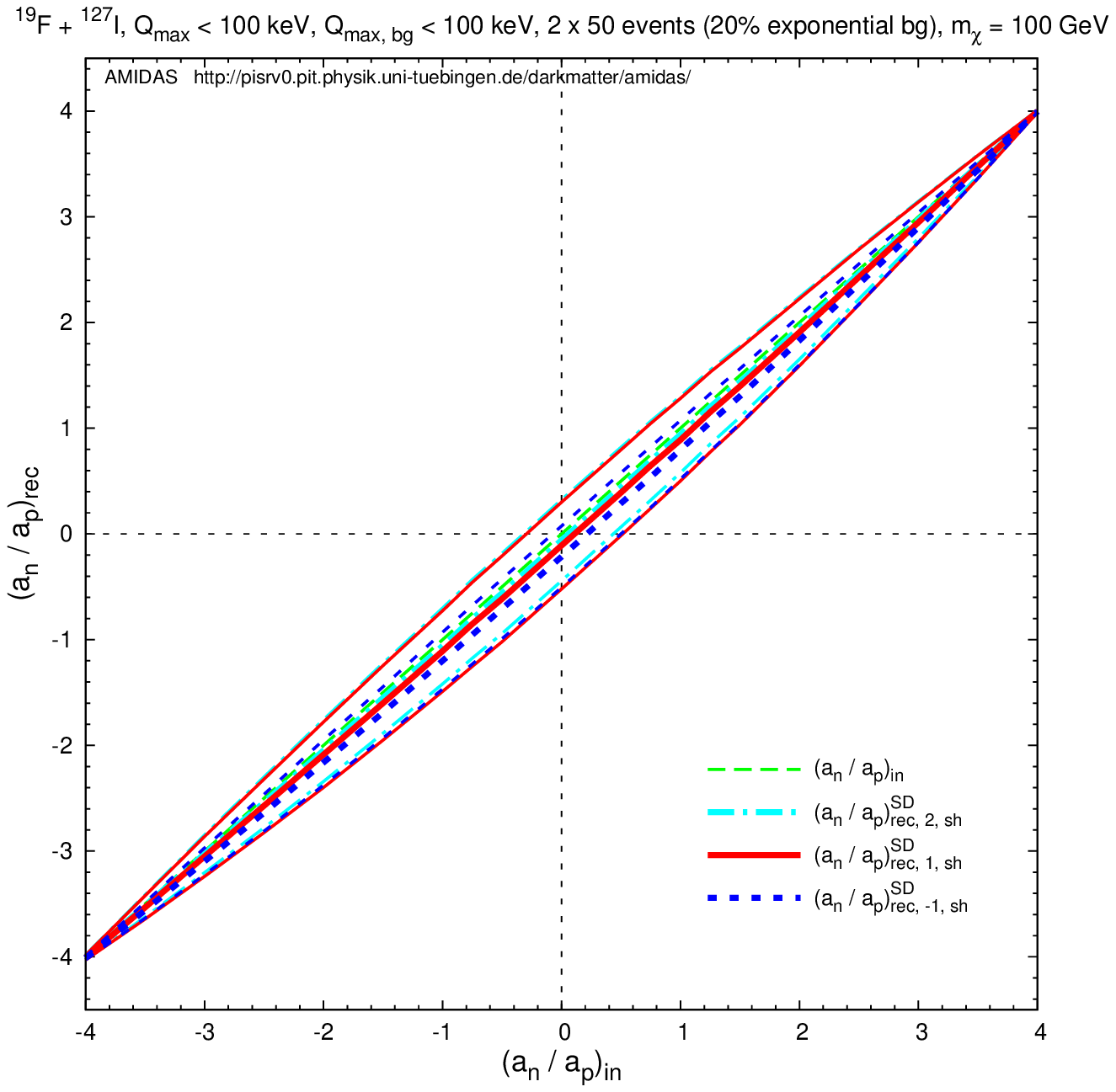}
\includegraphics[width=8.5cm]{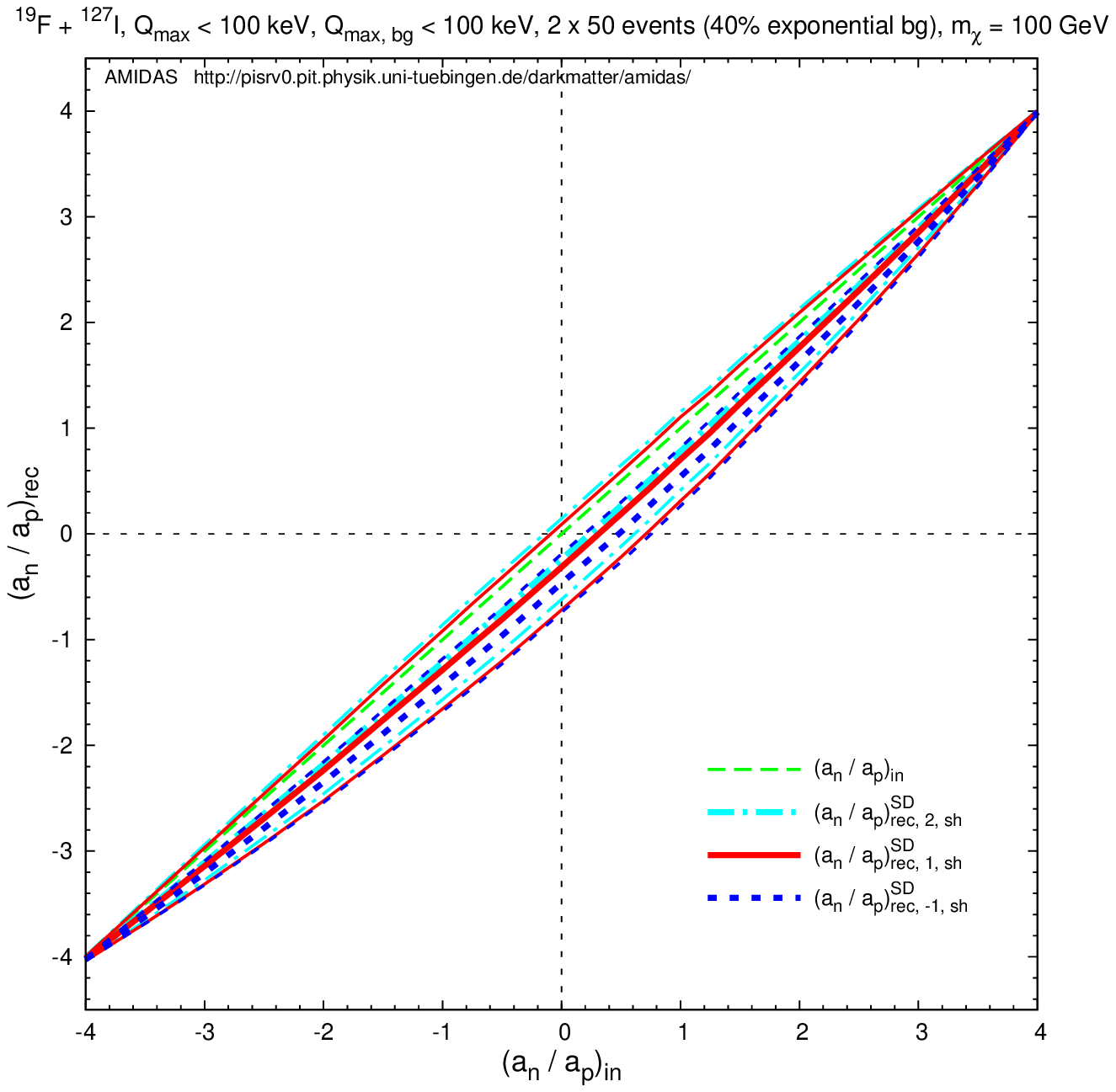} \\
\vspace{-0.25cm}
\end{center}
\caption{
 As in Figs.~\ref{fig:ranapSD-ranap-rec-ex},
 except that
 we estimate $(\armn / \armp)_{\pm, n}^{\rm SD}$ by Eq.~(\ref{eqn:ranapSD})
 with the counting rates
 at the shifted points of the first $Q-$bin,
 $r_{(X, Y)}(Q_{s, 1, (X, Y)}) = r_{(X, Y), 1}$.
}
\label{fig:ranapSD-ranap-sh-rec-ex}
\end{figure}

 In Figs.~\ref{fig:ranapSD-ranap-sh-rec-ex}
 I show the reconstructed $\armn / \armp$ ratios
 and the lower and upper bounds of
 their 1$\sigma$ statistical uncertainties
 estimated by Eqs.~(\ref{eqn:ranapSD})
 and (\ref{eqn:sigma_ranapSD})
 with the counting rates
 at the shifted points of the first $Q-$bin,
 $r_{(X, Y)}(Q_{s, 1, (X, Y)}) = r_{(X, Y), 1}$,
 as functions of the input $\armn / \armp$ ratio%
\footnote{
 Labeled hereafter with an ``sh'' in the subscript.
}.
 Different from the results
 shown in Figs.~\ref{fig:ranapSD-ranap-rec-ex},
 the statistical uncertainty on
 the reconstructed $\armn / \armp$ ratio with $n = -1$
 is a little bit {\em smaller} than
 those reconstructed with $n = 2$ and $n = 1$.%
\footnote{
 This is because
 we set simply the experimental threshold energies
 to be {\em negligible} in our simulations
 (see Ref.~\cite{DMDDranap}
  for cases with {\em non--negligible} threshold energies).
\label{footnote:Qmin}
}
 But,
 the same as shown in Figs.~\ref{fig:ranapSD-ranap-rec-ex},
 the larger the background ratio
 in the analyzed data sets,
 the more strongly
 the reconstructed $\armn / \armp$ ratios
 would be systematically underestimated,
 and the larger the $n$ value,
 the smaller this systematic deviation.
 However,
 the incompatibility between the $\armn / \armp$ ratios
 reconstructed with different $n$
 is not so significant as shown in Figs.~\ref{fig:ranapSD-ranap-rec-ex}.

\begin{figure}[t!]
\begin{center}
\includegraphics[width=8.5cm]{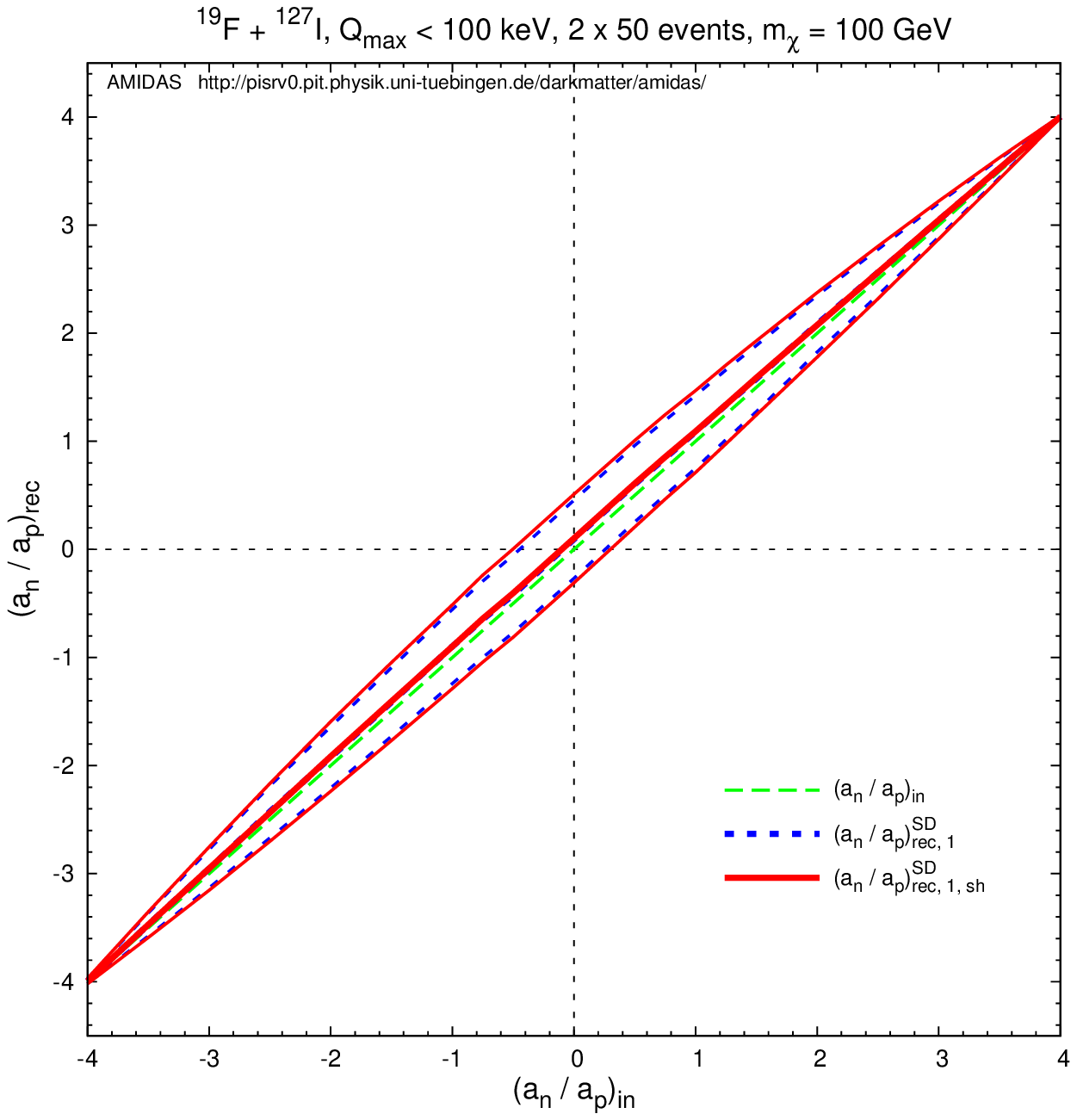}
\includegraphics[width=8.5cm]{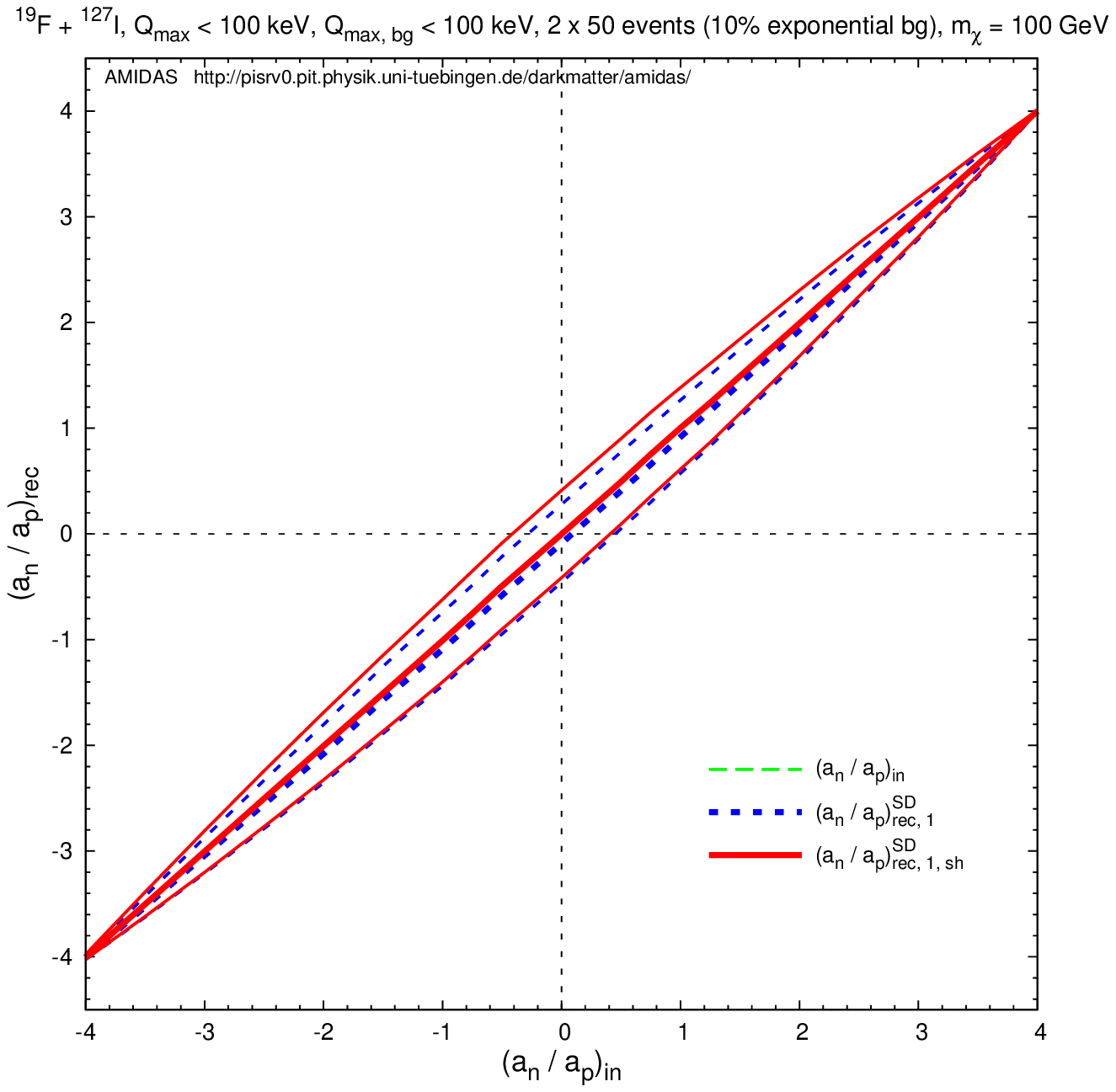} \\
\vspace{0.5cm}
\includegraphics[width=8.5cm]{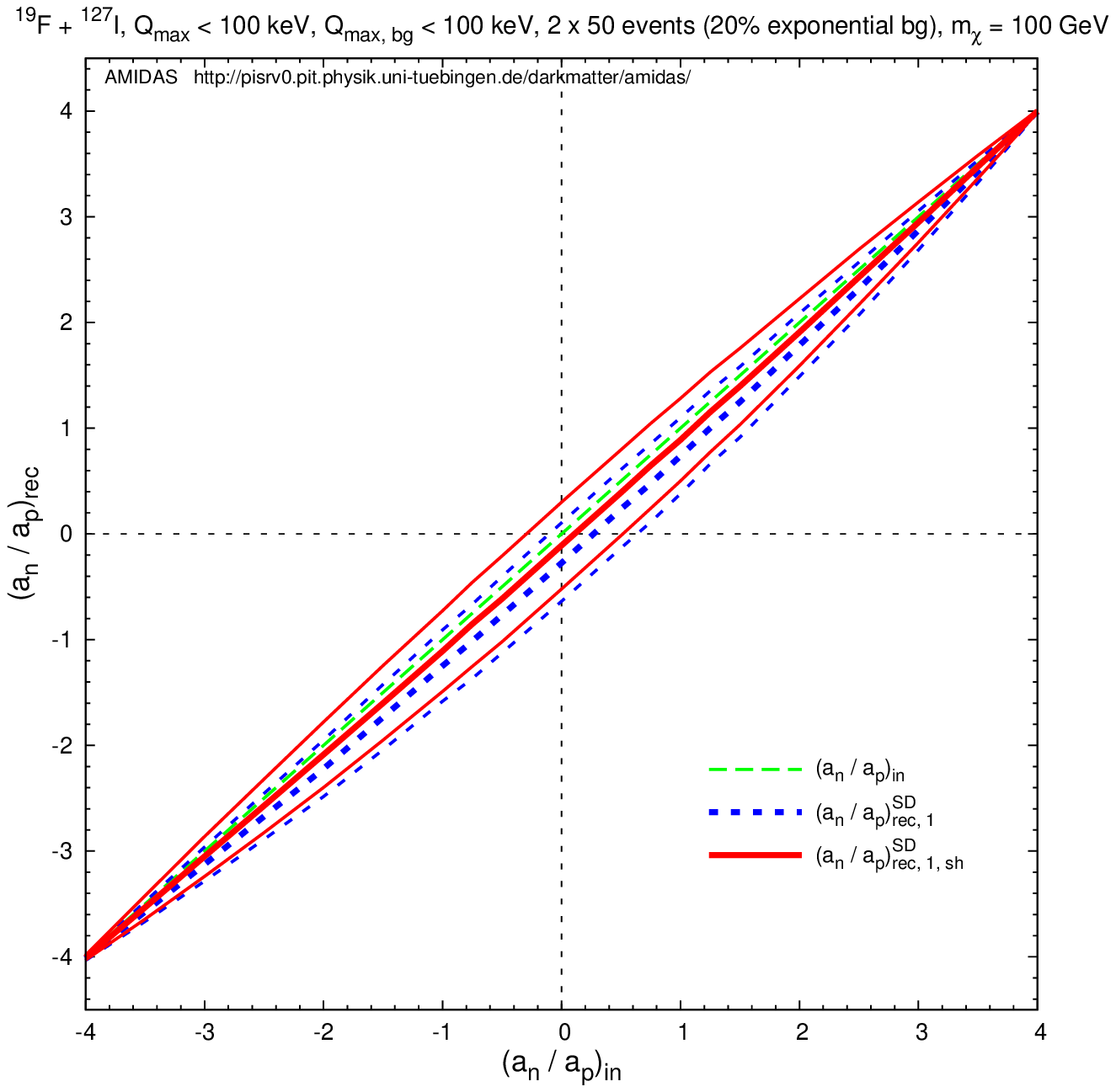}
\includegraphics[width=8.5cm]{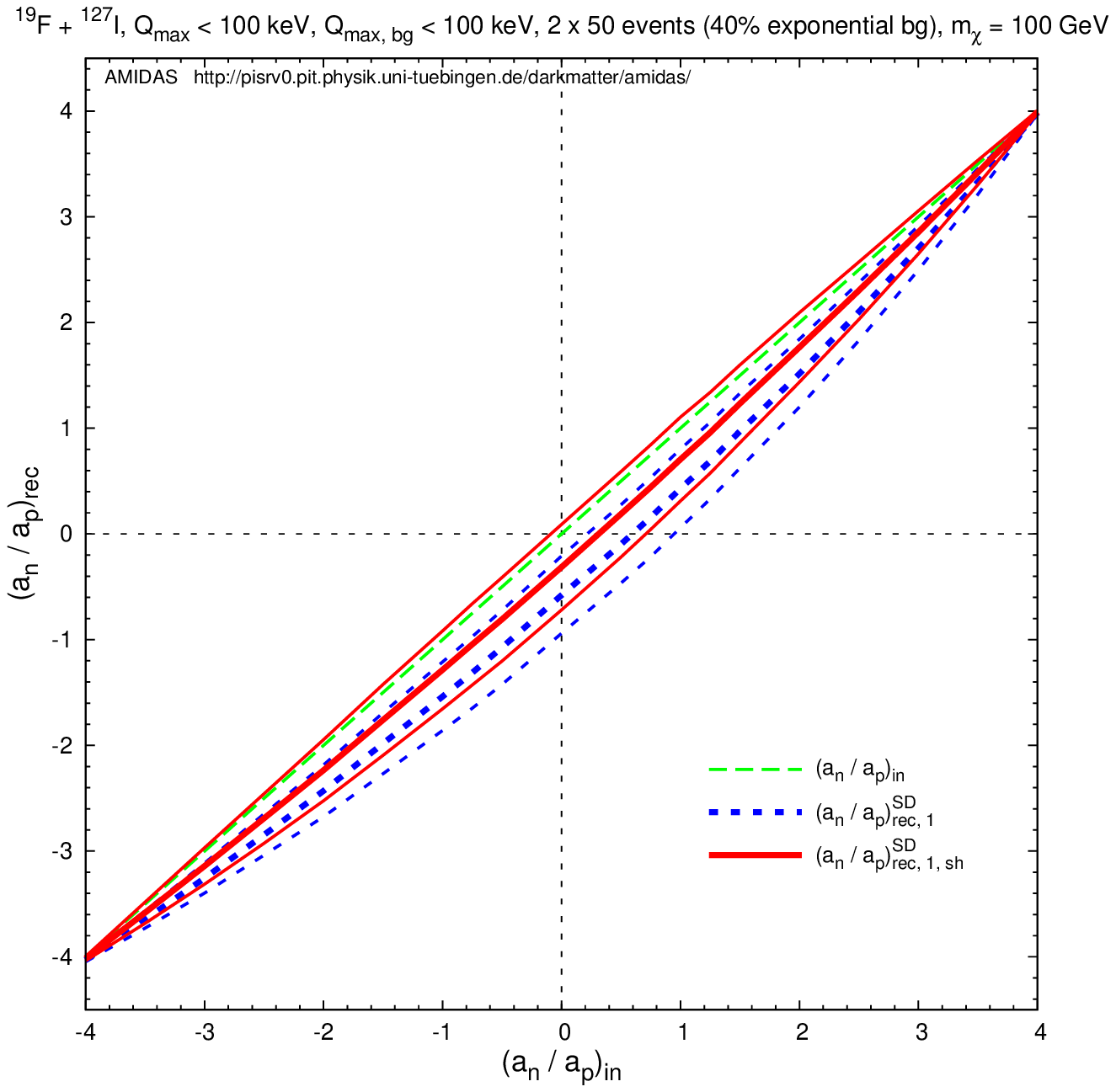} \\
\vspace{-0.25cm}
\end{center}
\caption{
 A comparison of
 the results shown in Figs.~\ref{fig:ranapSD-ranap-rec-ex}
 estimated with $r_{(X, Y)}(Q_{{\rm min}, (X, Y)} = 0)$ (dashed blue)
 and those shown in Figs.~\ref{fig:ranapSD-ranap-sh-rec-ex}
 with $r_{(X, Y)}(Q_{s, 1, (X, Y)})$ (solid red).
 Only the results estimated with $n = 1$
 are shown here.
}
\label{fig:ranapSD-ranap-ex}
\end{figure}
\begin{figure}[t!]
\begin{center}
\includegraphics[width=8.5cm]{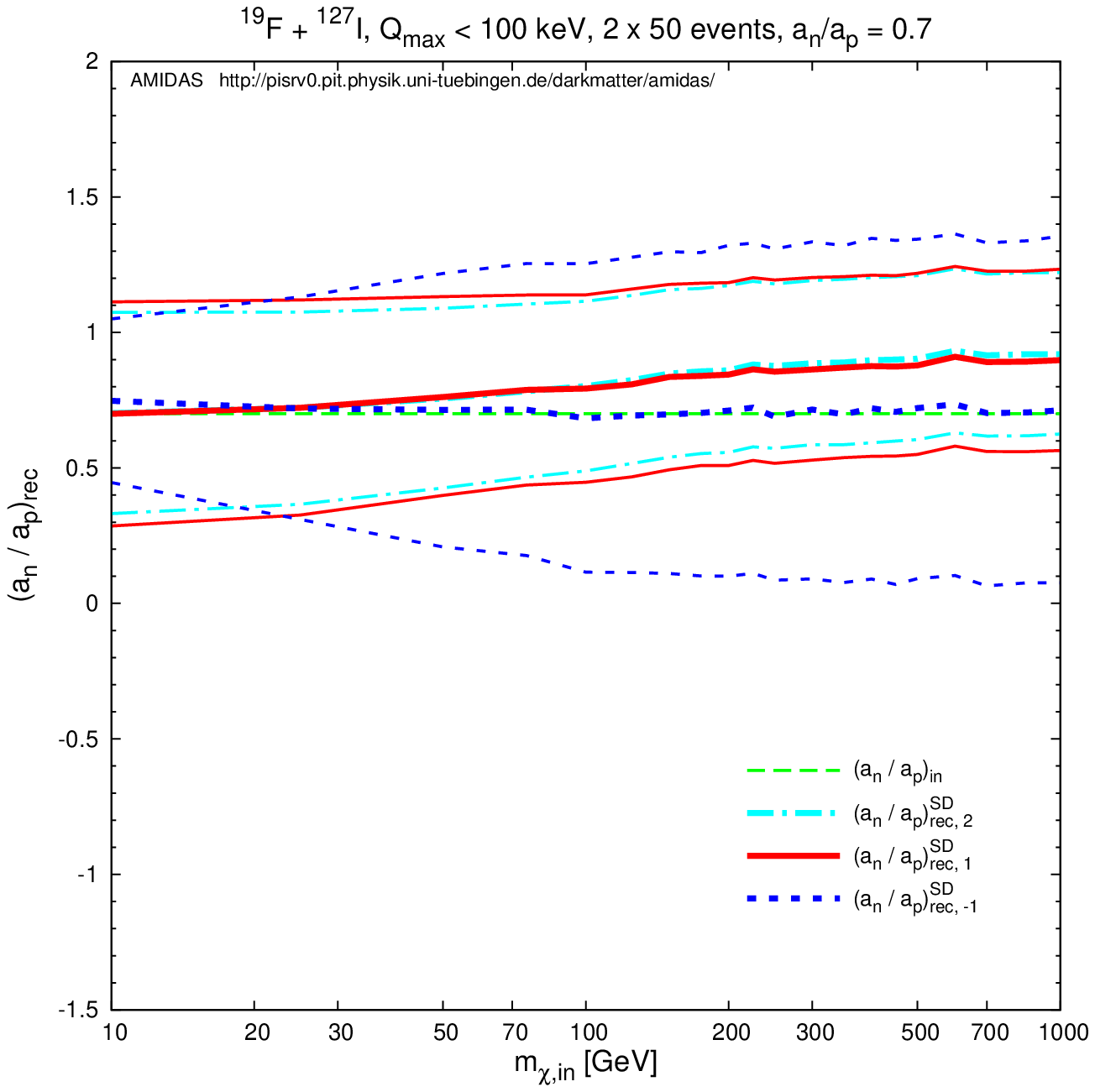}
\includegraphics[width=8.5cm]{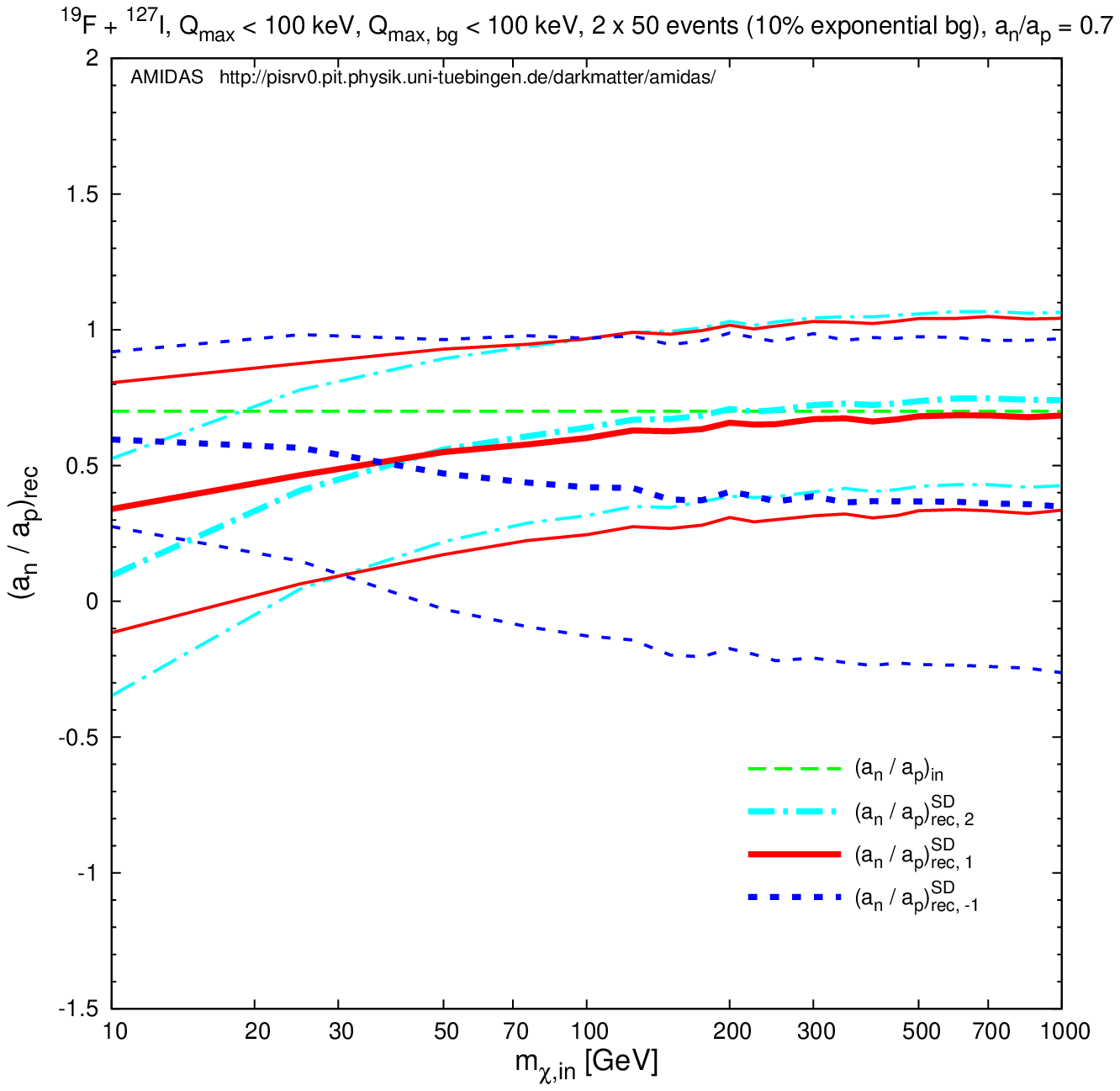} \\
\vspace{0.5cm}
\includegraphics[width=8.5cm]{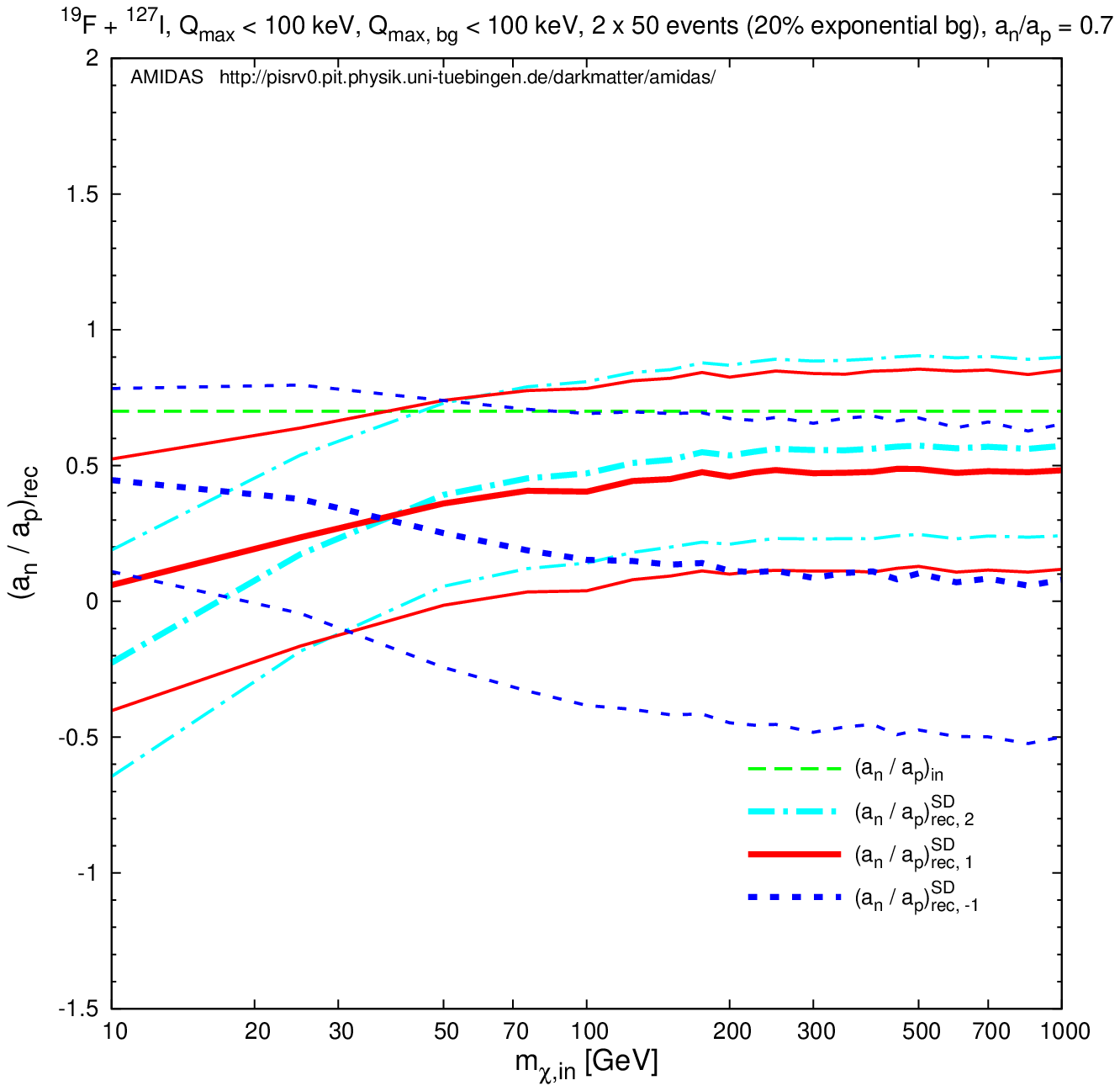}
\includegraphics[width=8.5cm]{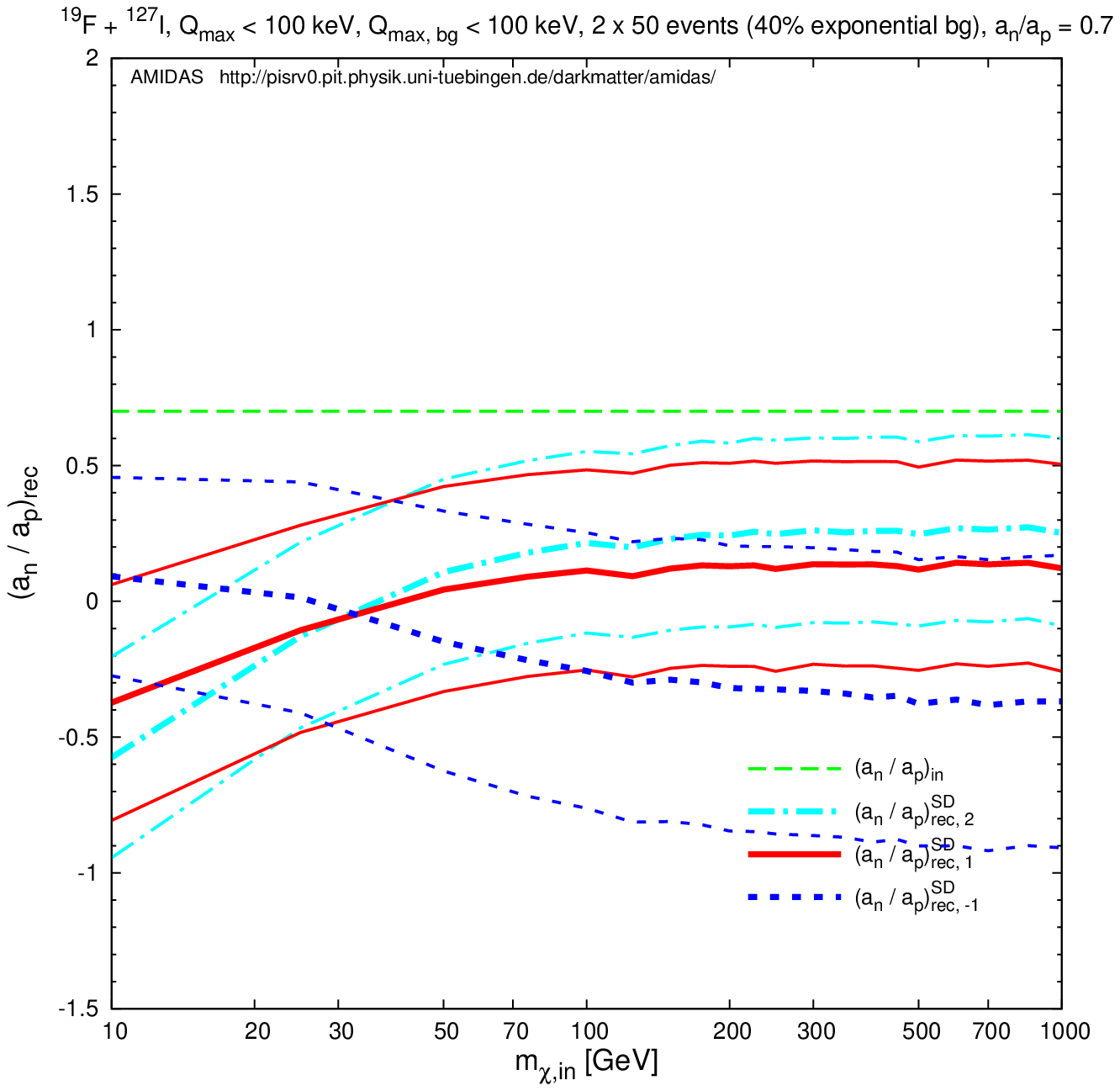} \\
\vspace{-0.25cm}
\end{center}
\caption{
 The reconstructed $\armn / \armp$ ratios
 estimated by Eq.~(\ref{eqn:ranapSD})
 and the lower and upper bounds of
 their 1$\sigma$ statistical uncertainties
 estimated by Eq.~(\ref{eqn:sigma_ranapSD})
 with $n = -1$ (dashed blue), 1 (solid red),
 and 2 (dash--dotted cyan)
 as functions of the input WIMP mass $\mchi$.
 The input $\armn / \armp$ ratio
 has been set as 0.7.
 The other parameters are as
 in Figs.~\ref{fig:ranapSD-ranap-rec-ex}.
}
\label{fig:ranapSD-mchi-rec-ex}
\end{figure}
\begin{figure}[t!]
\begin{center}
\includegraphics[width=8.5cm]{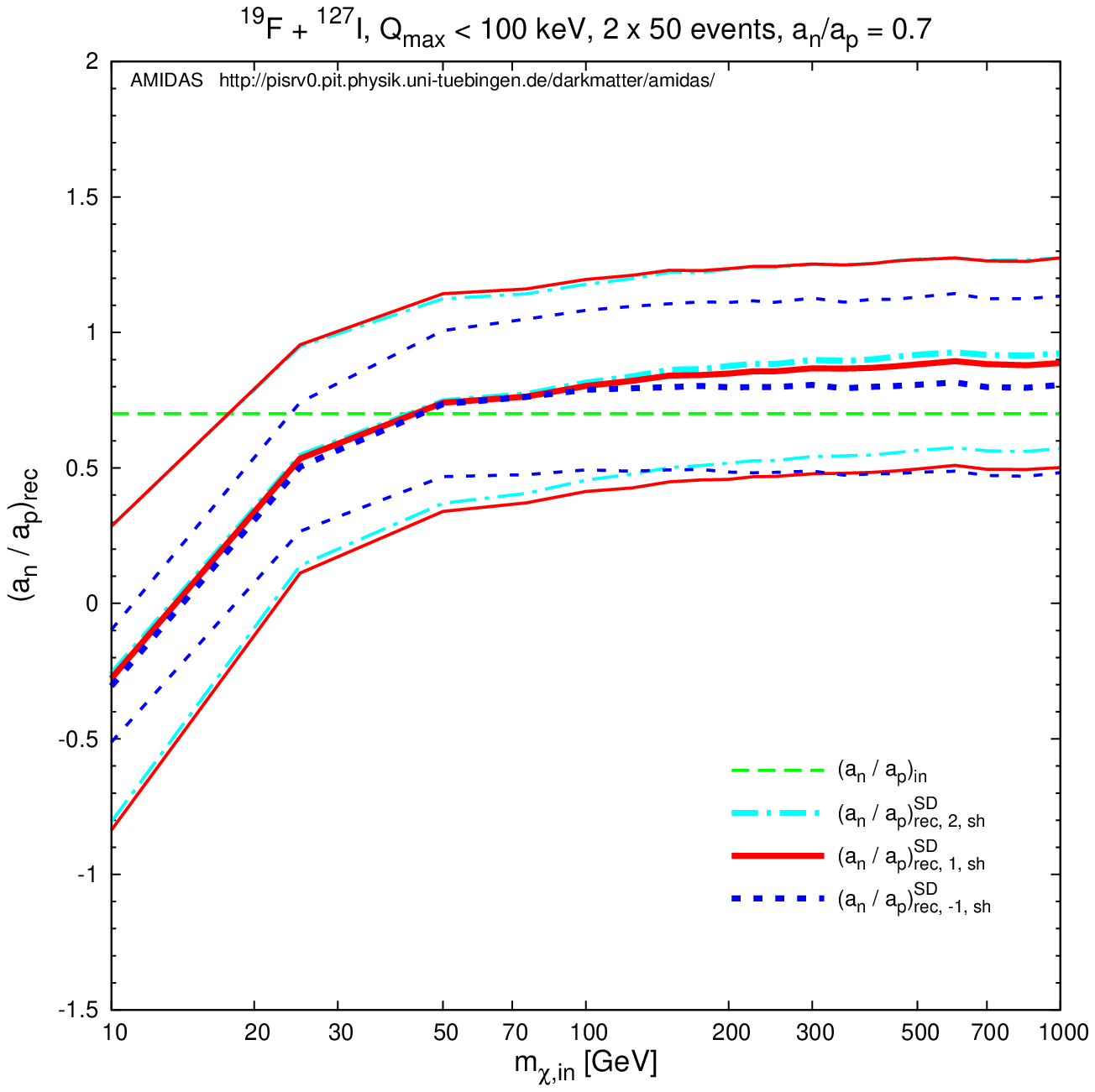}
\includegraphics[width=8.5cm]{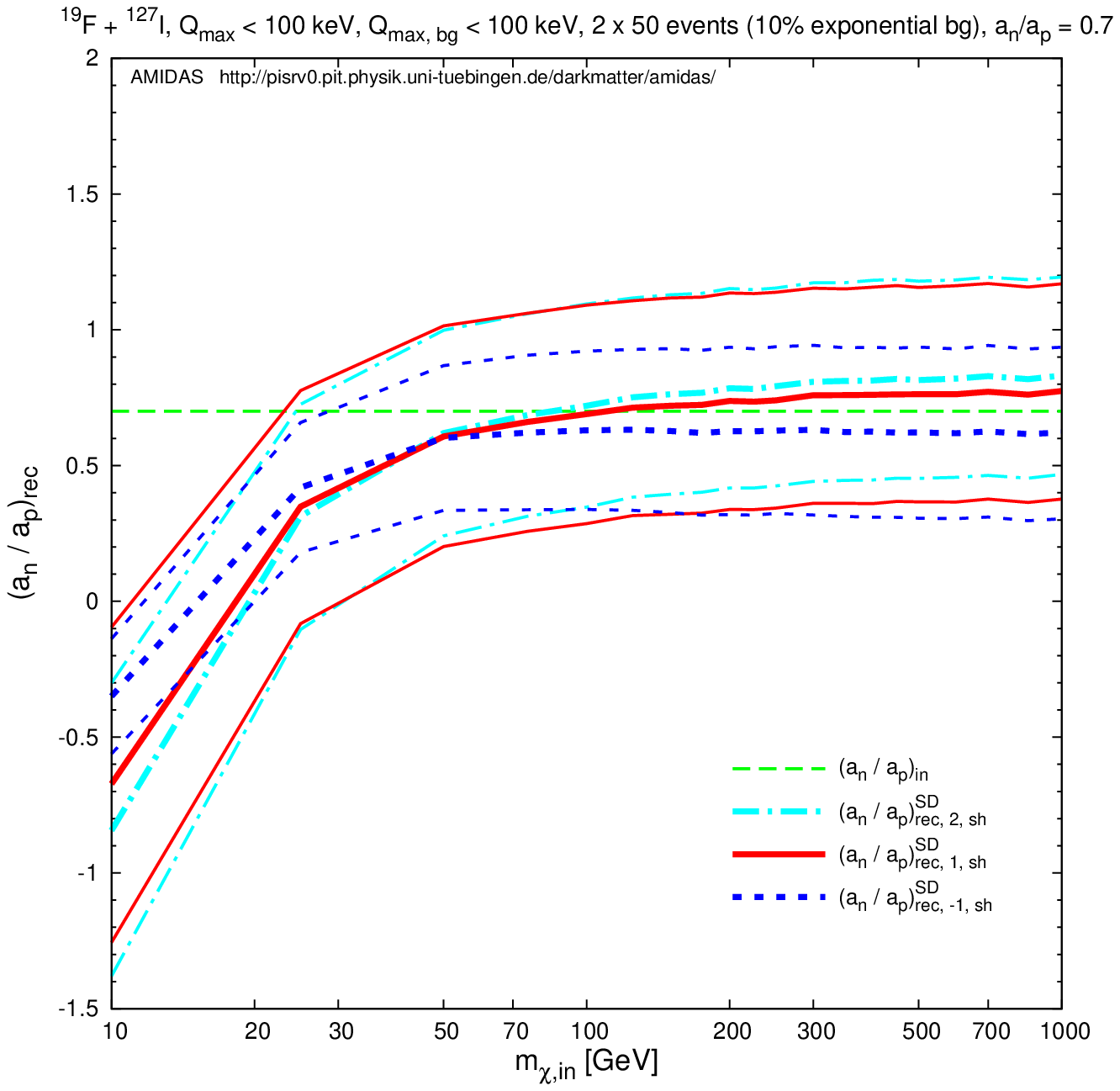} \\
\vspace{0.5cm}
\includegraphics[width=8.5cm]{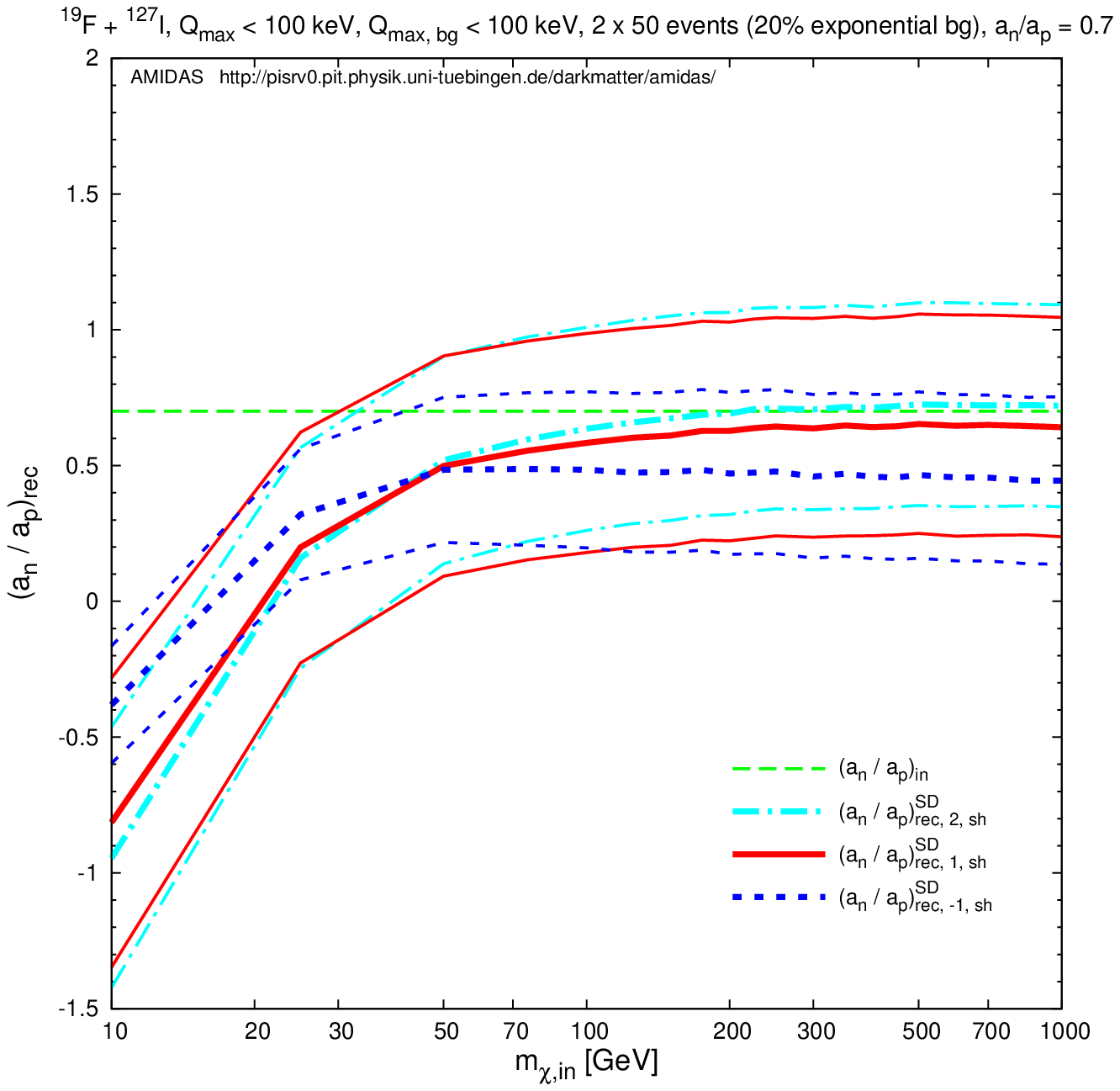}
\includegraphics[width=8.5cm]{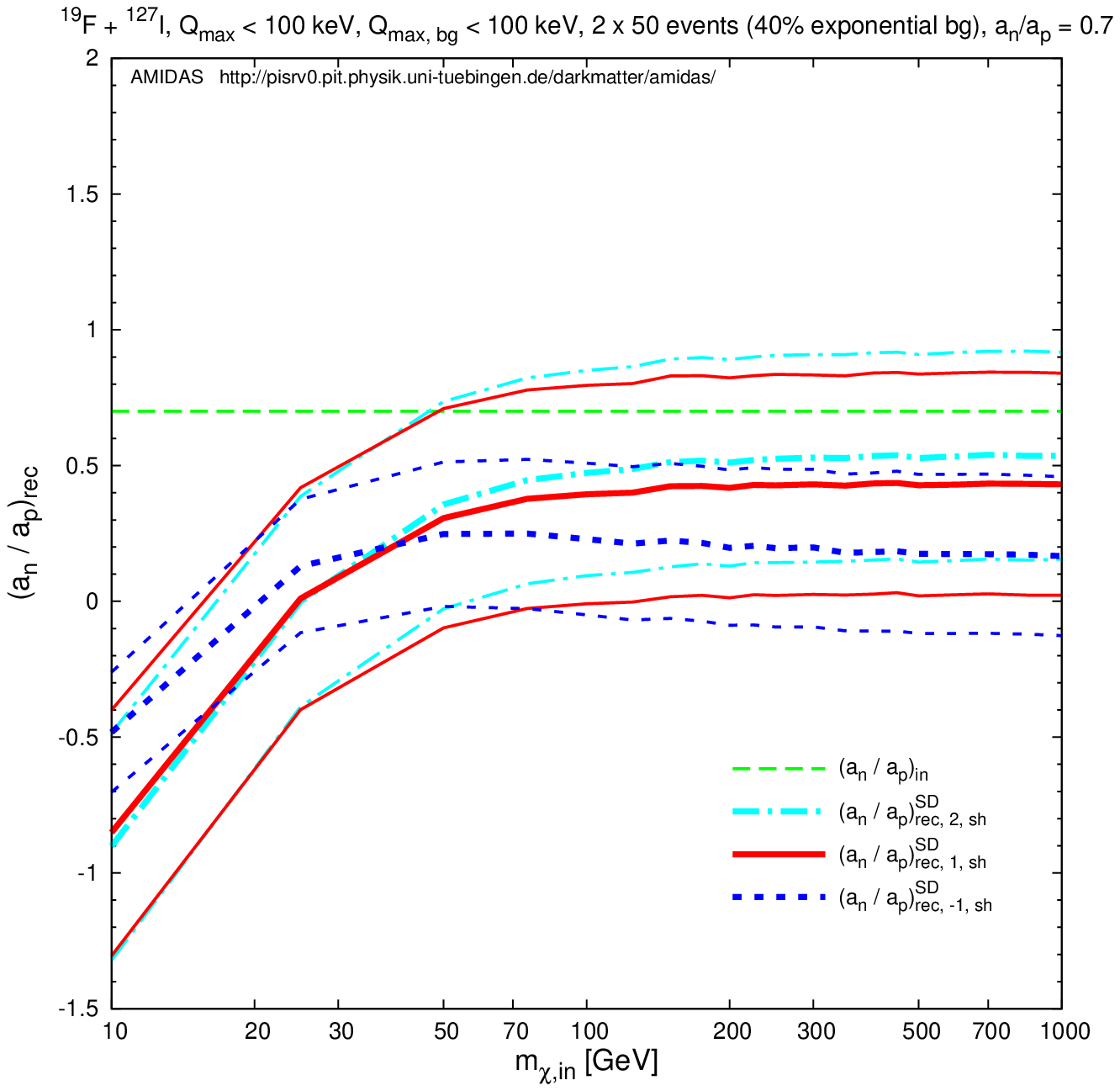} \\
\vspace{-0.25cm}
\end{center}
\caption{
 As in Figs.~\ref{fig:ranapSD-mchi-rec-ex},
 except that
 we estimate $(\armn / \armp)_{\pm, n}^{\rm SD}$ by Eq.~(\ref{eqn:ranapSD})
 with the counting rates
 at the shifted points of the first $Q-$bin,
 $r_{(X, Y)}(Q_{s, 1, (X, Y)}) = r_{(X, Y), 1}$.
}
\label{fig:ranapSD-mchi-sh-rec-ex}
\end{figure}
\begin{figure}[t!]
\begin{center}
\includegraphics[width=8.5cm]{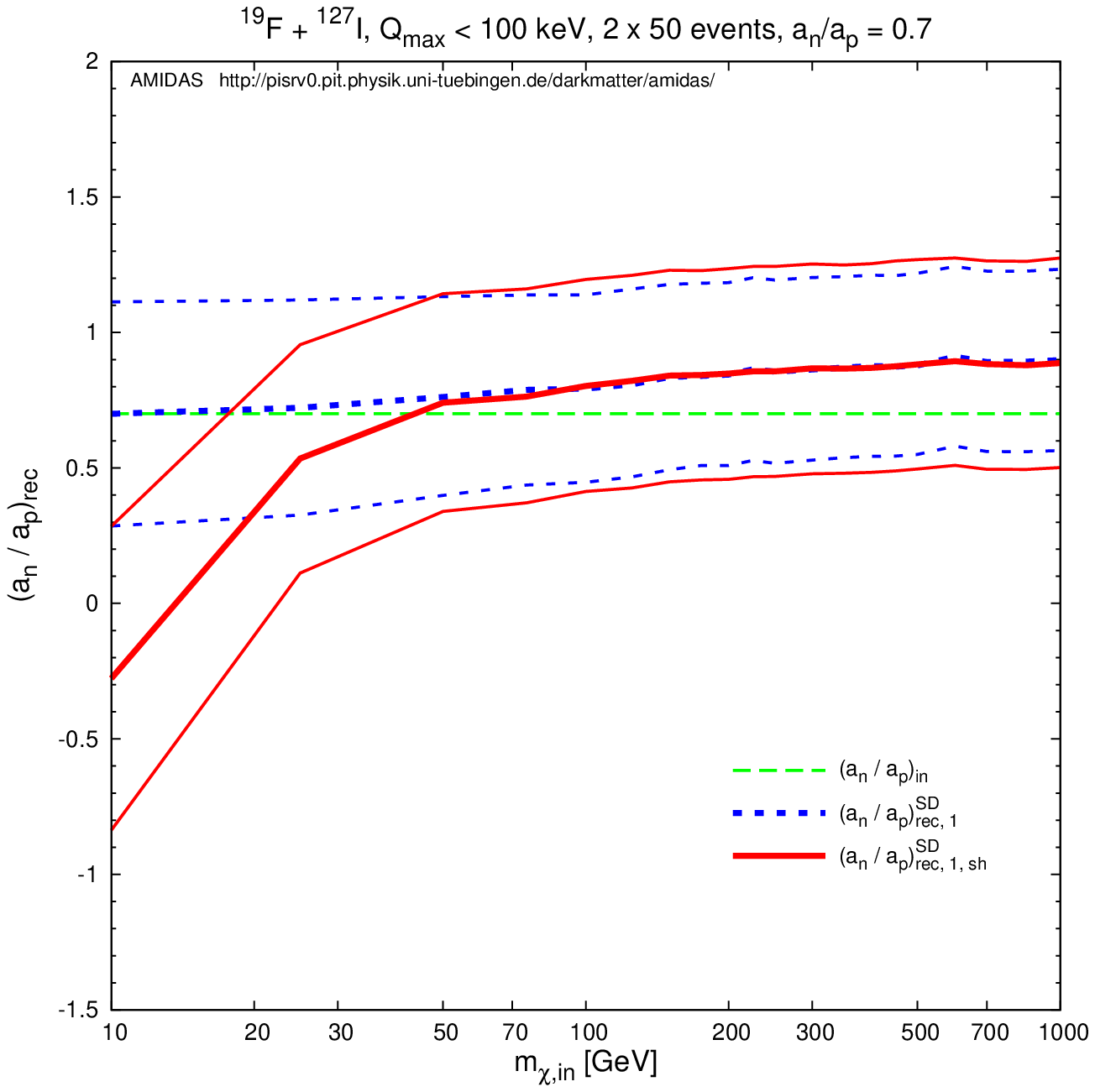}
\includegraphics[width=8.5cm]{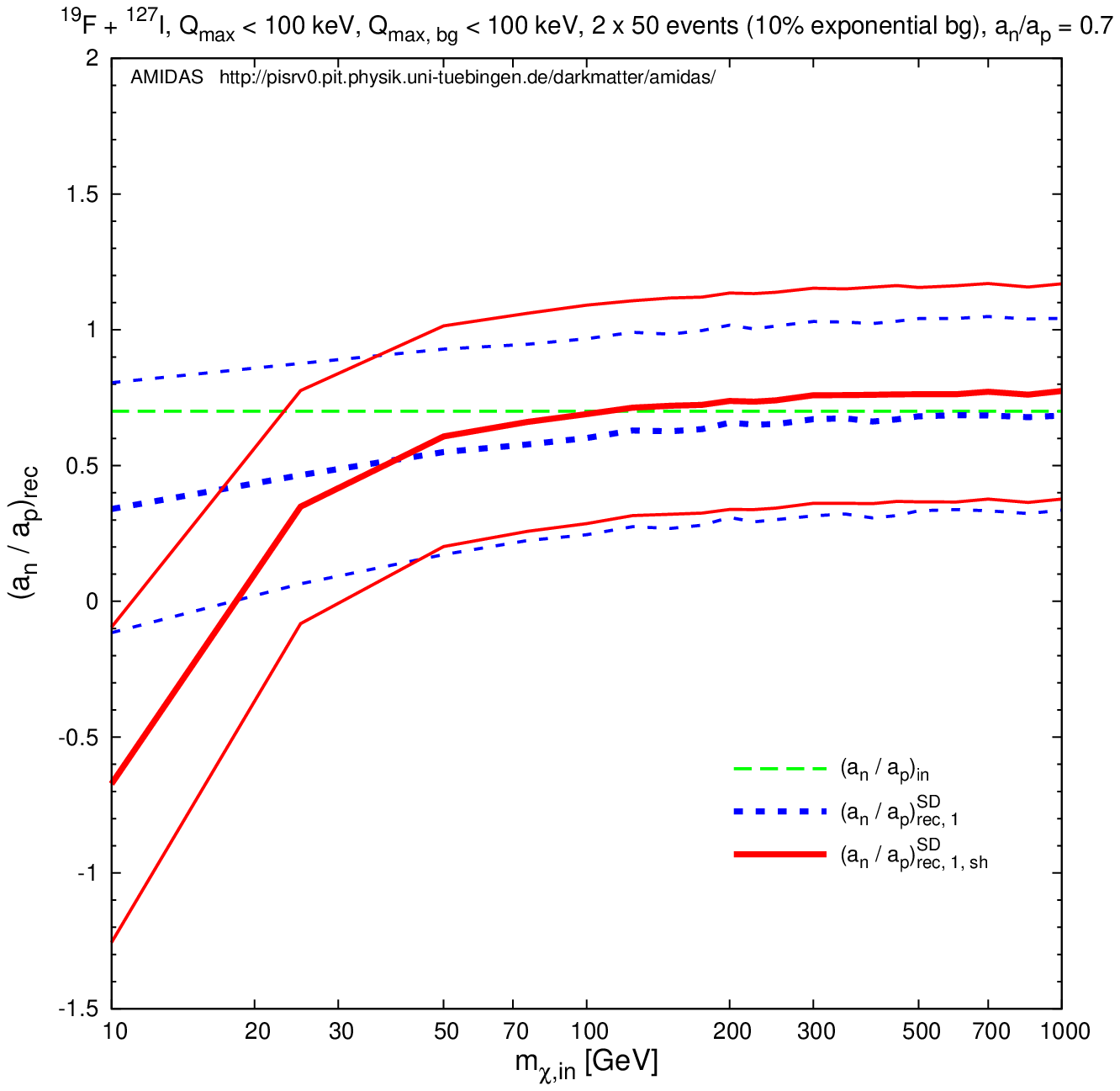} \\
\vspace{0.5cm}
\includegraphics[width=8.5cm]{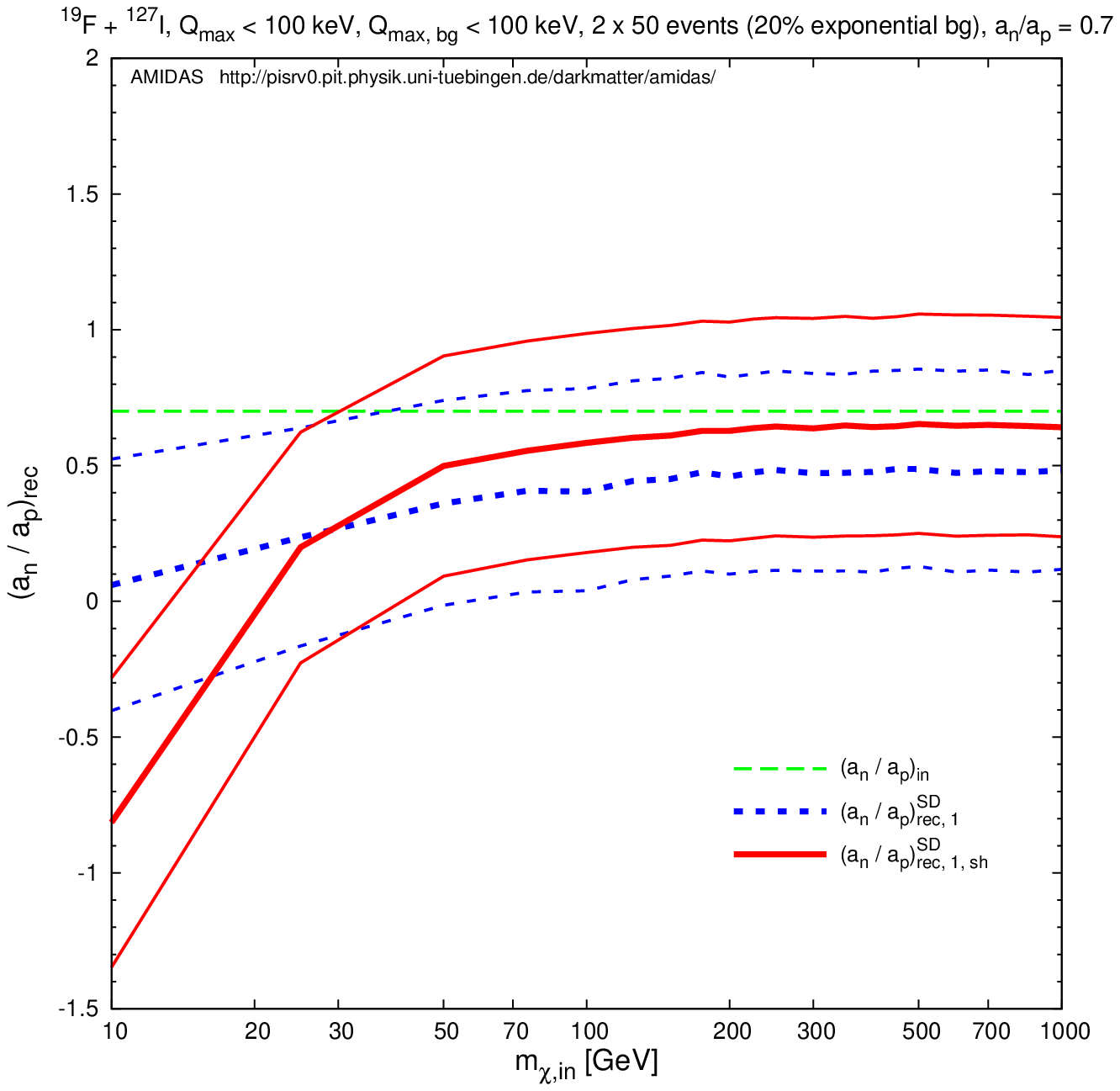}
\includegraphics[width=8.5cm]{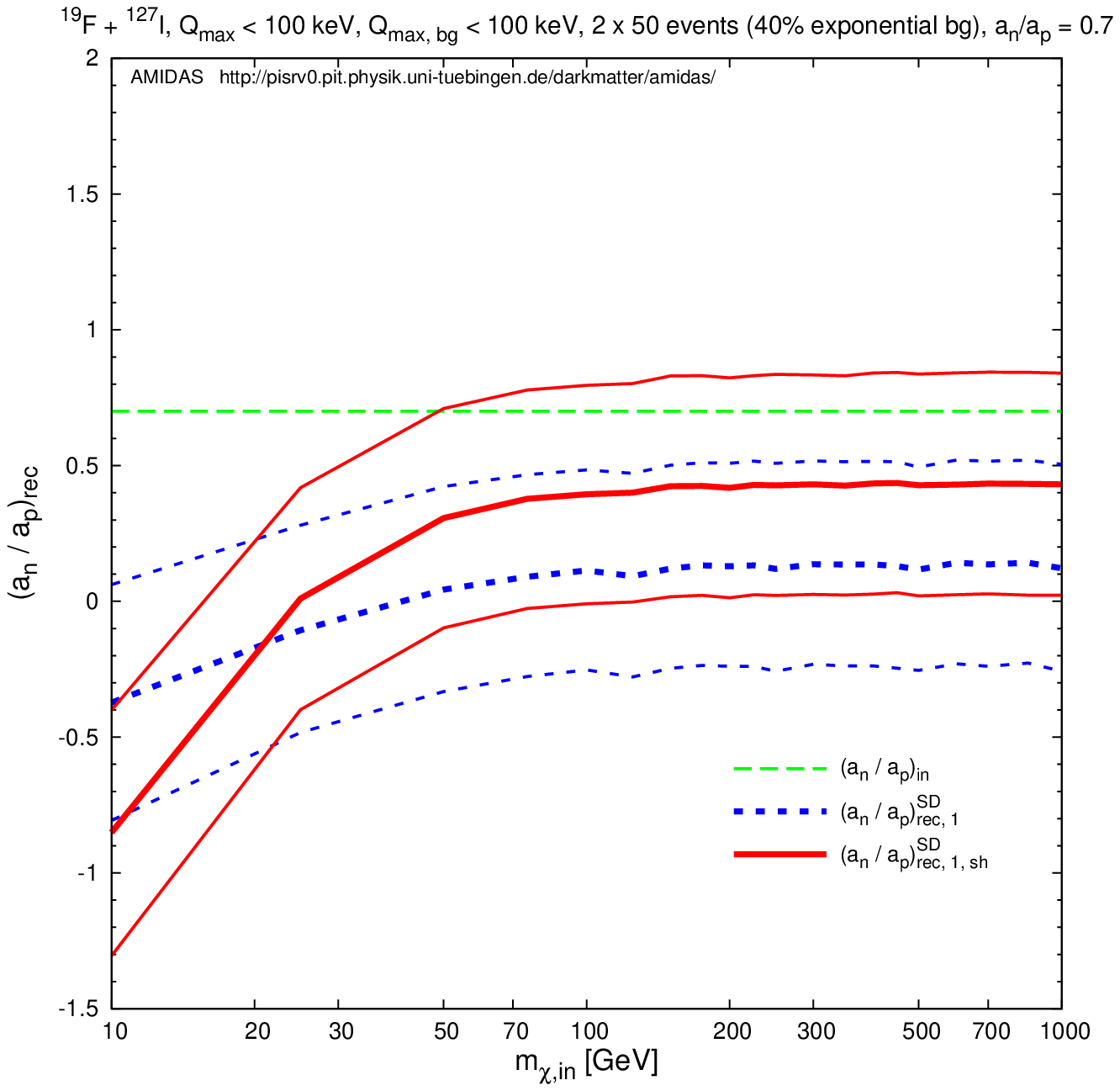} \\
\vspace{-0.25cm}
\end{center}
\caption{
 A comparison of
 the results shown in Figs.~\ref{fig:ranapSD-mchi-rec-ex}
 estimated with $r_{(X, Y)}(Q_{{\rm min}, (X, Y)} = 0)$ (dashed blue)
 and those shown in Figs.~\ref{fig:ranapSD-mchi-sh-rec-ex}
 with $r_{(X, Y)}(Q_{s, 1, (X, Y)})$ (solid red).
 Only the results estimated with $n = 1$
 are shown here.
}
\label{fig:ranapSD-mchi-ex}
\end{figure}

 As a comparison,
 I show
 the results shown in Figs.~\ref{fig:ranapSD-ranap-rec-ex}
 estimated with $r_{(X, Y)}(Q_{{\rm min}, (X, Y)} = 0)$ (dashed blue)
 and those shown in Figs.~\ref{fig:ranapSD-ranap-sh-rec-ex}
 with $r_{(X, Y)}(Q_{s, 1, (X, Y)})$ (solid red)
 together (only cases with $n = 1$)
 in Figs.~\ref{fig:ranapSD-ranap-ex}.
 It can be seen here clearly
 as well as from
 Figs.~\ref{fig:ranapSD-ranap-rec-ex}
 and \ref{fig:ranapSD-ranap-sh-rec-ex} that,
 because we set simply the experimental threshold energies
 to be {\em negligible} in our simulations,
 the statistical uncertainties on $\armn / \armp$
 estimated with $r_{(X, Y)}(Q_{s, 1, (X, Y)})$
 are a little bit larger.
 However,
 as shown in Ref.~\cite{DMDDranap},
 once threshold energies of analyzed data sets
 are {\em non--negligible},
 the statistical uncertainties on $\armn / \armp$
 estimated with $r_{(X, Y)}(Q_{s, 1, (X, Y)})$
 could be $\sim$ 16\% smaller than
 those estimated with
 $r_{(X, Y)}(Q_{{\rm min}, (X, Y)} = 5~{\rm keV})$.
 Moreover,
 as shown here as well as
 in Figs.~\ref{fig:ranapSD-ranap-rec-ex}
 and \ref{fig:ranapSD-ranap-sh-rec-ex},
 for the cases with $r_{(X, Y)}(Q_{s, 1, (X, Y)})$
 the systematic deviations
 caused by residue background events
 are much smaller.

 Quantitatively,
 Figs.~\ref{fig:ranapSD-ranap-rec-ex} to \ref{fig:ranapSD-ranap-ex}
 show that,
 with even $\sim$ 20\% -- 40\% residue background events
 in the analyzed data sets,
 one could in principle still reconstruct
 the ratio between the SD WIMP couplings on neutrons and that on protons
 pretty well;
 for a WIMP mass \mbox{$\mchi = 100$ GeV} and $\armn / \armp = 0.75$,
 by using Eq.~(\ref{eqn:ranapSD})
 with $r_{(X, Y)}(Q_{{\rm min}, (X, Y)} = 0)$
 (with $r_{(X, Y)}(Q_{s, 1, (X, Y)})$)
 and $n = 1$
 to analyze data sets of a 10\% (20\%) background ratio,
 the systematic deviation
 could be $\lsim -12\%$
 ($\lsim -14\%$)
 with an $\sim 55\%$ ($\sim 60\%$)
 statistical uncertainty%
\footnote{
 In Refs.~\cite{DMDDidentification-DARK2009, DMDDranap}
 another combination of detector materials:
 $\rmXA{Ge}{73}$ + $\rmXA{Cl}{37}$ has also been considered.
 With the same setup used here,
 expect non--negligible experimental threshold energies
 ($\Qmin = 5$ keV),
 the statistical uncertainties on
 the $\armn / \armp$ ratios reconstructed with Ge + Cl
 are only $\sim 1/5 - 1/3$ of those reconstructed with F + I.
 For detailed discussions about
 the reasons of this difference,
 see Ref.~\cite{DMDDranap}.
}.

 The expression (\ref{eqn:ranapSD}) for estimating
 the ratio between two SD WIMP--nucleon couplings
 is {\em independent} of the WIMP mass.
 However,
 as discussed in Sec.~3 and
 in Ref.~\cite{DMDDbg-mchi} in more detail,
 an exponential--like residue background spectrum
 could cause an over--/underestimate of the reconstructed WIMP mass
 for light/heavy WIMPs.
 Hence,
 in order to check the WIMP--mass independence
 of the reconstructed $\armn / \armp$ ratio
 with non--negligible background events,
 in Figs.~\ref{fig:ranapSD-mchi-rec-ex}
 and \ref{fig:ranapSD-mchi-sh-rec-ex}
 I show the reconstructed results
 with $n = -1$ (dashed blue), 1 (solid red),
 and 2 (dash--dotted cyan)
 as functions of the input WIMP mass $\mchi$.
 The input $\armn / \armp$ ratio
 has been set as 0.7.

 It can be seen here that,
 for WIMP masses $\mchi~\gsim$ 50 GeV,
 except the statistical uncertainty estimated
 with $r_{(X, Y)}(Q_{{\rm min}, (X, Y)})$ and $n = -1$
 (dashed blue curves
  labeled as $(\armn / \armp)_{\rm rec,~-1}^{\rm SD}$
  in Figs.~\ref{fig:ranapSD-mchi-rec-ex}),
 the reconstructed $\armn / \armp$ ratios
 as well as their statistical uncertainties
 are indeed (almost) independent of the WIMP mass.
 However,
 as discussed above,
 the larger the background ratios
 in our data sets,
 the more strongly {\em underestimated}
 the reconstructed $\armn / \armp$ ratios
 for all input WIMP masses.
 And,
 as shown in Figs.~\ref{fig:ranapSD-mchi-ex},
 the $\armn / \armp$ ratios reconstructed
 with $r_{(X, Y)}(Q_{{\rm min}, (X, Y)})$ (dashed blue)
 would be more strongly underestimated as
 those reconstructed
 with $r_{(X, Y)}(Q_{s, 1, (X, Y)})$ (solid red).
 Nevertheless,
 with data sets of $\lsim~20\%$ residue background events,
 the reconstructed 1$\sigma$ statistical uncertainty intervals
 could in principle always cover
 the input (true) $\armn / \armp$ ratios
 pretty well.

 On the other hand,
 for WIMP masses $\lsim~50$ GeV,
 interestingly,
 Figs.~\ref{fig:ranapSD-mchi-rec-ex} show that
 the $\armn / \armp$ ratio reconstructed
 with $r_{(X, Y)}(Q_{{\rm min}, (X, Y)})$
 and $n = -1$
 becomes the best result%
\footnote{
 Remind that
 this conclusion holds only
 when the experimental threshold energies of the analyzed data sets
 can be {\em negligible}.
 Otherwise,
 as shown in Ref.~\cite{DMDDranap},
 the $\armn / \armp$ ratio reconstructed
 with $r_{(X, Y)}(Q_{{\rm min}, (X, Y)})$ and $n = -1$
 would also be strongly underestimated.
\label{footnote:rmin}
}:
 with an $\lsim~20\%$ background ratio
 the 1$\sigma$ statistical uncertainty interval
 could still cover
 the input (true) $\armn / \armp$ ratios well.
 In contrast,
 all other five reconstructed $\armn / \armp$ ratios
 shown in Figs.~\ref{fig:ranapSD-mchi-rec-ex}
 and \ref{fig:ranapSD-mchi-sh-rec-ex}
 are (strongly) {\em underestimated}.

\begin{figure}[t!]
\begin{center}
\includegraphics[width=8.5cm]{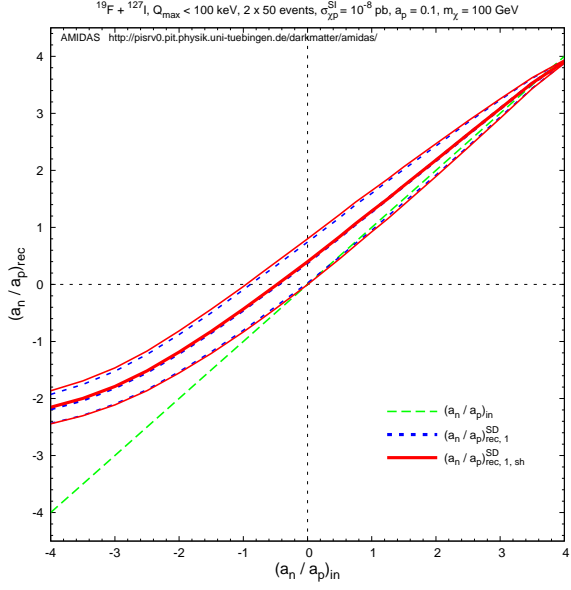}
\includegraphics[width=8.5cm]{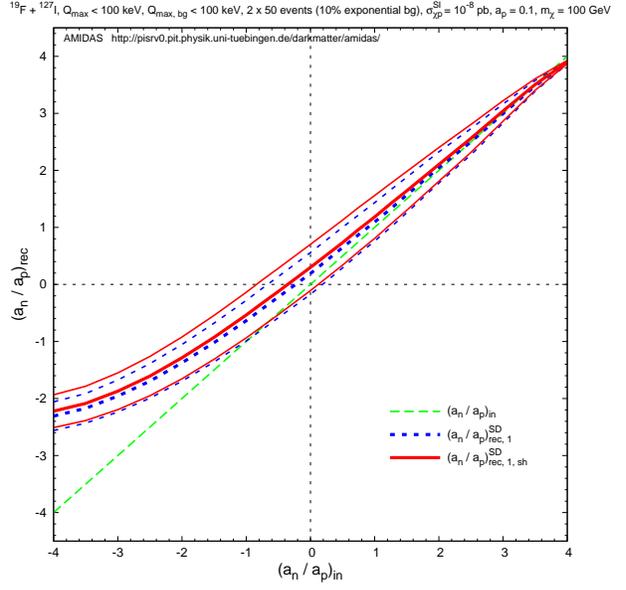} \\
\vspace{0.5cm}
\includegraphics[width=8.5cm]{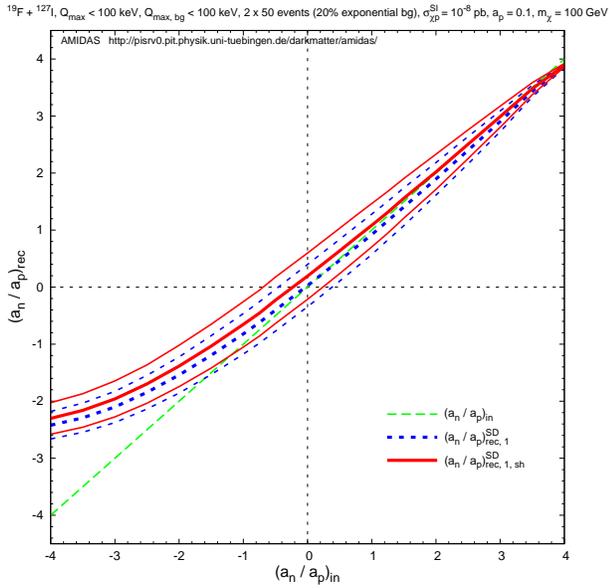}
\includegraphics[width=8.5cm]{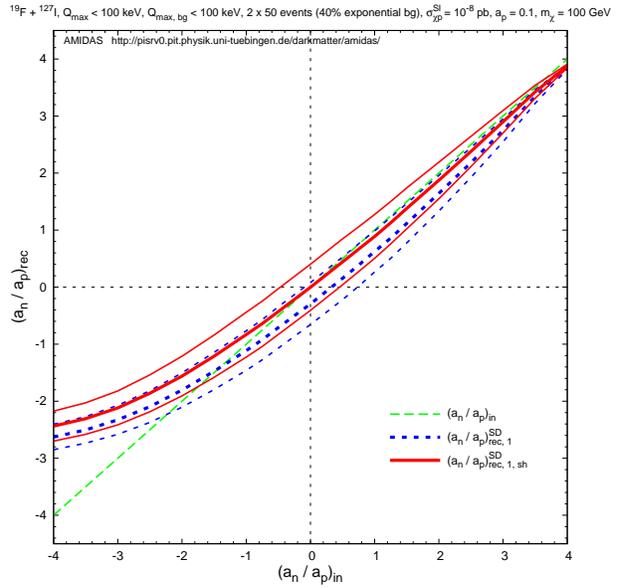} \\
\vspace{-0.25cm}
\end{center}
\caption{
 As in Figs.~\ref{fig:ranapSD-ranap-ex},
 except that
 the SI WIMP--nucleon cross section
 and the SD WIMP--proton coupling
 have been set as \mbox{$\sigmapSI = 10^{-8}$ pb} and $\armp = 0.1$,
 respectively.
}
\label{fig:ranapSD-08-ranap-ex}
\end{figure}
\begin{figure}[t!]
\begin{center}
\includegraphics[width=8.5cm]{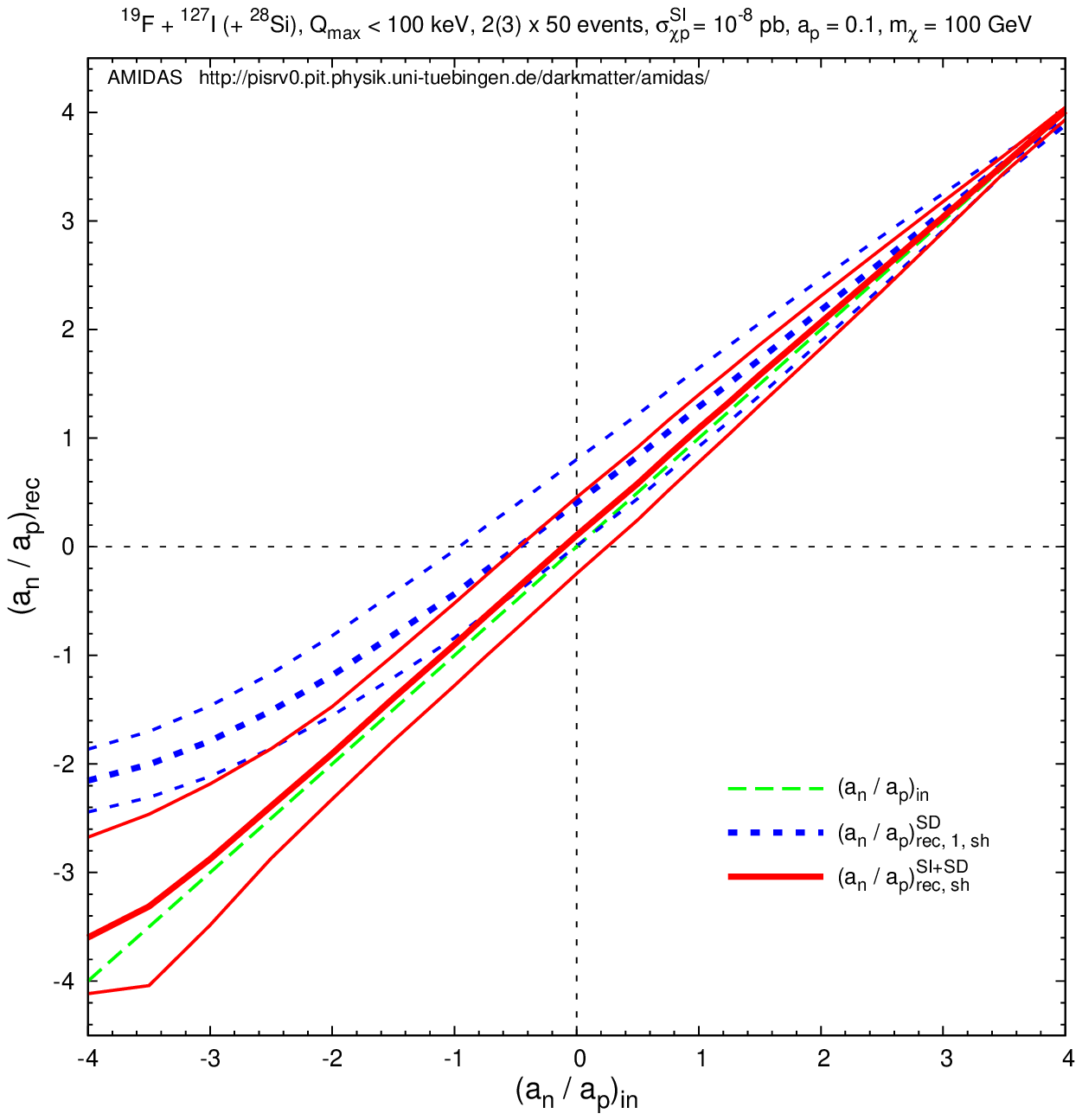}
\includegraphics[width=8.5cm]{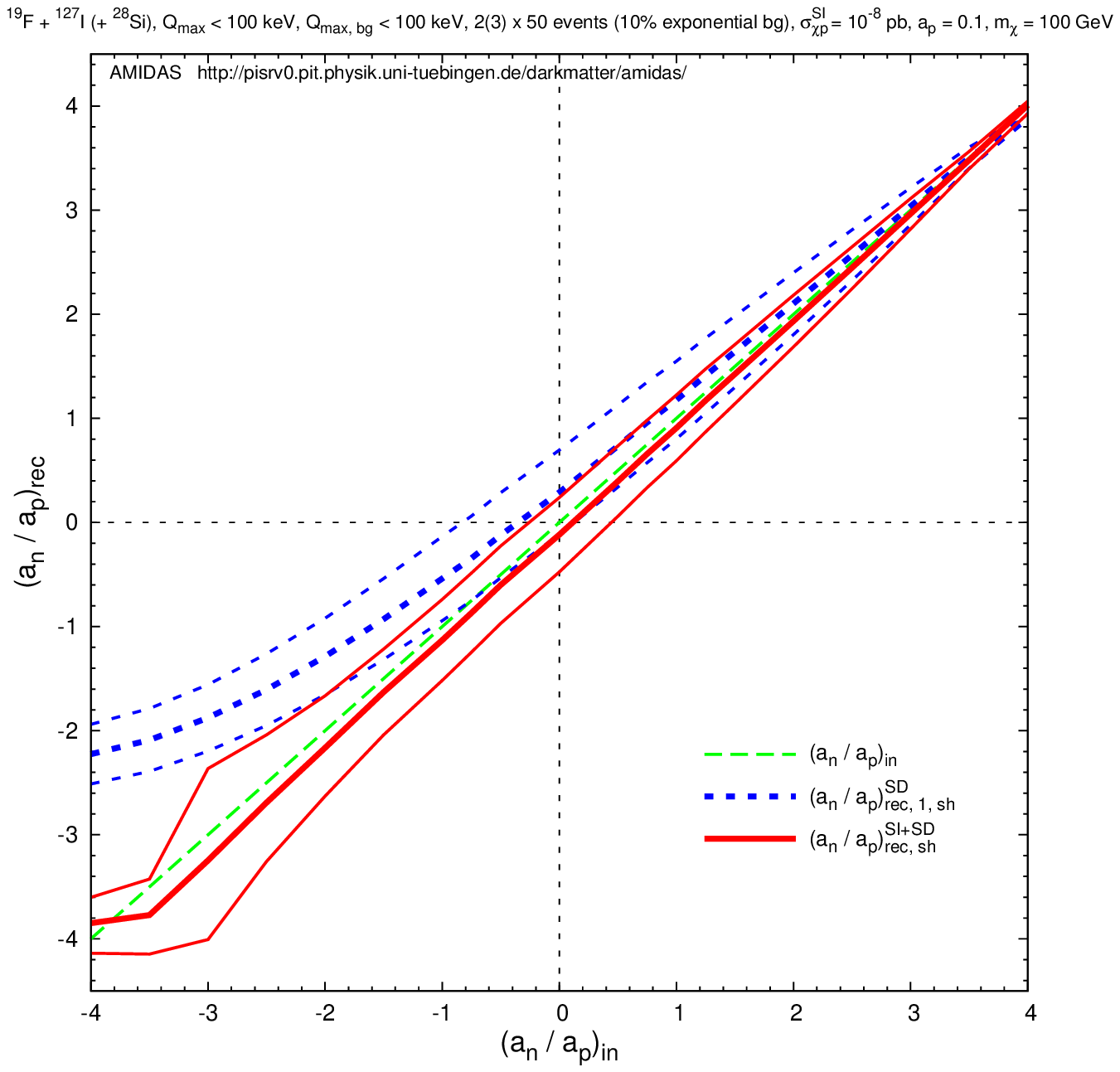} \\
\vspace{0.5cm}
\includegraphics[width=8.5cm]{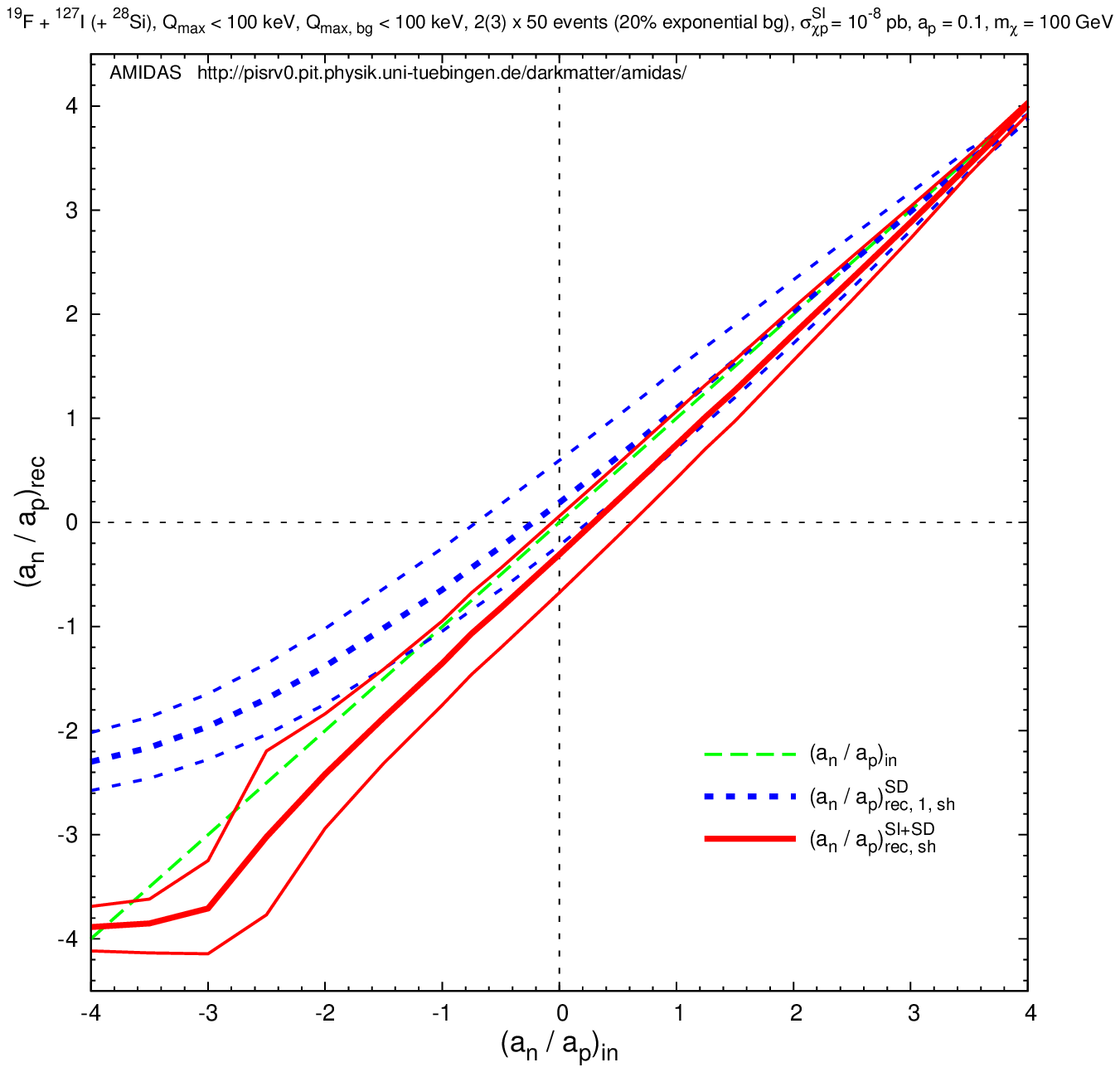}
\includegraphics[width=8.5cm]{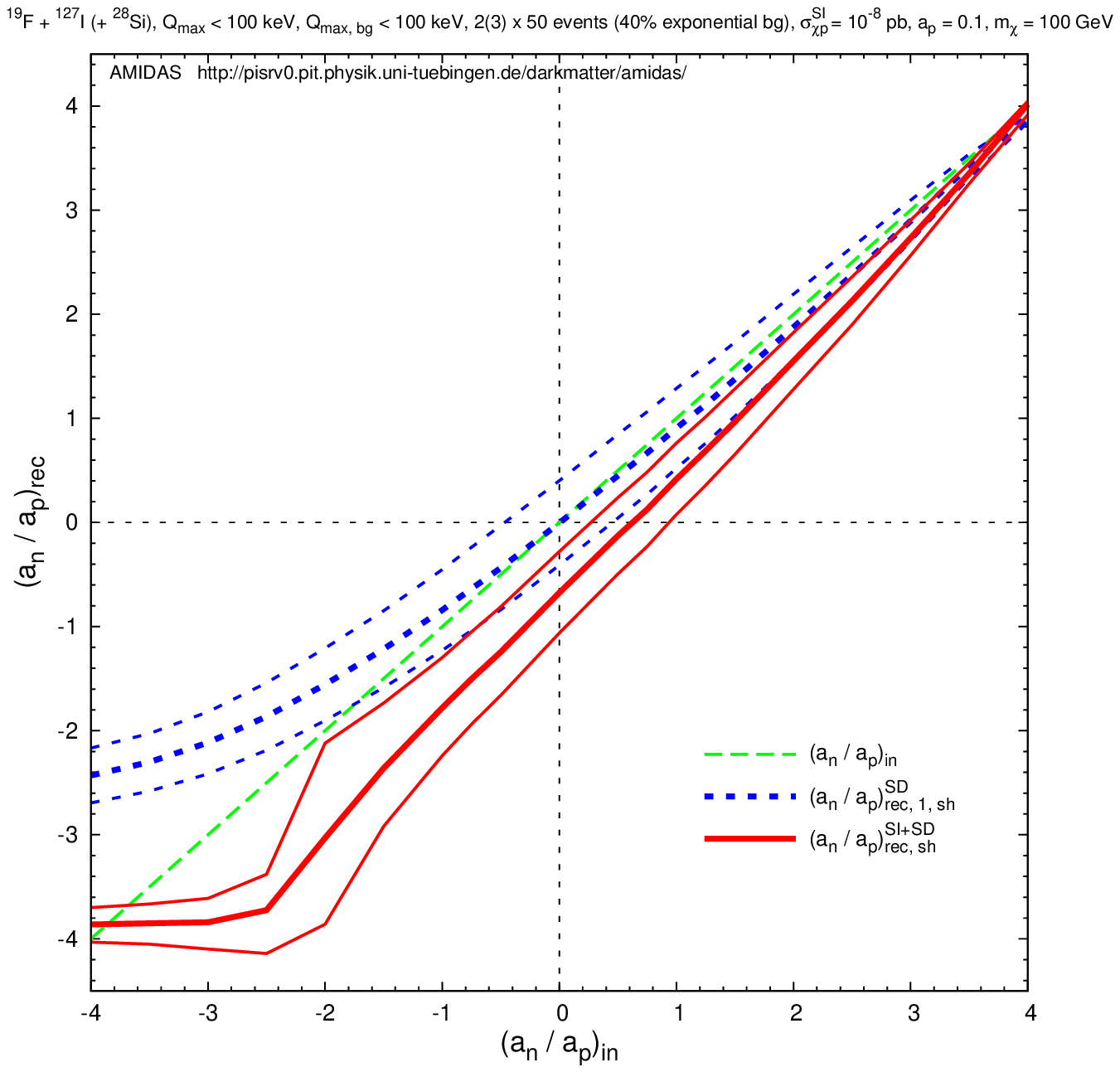} \\
\vspace{-0.25cm}
\end{center}
\caption{
 The reconstructed $\armn / \armp$ ratios
 estimated by Eqs.~(\ref{eqn:ranapSD}) (dashed blue, $n = 1$)
 and (\ref{eqn:ranapSISD}) (solid red)
 and the lower and upper bounds of
 their 1$\sigma$ statistical uncertainties
 estimated by Eqs.~(\ref{eqn:sigma_ranapSD})
 and (\ref{eqn:sigma_ranapSISD})
 with $r_{(X, Y, Z)}(Q_{s, 1, (X, Y, Z)})$
 as functions of the input $\armn / \armp$ ratio.
 Besides $\rmXA{F}{19}$ and $\rmXA{I}{127}$,
 $\rmXA{Si}{28}$ has been chosen as the third target
 for estimating $c_{{\rm p}, (X, Y)}$ by
 Eqs.~(\ref{eqn:cpX}) and (\ref{eqn:cpY}).
 The other parameters are as
 in Figs.~\ref{fig:ranapSD-08-ranap-ex}.
}
\label{fig:ranapSISD-08-ranap-sh-rec-ex}
\end{figure}
\begin{figure}[t!]
\begin{center}
\includegraphics[width=8.5cm]{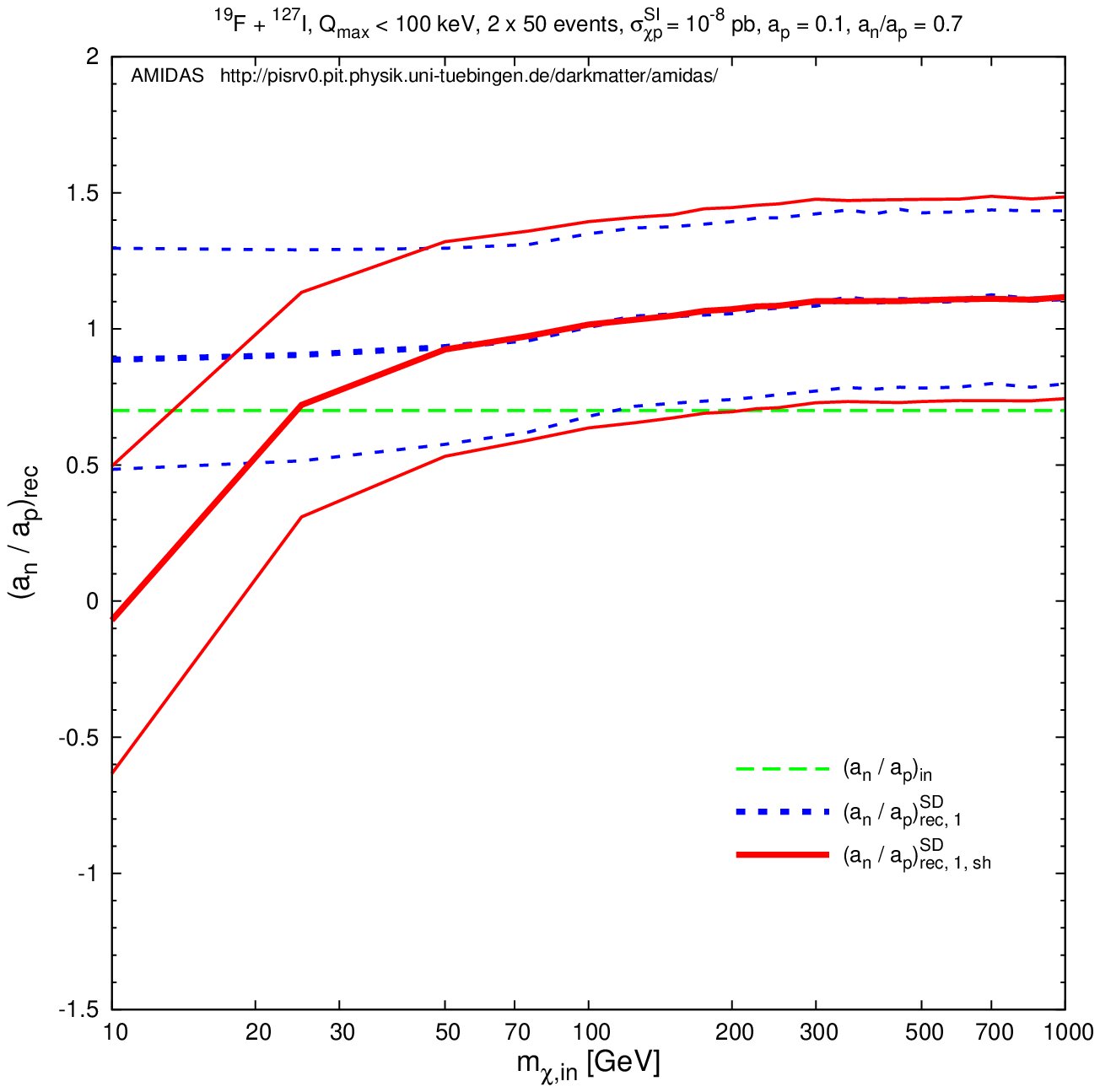}
\includegraphics[width=8.5cm]{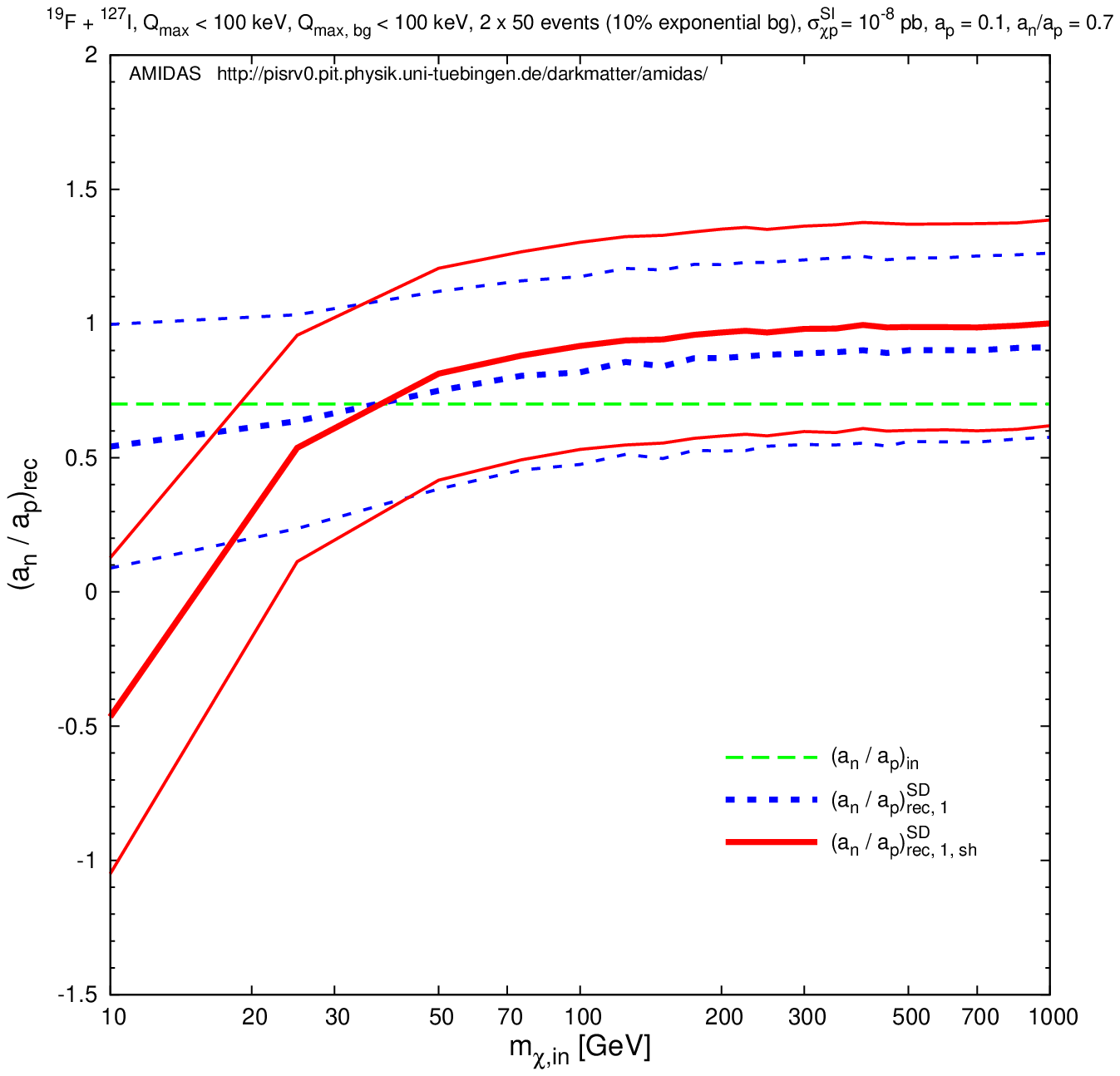} \\
\vspace{0.5cm}
\includegraphics[width=8.5cm]{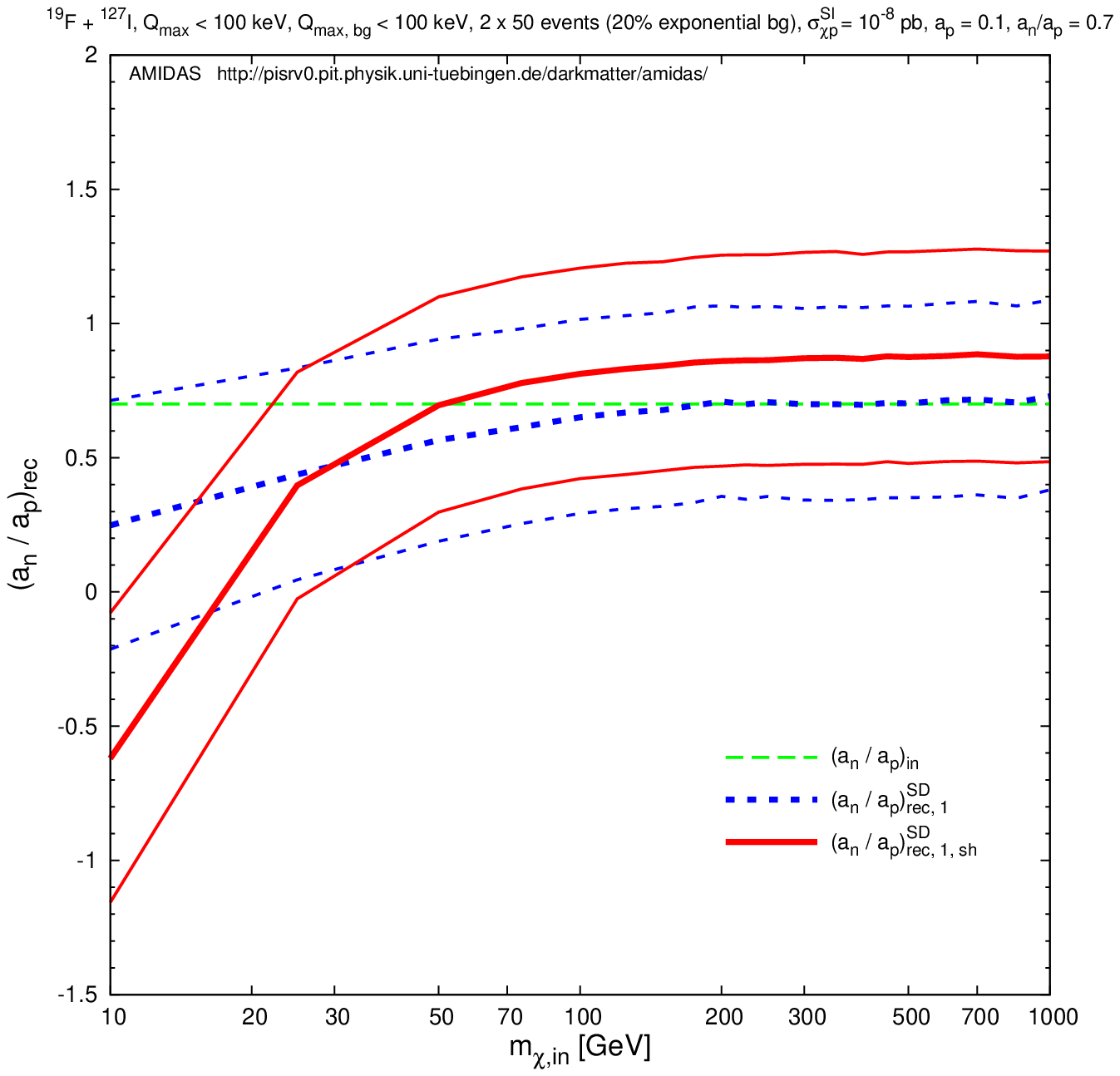}
\includegraphics[width=8.5cm]{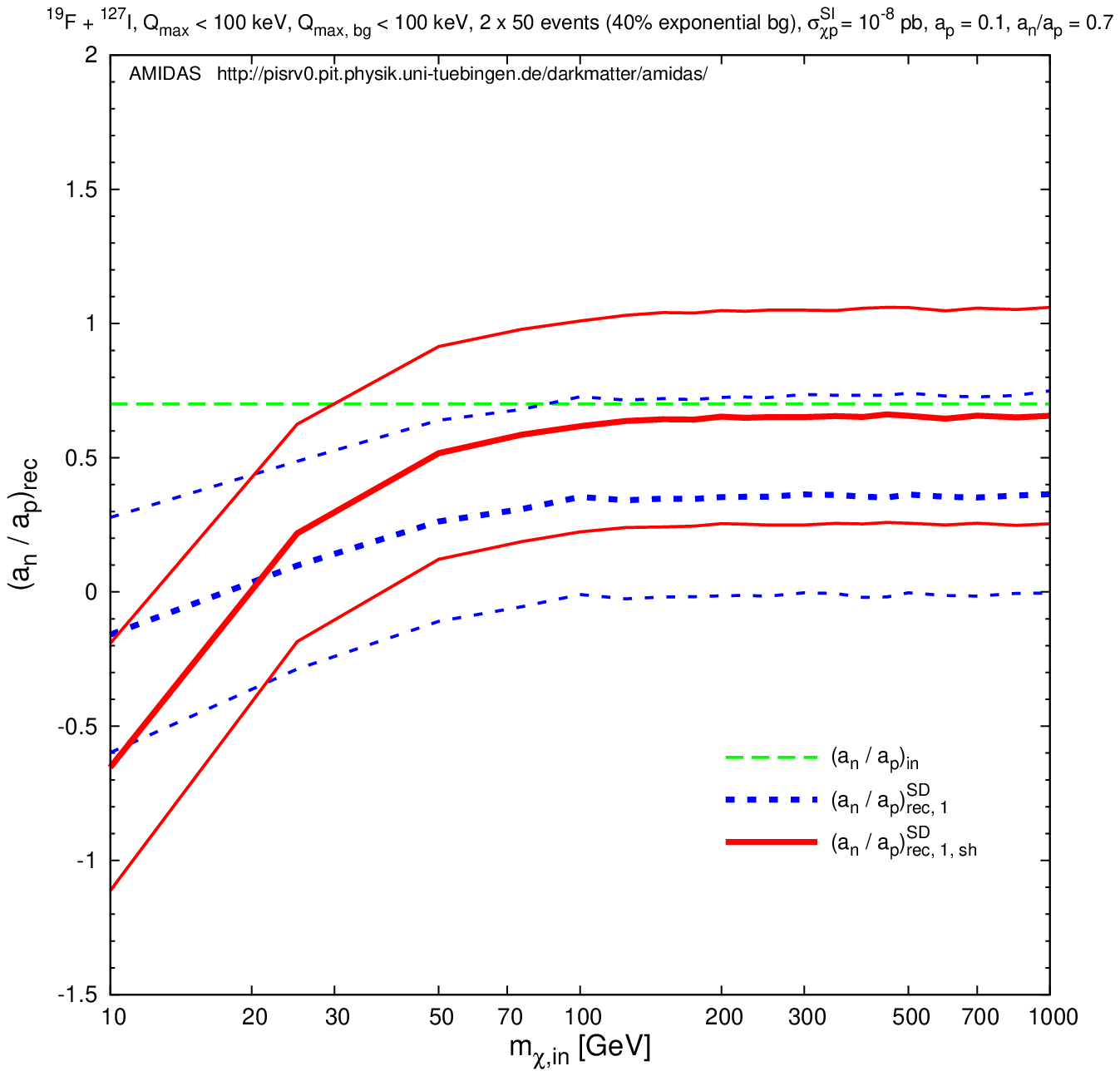} \\
\vspace{-0.25cm}
\end{center}
\caption{
 As in Figs.~\ref{fig:ranapSD-mchi-ex},
 except that
 the SI WIMP--nucleon cross section
 and the SD WIMP--proton coupling
 have been set as \mbox{$\sigmapSI = 10^{-8}$ pb} and $\armp = 0.1$,
 respectively.
}
\label{fig:ranapSD-08-mchi-ex}
\end{figure}
\begin{figure}[t!]
\begin{center}
\includegraphics[width=8.5cm]{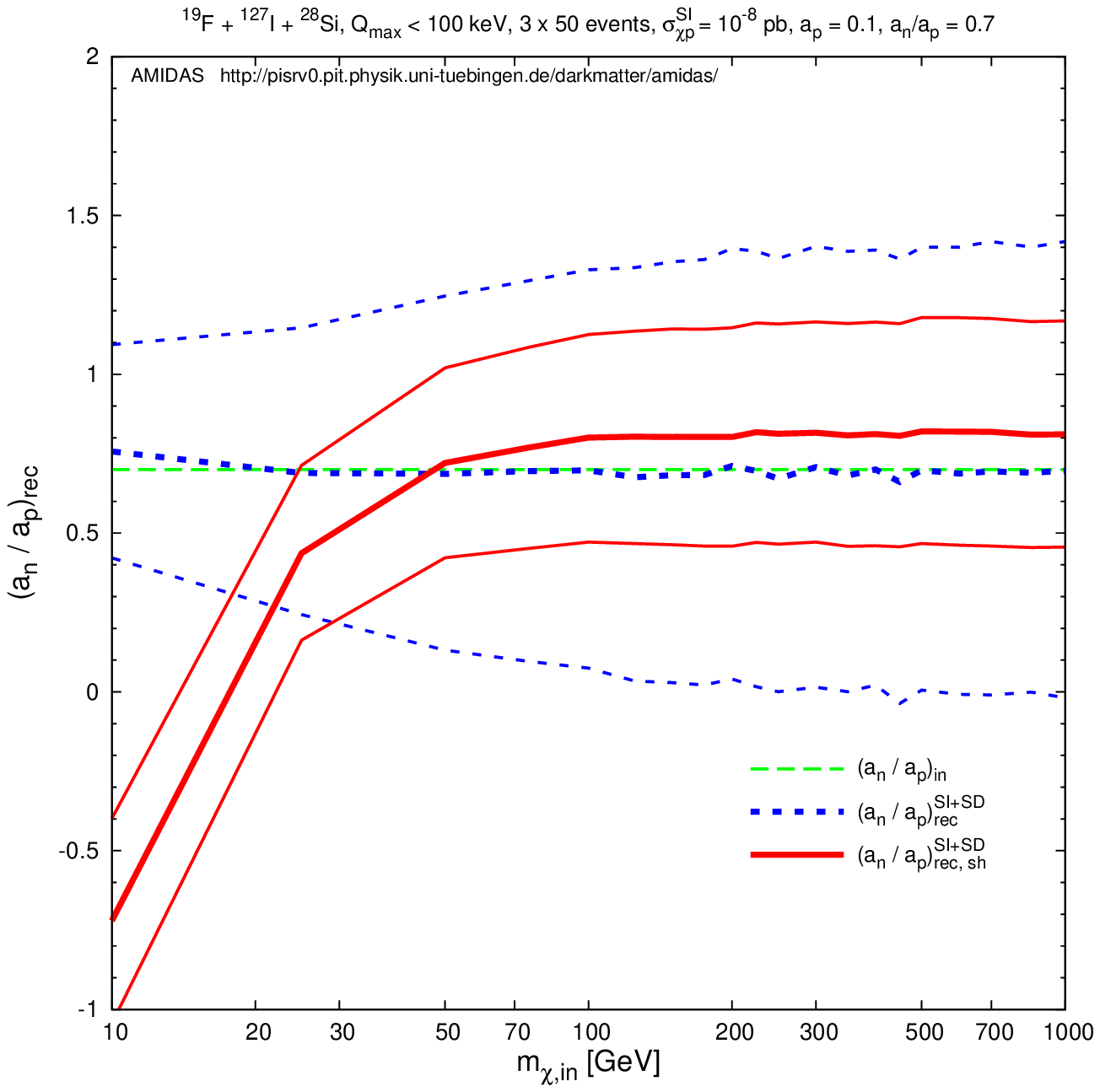}
\includegraphics[width=8.5cm]{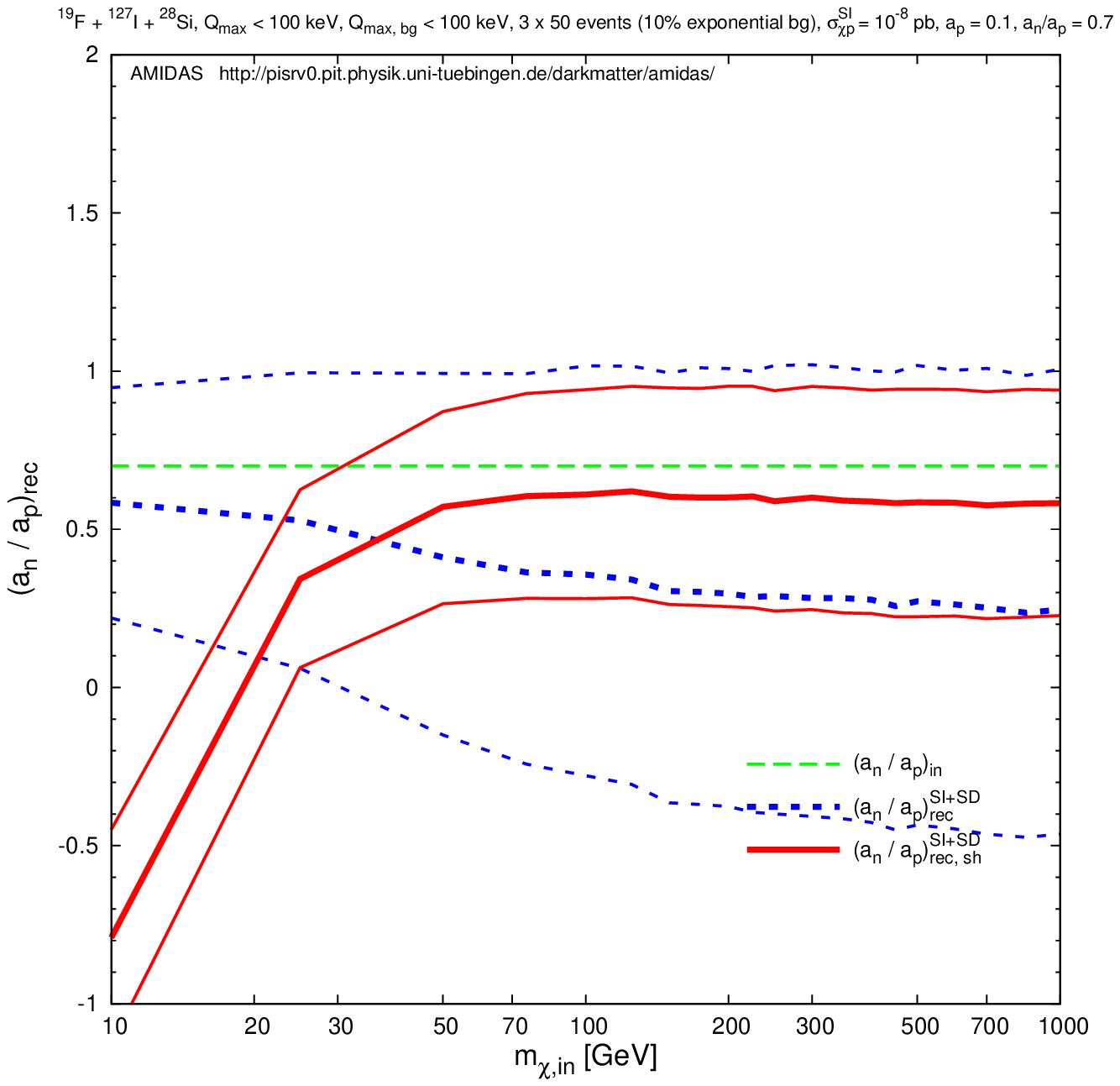} \\
\vspace{0.5cm}
\includegraphics[width=8.5cm]{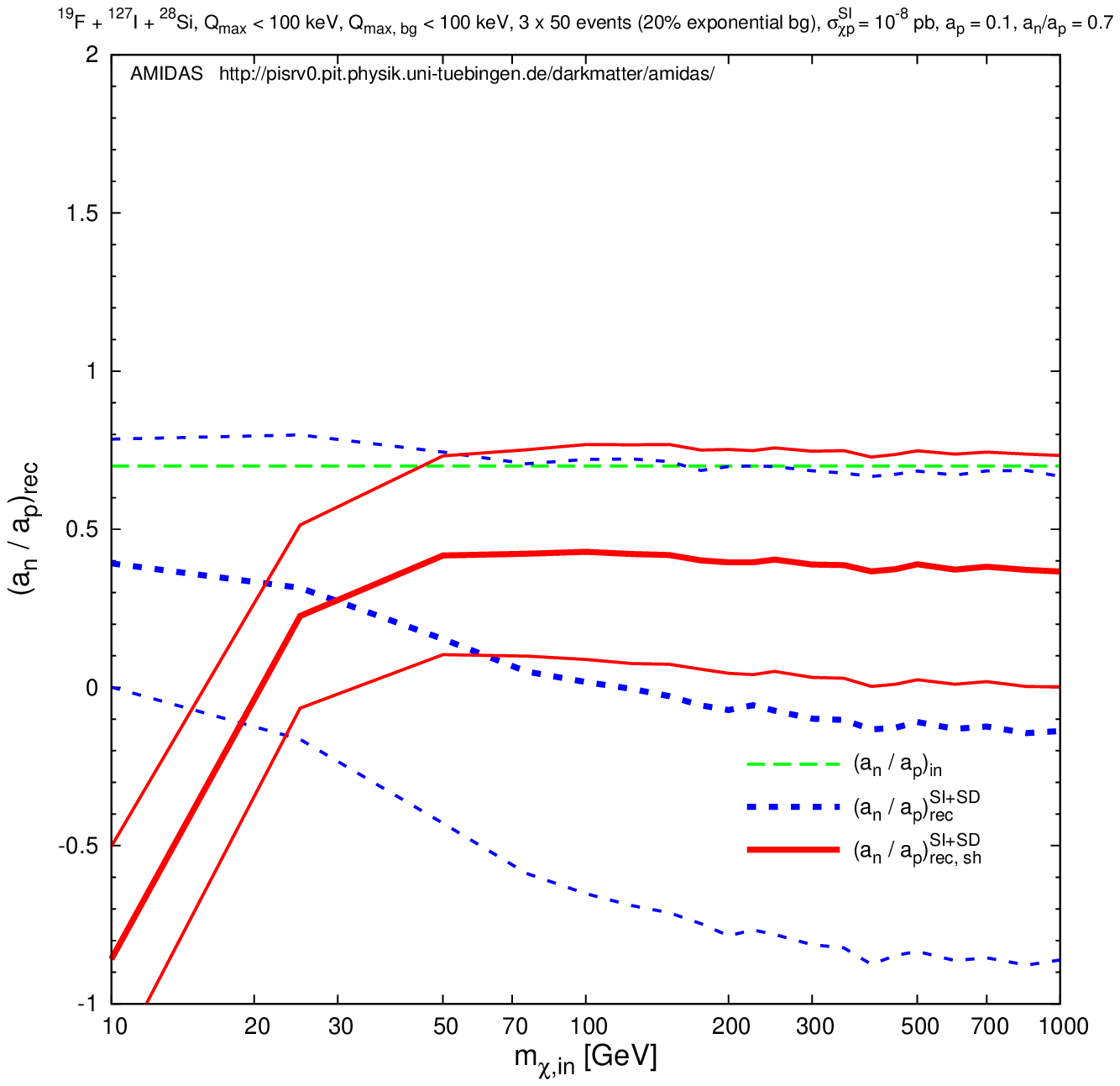}
\includegraphics[width=8.5cm]{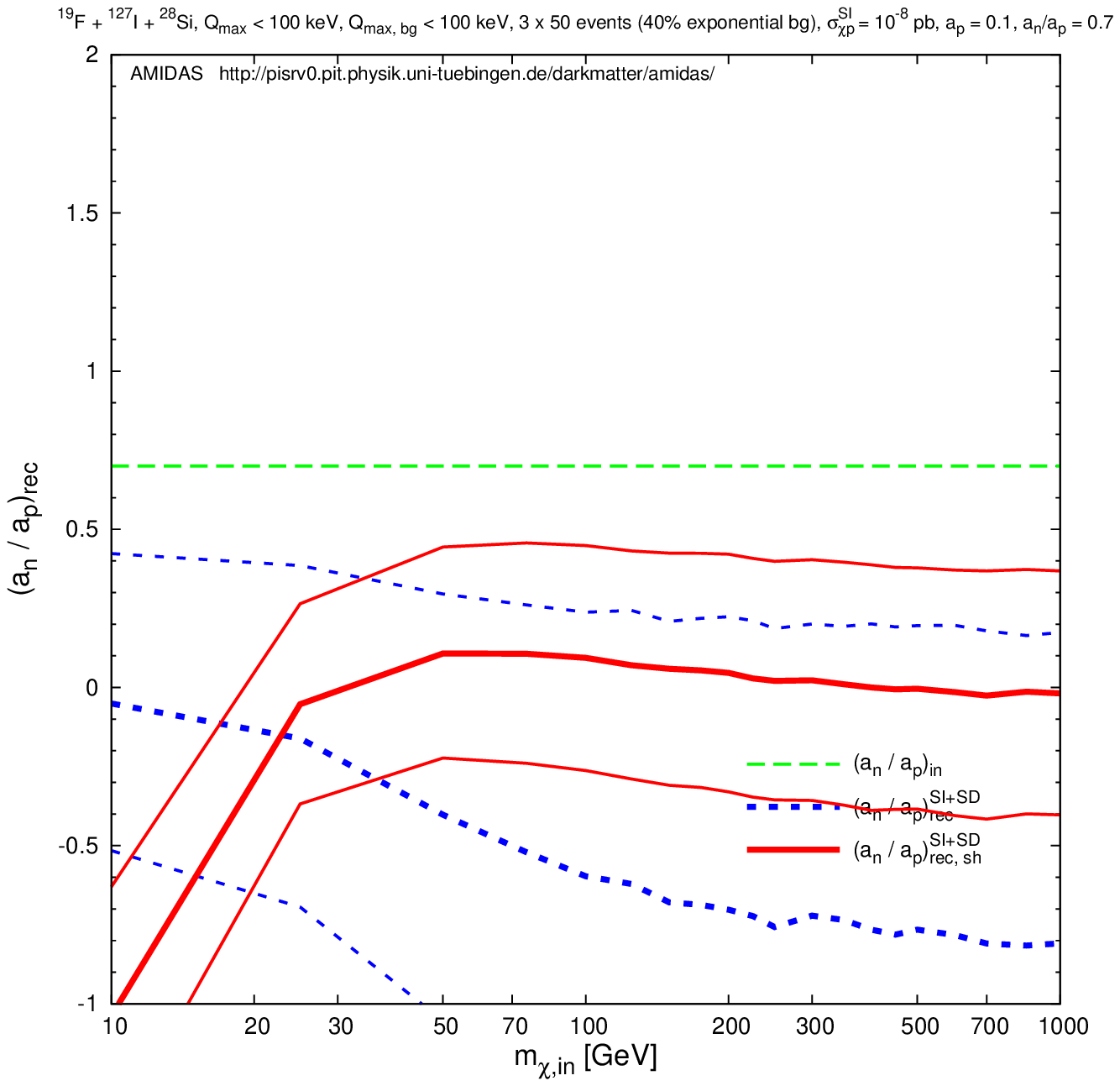} \\
\vspace{-0.25cm}
\end{center}
\caption{
 The reconstructed $\armn / \armp$ ratios
 estimated by Eq.~(\ref{eqn:ranapSISD})
 and the lower and upper bounds of
 their 1$\sigma$ statistical uncertainties
 estimated by Eq.~(\ref{eqn:sigma_ranapSISD})
 with $r_{(X, Y, Z)}(Q_{{\rm min}, (X, Y, Z)})$ (dashed blue)
 and with $r_{(X, Y, Z)}(Q_{s, 1, (X, Y, Z)})$ (solid red)
 as functions of the input WIMP mass $\mchi$.
 Besides $\rmXA{F}{19}$ and $\rmXA{I}{127}$,
 $\rmXA{Si}{28}$ has been chosen as the third target
 for estimating $c_{{\rm p}, (X, Y)}$ by
 Eqs.~(\ref{eqn:cpX}) and (\ref{eqn:cpY}).
 The other parameters are as
 in Figs.~\ref{fig:ranapSD-08-mchi-ex}.
}
\label{fig:ranapSISD-08-mchi-ex}
\end{figure}
\begin{figure}[t!]
\begin{center}
\includegraphics[width=8.5cm]{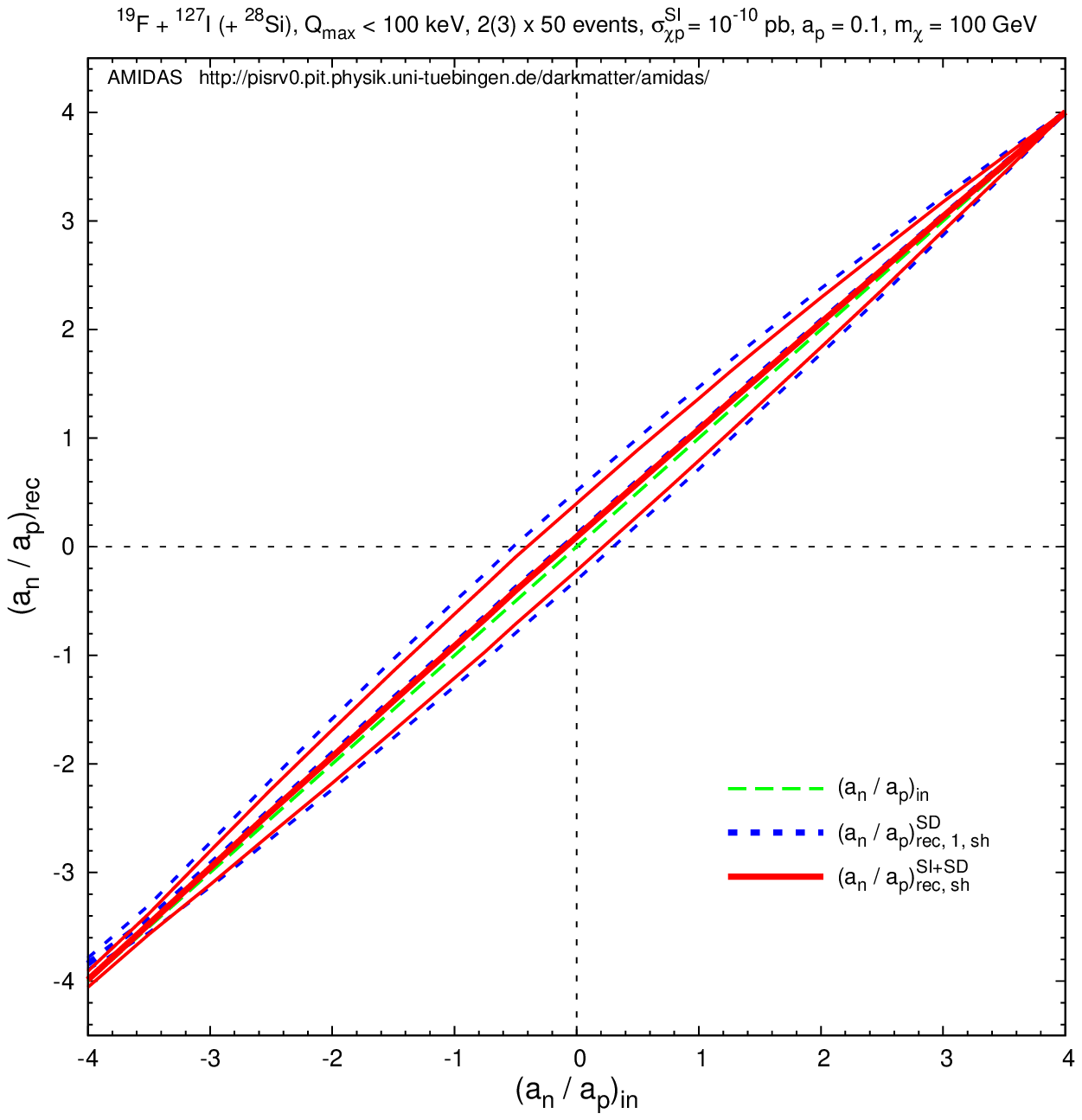}
\includegraphics[width=8.5cm]{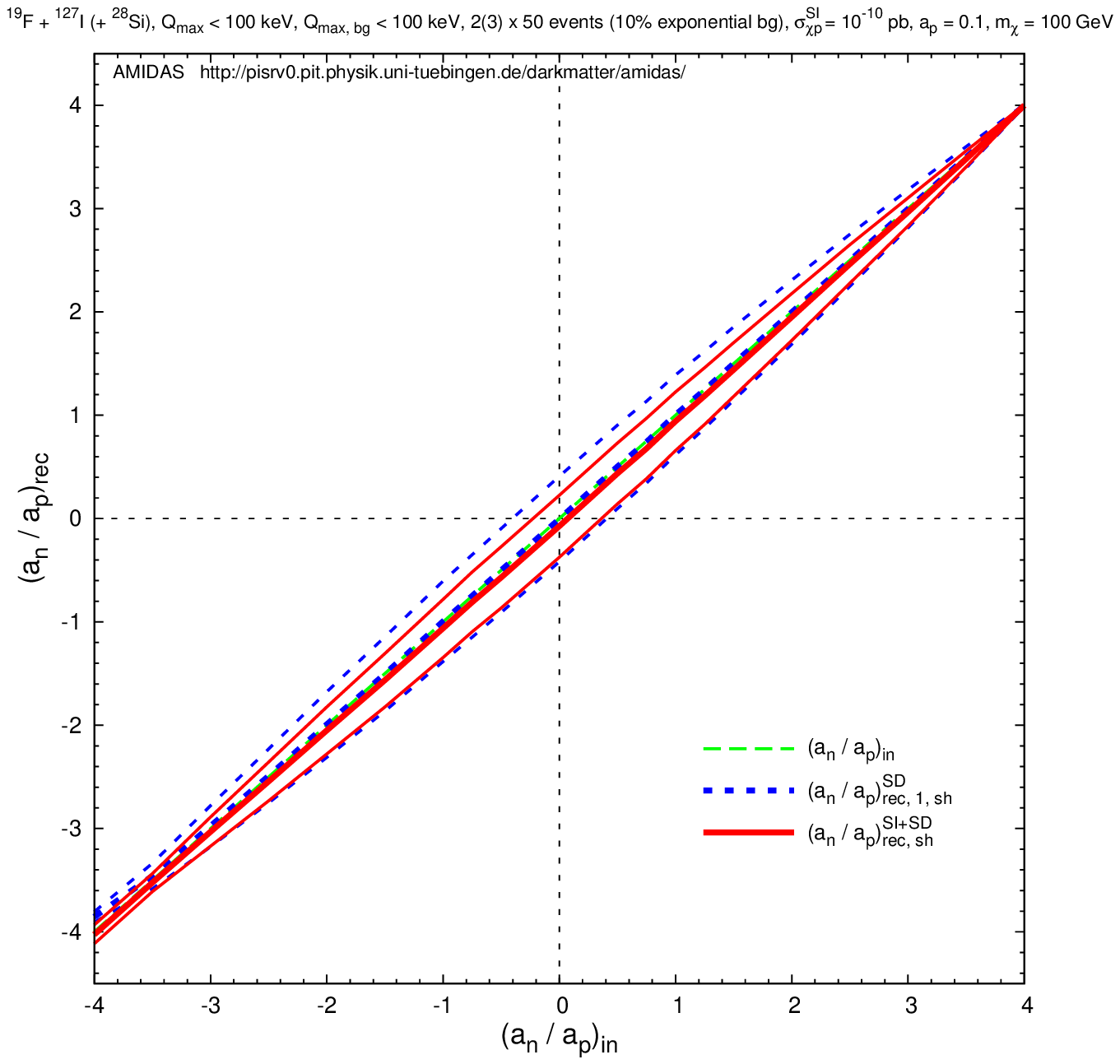} \\
\vspace{0.5cm}
\includegraphics[width=8.5cm]{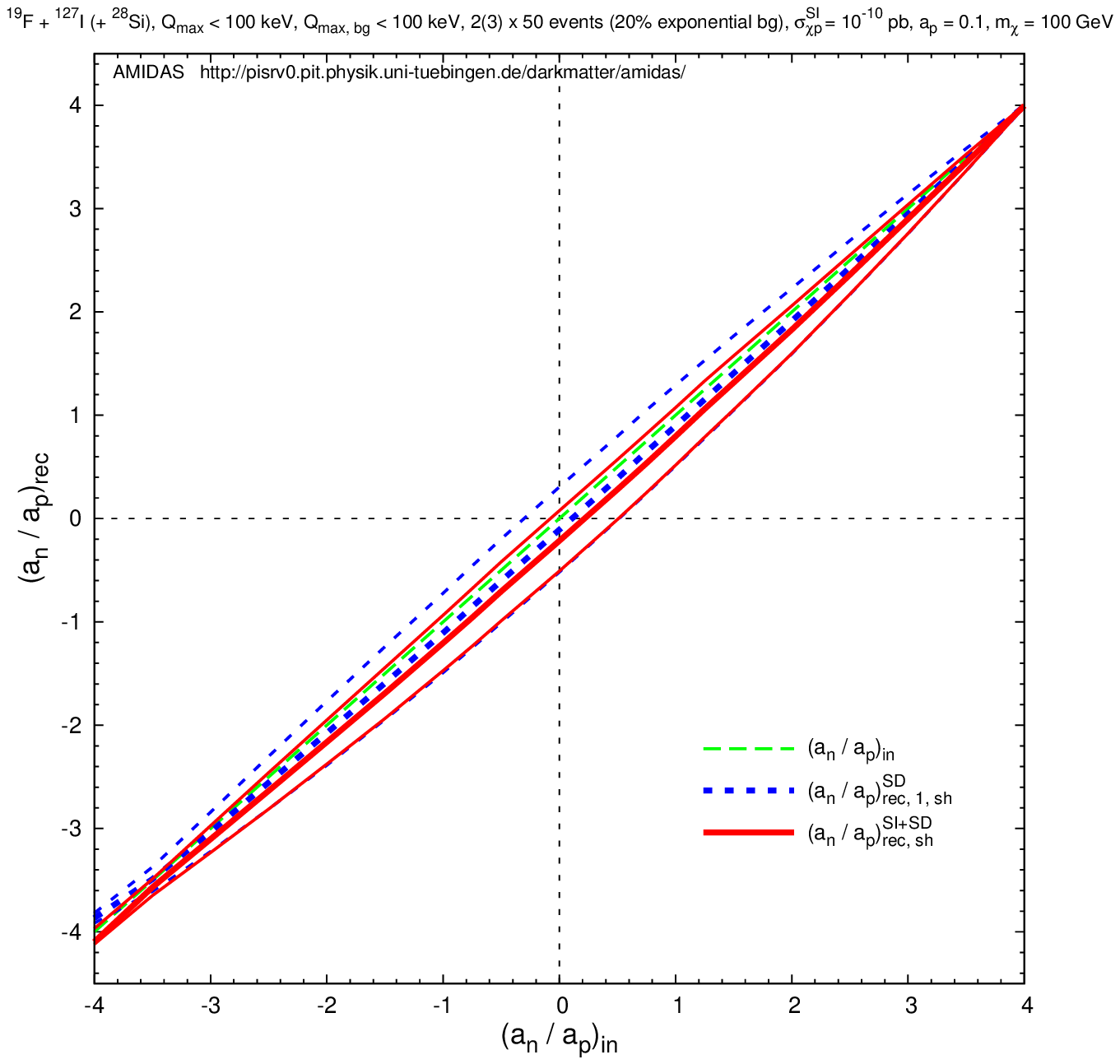}
\includegraphics[width=8.5cm]{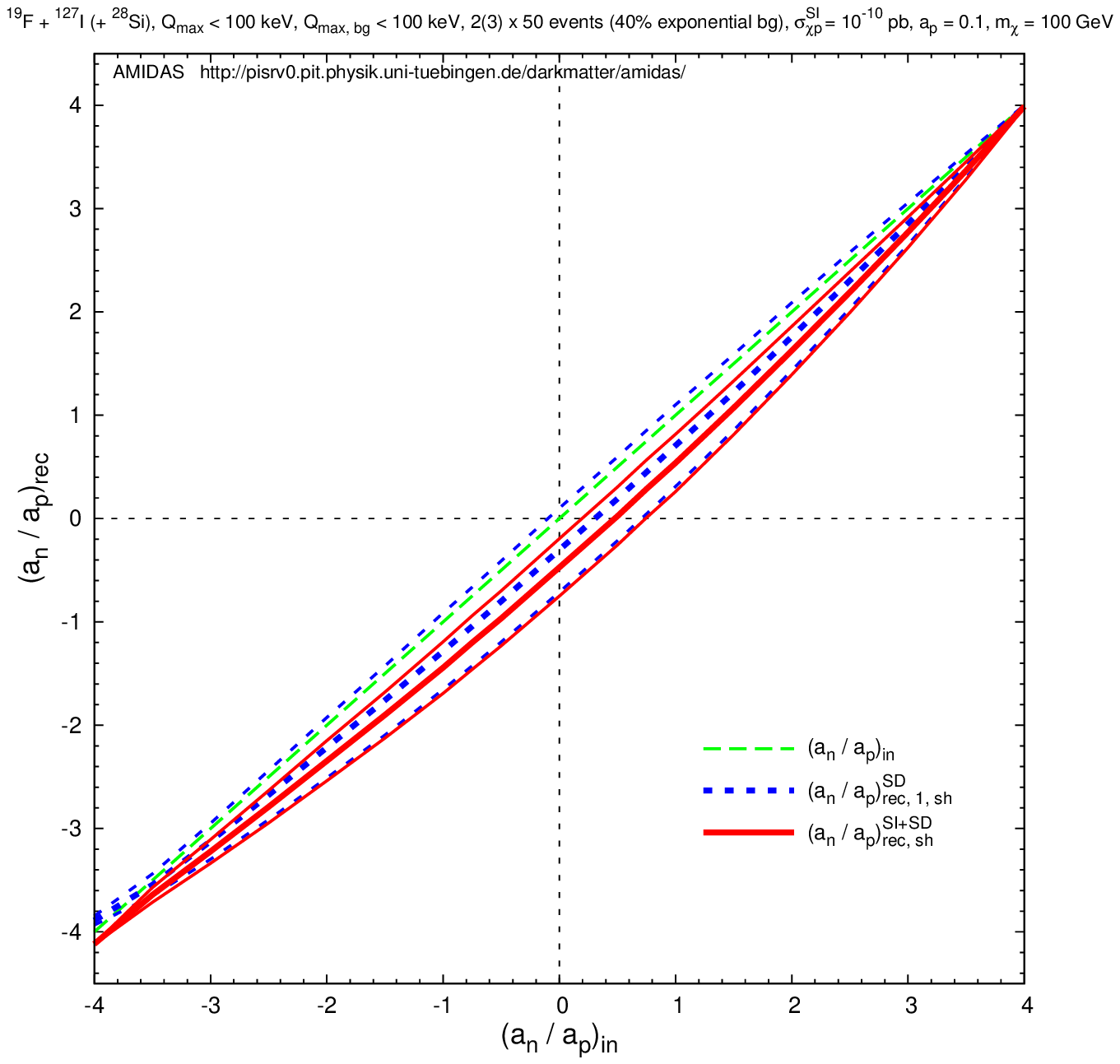} \\
\vspace{-0.25cm}
\end{center}
\caption{
 As in Figs.~\ref{fig:ranapSISD-08-ranap-sh-rec-ex},
 except that
 the SI WIMP--nucleon cross section
 has been set as \mbox{$\sigmapSI = 10^{-10}$ pb}.
}
\label{fig:ranapSISD-10-ranap-sh-rec-ex}
\end{figure}
\begin{figure}[t!]
\begin{center}
\includegraphics[width=8.5cm]{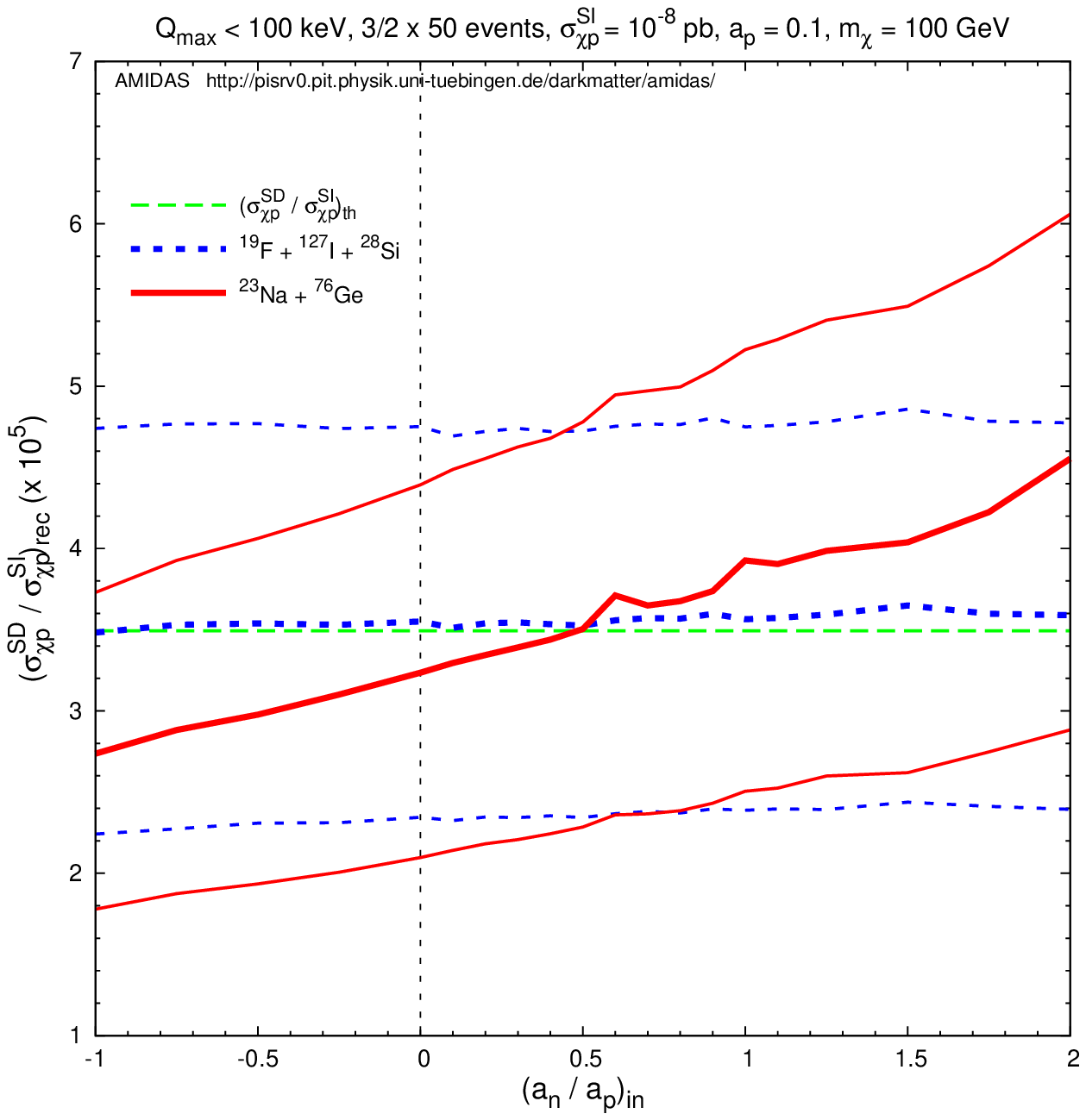}
\includegraphics[width=8.5cm]{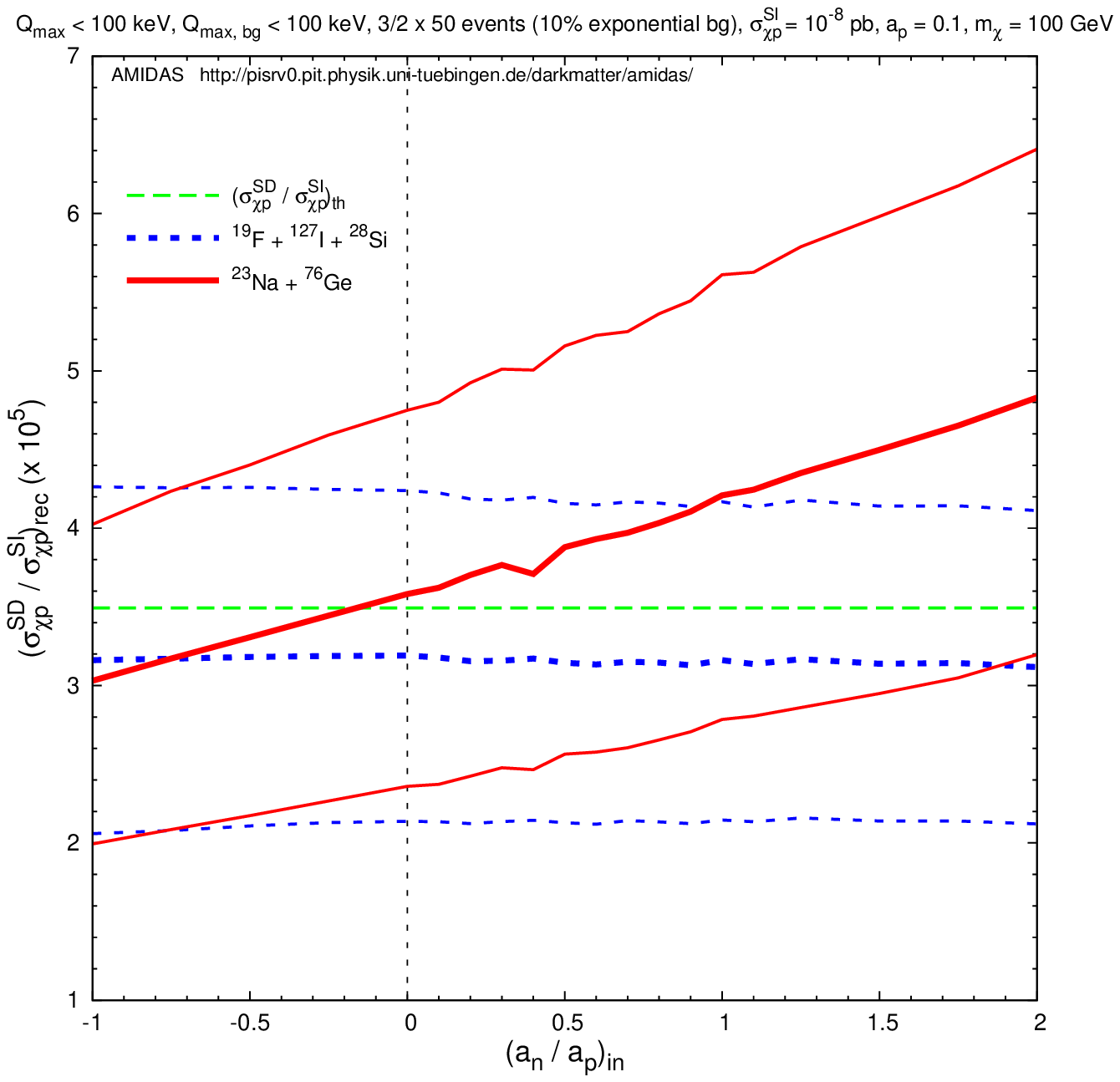} \\
\vspace{0.5cm}
\includegraphics[width=8.5cm]{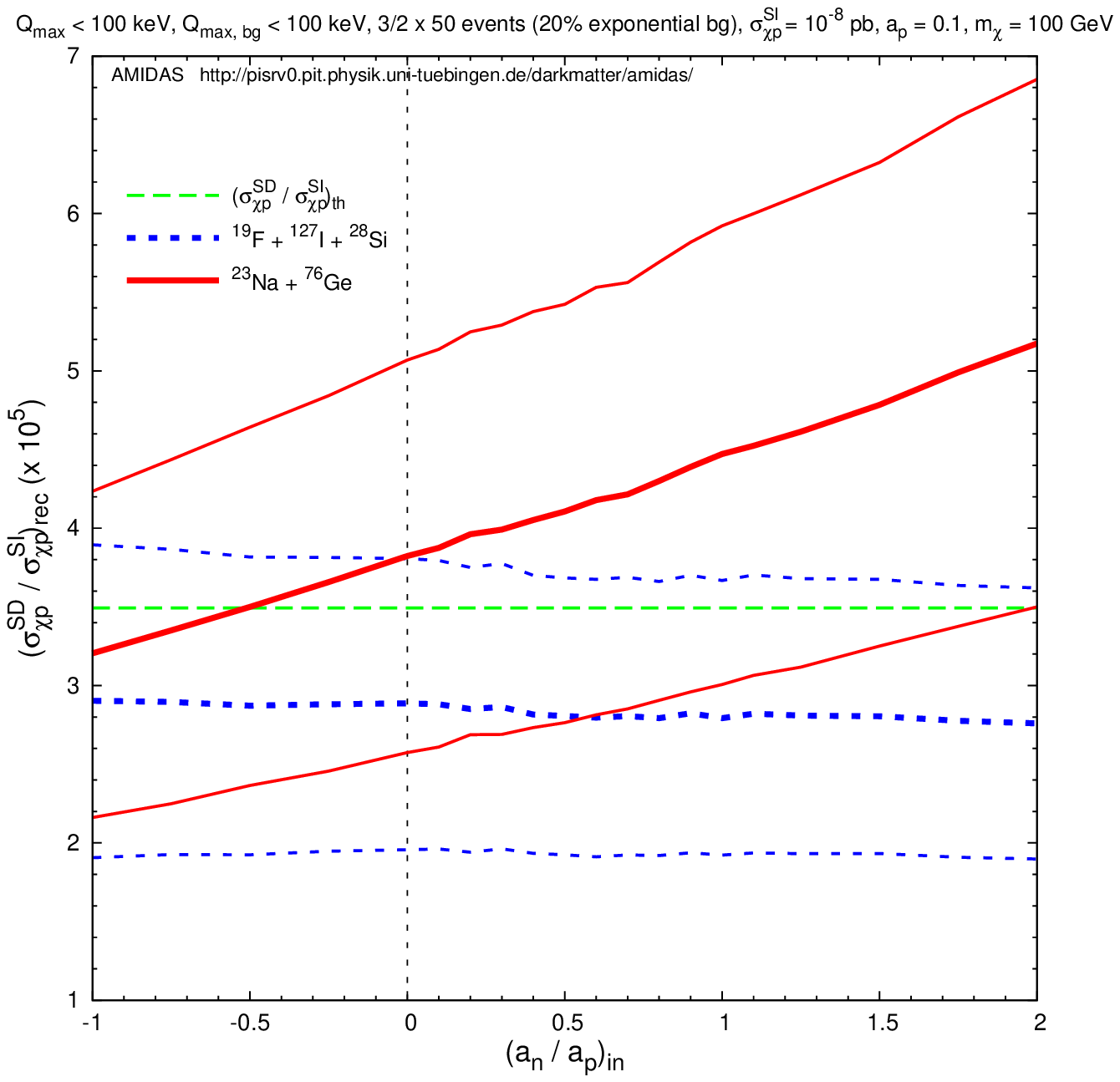}
\includegraphics[width=8.5cm]{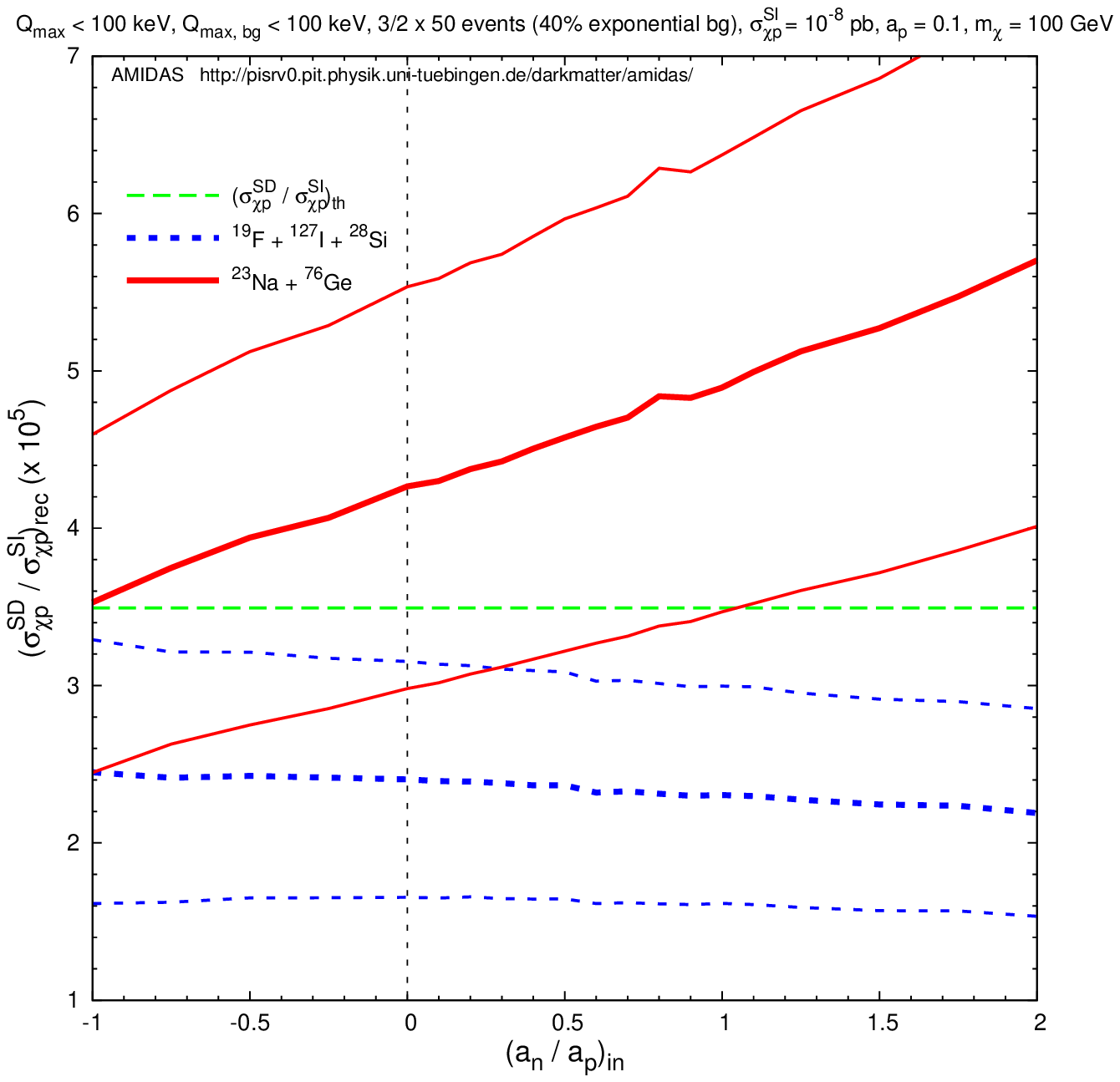} \\
\vspace{-0.25cm}
\end{center}
\caption{
 The reconstructed $\sigmapSD / \sigmapSI$ ratios
 and the lower and upper bounds of
 their 1$\sigma$ statistical uncertainties
 as functions of the input $\armn / \armp$ ratio.
 The dashed blue curves indicate the ratios
 estimated by Eq.~(\ref{eqn:rsigmaSDpSI})
 with $\armn / \armp$ estimated by Eq.~(\ref{eqn:ranapSISD})
 ({\em not} by Eq.~(\ref{eqn:ranapSD})),
 whereas the solid red curves indicate the ratios
 estimated by Eq.~(\ref{eqn:rsigmaSDpSI_even}).
 $\rmXA{Ge}{76}$ has been chosen as the second target
 having only the SI interaction with WIMPs and combined with
 $\rmXA{Na}{23}$ for using Eq.~(\ref{eqn:rsigmaSDpSI_even}).
 Parameters are as
 in Figs.~\ref{fig:ranapSISD-08-ranap-sh-rec-ex}.
 Note that
 the input $\armn / \armp$ ratio
 ranges only between $-1$ and 2.
}
\label{fig:rsigmaSDpSI-08-ranap-ex}
\end{figure}
\begin{figure}[t!]
\begin{center}
\includegraphics[width=8.5cm]{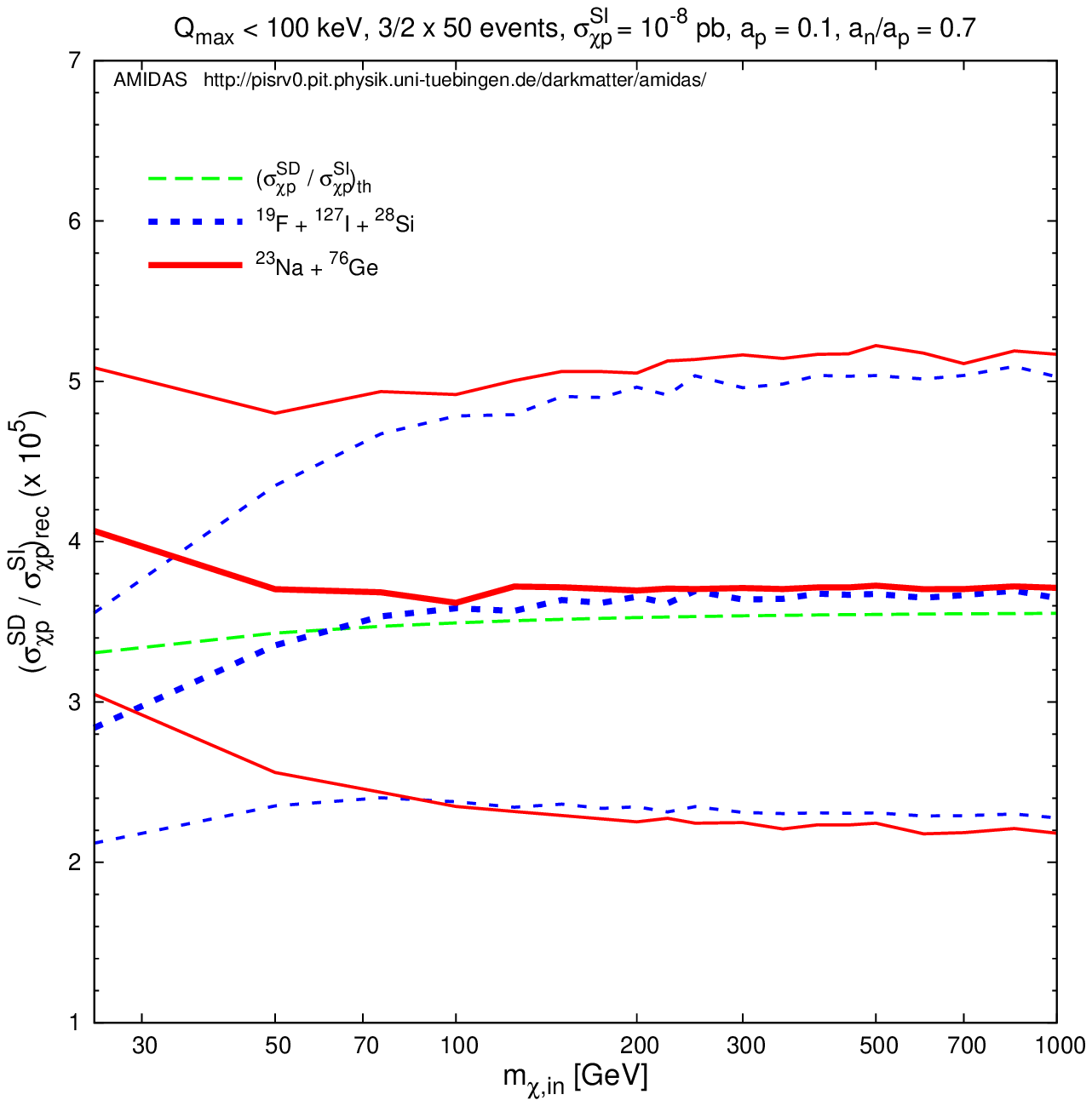}
\includegraphics[width=8.5cm]{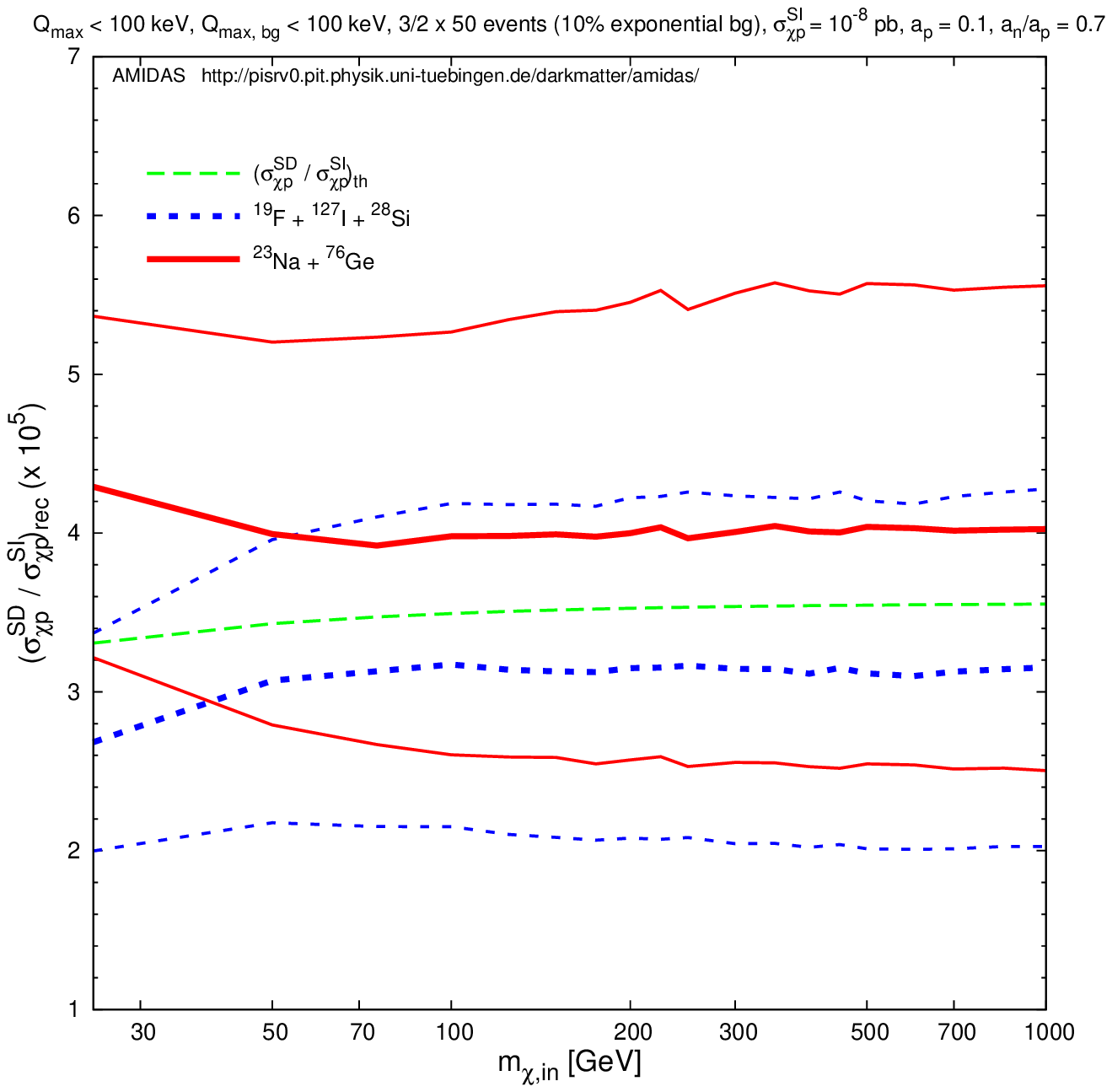} \\
\vspace{0.5cm}
\includegraphics[width=8.5cm]{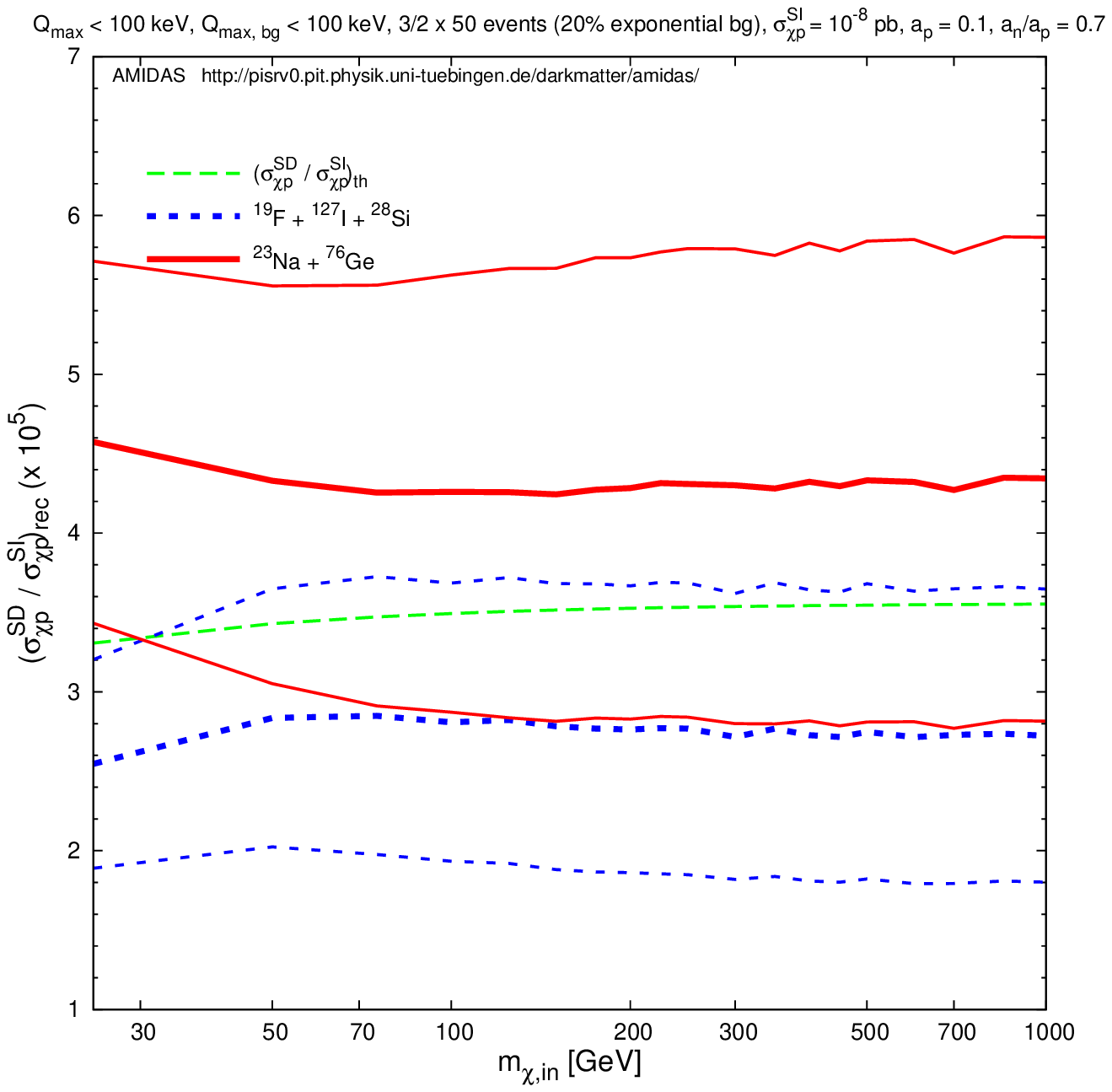}
\includegraphics[width=8.5cm]{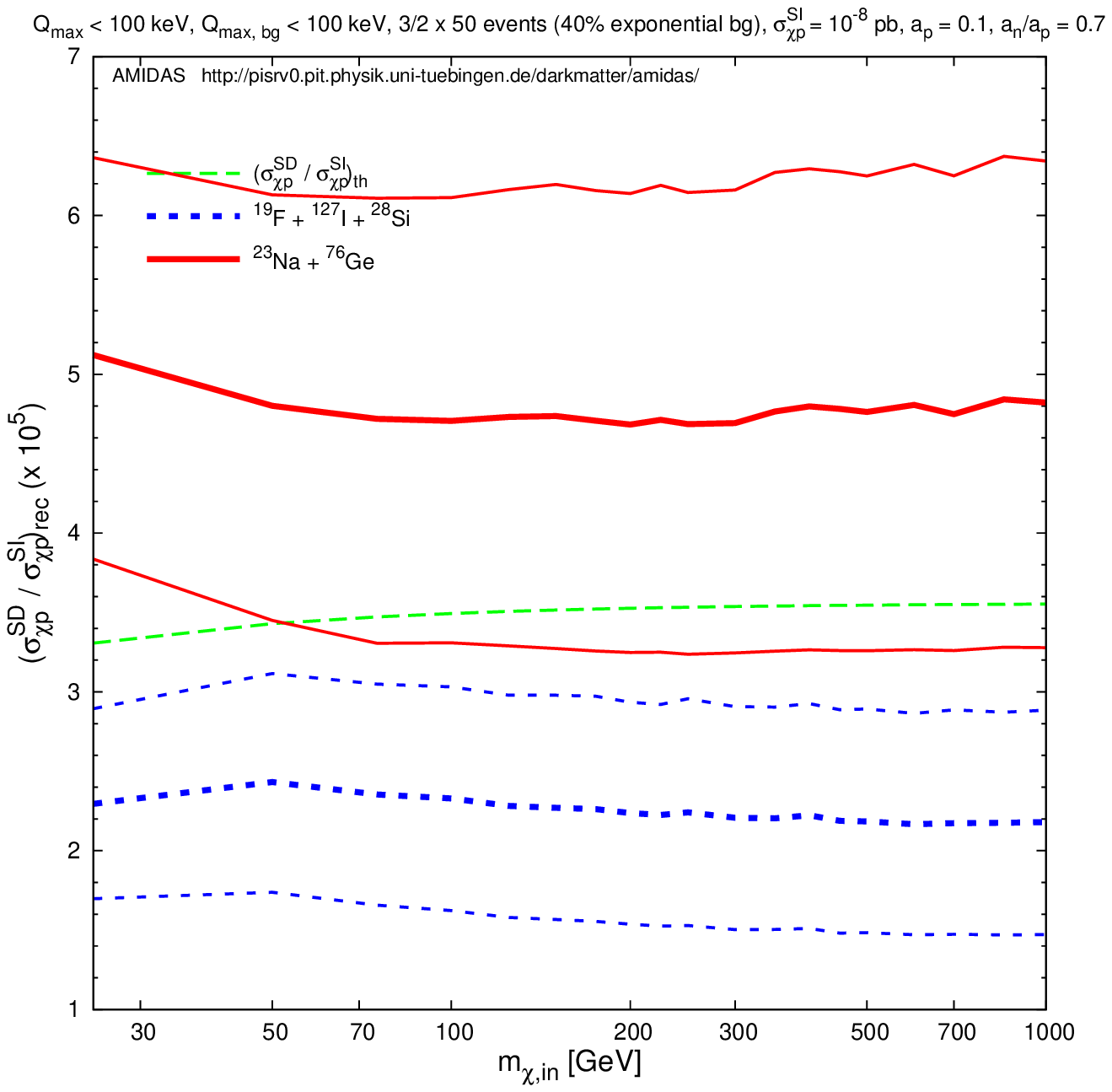} \\
\vspace{-0.25cm}
\end{center}
\caption{
 The reconstructed $\sigmapSD / \sigmapSI$ ratios
 and the lower and upper bounds of
 their 1$\sigma$ statistical uncertainties
 as functions of the input WIMP mass $\mchi$.
 The input $\armn / \armp = 0.7$,
 the other parameters and notations
 are as in Figs.~\ref{fig:rsigmaSDpSI-08-ranap-ex}.
 Note that
 the input WIMP mass starts from \mbox{25 GeV}.
}
\label{fig:rsigmaSDpSI-08-mchi-ex}
\end{figure}
\begin{figure}[t!]
\begin{center}
\includegraphics[width=8.5cm]{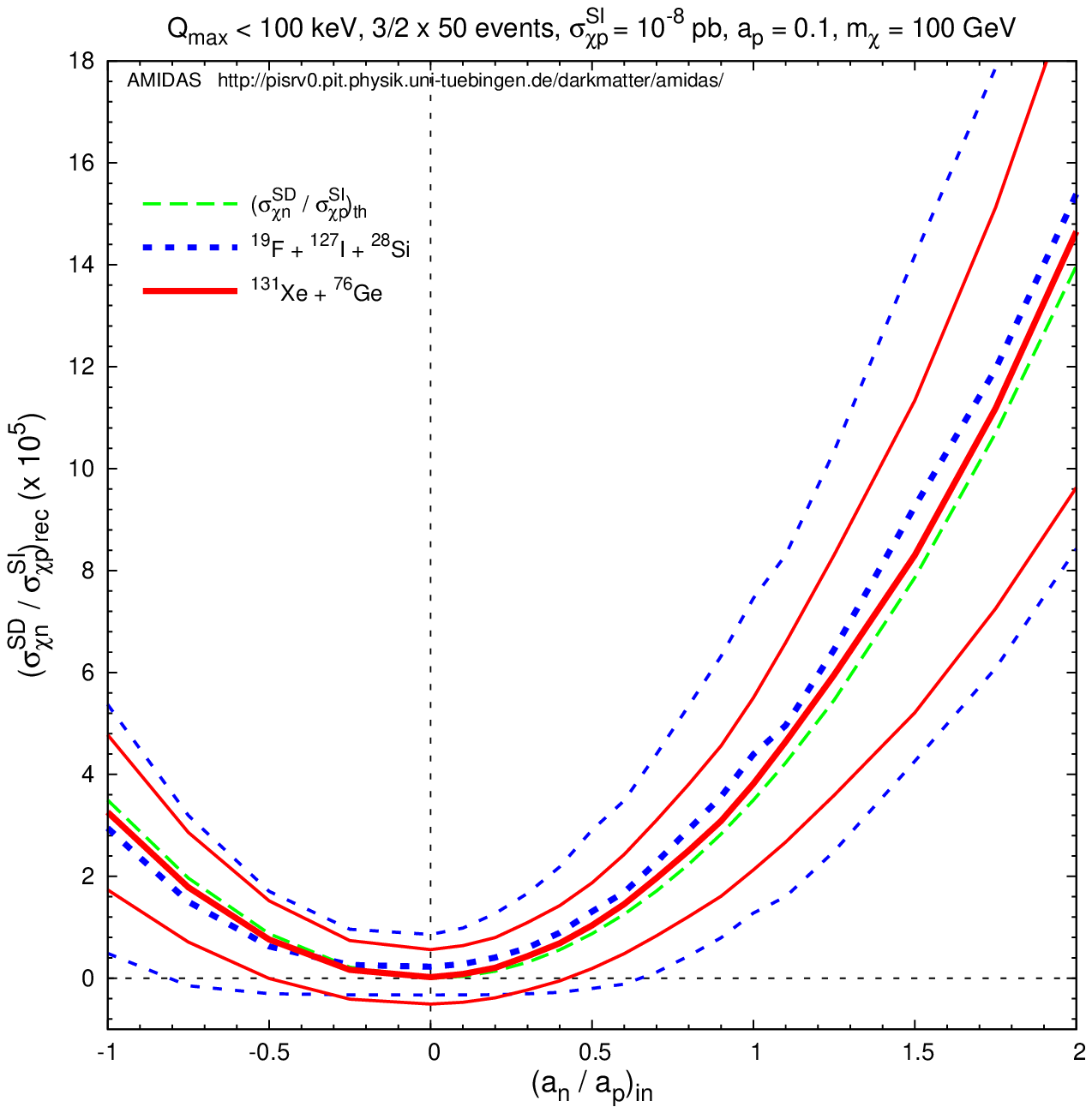}
\includegraphics[width=8.5cm]{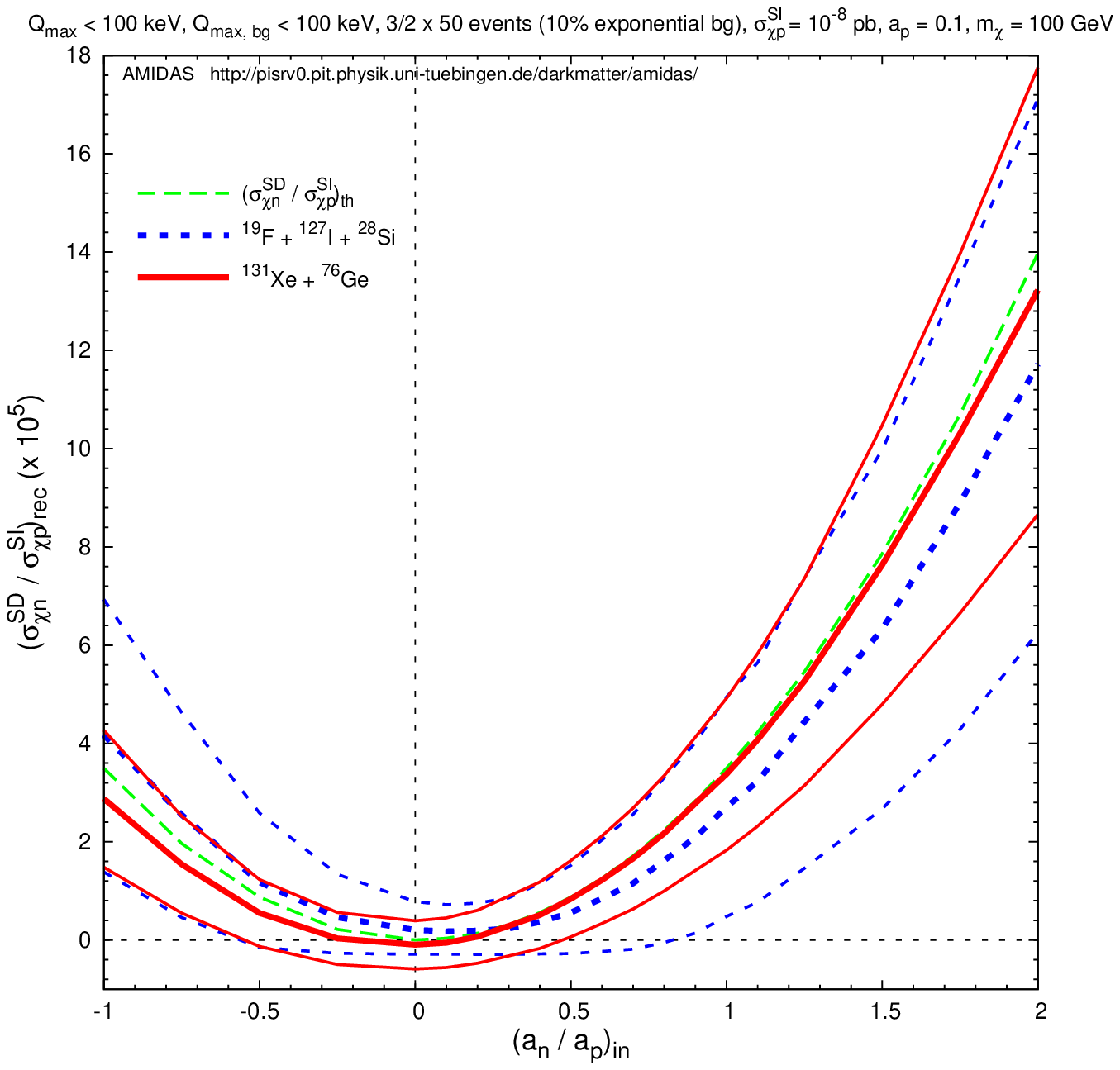} \\
\vspace{0.5cm}
\includegraphics[width=8.5cm]{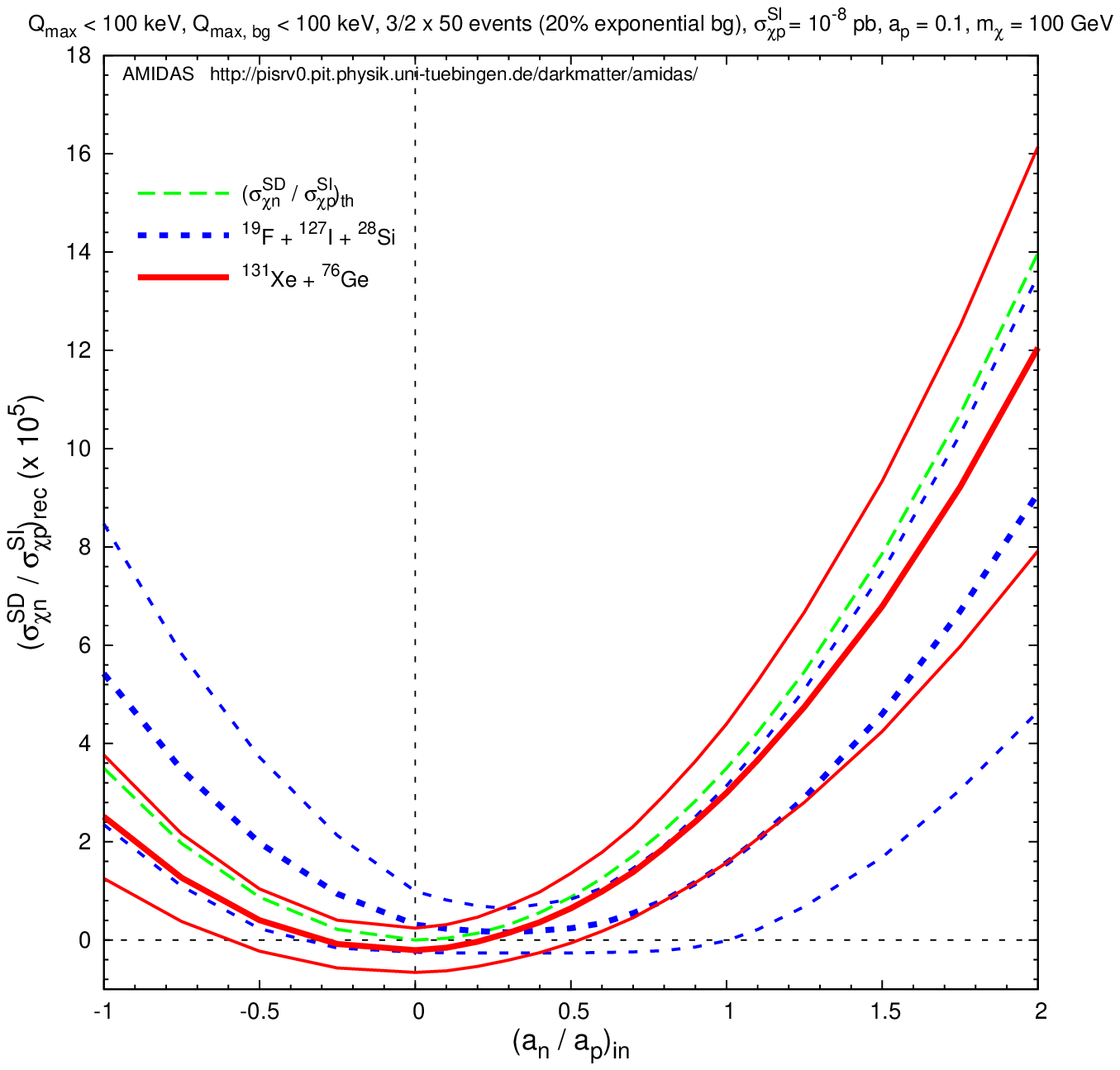}
\includegraphics[width=8.5cm]{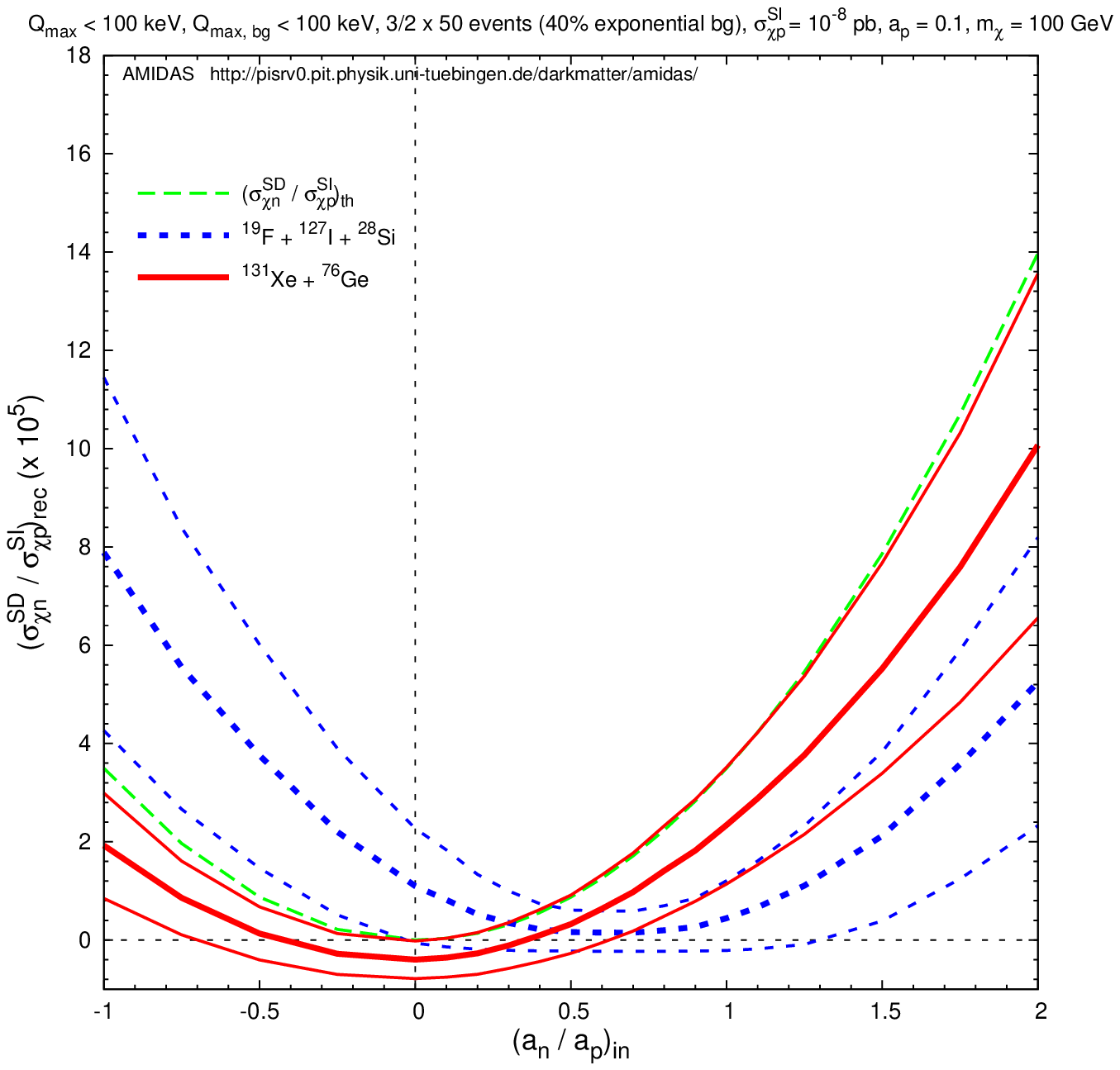} \\
\vspace{-0.25cm}
\end{center}
\caption{
 The reconstructed $\sigmanSD / \sigmapSI$ ratios
 and the lower and upper bounds of
 their 1$\sigma$ statistical uncertainties
 as functions of the input $\armn / \armp$ ratio.
 The dashed blue curves indicate the ratios
 estimated by Eq.~(\ref{eqn:rsigmaSDpSI})
 with $\armn / \armp$ estimated by Eq.~(\ref{eqn:ranapSISD}),
 whereas the solid red curves indicate the ratios
 estimated by Eq.~(\ref{eqn:rsigmaSDpSI_even}).
 $\rmXA{Ge}{76}$ has been chosen as the second target
 having only the SI interaction with WIMPs and combined with
 $\rmXA{Xe}{131}$ for using Eq.~(\ref{eqn:rsigmaSDpSI_even}).
 Parameters are as in Figs.~\ref{fig:rsigmaSDpSI-08-ranap-ex}.
 Note that
 the input $\armn / \armp$ ratio
 ranges only between $-1$ and 2.
}
\label{fig:rsigmaSDnSI-08-ranap-ex}
\end{figure}
\begin{figure}[t!]
\begin{center}
\includegraphics[width=8.5cm]{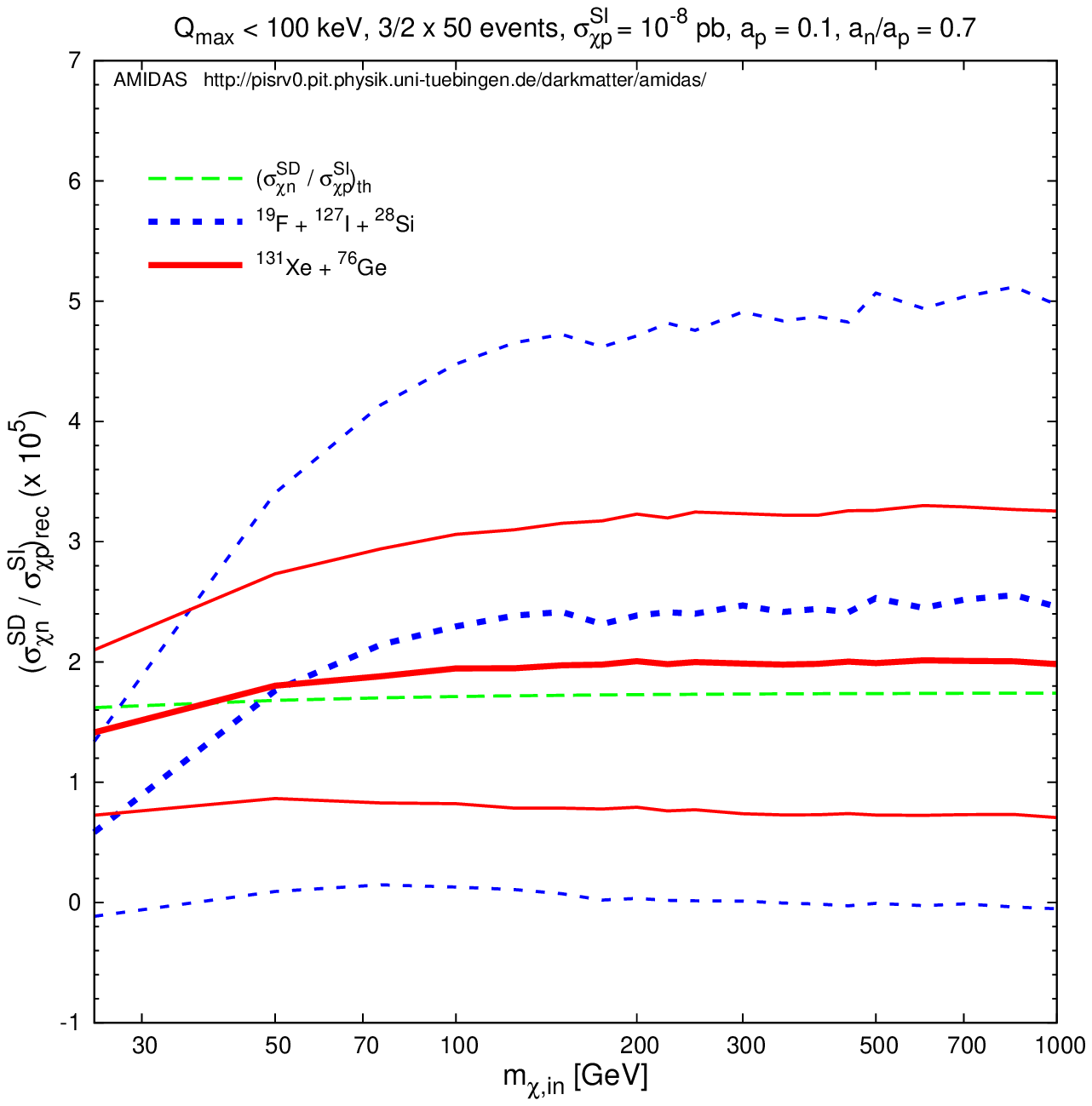}
\includegraphics[width=8.5cm]{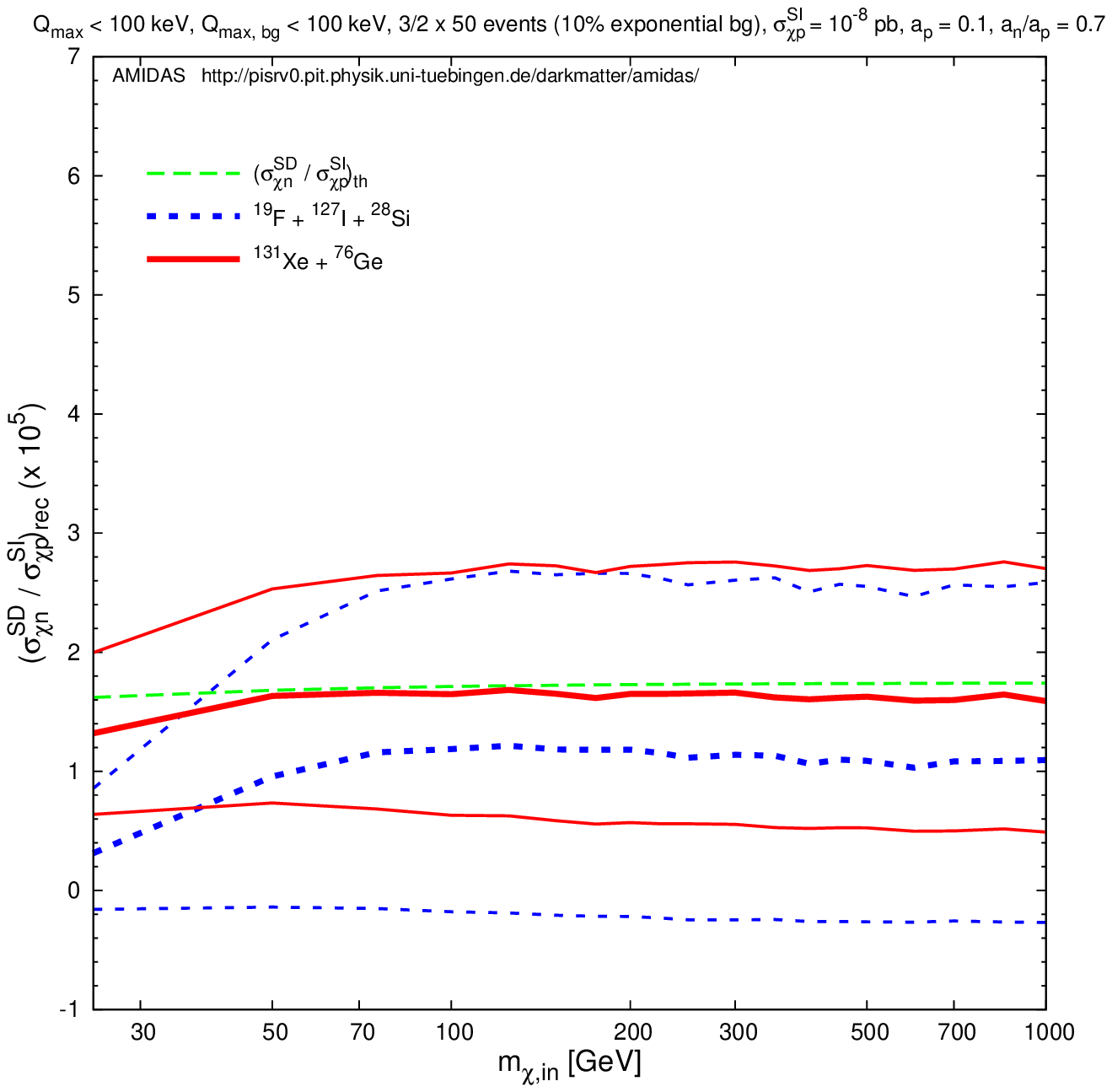} \\
\vspace{0.5cm}
\includegraphics[width=8.5cm]{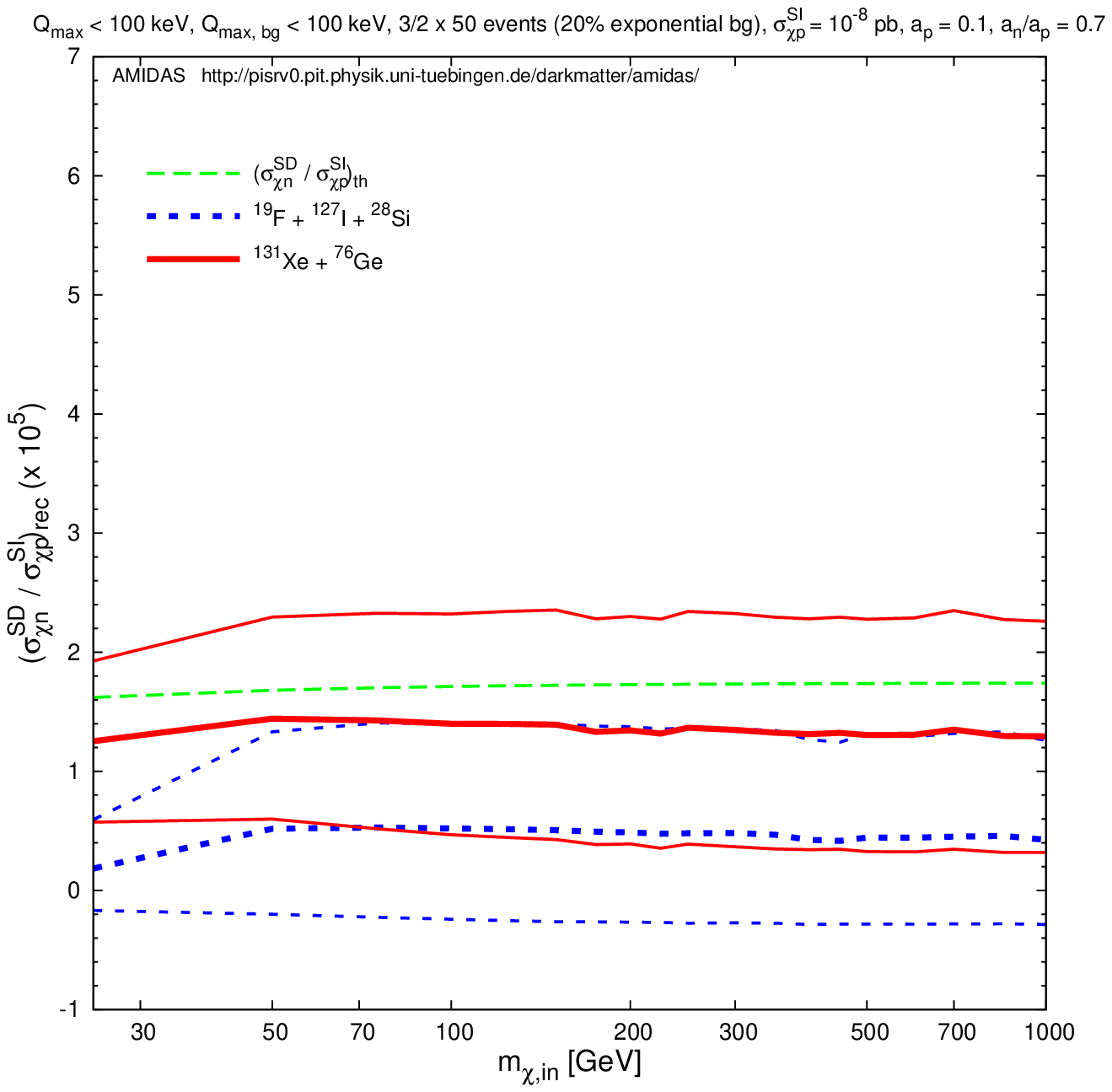}
\includegraphics[width=8.5cm]{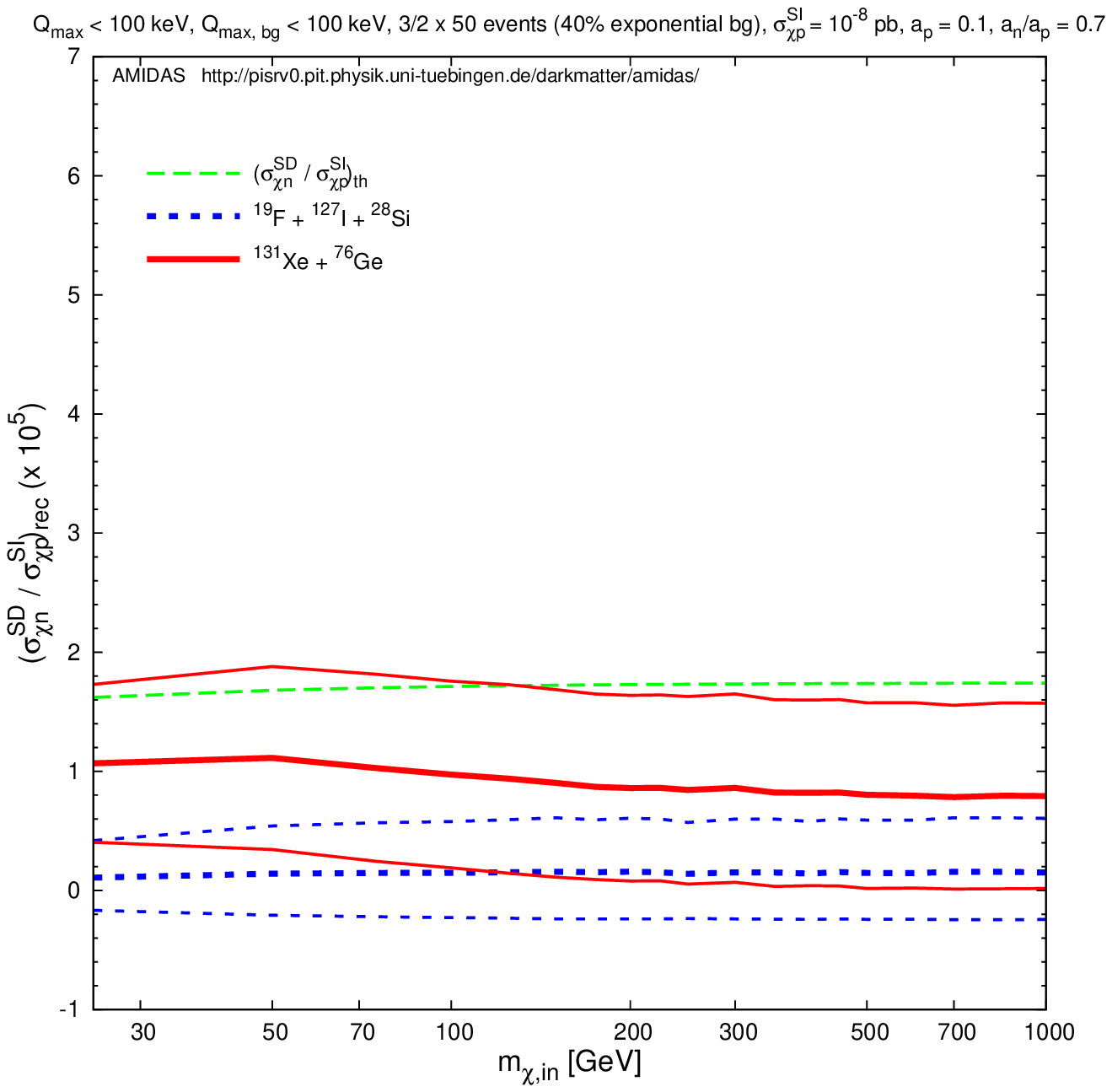} \\
\vspace{-0.25cm}
\end{center}
\caption{
 The reconstructed $\sigmanSD / \sigmapSI$ ratios
 and the lower and upper bounds of
 their 1$\sigma$ statistical uncertainties
 as functions of the input WIMP mass $\mchi$.
 The input $\armn / \armp = 0.7$,
 the other parameters and notations
 are as in Figs.~\ref{fig:rsigmaSDnSI-08-ranap-ex}.
 Note that
 the input WIMP mass starts from \mbox{25 GeV}.
}
\label{fig:rsigmaSDnSI-08-mchi-ex}
\end{figure}
\subsection{Reconstructed \boldmath$(\armn / \armp)_{\pm}^{\rm SI + SD}$}
\begin{figure}[t!]
\begin{center}
\includegraphics[width=8.5cm]{ranapSD-ranap-sh-rec-ex-00}
\includegraphics[width=8.5cm]{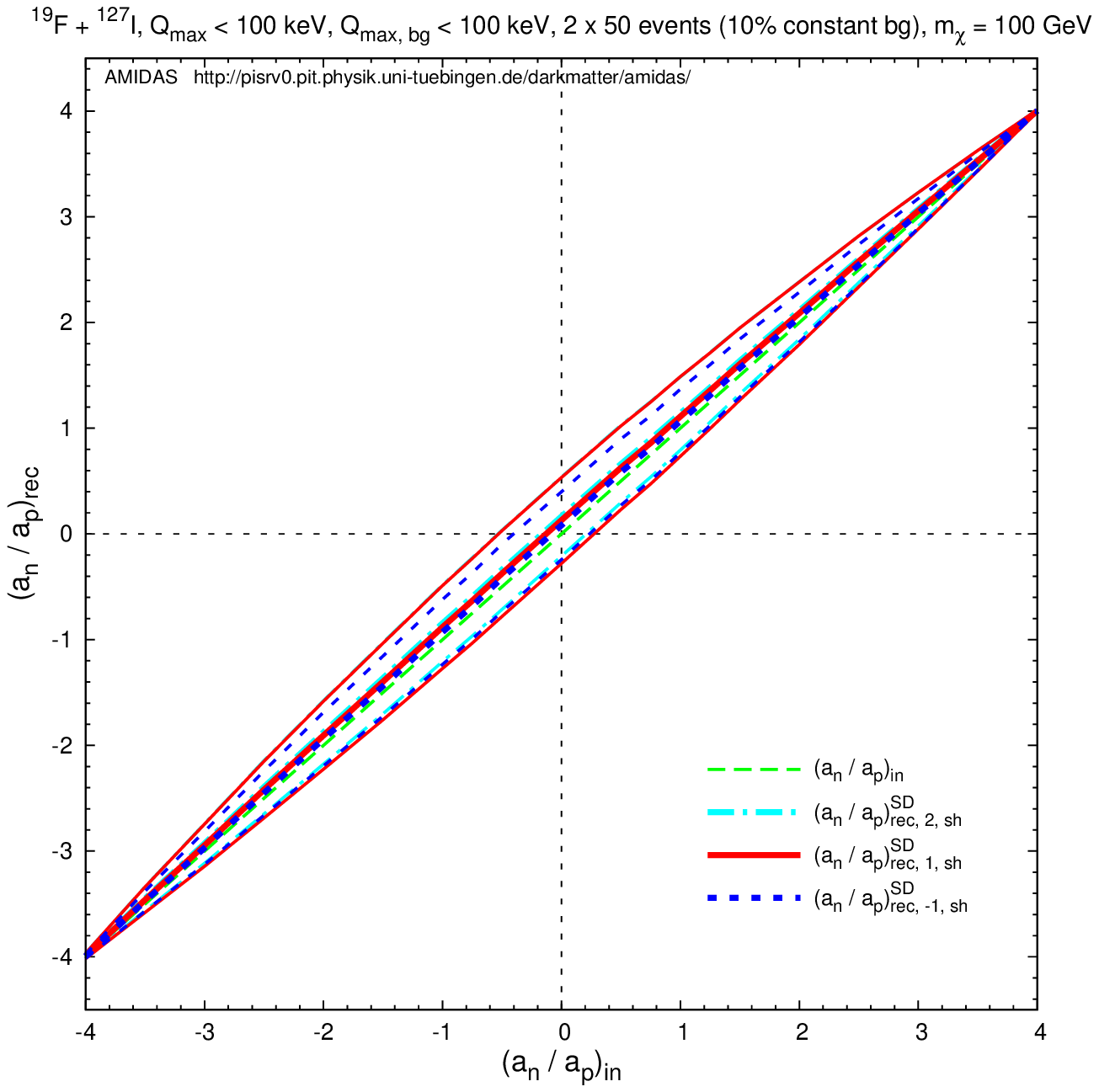} \\
\vspace{0.5cm}
\includegraphics[width=8.5cm]{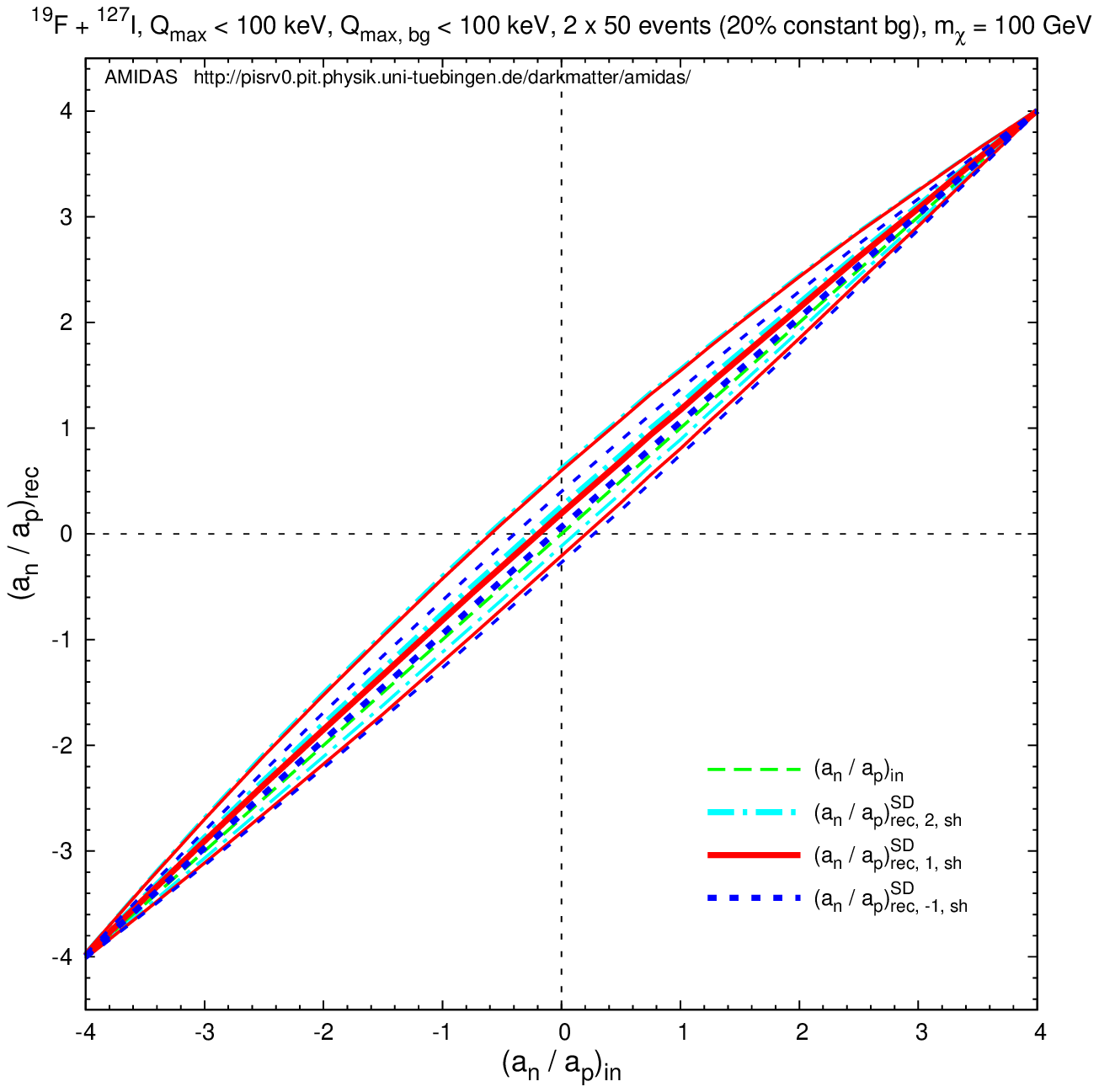}
\includegraphics[width=8.5cm]{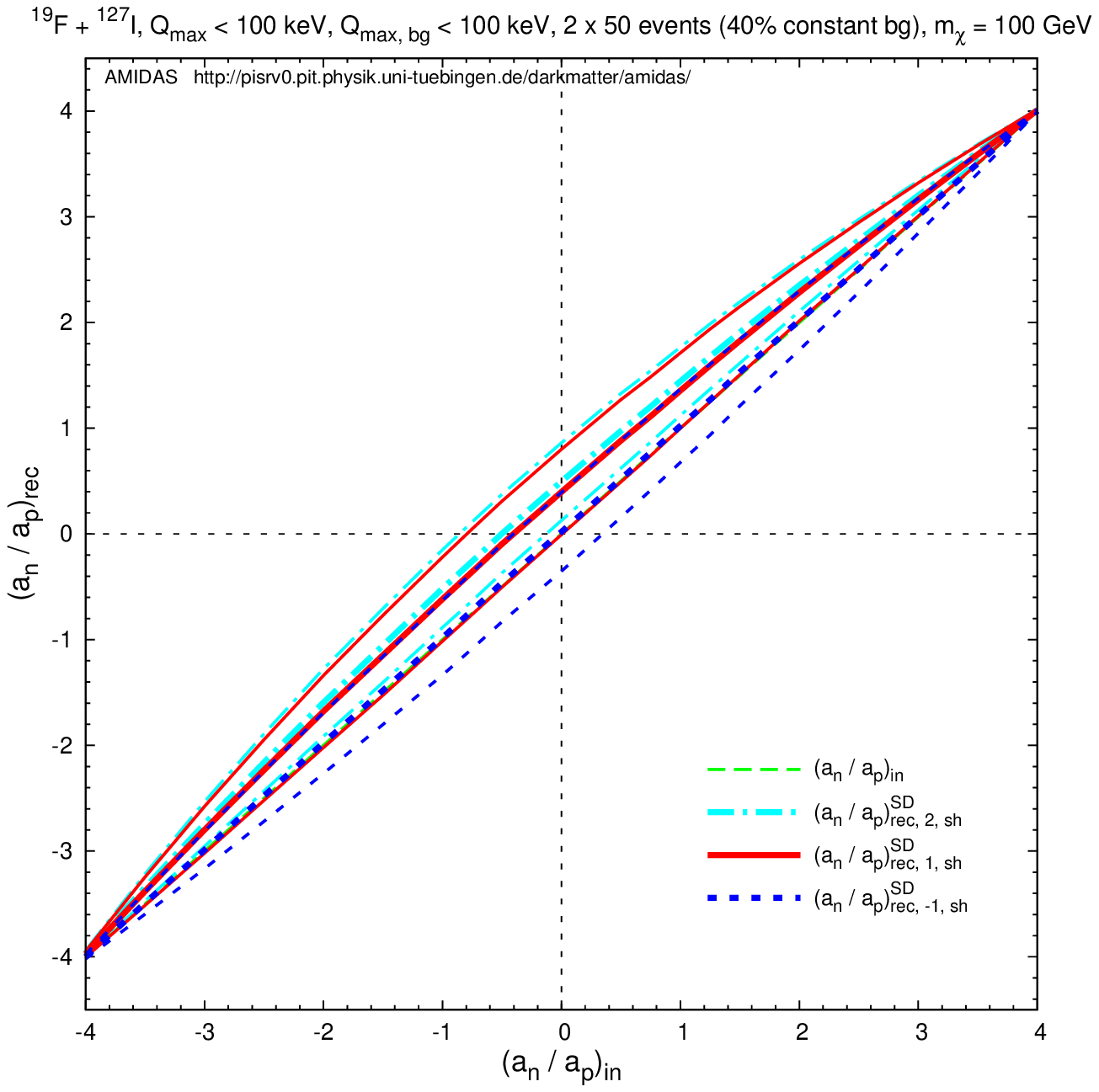} \\
\vspace{-0.25cm}
\end{center}
\caption{
 As in Figs.~\ref{fig:ranapSD-ranap-sh-rec-ex},
 except that the constant background spectrum
 has been used here.
}
\label{fig:ranapSD-ranap-sh-rec-const}
\end{figure}
\begin{figure}[t!]
\begin{center}
\includegraphics[width=8.5cm]{ranapSD-mchi-rec-ex-00}
\includegraphics[width=8.5cm]{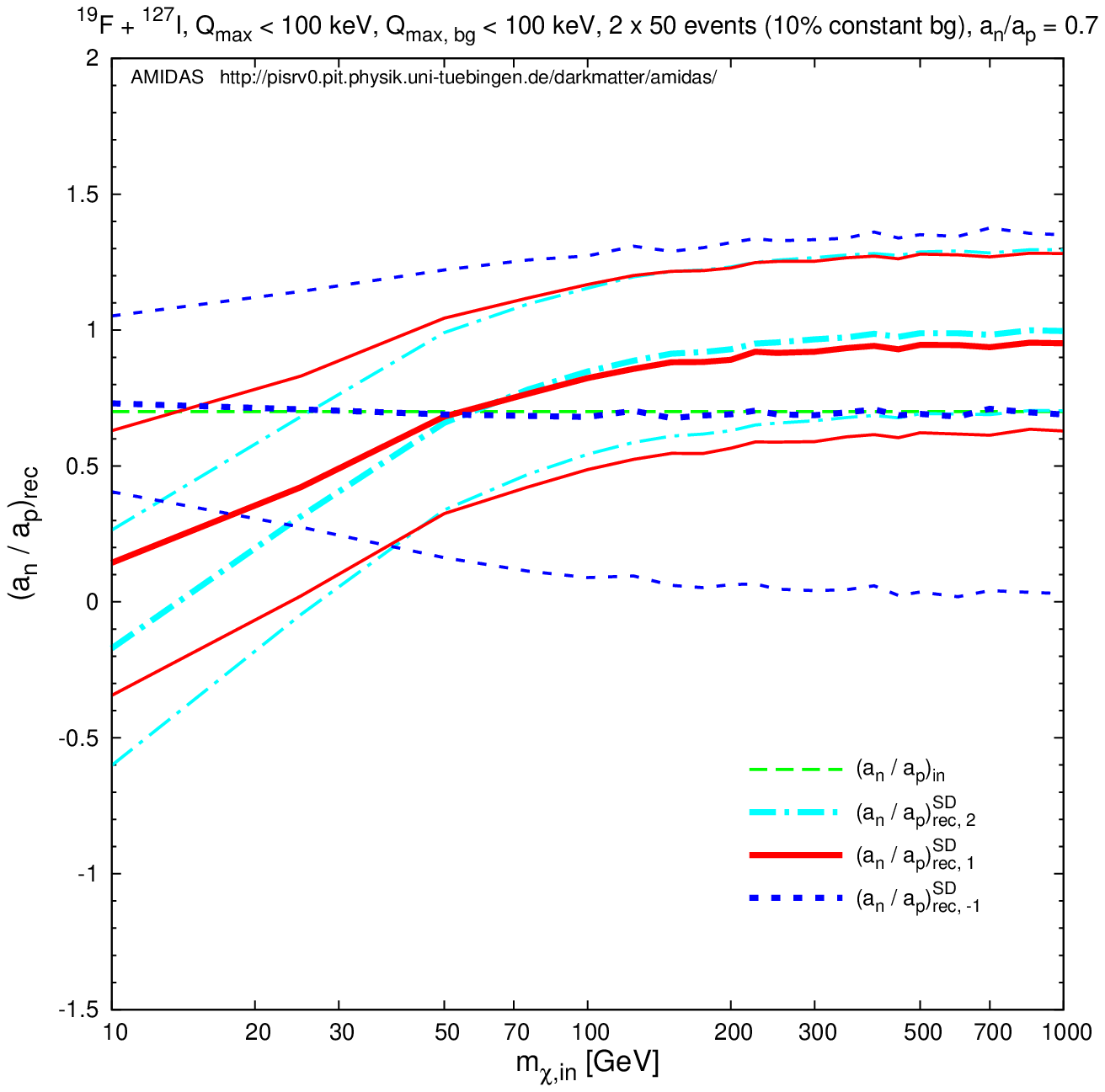} \\
\vspace{0.5cm}
\includegraphics[width=8.5cm]{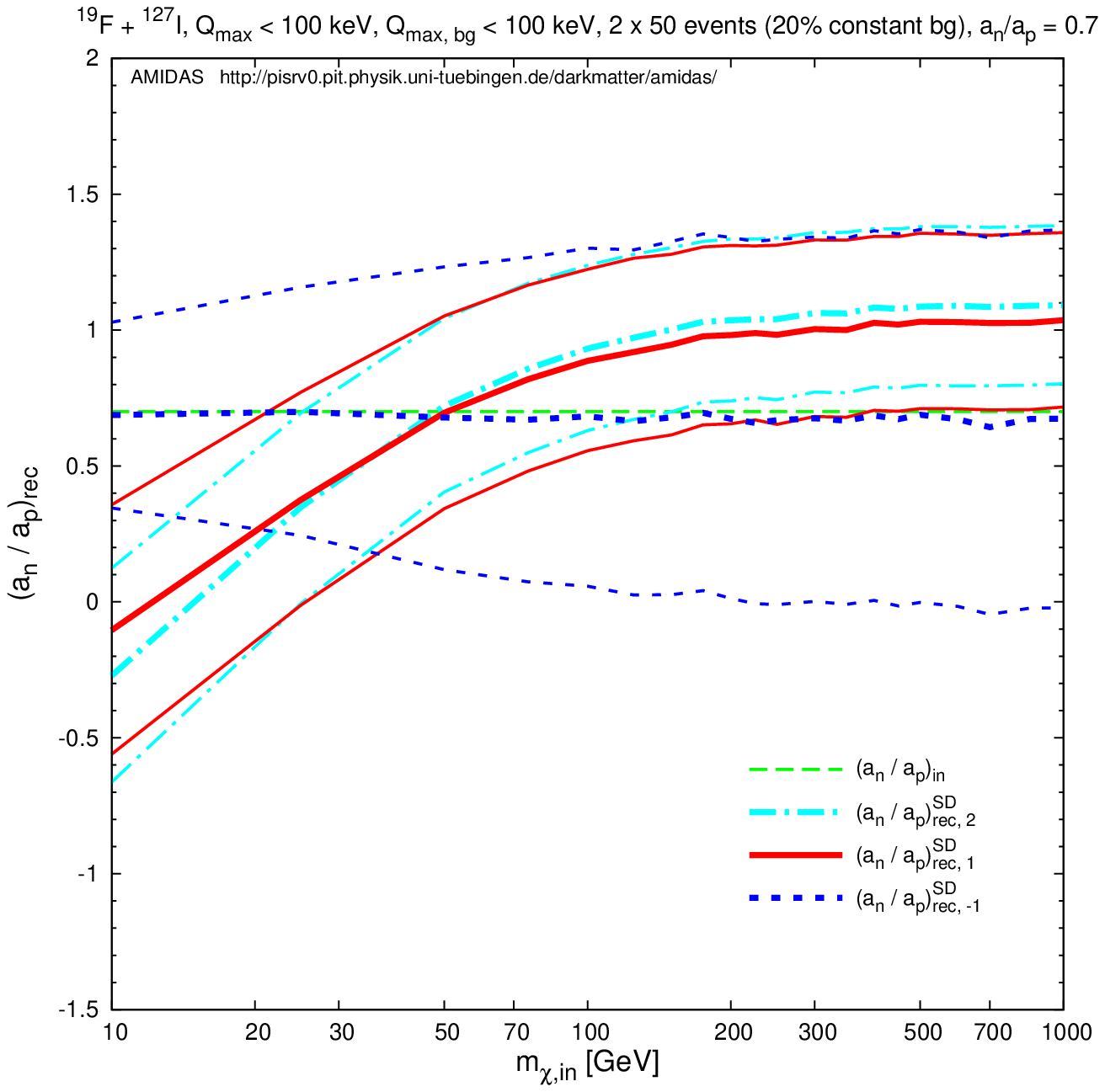}
\includegraphics[width=8.5cm]{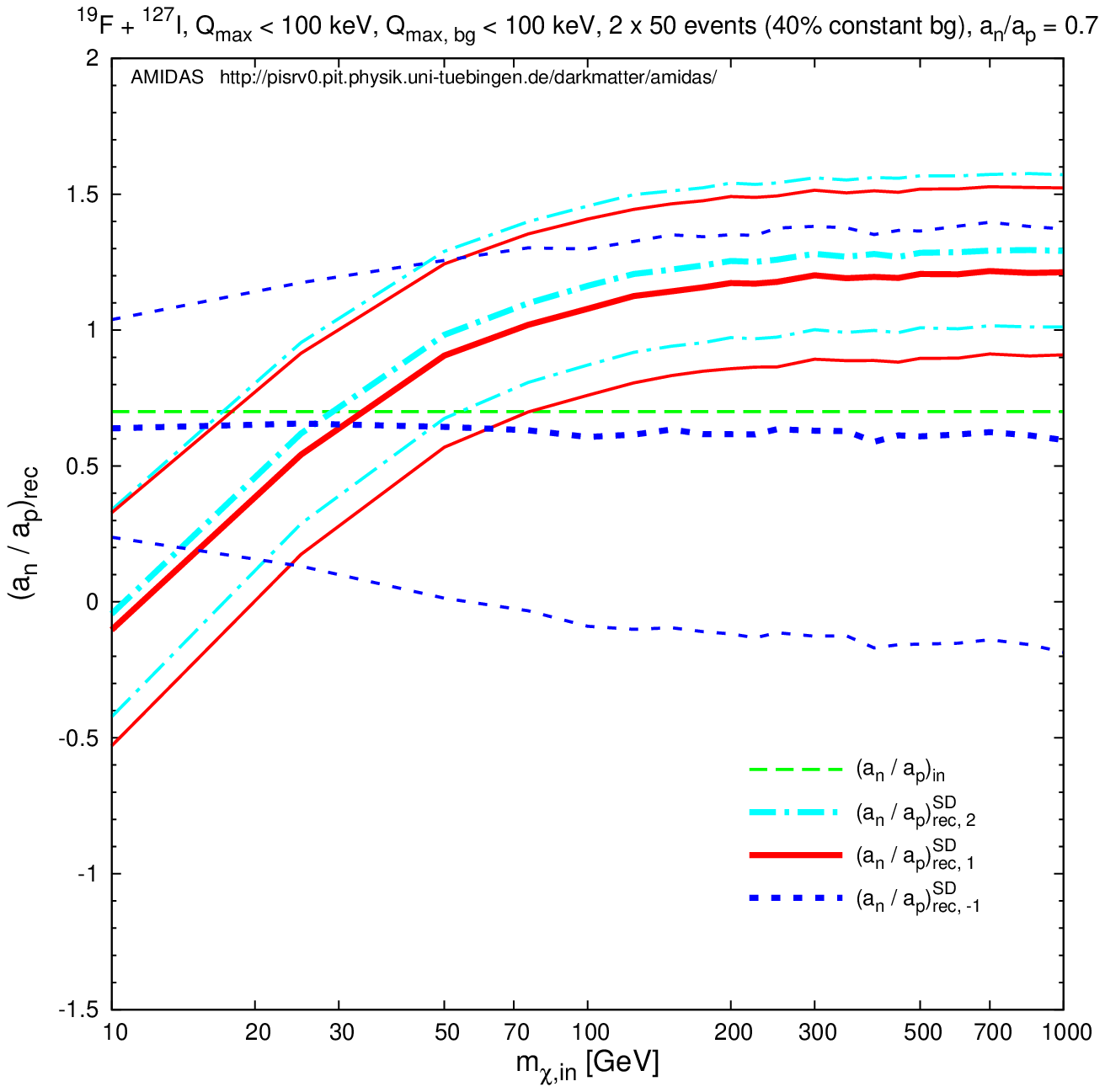} \\
\vspace{-0.25cm}
\end{center}
\caption{
 As in Figs.~\ref{fig:ranapSD-mchi-rec-ex},
 except that the constant background spectrum
 has been used here.
}
\label{fig:ranapSD-mchi-rec-const}
\end{figure}
\begin{figure}[t!]
\begin{center}
\includegraphics[width=8.5cm]{ranapSD-mchi-sh-rec-ex-00}
\includegraphics[width=8.5cm]{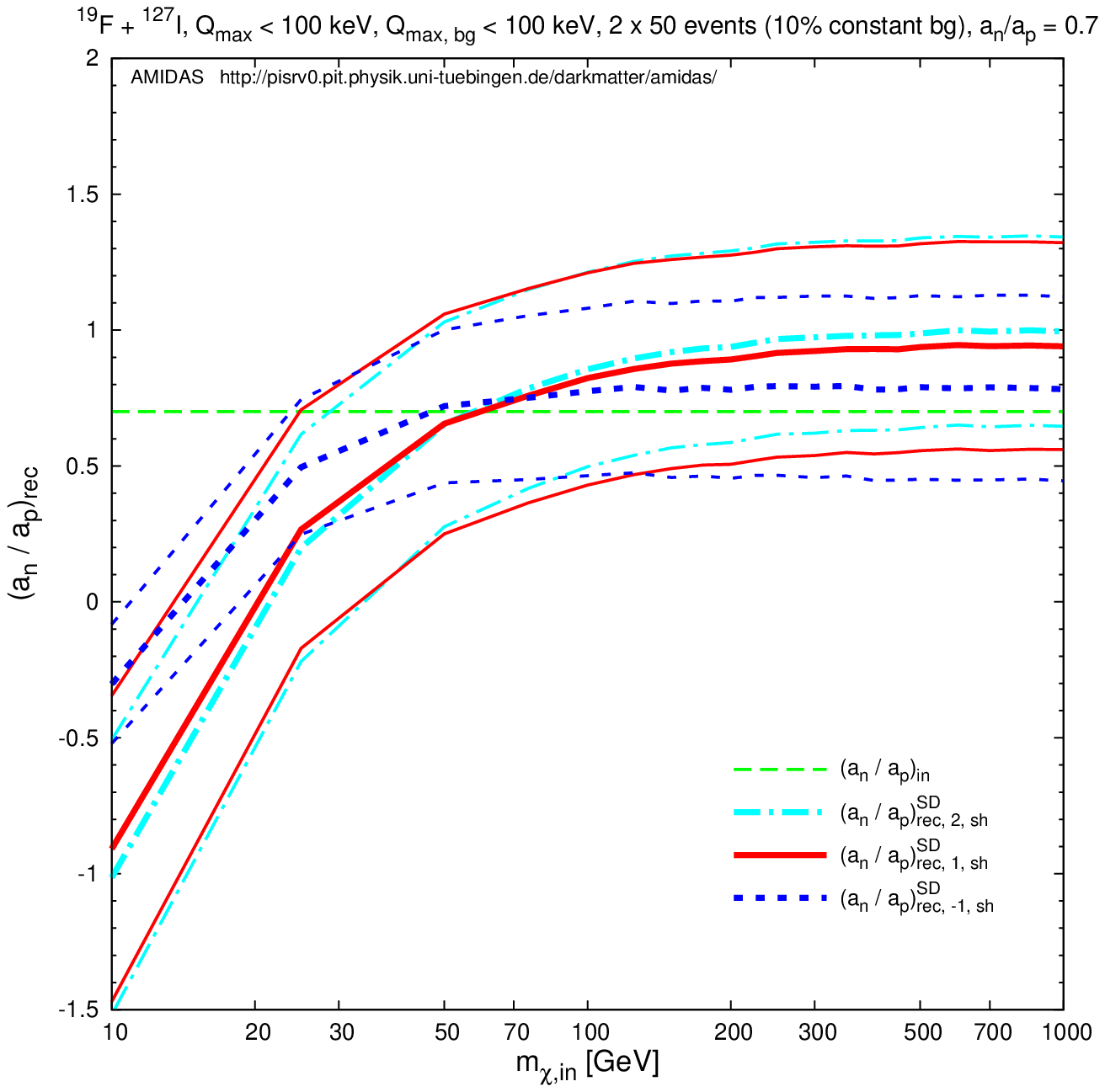} \\
\vspace{0.5cm}
\includegraphics[width=8.5cm]{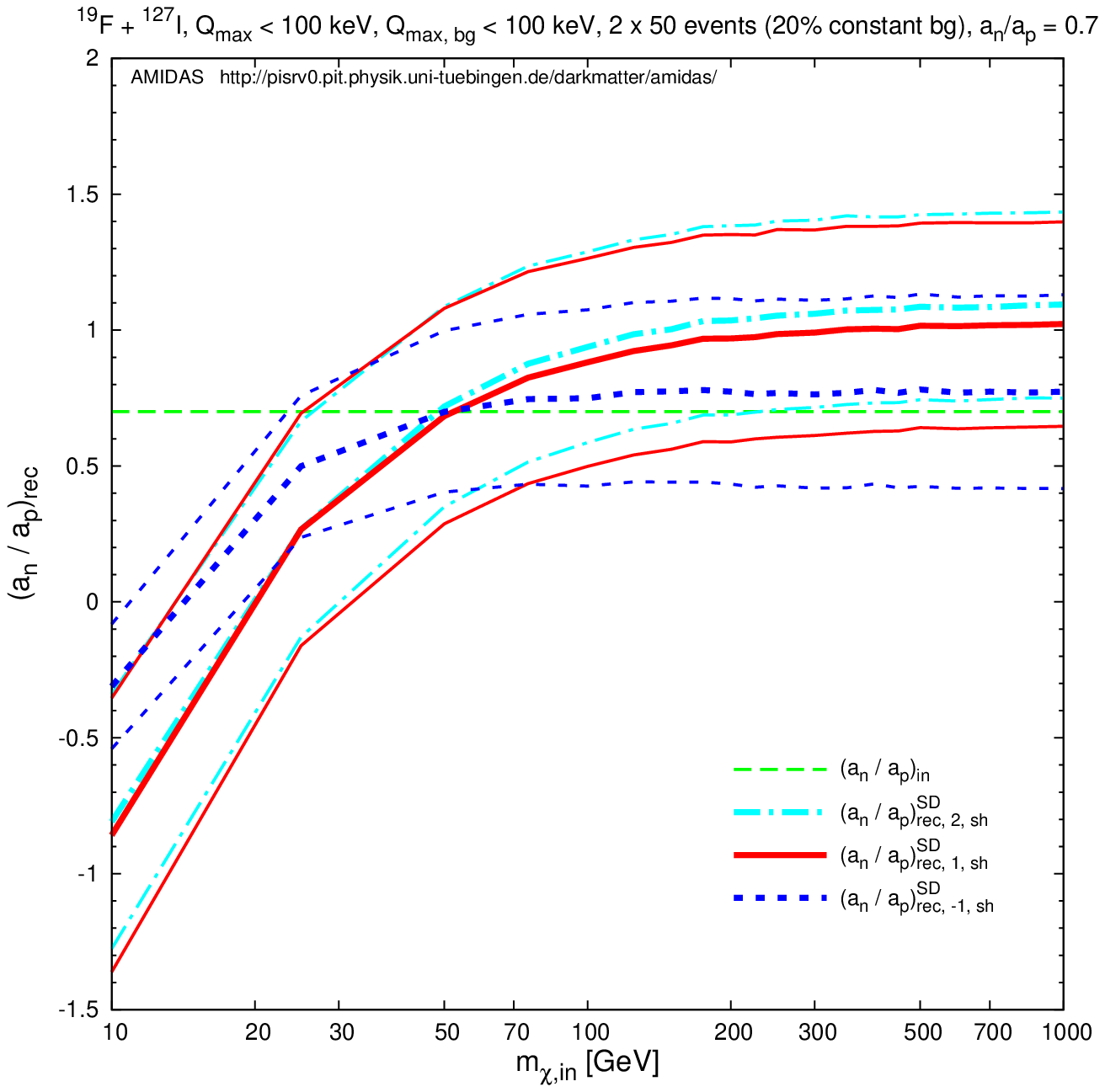}
\includegraphics[width=8.5cm]{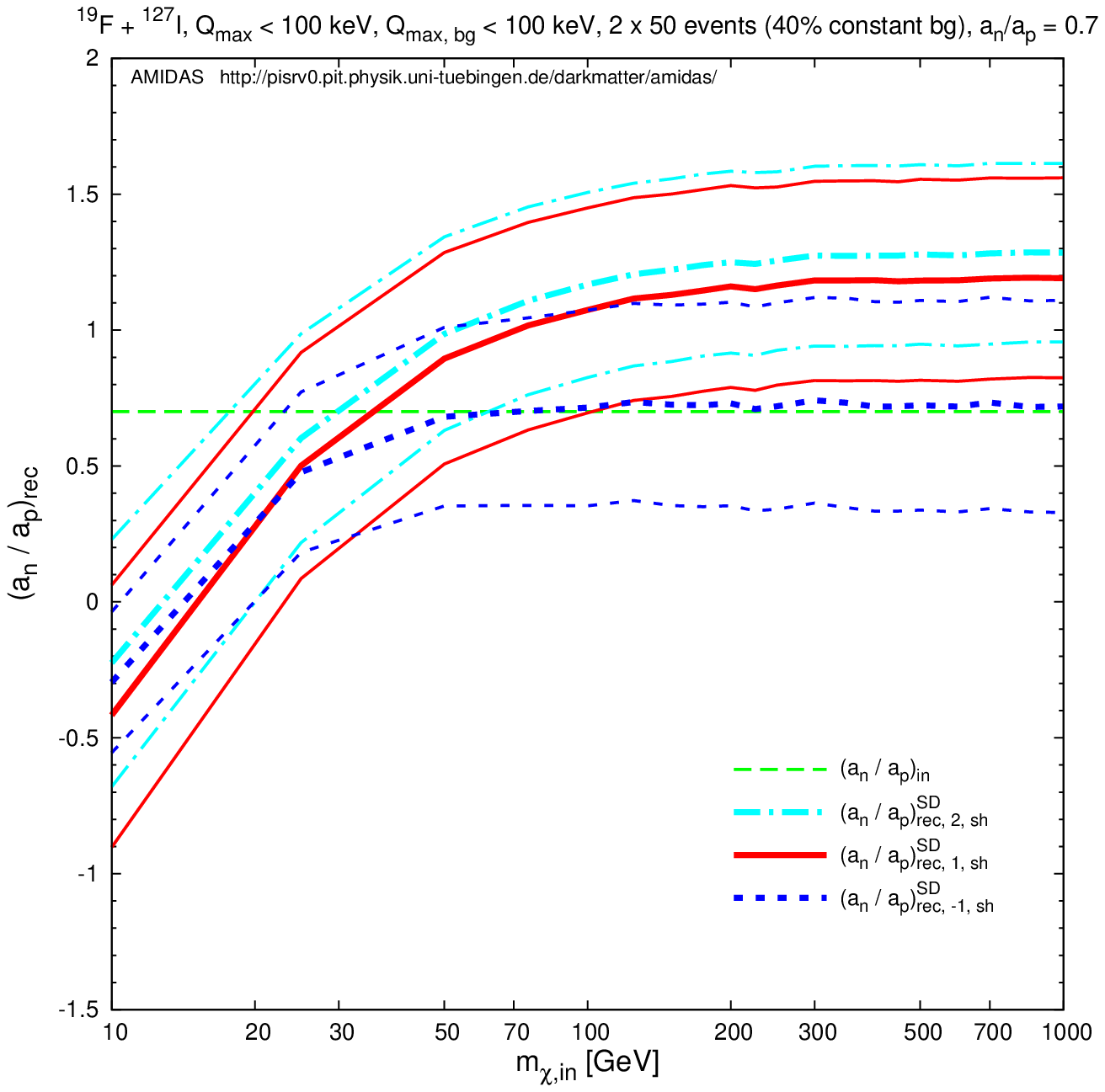} \\
\vspace{-0.25cm}
\end{center}
\caption{
 As in Figs.~\ref{fig:ranapSD-mchi-sh-rec-ex},
 except that the constant background spectrum
 has been used here.
}
\label{fig:ranapSD-mchi-sh-rec-const}
\end{figure}
\begin{figure}[t!]
\begin{center}
\includegraphics[width=8.5cm]{ranapSISD-08-ranap-sh-rec-ex-00}
\includegraphics[width=8.5cm]{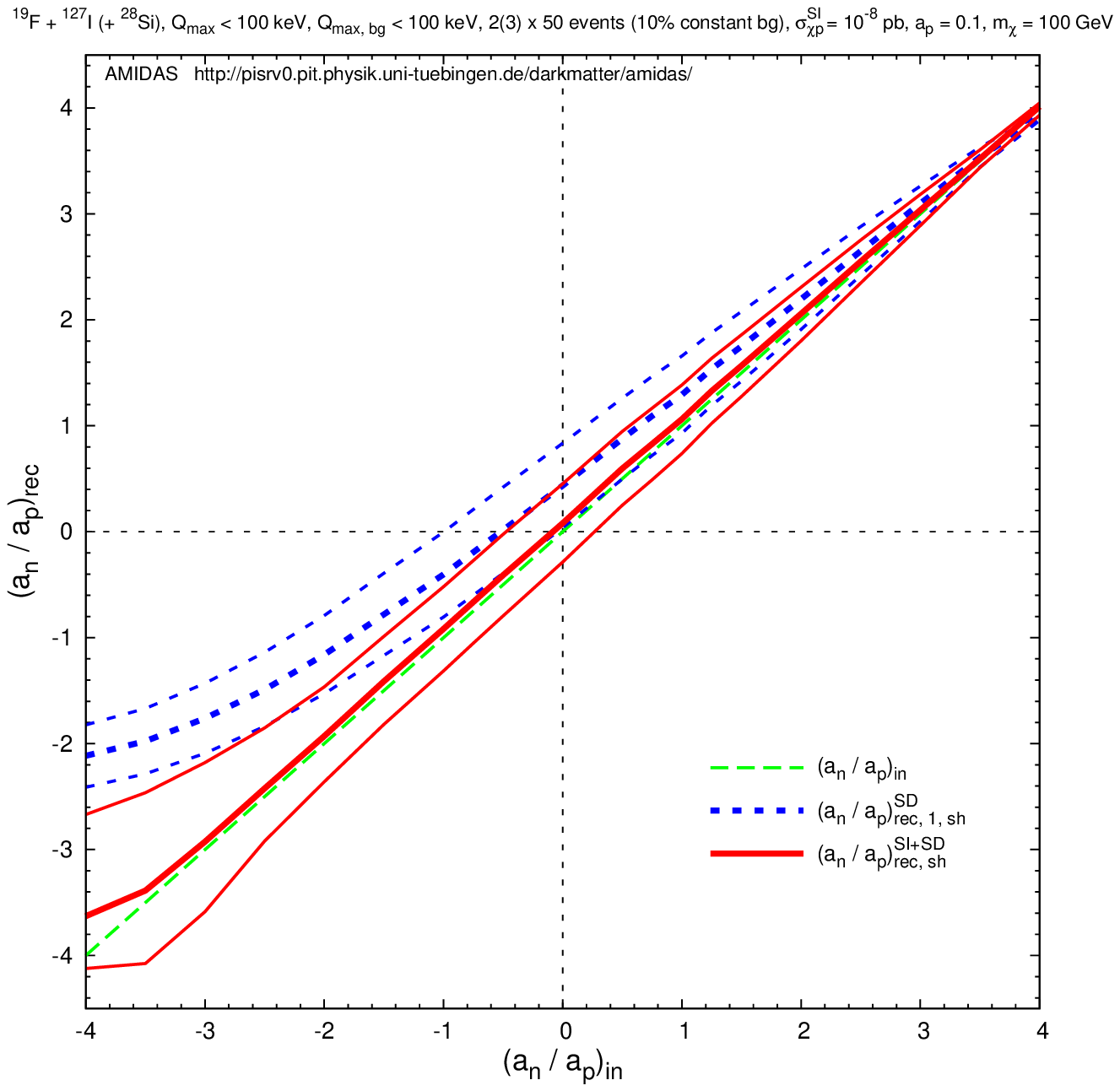} \\
\vspace{0.5cm}
\includegraphics[width=8.5cm]{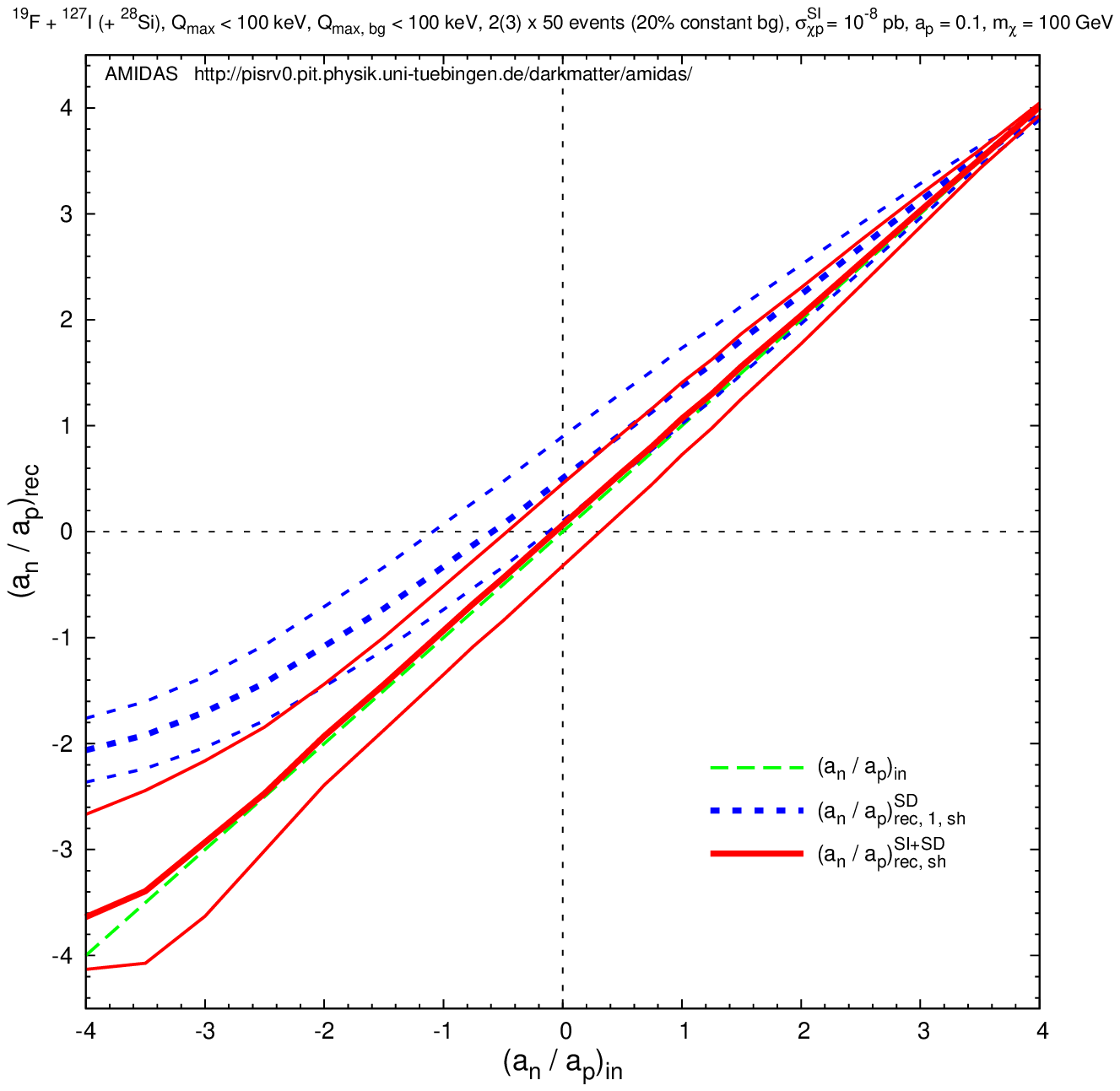}
\includegraphics[width=8.5cm]{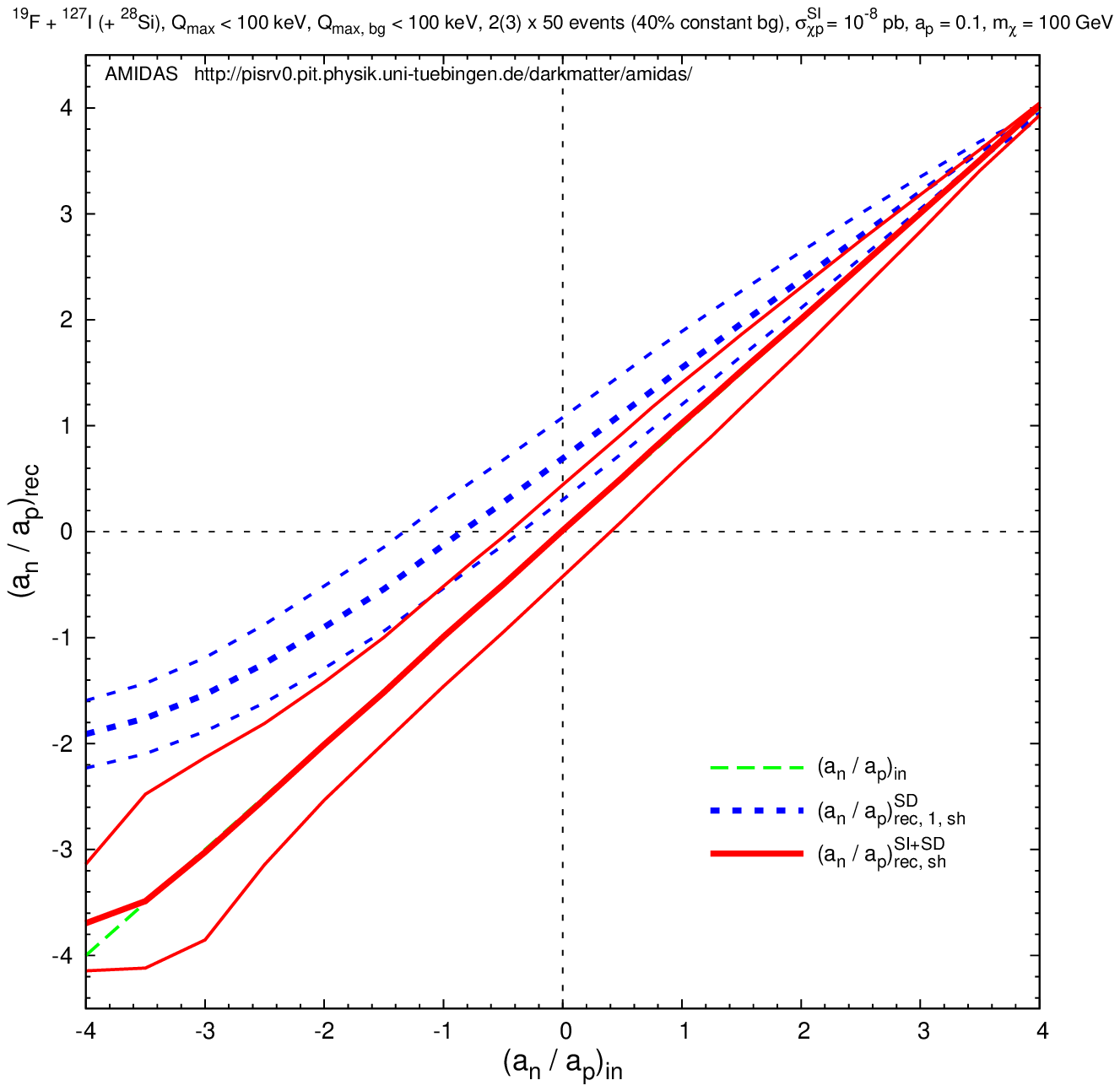} \\
\vspace{-0.25cm}
\end{center}
\caption{
 As in Figs.~\ref{fig:ranapSISD-08-ranap-sh-rec-ex},
 except that the constant background spectrum
 has been used here.
}
\label{fig:ranapSISD-08-ranap-sh-rec-const}
\end{figure}
\begin{figure}[t!]
\begin{center}
\includegraphics[width=8.5cm]{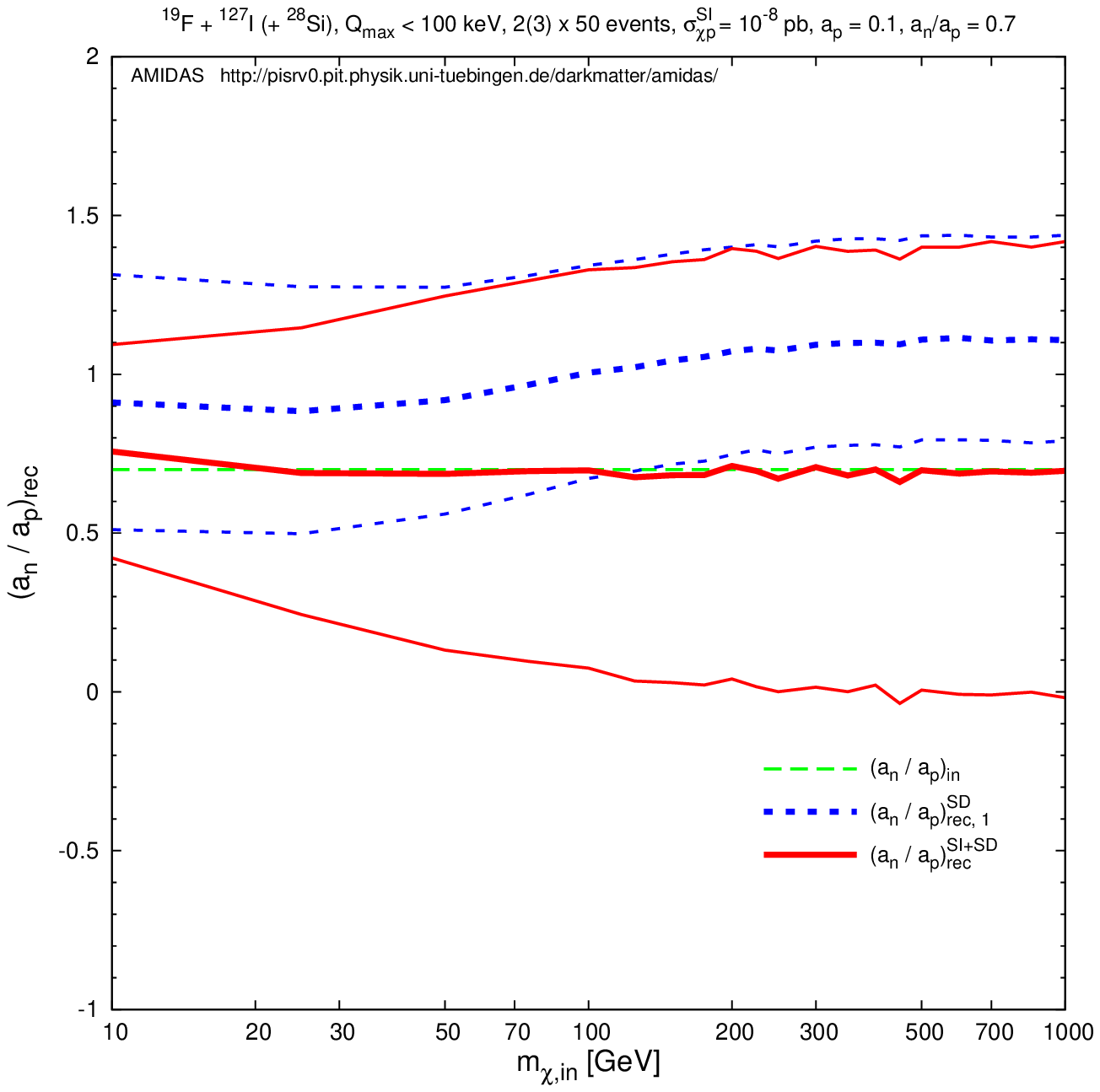}
\includegraphics[width=8.5cm]{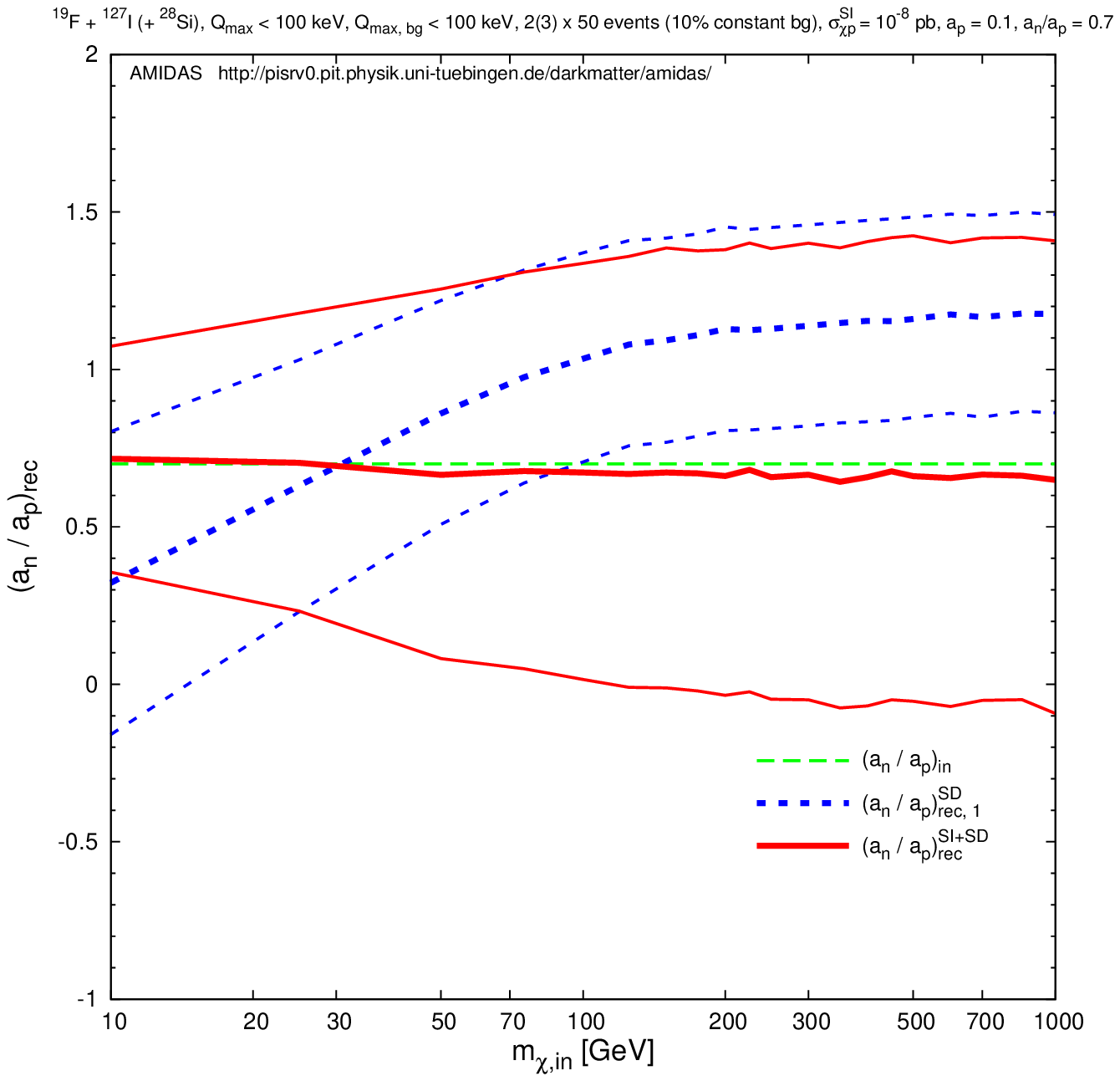} \\
\vspace{0.5cm}
\includegraphics[width=8.5cm]{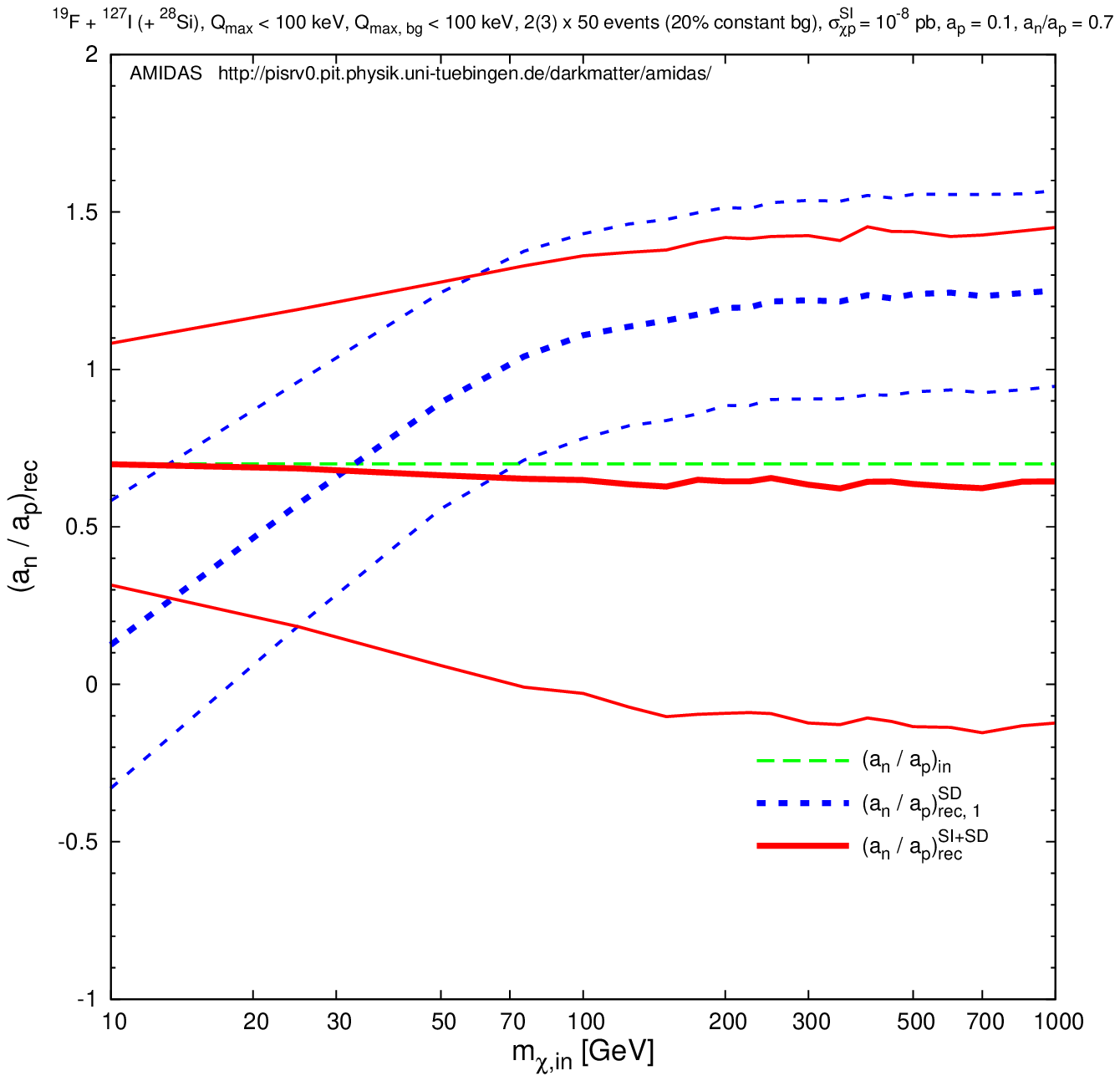}
\includegraphics[width=8.5cm]{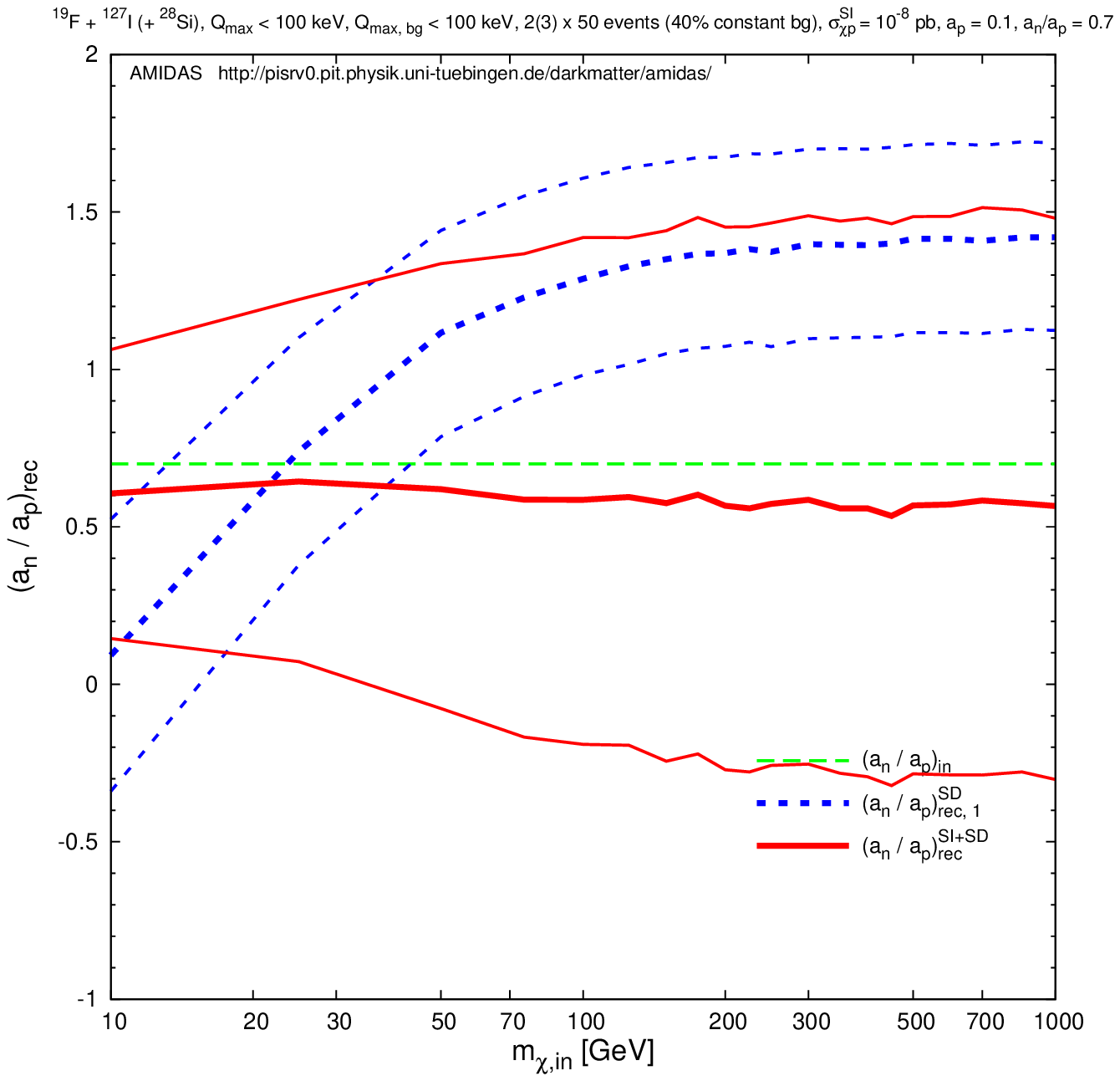} \\
\vspace{-0.25cm}
\end{center}
\caption{
 As in Figs.~\ref{fig:ranapSD-08-mchi-ex}
 and \ref{fig:ranapSISD-08-mchi-ex},
 except that the constant background spectrum
 has been used here.
 The dashed blue (solid red) curves indicate
 the reconstructed $\armn / \armp$ ratios
 estimated by Eq.~(\ref{eqn:ranapSD}) with $n = 1$
 (Eq.~(\ref{eqn:ranapSISD}))
 with $r_{(X, Y)}(Q_{{\rm min}, (X, Y)})$
 as functions of the input $\armn / \armp$ ratio.
}
\label{fig:ranapSISD-08-mchi-rec-const}
\end{figure}
\begin{figure}[t!]
\begin{center}
\includegraphics[width=8.5cm]{rsigmaSDpSI-08-ranap-ex-00}
\includegraphics[width=8.5cm]{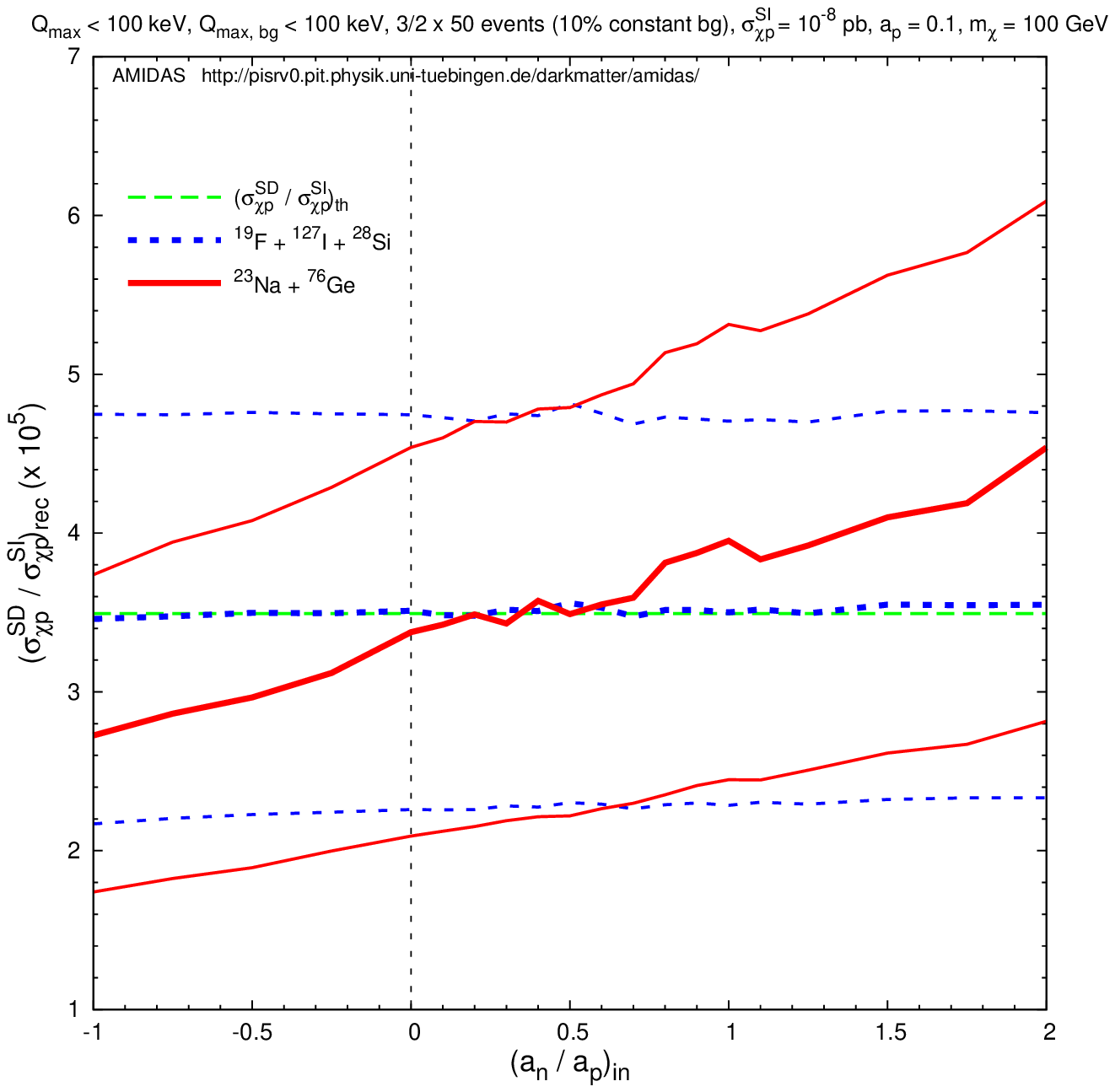} \\
\vspace{0.5cm}
\includegraphics[width=8.5cm]{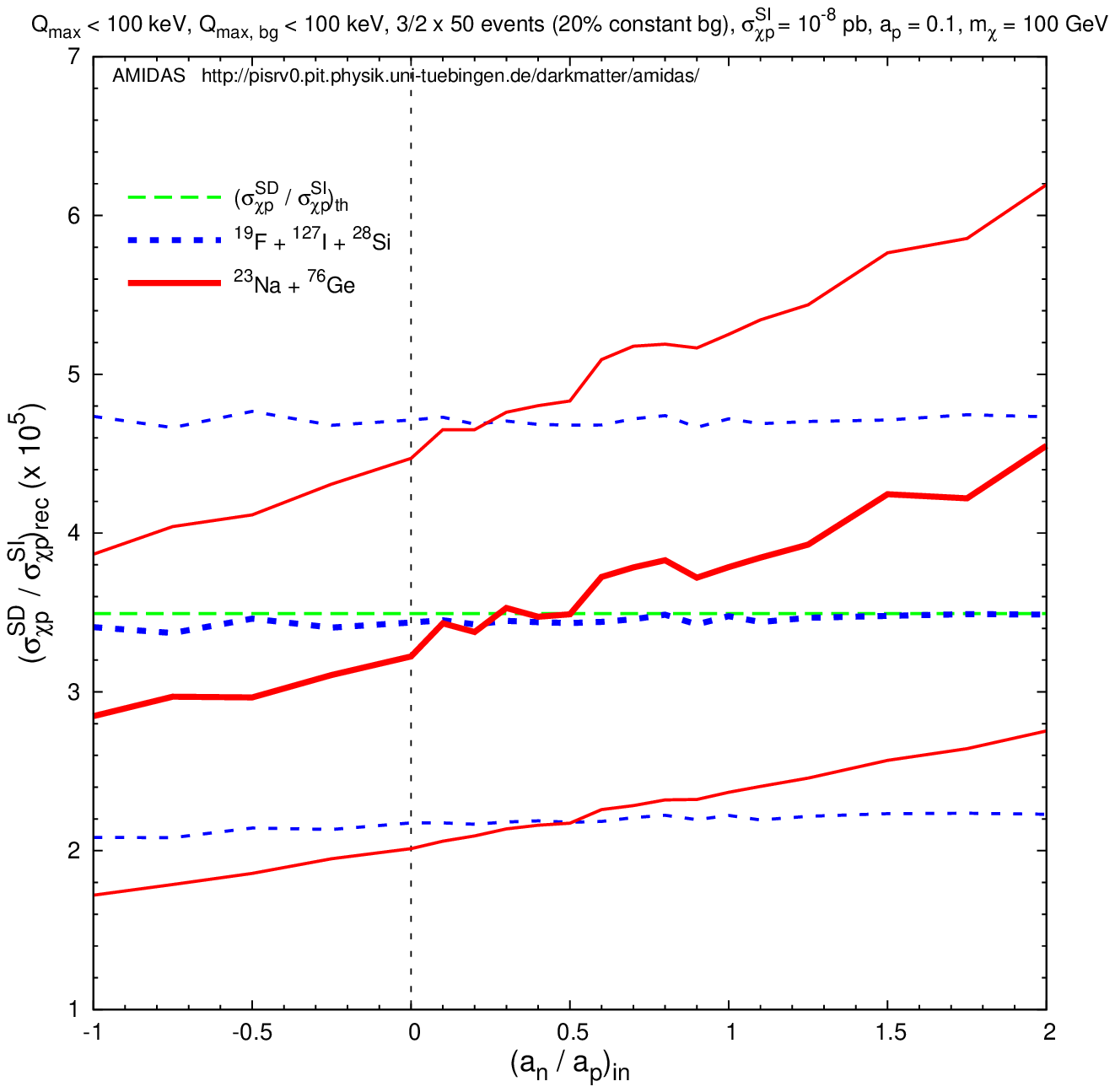}
\includegraphics[width=8.5cm]{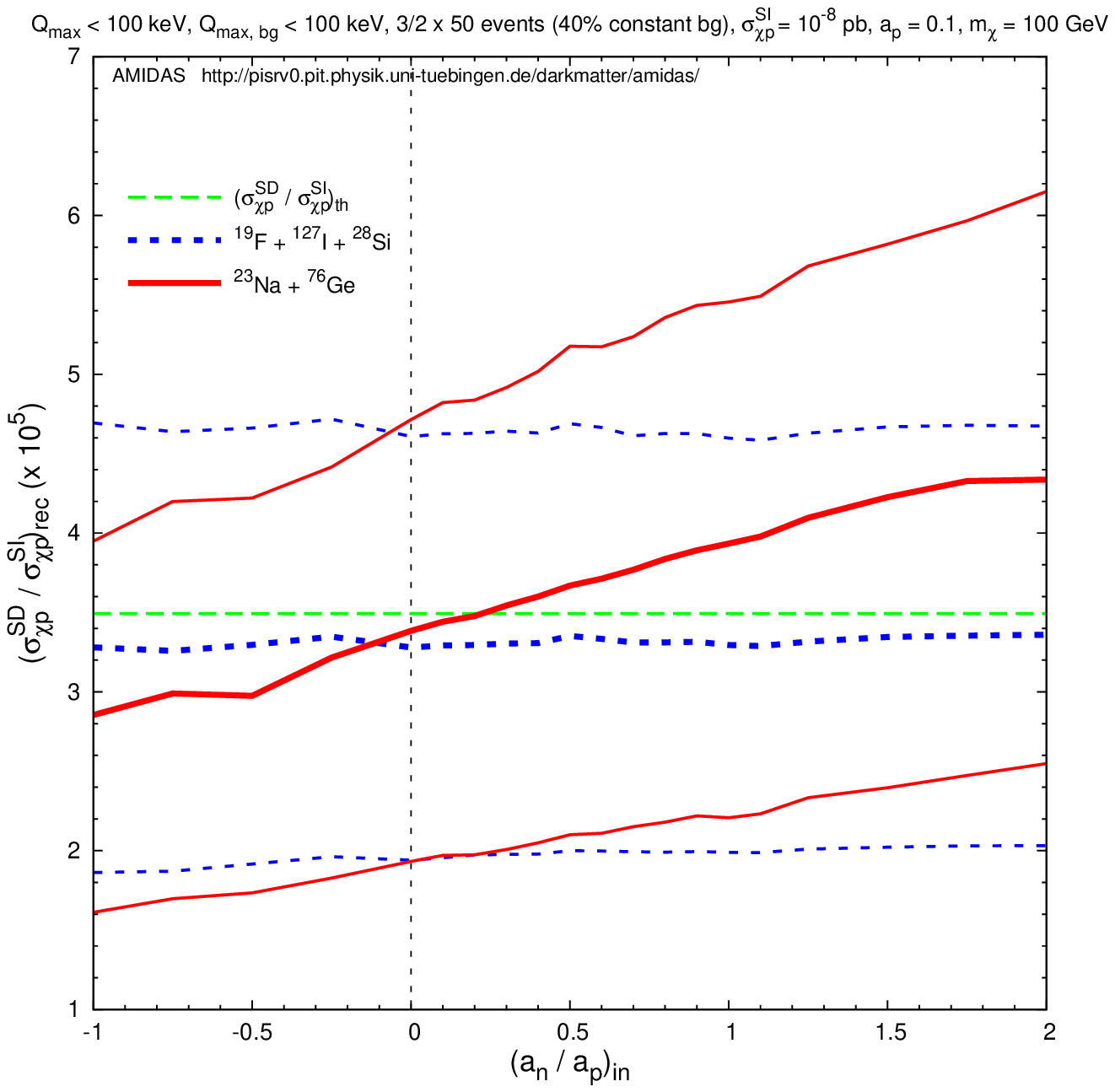} \\
\vspace{-0.25cm}
\end{center}
\caption{
 As in Figs.~\ref{fig:rsigmaSDpSI-08-ranap-ex},
 except that the constant background spectrum
 has been used here.
}
\label{fig:rsigmaSDpSI-08-ranap-const}
\end{figure}
\begin{figure}[t!]
\begin{center}
\includegraphics[width=8.5cm]{rsigmaSDpSI-08-mchi-ex-00}
\includegraphics[width=8.5cm]{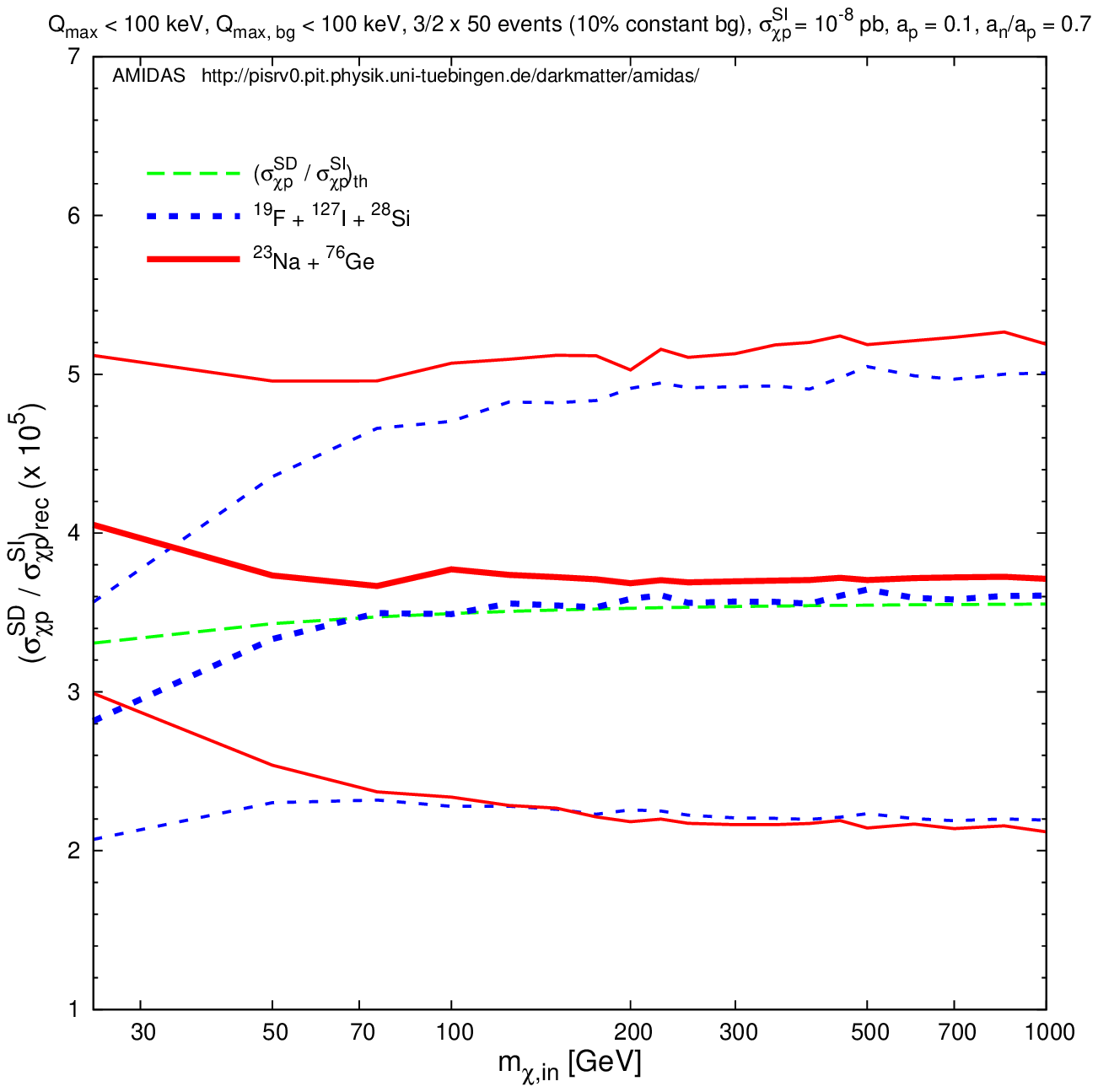} \\
\vspace{0.5cm}
\includegraphics[width=8.5cm]{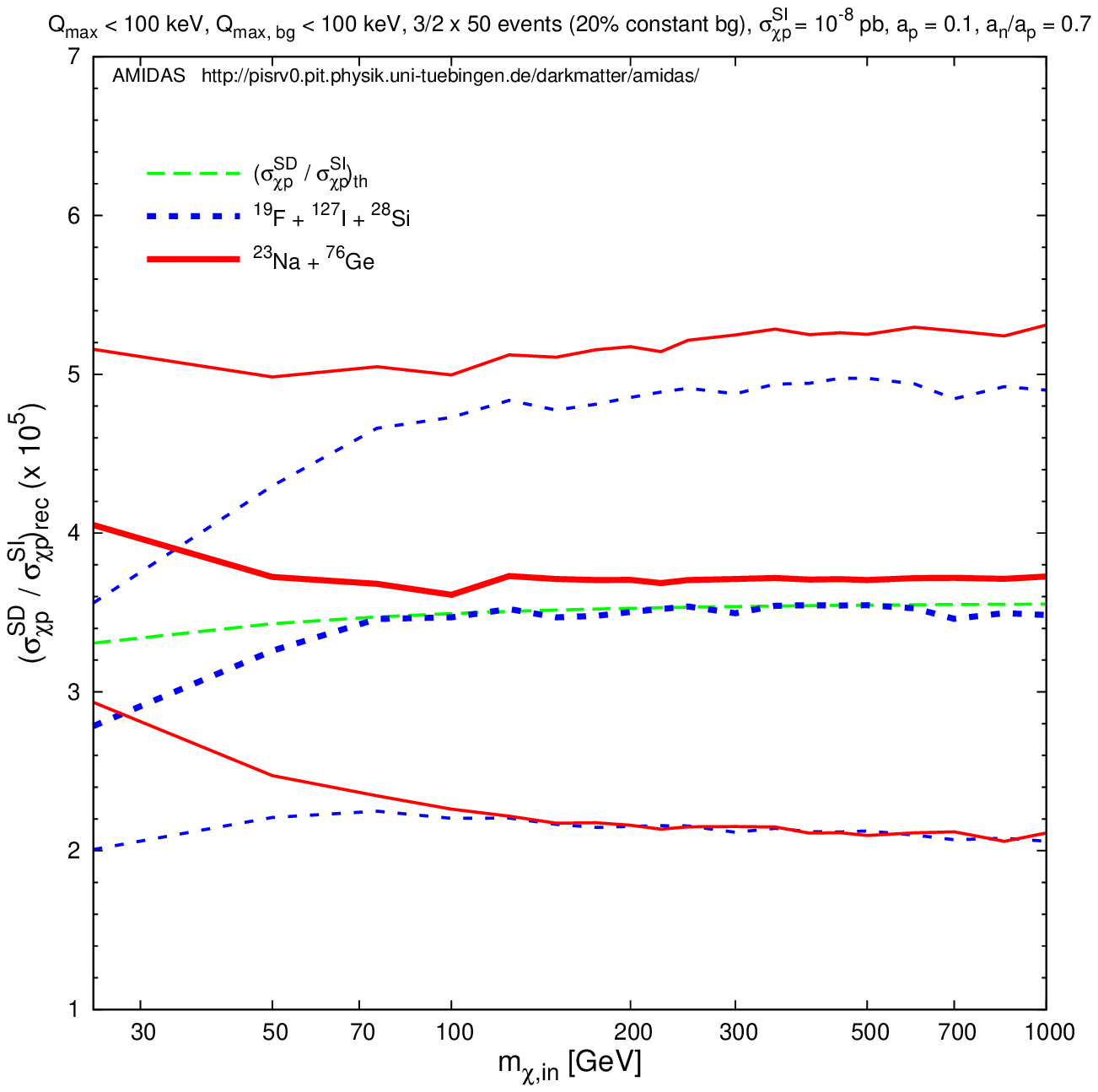}
\includegraphics[width=8.5cm]{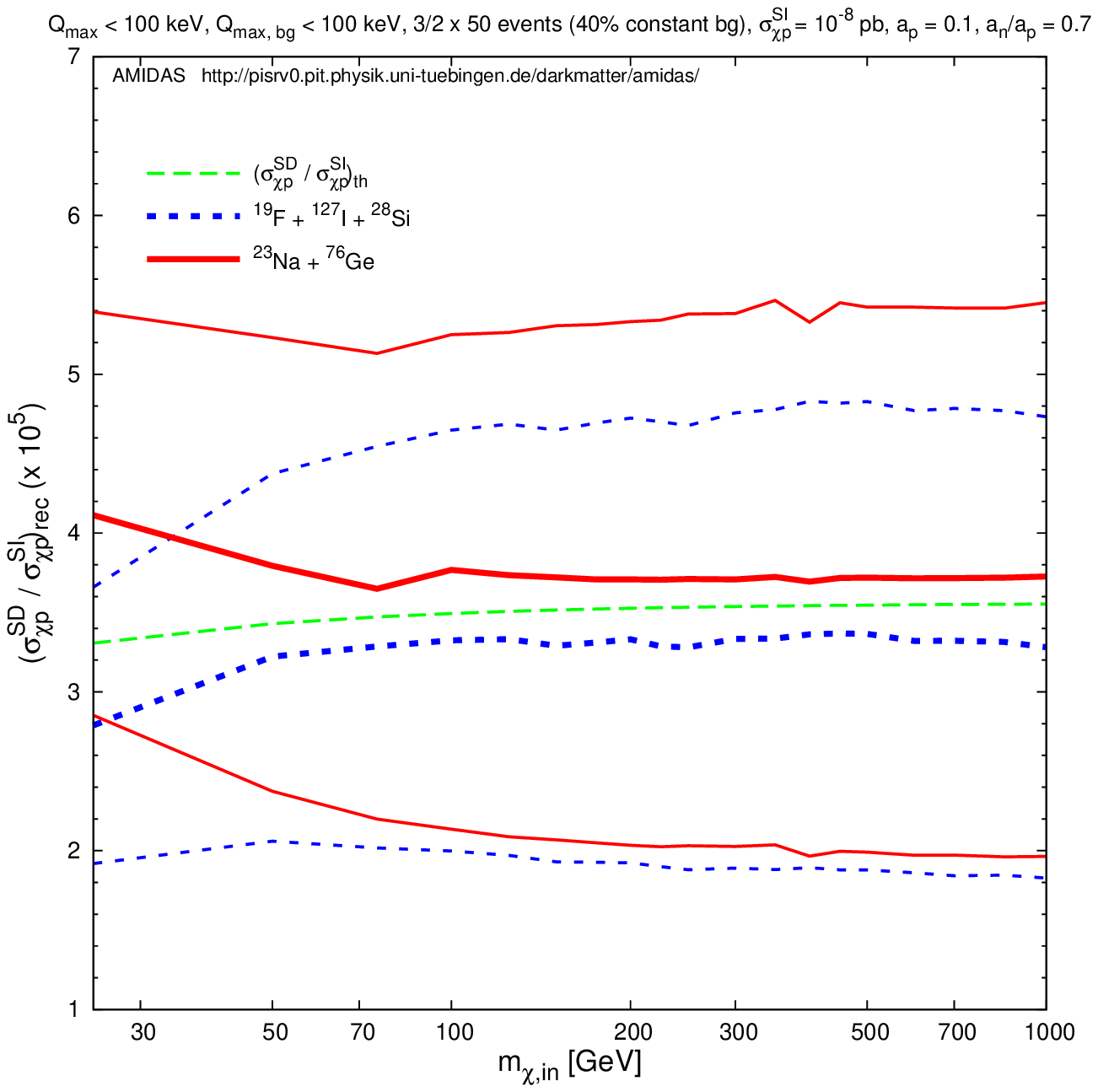} \\
\vspace{-0.25cm}
\end{center}
\caption{
 As in Figs.~\ref{fig:rsigmaSDpSI-08-mchi-ex},
 except that the constant background spectrum
 has been used here.
}
\label{fig:rsigmaSDpSI-08-mchi-const}
\end{figure}
\begin{figure}[t!]
\begin{center}
\includegraphics[width=8.5cm]{rsigmaSDnSI-08-ranap-ex-00}
\includegraphics[width=8.5cm]{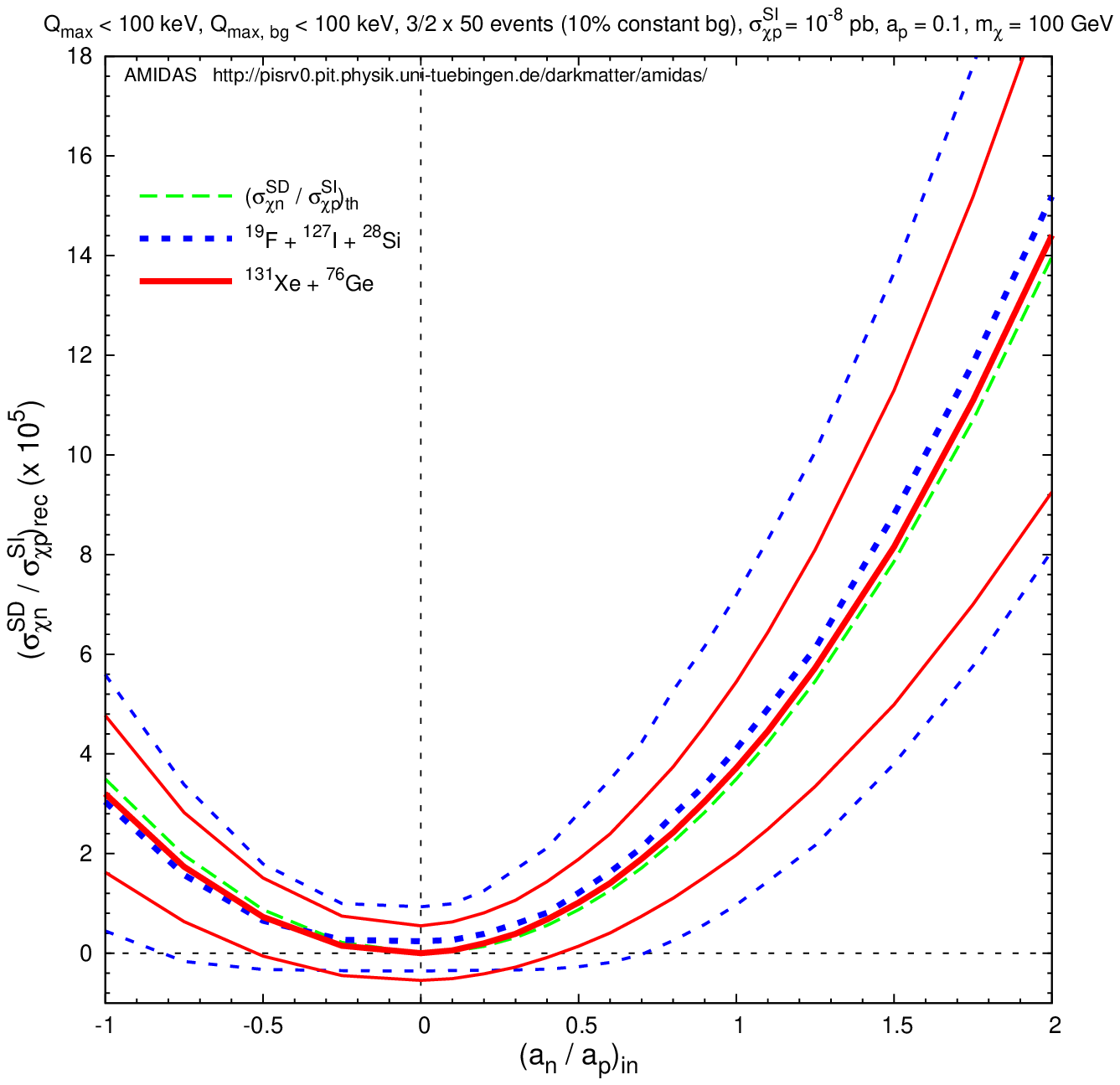} \\
\vspace{0.5cm}
\includegraphics[width=8.5cm]{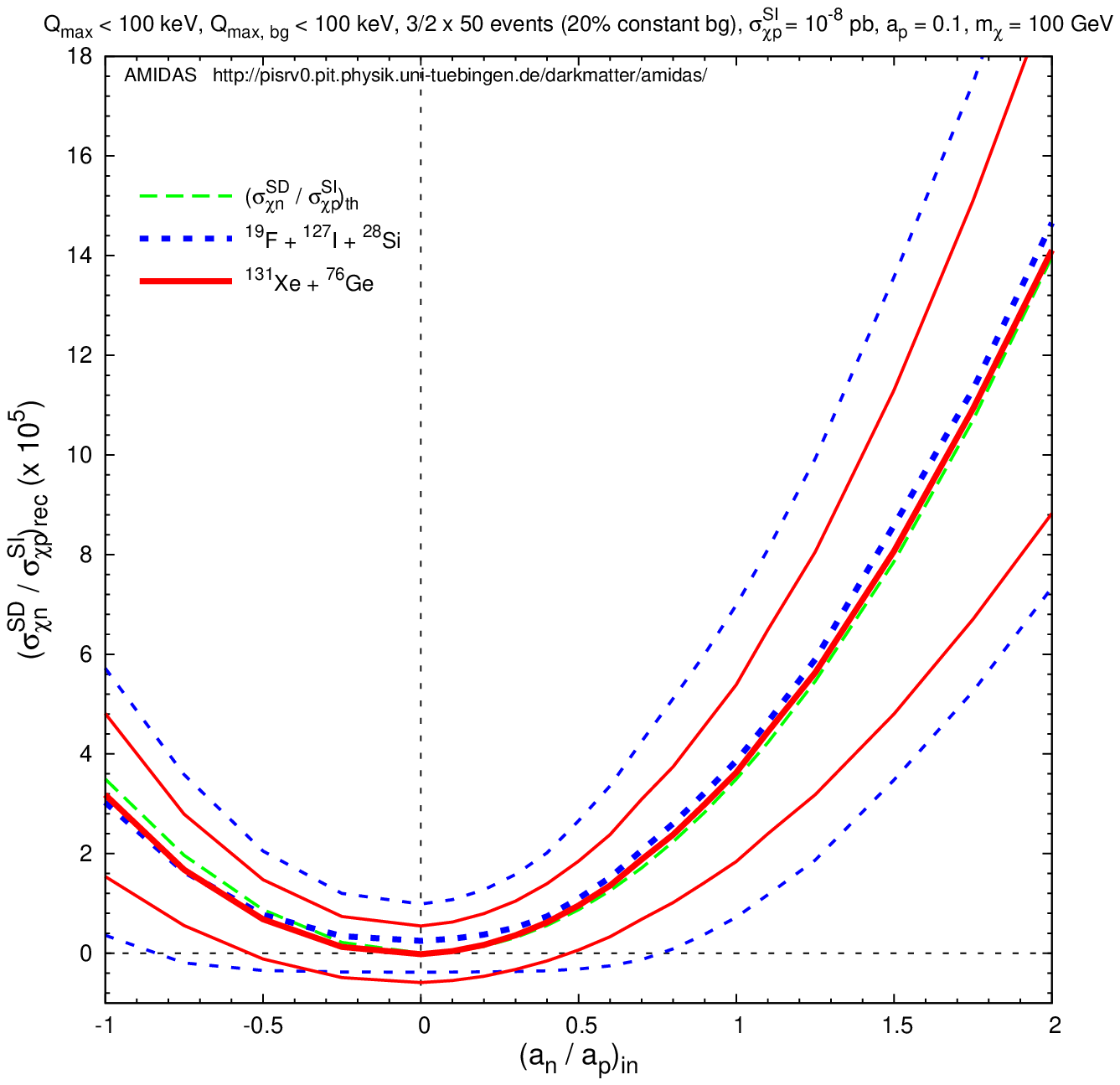}
\includegraphics[width=8.5cm]{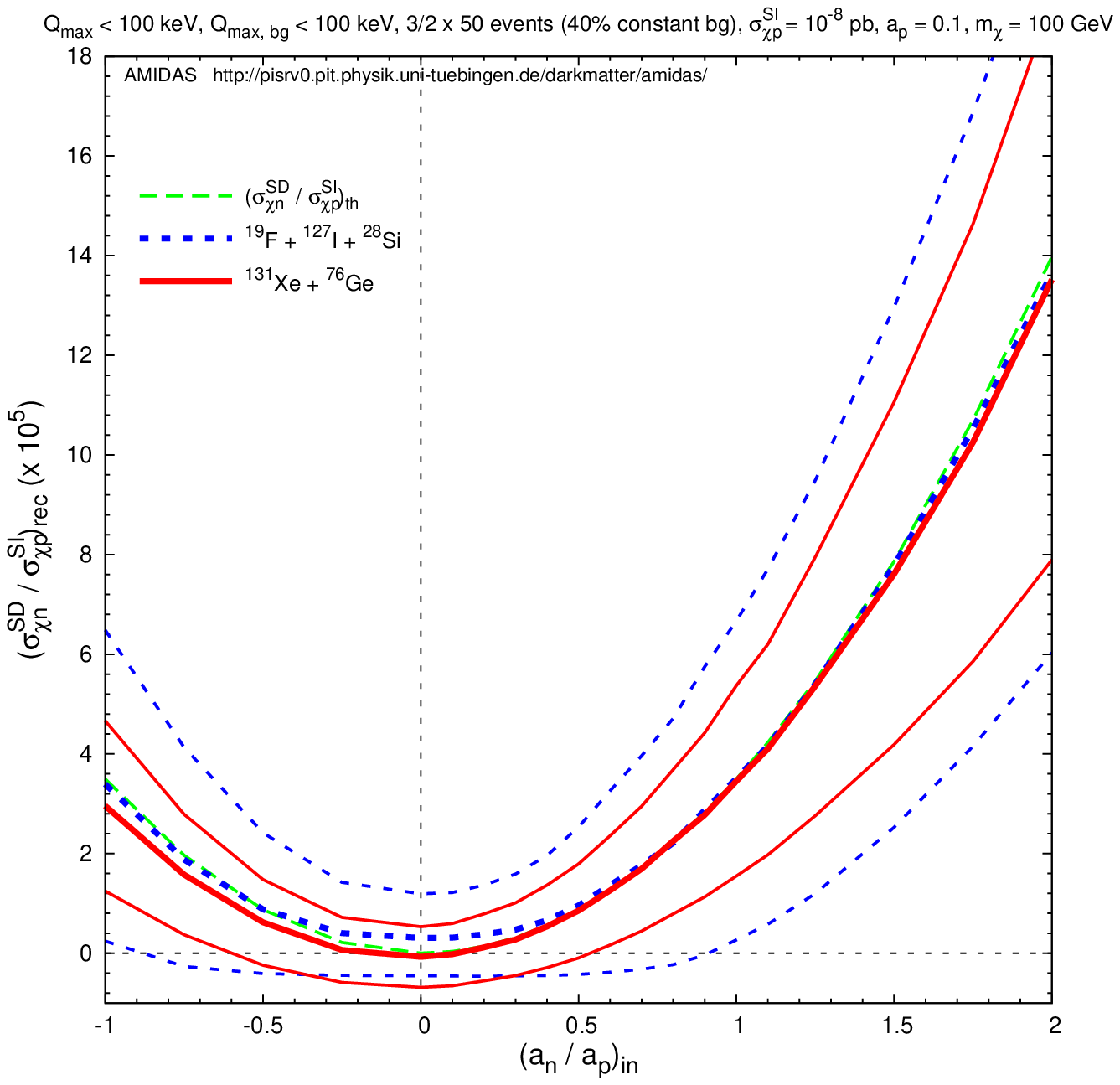} \\
\vspace{-0.25cm}
\end{center}
\caption{
 As in Figs.~\ref{fig:rsigmaSDnSI-08-ranap-ex},
 except that the constant background spectrum
 has been used here.
}
\label{fig:rsigmaSDnSI-08-ranap-const}
\end{figure}
\begin{figure}[t!]
\begin{center}
\includegraphics[width=8.5cm]{rsigmaSDnSI-08-mchi-ex-00}
\includegraphics[width=8.5cm]{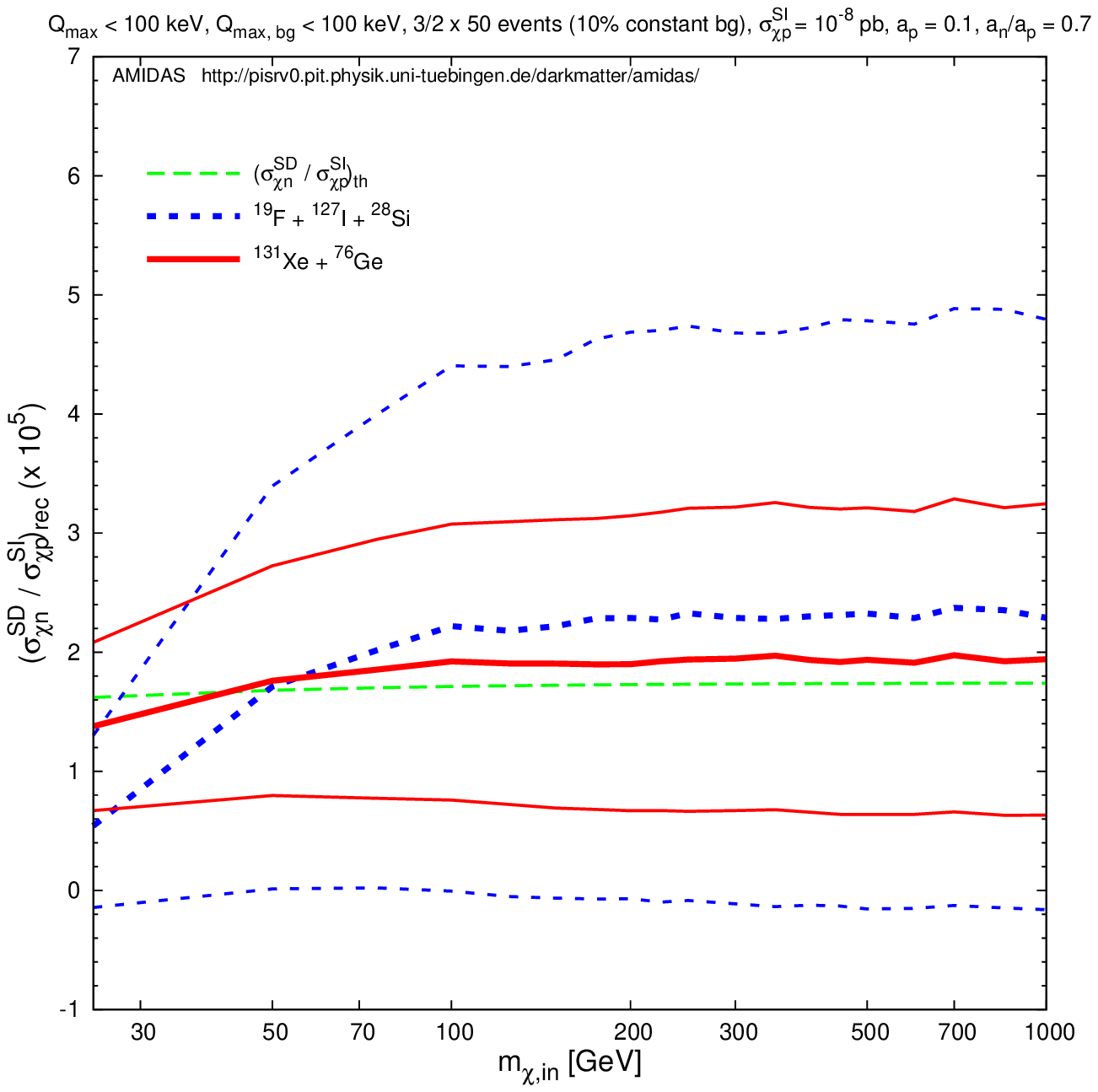} \\
\vspace{0.5cm}
\includegraphics[width=8.5cm]{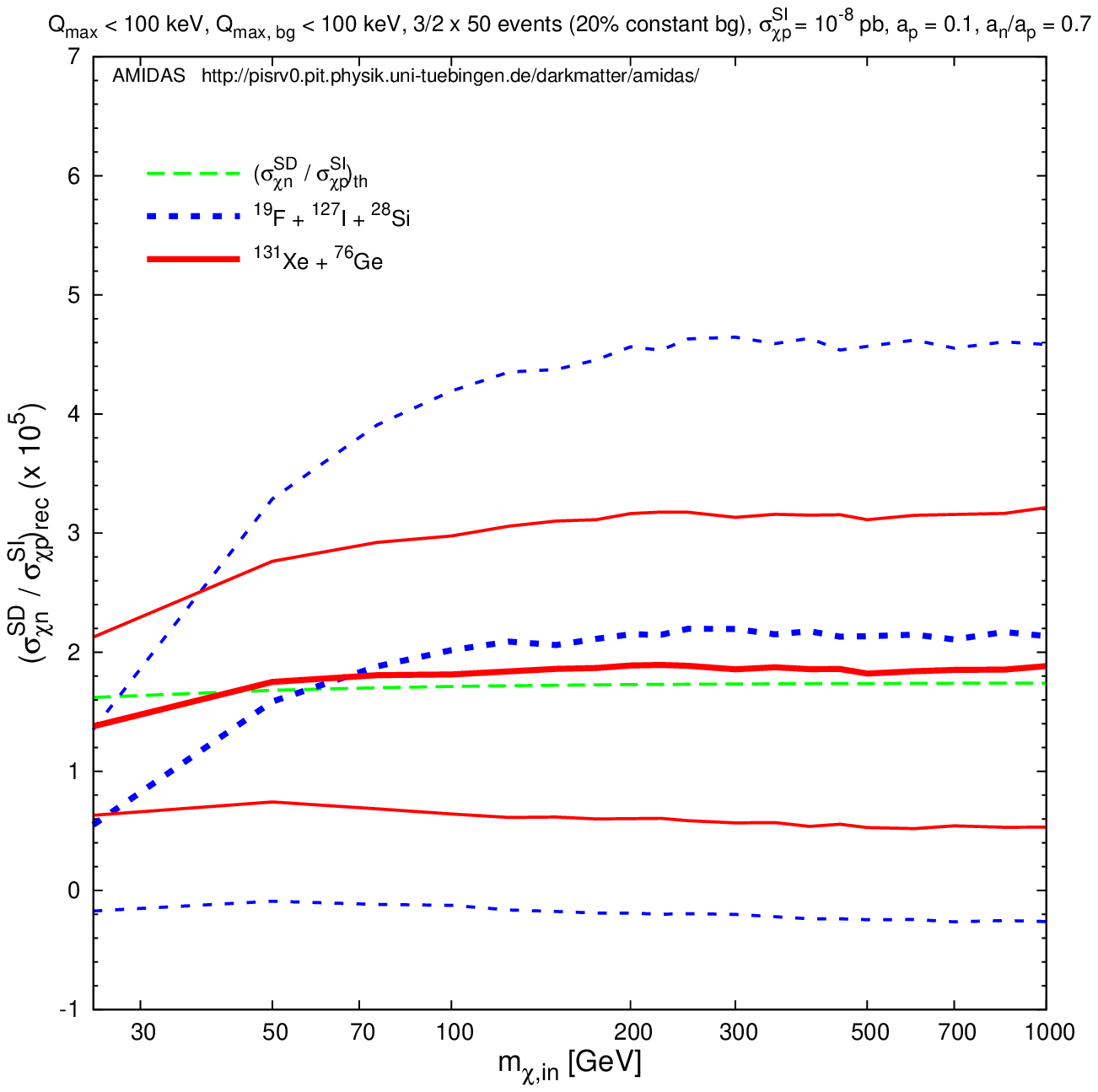}
\includegraphics[width=8.5cm]{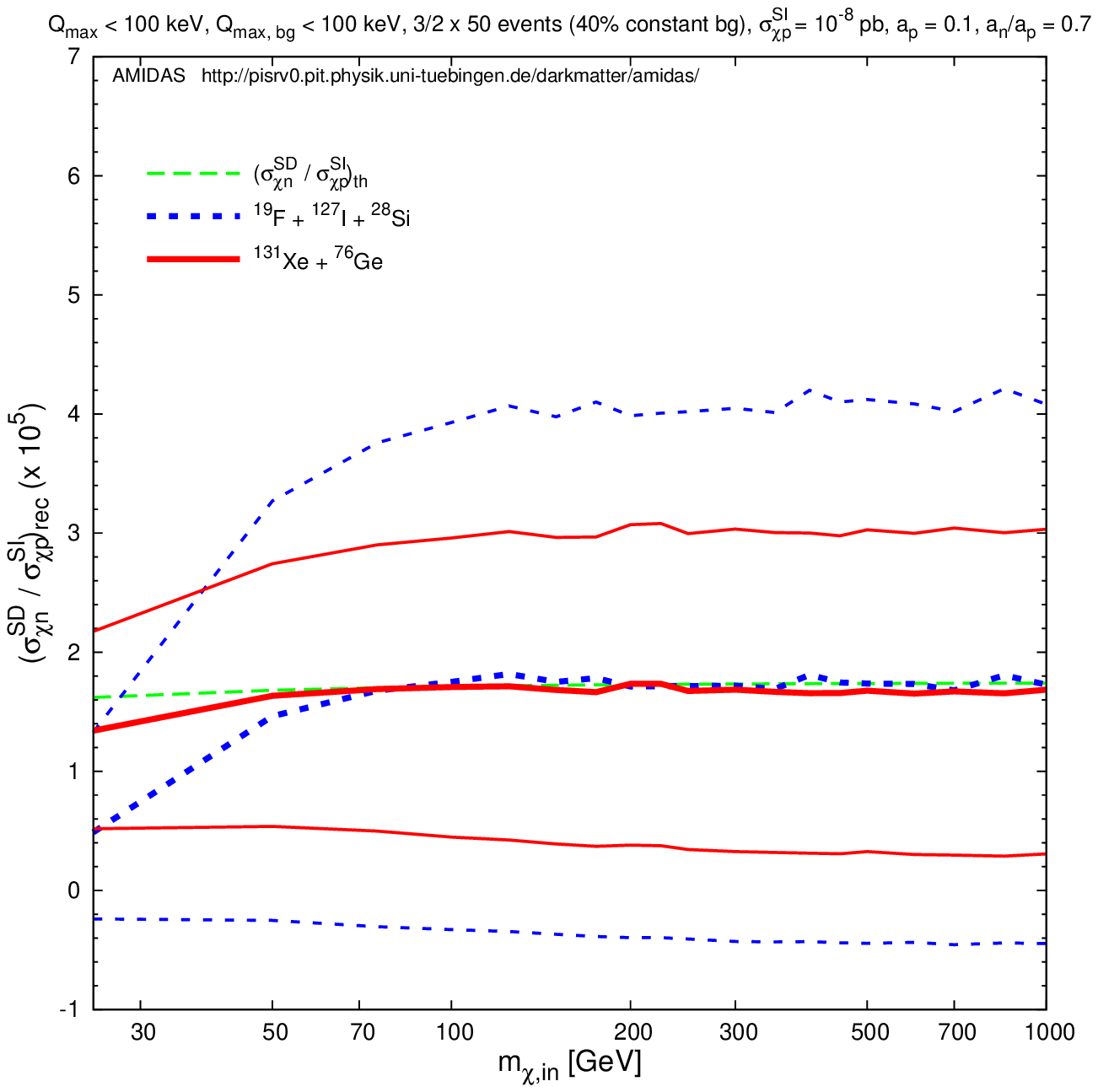} \\
\vspace{-0.25cm}
\end{center}
\caption{
 As in Figs.~\ref{fig:rsigmaSDnSI-08-mchi-ex},
 except that the constant background spectrum
 has been used here.
}
\label{fig:rsigmaSDnSI-08-mchi-const}
\end{figure}

 In this and the next subsections
 I consider the case
 with a {\em non--negligible} SI WIMP--nucleus interaction.
 The input SI WIMP--nucleon cross section
 and the input SD WIMP--proton coupling
 have been set as \mbox{$\sigmapSI = 10^{-8}$ pb} and $\armp = 0.1$,
 respectively.

 At first I show in Figs.~\ref{fig:ranapSD-08-ranap-ex}
 the reconstructed $\armn / \armp$ ratios
 estimated by Eq.~(\ref{eqn:ranapSD})
 and the lower and upper bounds of
 their 1$\sigma$ statistical uncertainties
 estimated by Eq.~(\ref{eqn:sigma_ranapSD})
 with $r_{(X, Y)}(Q_{{\rm min}, (X, Y)})$ (dashed blue)
 and with $r_{(X, Y)}(Q_{s, 1, (X, Y)})$ (solid red)
 as functions of the input $\armn / \armp$ ratio
 together (only cases with $n = 1$).
 Since for this simulation setup
 the SD WIMP--nucleus cross section
 doesn't really dominate over the SI one,
 by using data sets with pure WIMP signals
 (no background events, top left frame),
 the reconstructed $\armn / \armp$ ratios
 are (strongly) {\em overestimated},
 especially for the input $\armn / \armp \le 0$.
 However,
 once some background events mix into our data sets,
 the reconstructed $\armn / \armp$ ratios
 become smaller
 (and even closer to the input (true) values,
  cf.~Figs.~\ref{fig:ranapSD-ranap-rec-ex}
  to \ref{fig:ranapSD-ranap-ex}).
 This seems to indicate that,
 with data sets of $\lsim~40\%$ residue background events
 in the analyzed data sets,
 one could in principle reconstruct
 the ratio between two SD WIMP--nucleon couplings
 by simply assuming a dominant SD WIMP--nucleus interaction
 (even though this assumption is not correct)
 and using Eq.~(\ref{eqn:ranapSD}).

 Nevertheless,
 once we consider both SI and SD WIMP interactions
 and thus use Eqs.~(\ref{eqn:ranapSISD})
 and (\ref{eqn:sigma_ranapSISD})
 to analyze the same data sets,
 as shown in Figs.~\ref{fig:ranapSISD-08-ranap-sh-rec-ex},
 the $\armn / \armp$ ratios
 could be reconstructed (much) better
 with data sets of $\lsim~20\%$ residue background events,
 even though the statistical uncertainties
 are pretty large
 for input $\armn / \armp \le 0$.
 For a WIMP mass \mbox{$\mchi = 100$ GeV} and $\armn / \armp = 0.75$,
 by using Eq.~(\ref{eqn:ranapSISD})
 with $r_{(X, Y, Z)}(Q_{s, 1, (X, Y, Z)})$
 to analyze data sets with a 10\% background ratio,
 the systematic deviation
 could be $\sim -12\%$
 with an $\sim 50\%$ statistical uncertainty.
 Moreover,
 with an increased background ratio,
 the (in)compatibility
 between the reconstructed $\armn / \armp$ ratios
 estimated by Eqs.~(\ref{eqn:ranapSD}) (dashed blue, $n = 1$)
 and (\ref{eqn:ranapSISD}) (solid red)
 becomes larger.
 Hence,
 as mentioned in the previous subsection,
 one could/should compare results
 reconstructed under different assumptions,
 with both $r_{(X, Y, Z)}(Q_{{\rm min}, (X, Y, Z)})$
 and $r_{(X, Y, Z)}(Q_{s, 1, (X, Y, Z)})$
 and with different $n$ values,
 in order to check the purity of the analyzed data sets
 (as well as the dominance of the SI or the SD WIMP interaction
  \cite{DMDDranap}).

 On the other hand,
 as done in the previous subsection,
 in order to check the WIMP--mass independence
 of the reconstructed results
 with non--negligible background events,
 I show the $\armn / \armp$ ratios
 estimated by Eq.~(\ref{eqn:ranapSD})
 with $r_{(X, Y)}(Q_{{\rm min}, (X, Y)})$ (dashed blue)
 and with $r_{(X, Y)}(Q_{s, 1, (X, Y)})$ (solid red)
 as functions of the input WIMP mass $\mchi$
 together (only cases with $n = 1$)
 in Figs.~\ref{fig:ranapSD-08-mchi-ex}
 as well as those
 estimated by Eq.~(\ref{eqn:ranapSISD})
 with $r_{(X, Y, Z)}(Q_{{\rm min}, (X, Y, Z)})$ (dashed blue)
 and with $r_{(X, Y, Z)}(Q_{s, 1, (X, Y, Z)})$ (solid red)
 as functions of the input WIMP mass $\mchi$
 in Figs.~\ref{fig:ranapSISD-08-mchi-ex}.

 As shown earlier,
 for WIMP masses \mbox{$\mchi~\gsim~50$ GeV},
 the reconstructed $\armn / \armp$ ratios
 are indeed (almost) independent of $\mchi$.
 And,
 with data sets of $\lsim~20\%$ background ratios,
 using Eq.~(\ref{eqn:ranapSISD})
 (solid red in Figs.~\ref{fig:ranapSISD-08-mchi-ex})
 and estimating with $r_{(X, Y, Z)}(Q_{s, 1, (X, Y, Z)})$
 could offer the best reconstructed $\armn / \armp$ ratios.
 Meanwhile,
 for WIMP masses \mbox{$\mchi~\lsim~50$ GeV},
 Figs.~\ref{fig:ranapSISD-08-mchi-ex} show also that
 the $\armn / \armp$ ratios reconstructed
 by Eq.~(\ref{eqn:ranapSISD})
 with $r_{(X, Y, Z)}(Q_{{\rm min}, (X, Y, Z)})$
 could be the best results%
\footnote{
 See footnote \ref{footnote:rmin}.
}
 (with however a pretty large statistical uncertainty):
 for a WIMP mass \mbox{$\mchi = 25$ GeV}
 and an input $\armn / \armp = 0.7$,
 by using Eq.~(\ref{eqn:ranapSISD})
 with $r_{(X, Y, Z)}(Q_{{\rm min}, (X, Y, Z)})$
 to analyze data sets of a 10\% background ratio,
 the systematic deviation
 could be $\sim -25\%$
 with an $\sim 88\%$ statistical uncertainty.

 Furthermore,
 in Figs.~\ref{fig:ranapSISD-10-ranap-sh-rec-ex}
 we reduce the SI WIMP--nucleon cross section
 two orders of magnitude smaller:
 \mbox{$\sigmapSI = 10^{-10}$ pb}.
 It can be found here that,
 although the $\armn / \armp$ ratios
 reconstructed by Eqs.~(\ref{eqn:ranapSD}) (dashed blue, $n = 1$)
 and (\ref{eqn:ranapSISD}) (solid red)
 with $r_{(X, Y, Z)}(Q_{s, 1, (X, Y, Z)})$
 match much better than those shown
 in Figs.~\ref{fig:ranapSISD-08-ranap-sh-rec-ex},
 with an increased background ratio
 the $\armn / \armp$ ratios could be
 a bit more strongly underestimated
 by using Eq.~(\ref{eqn:ranapSISD}).
 Nevertheless,
 Figs.~\ref{fig:ranapSISD-08-ranap-sh-rec-ex},
 Figs.~\ref{fig:ranapSISD-08-mchi-ex}, and
 Figs.~\ref{fig:ranapSISD-10-ranap-sh-rec-ex} show that,
 it doesn't matter whether
 the SD WIMP--nucleus interaction really dominates
 over the SI one or not,
 by using Eq.~(\ref{eqn:ranapSISD})
 one could in principle always reconstruct
 the ratio of the SD WIMP coupling
 on neutrons to that on protons
 with data sets of $\lsim~20\%$ residue background events
 pretty well.
 But,
 the larger the relative strength between
 the SD WIMP--nucleus interaction to the SI one,
 the smaller the systematic deviations
 as well as the statistical uncertainties.
 For $\sigmapSI = 10^{-10}$ pb
 with a WIMP mass \mbox{$\mchi = 100$ GeV}
 and $\armn / \armp = 0.75$,
 by using data sets of a 10\% background ratio,
 the systematic deviation
 could be $\sim -10\%$
 with an $\sim 43\%$ statistical uncertainty.

\subsection{Reconstructed \boldmath$\sigma_{\chi ({\rm p, n})}^{\rm SD} / \sigmapSI$}

 In Figs.~\ref{fig:rsigmaSDpSI-08-ranap-ex}
 and \ref{fig:rsigmaSDpSI-08-mchi-ex}
 I show the reconstructed $\sigmapSD / \sigmapSI$ ratios
 and the lower and upper bounds of
 their 1$\sigma$ statistical uncertainties
 as functions of the input $\armn / \armp$ ratio
 as well as of the input WIMP mass $\mchi$,
 respectively.
 The dashed blue curves indicate the ratios
 estimated by Eq.~(\ref{eqn:rsigmaSDpSI})
 with $\armn / \armp$ estimated by Eq.~(\ref{eqn:ranapSISD})
 ({\em not} by Eq.~(\ref{eqn:ranapSD})),
 whereas the solid red curves indicate the ratios
 estimated by Eq.~(\ref{eqn:rsigmaSDpSI_even}).
 $\rmXA{Ge}{76}$ has been chosen as the second target
 having only the SI interaction with WIMPs and combined with
 $\rmXA{Na}{23}$ for using Eq.~(\ref{eqn:rsigmaSDpSI_even}).

 It can be seen here that,
 interestingly,
 while the $\sigmapSD / \sigmapSI$ ratios
 reconstructed with $\rm F + I + Si$ targets
 become as usual more and more strongly {\em underestimated}
 with an increased background ratio,
 those reconstructed with $\rm Na + Ge$ targets
 become in contrast more and more strongly {\em overestimated}.
 Nevertheless,
 for WIMP masses \mbox{$\mchi~\gsim~25$ GeV},
 with $\lsim~20\%$ residue background events,
 the overlap of the 1$\sigma$ statistical uncertainty intervals
 estimated by two target combinations
 could cover the input (true) $\sigmapSD / \sigmapSI$ ratio well.
 For a WIMP mass \mbox{$\mchi = 100$ GeV} and $\armn / \armp = 0.7$
 (the theoretical value of \mbox{$\sigmapSD / \sigmapSI = 3.49 \times 10^5$}),
 by using $\rm F + I + Si$ ($\rm Na + Ge$) targets
 with data sets of a 20\% background ratio,
 one could in principle reconstruct
 the $\sigmapSD / \sigmapSI$ ratio
 with an $\sim -20\%$ ($\sim +21\%$) systematic deviation
 and an $\sim 31\%$ ($\sim 32\%$) statistical uncertainty.

 On the other hand,
 in Figs.~\ref{fig:rsigmaSDnSI-08-ranap-ex}
 and \ref{fig:rsigmaSDnSI-08-mchi-ex}
 I show the reconstructed $\sigmanSD / \sigmapSI$ ratios
 and the lower and upper bounds of
 their 1$\sigma$ statistical uncertainties
 as functions of the input $\armn / \armp$ ratio
 as well as of the input WIMP mass $\mchi$,
 respectively.
 The dashed blue curves indicate the ratios
 estimated by Eq.~(\ref{eqn:rsigmaSDpSI})
 with $\armn / \armp$ estimated by Eq.~(\ref{eqn:ranapSISD}),
 whereas the solid red curves indicate the ratios
 estimated by Eq.~(\ref{eqn:rsigmaSDpSI_even}).
 $\rmXA{Ge}{76}$ has been chosen as the second target
 having only the SI interaction with WIMPs and combined with
 $\rmXA{Xe}{131}$ for using Eq.~(\ref{eqn:rsigmaSDpSI_even}).

 It can be found here that,
 more interestingly,
 while the $\sigmanSD / \sigmapSI$ ratios
 reconstructed with $\rm F + I + Si$ targets
 become more and more strongly underestimated
 with an increased background ratio
 for all input $\armn / \armp$ values,
 those reconstructed with $\rm Xe + Ge$ targets
 become more and more strongly {\em underestimated}
 for $\armn / \armp~\gsim~0$
 and more and more strongly {\em overestimated}
 for $\armn / \armp~\lsim~0$.
 Nevertheless,
 for WIMP masses \mbox{$\mchi~\gsim~25$ GeV},
 with $\lsim~20\%$ residue background events,
 the (overlap of the) 1$\sigma$ statistical uncertainty intervals
 estimated by two target combinations
 could cover the input (true) $\sigmanSD / \sigmapSI$ ratio well.
 For a WIMP mass \mbox{$\mchi = 100$ GeV} and $\armn / \armp = 0.7$
 (the theoretical value of \mbox{$\sigmanSD / \sigmapSI = 1.71 \times 10^5$}),
 by using $\rm F + I + Si$ ($\rm Xe + Ge$) targets
 with data sets of a 10\% background ratio,
 one could in principle reconstruct
 the $\sigmapSD / \sigmapSI$ ratio
 with an $\sim -32\%$ ($\sim -3.2\%$) systematic deviation
 and an $\sim 120\%$ ($\sim 62\%$) statistical uncertainty.

\section{Results of the reconstructed ratios of
         WIMP--nucleon couplings/cross sections II:
         with constant background spectra}

 In this section
 I show simulation results
 with residue background events
 generated by the constant spectrum
 given in Eq.~(\ref{eqn:dRdQ_bg_const})
 and compare them
 with those shown in the previous section.
 Some general rules
 about the effects of different background sources
 will also be discussed.

\subsection{Reconstructed \boldmath$(\armn / \armp)_{\pm, n}^{\rm SD}$}

 As in the previous section,
 I consider at first
 the case of a dominant SD WIMP--nucleus interaction.

 In Figs.~\ref{fig:ranapSD-ranap-sh-rec-const}
 I show the reconstructed $\armn / \armp$ ratios
 and the lower and upper bounds of
 their 1$\sigma$ statistical uncertainties
 estimated by Eqs.~(\ref{eqn:ranapSD})
 and (\ref{eqn:sigma_ranapSD})
 with $n = -1$ (dashed blue), 1 (solid red),
 and 2 (dash--dotted cyan)
 as functions of the input $\armn / \armp$ ratio.
 Note that
 only results with the counting rates
 at the shifted points of the first $Q-$bin,
 $r_{(X, Y)}(Q_{s, 1, (X, Y)})$,
 are shown here,
 because their systematic deviations
 due to the non--negligible background events
 as well as their statistical uncertainties
 (for non--zero experimental threshold energies)
 are (much) smaller%
\footnote{
 See Figs.~\ref{fig:ranapSD-ranap-ex}
 and discussions there.
}.

 It can be found here that,
 firstly,
 in contrast to the results
 shown in Figs.~\ref{fig:ranapSD-ranap-rec-ex}
 to \ref{fig:ranapSD-ranap-ex},
 the $\armn / \armp$ ratios
 reconstructed with $n = 1$ (solid red)
 and 2 (dash--dotted cyan)
 are now {\em overestimated}.
 This should be caused by the background contribution
 to {\em high} energy ranges.
 Remind that,
 as shown in Figs.~\ref{fig:dRdQ-bg-ex-F-000-100-100}
 and \ref{fig:dRdQ-bg-const-I-000-100-100},
 while an exponential(--like) background spectrum
 contributes (almost) only/mainly to low energy ranges,
 a constant/(approximately) flat one
 contributes mainly to high energy ranges.
 This indicates in turn that,
 while the reconstructed $\armn / \armp$ ratio
 would be {\em underestimated}
 by using data sets with residue background events
 existing mostly in low energy ranges,
 e.g., electronic noise or
 incompletely charged surface events,
 the $\armn / \armp$ ratio
 would be {\em overestimated}
 by using data sets with backgrounds
 relatively mainly in high energy ranges,
 e.g., cosmic rays and cosmic--ray induced $\gamma$-rays
 with energies of \mbox{$\cal O$(100) keV}.

 However,
 interestingly and importantly,
 the $\armn / \armp$ ratio
 reconstructed with $n = -1$ (dashed blue)
 in Figs.~\ref{fig:ranapSD-ranap-sh-rec-const}
 is almost {\em not} affected
 by the {\em constant} background spectrum!
 The reason can be understood as follows.
 The $(\armn / \armp)_{-, -1, {\rm sh}}^{\rm SD}$ ratio
 has been estimated with $\calR_{J, -1, (X, Y)}$
 given in Eq.~(\ref{eqn:JmaX})
 and is in fact a function of only
 $r_{(X, Y)}(Q_{s, 1, (X, Y)})$.
 Since the constant background spectra contribute
 only (very) small amounts to the lowest energy ranges
 (see Figs.~\ref{fig:dRdQ-bg-const-I-000-100-100}),
 $r_{(X, Y)}(Q_{s, 1, (X, Y)})$
 estimated by using events recorded in the first $Q-$bins
 would {\em not} change or only {\em very slightly},
 and in turn also the reconstructed $\armn / \armp$ ratio.
 
 In Figs.~\ref{fig:ranapSD-mchi-rec-const}
 and \ref{fig:ranapSD-mchi-sh-rec-const},
 I show the reconstructed $\armn / \armp$ ratios
 as functions of the input WIMP mass.
 Both of them show that,
 while the $\armn / \armp$ ratios
 reconstructed with $n = 1$ (solid red)
 and 2 (dash--dotted cyan)
 are more and more strongly overestimated
 with an increased background ratio,
 the ratios reconstructed with $n = -1$ (dashed blue)
 just become a little bit smaller
 (\mbox{$\sim$ 15\%} smaller
  for a background ratio of 40\%
  and an input WIMP mass of 1 TeV,
  compared to the results
  with no background events)
 and the statistical uncertainties
 also grow only slightly
 ($\lsim~45\%$ larger).

\subsection{Reconstructed \boldmath$(\armn / \armp)_{\pm}^{\rm SI + SD}$}

 In this subsection
 I consider the reconstruction of the $\armn / \armp$ ratio
 with a non--zero SI WIMP--nucleus cross section.

 As shown in the previous subsection,
 Figs.~\ref{fig:ranapSISD-08-ranap-sh-rec-const}
 and \ref{fig:ranapSISD-08-mchi-rec-const}
 show that,
 while with an increased background ratio
 the $\armn / \armp$ ratios
 reconstructed by Eq.~(\ref{eqn:ranapSD})
 (dashed blue, with $n = 1$ and
  $r_{(X, Y)}(Q_{{\rm min}, (X, Y)} = 0)$
  or $r_{(X, Y)}(Q_{s, 1, (X, Y)})$)
 are more and more strongly overestimated
 caused by contributions of the constant background spectrum
 to high energy ranges,
 those reconstructed by Eq.~(\ref{eqn:ranapSISD})
 (solid red)
 just become a little bit smaller
 ($\sim$ 9\% smaller
  for a background ratio of 40\%
  and an input WIMP mass of 100 GeV,
  compared to the results
  with no background events)
 and the statistical uncertainties
 also grow only slightly
 ($\sim$ 34\% larger).

 This is simply because that
 $(\armn / \armp)_{-}^{\rm SI + SD}$
 given in Eq.~(\ref{eqn:ranapSISD})
 depends only on $c_{{\rm p}, (X, Y)}$,
 which are in turn just the functions of
 $r_{(X, Y, Z)}(Q_{{\rm min}, (X, Y, Z)})$
 or $r_{(X, Y, Z)}(Q_{s, 1, (X, Y, Z)})$
 and can be estimated by using events
 in the lowest energy ranges.
 Hence,
 once residue background events
 exist regularly between the experimental
 minimal and maximal cut--off energies
 or (even better) (mostly) in high energy ranges,
 one can in principle estimate
 $r_{(X, Y, Z)}(Q_{{\rm min}, (X, Y, Z)})$
 and $r_{(X, Y, Z)}(Q_{s, 1, (X, Y, Z)})$
 as well as
 the ratio between two SD WIMP--nucleon couplings
 (pretty) precisely
 by using Eq.~(\ref{eqn:ranapSISD})
 {\em without} worrying about
 the non--negligible backgrounds.

\subsection{Reconstructed \boldmath$\sigma_{\chi ({\rm p, n})}^{\rm SD} / \sigmapSI$}

 Finally,
 we check the reconstruction of the ratios
 between the SD and SI WIMP--proton(neutron) cross sections
 by using Eqs.~(\ref{eqn:rsigmaSDpSI}) and
 (\ref{eqn:rsigmaSDpSI_even})
 with the constant background spectrum.

 In Figs.~\ref{fig:rsigmaSDpSI-08-ranap-const}
 and \ref{fig:rsigmaSDpSI-08-mchi-const}
 we can find that,
 as observed in Sec.~4.3,
 while the $\sigmapSD / \sigmapSI$ ratios
 reconstructed with $\rm F + I + Si$ targets
 by Eq.~(\ref{eqn:rsigmaSDpSI})
 with $\armn / \armp$ reconstructed by Eq.~(\ref{eqn:ranapSISD})
 (dashed blue)
 become more and more strongly underestimated
 with an increased background ratio,
 those reconstructed with $\rm Na + Ge$ targets
 by Eq.~(\ref{eqn:rsigmaSDpSI_even})
 (solid red)
 become more and more strongly overestimated.
 Meanwhile,
 Figs.~\ref{fig:rsigmaSDnSI-08-ranap-const}
 and \ref{fig:rsigmaSDnSI-08-mchi-const}
 show that,
 while the $\sigmanSD / \sigmapSI$ ratios
 reconstructed with $\rm F + I + Si$ targets
 (dashed blue)
 become more and more strongly underestimated
 with an increased background ratio
 for all $\armn / \armp$ values,
 those reconstructed with $\rm Xe + Ge$ targets
 (solid red)
 become more and more strongly underestimated
 for $\armn / \armp~\gsim~0$
 and more and more strongly overestimated
 for $\armn / \armp~\lsim~0$.

 However,
 the shifts caused by events with
 the constant background spectrum
 are much smaller than those
 caused with the exponential one.
 Quantitatively,
 for an input WIMP mass of 100 GeV,
 an input SI WIMP--nucleon cross section of $10^{-8}$ pb,
 an input SD WIMP--proton coupling $\armp = 0.1$,
 and an input $\armn / \armp = 0.7$,
 by using 2 (3) data sets of a 40\% background ratio,
 the $\sigmapSD / \sigmapSI$ ratios
 would be reconstructed 4\% larger (7\% smaller)
 with $<$ 20\% larger statistical uncertainties,
 whereas the $\sigmanSD / \sigmapSI$ ratios
 would be reconstructed 14\% (30\%) smaller
 with \mbox{$\sim$ 30\%} larger statistical uncertainties,
 compared to the results
 with no background events.

 These results indicate that,
 once residue background events
 exist regularly between the experimental
 minimal and maximal cut--off energies
 or (mostly) in high energy ranges
 and one can therefore in principle estimate
 $r_{(X, Y, Z)}(Q_{{\rm min}, (X, Y, Z)})$
 and $r_{(X, Y, Z)}(Q_{s, 1, (X, Y, Z)})$
 (pretty) well,
 the ratios between the SD and SI WIMP--nucleon cross sections
 could then be estimated
 (pretty) precisely {\em without} worrying about
 the non--negligible backgrounds.

\section{Summary and conclusions}
 In this paper
 I reexamine the model--independent data analysis methods
 introduced in Refs.~\cite{DMDDidentification-DARK2009, DMDDranap}
 for the determinations of
 ratios between different WIMP--nucleon couplings/cross sections
 from data (measured recoil energies) of
 direct Dark Matter detection experiments directly
 by taking into account a fraction of residue background events,
 which pass all discrimination criteria and
 then mix with other real WIMP--induced events
 in the analyzed data sets.
 These methods require {\em neither} prior knowledge
 about the WIMP scattering and
 different possible background spectra
 {\em nor} about the WIMP mass;
 the unique needed information is the recoil energies
 recorded in two (or more) direct detection experiments.

 I considered at first
 the case of a dominant SD WIMP--nucleus interaction.
 Our simulations show that,
 due to the contribution of
 non--negligible residue background events
 in the analyzed data sets
 to low/high energy ranges,
 the reconstructed $\armn / \armp$ ratios
 would be {\em under-/overestimated};
 the larger the background ratio
 the larger these systematic deviations of
 the reconstructed $\armn / \armp$.
 But,
 by estimating the counting rates
 at the shifted points,
 instead of at the experimental minimal cut--off energies,
 one could (strongly) alleviate these systematic deviations
 as well as reduce the statistical uncertainties
 (with non--negligible
  experimental threshold energies).
 By using data sets of $\sim$ 20\% -- 40\% residue background events,
 the ratio between the SD WIMP couplings on neutrons and on protons
 could in principle still be reconstructed pretty well:
 for a WIMP mass \mbox{$\mchi = 100$ GeV} and $\armn / \armp = 0.75$,
 with data sets of a 10\% (20\%) background ratio,
 the systematic deviation
 could be $\lsim -12\%$
 ($\lsim -14\%$)
 with an $\sim 55\%$ ($\sim 60\%$)
 statistical uncertainty.

 Then I turned to consider
 the general combination of the SD WIMP--nucleus interaction
 with a non--negligible SI one.
 Our simulations show that,
 by combining three (two spin--sensitive) target nuclei,
 our method can be used to reconstruct
 the ratio of the SD WIMP coupling on neutrons to that on protons
 with data sets of $\lsim~20\%$ residue background events.
 And,
 more importantly,
 it doesn't matter whether
 the SD WIMP--nucleus cross section
 really dominates over the SI one or not.
 But,
 the larger the relative strength between
 the SD WIMP--nucleus interaction to the SI one,
 the smaller the systematic deviations
 as well as the statistical uncertainties.

 I considered also the reconstruction of
 the ratios between the SD and SI WIMP--nucleon cross sections.
 For WIMP masses \mbox{$\mchi~\gsim~25$ GeV},
 by using either the reconstructed $\armn / \armp$ ratio
 or a combination of a spin--sensitive nucleus with
 an only SI--sensitive one
 with data sets of $\lsim~20\%$ background ratios,
 while the $\sigmapSD / \sigmapSI$ ratio could be reconstructed
 with a $\sim$ 30\% statistical uncertainty,
 one could (only) estimate
 the order of magnitude of the $\sigmanSD / \sigmapSI$ ratio
 (because of an \mbox{$\sim$ 120\%} statistical uncertainty).

 Moreover,
 our simulations show also
 the WIMP--mass independence of
 the reconstructed $\armn / \armp$ ratios
 with non--negligible background events,
 especially for WIMP masses \mbox{$\mchi~\gsim$ 50 GeV}.
 For WIMP masses $\lsim~50$ GeV,
 the $\armn / \armp$ ratios
 could be (strongly) {\em underestimated},
 even with zero background events.
 However,
 this underestimate could be alleviated/corrected
 by decrease the experimental minimal cut--off energies
 of the analyzed date sets
 (to be negligible).
 Then,
 with data sets of $\lsim~20\%$ residue background events,
 one could still reconstruct the $\armn / \armp$ ratios
 pretty well,
 by either assuming a dominant SD WIMP interaction
 or using the general combination of the SI and SD cross sections.
 But,
 the statistical uncertainty
 could be pretty large,
 once the SD WIMP--nucleus interaction
 doesn't dominate over the SI one.

 Furthermore,
 it has also been found that,
 firstly,
 by taking different assumptions about the relative strength
 between the SI and SD WIMP--nucleus interactions,
 and/or using different moments of
 the one--dimensional WIMP velocity distribution function,
 at either the experimental minimal cut--off energies
 or the shifted energy points,
 there would be an (in)compatibility
 between different reconstructed $\armn / \armp$ ratios;
 with an increased background ratio,
 the incompatibility
 between the reconstructed results
 would become larger.
 Hence,
 this (in)compatibility could allow us to check
 the purity/availability of the analyzed data sets
 (as well as the dominance of the SI or SD WIMP interaction).

 Secondly and more importantly,
 our simulations with a constant background spectrum
 indicate that,
 once residue background events
 exist regularly between the experimental
 minimal and maximal cut--off energies
 or (even better) (mostly) in high energy ranges,
 one could in principle estimate
 the counting rates of the recoil spectrum
 of {\em only} WIMP--induced events
 at the experimental threshold energies
 (pretty) precisely.
 Then the ratio between two SD WIMP--nucleon couplings
 as well as the ratios between the SD and SI WIMP--nucleon cross sections
 could be estimated (pretty) precisely
 without worrying about
 the non--negligible backgrounds.

 In summary,
 as the forth part of
 the study of the effects of residue background events
 in direct Dark Matter detection experiments,
 we considered the determinations of ratios
 between different WIMP--nucleon couplings/cross sections.
 Our results show that,
 with currently running and projected experiments
 using detectors with $10^{-9}$ to $10^{-11}$ pb sensitivities
 \cite{Baudis07a, Aprile09a, Gascon09, Drees10}
 and $< 10^{-6}$ background rejection ability
 \cite{CRESST-bg, EDELWEISS-bg, Lang09b, Ahmed09b},
 once two or more experiments with different {\em spin--sensitive} target nuclei
 could accumulate a few tens events
 (in one experiment),
 we could in principle already estimate
 the relative strengths of couplings/cross sections of Dark Matter particles
 on ordinary matter
 with a reasonable precession,
 even though there could be some background events
 mixed in our data sets for analyses.
 Moreover,
 although two forms for background spectrum
 considered in this work is rather naive,
 the nuclear form factors for the SD WIMP interaction
 with different target nuclei
 are also more complicated as
 the simple thin--shell form used in our simulations,
 and the relative signs of
 the (ratios of the) expected/measured proton/neutron group spins
 of the used target nuclei
 could also change the reconstructed results
 (to be larger or smaller, underestimated or overestimated),
 one should be able to extend our observations/discussions
 to predict the effects of possible background events
 in their own experiments.
 Hopefully,
 this will not only encourage our experimental colleagues
 to present their (future) results
 in the parameter space of Dark Matter particles,
 but also help them
 to check the purity of their data sets,
 to understand (residue) background events in their experiments,
 as well as to improve their background discrimination techniques.
\subsubsection*{Acknowledgments}
 The author would like to thank
 the Physikalisches Institut der Universit\"at T\"ubingen
 for the technical support of the computational work
 demonstrated in this article.
 This work
 was partially supported by
 the National Science Council of R.O.C.~%
 under contract no.~NSC-99-2811-M-006-031
 as well as by
 the LHC Physics Focus Group,
 National Center of Theoretical Sciences, R.O.C..
\appendix
\setcounter{equation}{0}
\setcounter{figure}{0}
\renewcommand{\theequation}{A\arabic{equation}}
\renewcommand{\thefigure}{A\arabic{figure}}
%
%
\section{Formulae needed in Sec.~2}
 Here I list all formulae needed
 for the model--independent data analysis procedures
 used in Sec.~2.
 Detailed derivations and discussions
 can be found in Refs.~\cite{DMDDf1v, DMDDranap}.
\subsection{Estimating \boldmath$r(\Qmin)$ and $I_n(\Qmin, \Qmax)$}
 First,
 consider experimental data described by
\beq
     {\T Q_n - \frac{b_n}{2}}
 \le \Qni
 \le {\T Q_n + \frac{b_n}{2}}
\~,
     ~~~~~~~~~~~~ 
     i
 =   1,~2,~\cdots,~N_n,~
     n
 =   1,~2,~\cdots,~B.
\label{eqn:Qni}
\eeq
 Here the total energy range between $\Qmin$ and $\Qmax$
 has been divided into $B$ bins
 with central points $Q_n$ and widths $b_n$.
 In each bin,
 $N_n$ events will be recorded.
 Since the recoil spectrum $dR / dQ$ is expected
 to be approximately exponential, 
 the following ansatz for the {\em measured} recoil spectrum
 ({\em before} normalized by the experimental exposure $\calE$)
 in the $n$th bin has been introduced \cite{DMDDf1v}:
\beq
        \adRdQ_{{\rm expt}, \~ n}
 \equiv \adRdQ_{{\rm expt}, \~ Q \simeq Q_n}
 \equiv \rn  \~ e^{k_n (Q - Q_{s, n})}
\~.
\label{eqn:dRdQn}
\eeq
 Here $r_n$ is the standard estimator
 for $(dR / dQ)_{\rm expt}$ at $Q = Q_n$:
\beq
   r_n
 = \frac{N_n}{b_n}
\~,
\label{eqn:rn}
\eeq
 $k_n$ is the logarithmic slope of
 the recoil spectrum in the $n$th $Q-$bin,
 which can be computed numerically
 from the average value of the measured recoil energies
 in this bin:
\beq
   \bQn
 = \afrac{b_n}{2} \coth\afrac{k_n b_n}{2}-\frac{1}{k_n}
\~,
\label{eqn:bQn}
\eeq
 where
\beq
        \bQxn{\lambda}
 \equiv \frac{1}{N_n} \sumiNn \abrac{\Qni - Q_n}^{\lambda}
\~.
\label{eqn:bQn_lambda}
\eeq
 The error on the logarithmic slope $k_n$
 can be estimated from Eq.~(\ref{eqn:bQn}) directly as
\beq
   \sigma^2(k_n)
 = k_n^4
   \cbrac{  1
          - \bfrac{k_n b_n / 2}{\sinh (k_n b_n / 2)}^2}^{-2}
            \sigma^2\abrac{\bQn}
\~,
\label{eqn:sigma_kn}
\eeq
 with
\beq
   \sigma^2\abrac{\bQn}
 = \frac{1}{N_n - 1} \bbigg{\bQQn - \bQn^2}
\~.
\label{eqn:sigma_bQn}
\eeq
 $Q_{s, n}$ in the ansatz (\ref{eqn:dRdQn})
 is the shifted point at which
 the leading systematic error due to the ansatz
 is minimal \cite{DMDDf1v},
\beq
   Q_{s, n}
 = Q_n + \frac{1}{k_n} \ln\bfrac{\sinh(k_n b_n/2)}{k_n b_n/2}
\~.
\label{eqn:Qsn}
\eeq
 Note that $Q_{s, n}$ differs from
 the central point of the $n$th bin, $Q_n$.
 From the ansatz (\ref{eqn:dRdQn}),
 the counting rate at $Q = \Qmin$ can be calculated by
\beq
   r(\Qmin)
 = r_1 e^{k_1 (\Qmin - Q_{s, 1})}
\~,
\label{eqn:rmin_eq}
\eeq
 and its statistical error can be expressed as
\beq
   \sigma^2(r(\Qmin))
 = r^2(\Qmin) 
   \cbrac{  \frac{1}{N_1}
          + \bbrac{  \frac{1}{k_1}
                   - \afrac{b_1}{2} 
                     \abrac{  1
                            + \coth\afrac{b_1 k_1}{2}}}^2
            \sigma^2(k_1)}
\~,
\label{eqn:sigma_rmin}
\eeq
 since
\beq
   \sigma^2(r_n)
 = \frac{N_n}{b_n^2}
\~.
\label{eqn:sigma_rn}
\eeq
 Finally,
 since all $I_n$ are determined from the same data,
 they are correlated with
\beq
   {\rm cov}(I_n, I_m)
 = \sum_{a = 1}^{N_{\rm tot}} \frac{Q_a^{(n+m-2)/2}}{F^4(Q_a)}
\~,
\label{eqn:cov_In}
\eeq
 where the sum runs over all events
 with recoil energy between $\Qmin$ and $\Qmax$. 
 And the correlation between the errors on $r(\Qmin)$,
 which is calculated entirely
 from the events in the first bin,
 and on $I_n$ is given by
\beqn
 \conti {\rm cov}(r(\Qmin), I_n)
        \non\\
 \=     r(\Qmin) \~ I_n(\Qmin, \Qmin + b_1)
        \non\\
 \conti ~~~~ \times 
        \cleft{  \frac{1}{N_1} 
               + \bbrac{  \frac{1}{k_1}
                        - \afrac{b_1}{2} \abrac{1 + \coth\afrac{b_1 k_1}{2}}}}
        \non\\
 \conti ~~~~~~~~~~~~~~ \times 
        \cright{ \bbrac{  \frac{I_{n+2}(\Qmin, \Qmin + b_1)}
                               {I_{n  }(\Qmin, \Qmin + b_1)}
                        - Q_1
                        + \frac{1}{k_1}
                        - \afrac{b_1}{2} \coth\afrac{b_1 k_1} {2}}
        \sigma^2(k_1)}
\~;
\label{eqn:cov_rmin_In}
\eeqn
 note that
 the sums $I_i$ here only count in the first bin,
 which ends at $Q = \Qmin + b_1$.

 On the other hand,
 with a functional form of the recoil spectrum
 (e.g., fitted to experimental data),
 $(dR / dQ)_{\rm expt}$,
 one can use the following integral forms
 to replace the summations given above.
 Firstly,
 the average $Q-$value in the $n$th bin
 defined in Eq.~(\ref{eqn:bQn_lambda})
 can be calculated by
\beq
   \bQxn{\lambda}
 = \frac{1}{N_n} \intQnbn \abrac{Q - Q_n}^{\lambda} \adRdQ_{\rm expt} dQ
\~.
\label{eqn:bQn_lambda_int}
\eeq
 For $I_n(\Qmin, \Qmax)$ given in Eq.~(\ref{eqn:In_sum}),
 we have
\beq
   I_n(\Qmin, \Qmax)
 = \int_{\Qmin}^{\Qmax} \frac{Q^{(n-1)/2}}{F^2(Q)} \adRdQ_{\rm expt} dQ
\~,
\label{eqn:In_int}
\eeq 
 and similarly for the covariance matrix for $I_n$
 in Eq.~(\ref{eqn:cov_In}),
\beq
   {\rm cov}(I_n, I_m)
 = \int_{\Qmin}^{\Qmax} \frac{Q^{(n+m-2)/2}}{F^4(Q)} \adRdQ_{\rm expt} dQ
\~.
\label{eqn:cov_In_int}
\eeq 
 Remind that
 $(dR / dQ)_{\rm expt}$ is the {\em measured} recoil spectrum
 {\em before} normalized by the exposure.
 Finally,
 $I_i(\Qmin, \Qmin + b_1)$ needed in Eq.~(\ref{eqn:cov_rmin_In})
 can be calculated by
\beq
   I_n(\Qmin, \Qmin + b_1)
 = \int_{\Qmin}^{\Qmin + b_1}
   \frac{Q^{(n-1)/2}}{F^2(Q)} \bbigg{r_1 \~ e^{k_1 (Q - Q_{s, 1})}} dQ
\~.
\label{eqn:In_1_int}
\eeq 
 Note that,
 firstly,
 $r(\Qmin)$ and $I_n(\Qmin, \Qmin + b_1)$ should be
 estimated by Eqs.~(\ref{eqn:rmin_eq}) and (\ref{eqn:In_1_int})
 with $r_1$, $k_1$ and $Q_{s, 1}$
 estimated by Eqs.~(\ref{eqn:rn}), (\ref{eqn:bQn}), and (\ref{eqn:Qsn})
 in order to use the other formulae for estimating
 the (correlations between the) statistical errors
 without any modification.
 Secondly,
 $r(\Qmin)$ and $I_n(\Qmin, \Qmax)$ estimated
 from a scattering spectrum fitted to experimental data
 are usually not model--independent any more.
 Moreover,
 for the use of Eqs.~(\ref{eqn:In_sum}),
 (\ref{eqn:cov_In}), (\ref{eqn:In_int}),
 (\ref{eqn:cov_In_int}), and (\ref{eqn:In_1_int})
 the elastic nuclear form factor $\FQ$
 should be understood to be chosen
 for the SI and SD WIMP--nucleon cross section correspondingly.
\subsection{Statistical uncertainty on
            \boldmath$\abrac{\armn / \armp}_{\pm, n}^{\rm SD}$}
 By using the standard Gaussian error propagation,
 the statistical uncertainty on
 $\abrac{\armn / \armp}_{\pm, n}^{\rm SD}$
 estimated by Eq.~(\ref{eqn:ranapSD})
 can be expressed as
\beqn
        \sigma\abrac{\afrac{\armn}{\armp}_{\pm, n}^{\rm SD}}
 \=     \frac{\vBig{\SpY \SnX - \SpX \SnY}}
             {\bBig{\SnX \pm \SnY (\calR_{J, n, X} / \calR_{J, n, Y})}^2}
        \abrac{\frac{1}{2} \cdot \frac{\calR_{J, n, X}}{\calR_{J, n, Y}}}
        \non\\
 \conti ~~ \times
        \cBiggl{  \sum_{i, j = 1}^3
                  \bbrac{  \frac{1}{\calR_{n     , X}} \aPp{\calR_{n     , X}}{c_{i, X}}
                         - \frac{1}{\calR_{\sigma, X}} \aPp{\calR_{\sigma, X}}{c_{i, X}} } }
        \non\\
 \conti ~~~~~~~~~~~~~~~~ \times 
                  \bbrac{  \frac{1}{\calR_{n     , X}} \aPp{\calR_{n     , X}}{c_{j, X}}
                         - \frac{1}{\calR_{\sigma, X}} \aPp{\calR_{\sigma, X}}{c_{j, X}} }
                  {\rm cov}(c_{i, X}, c_{j, X})
        \non\\
 \conti ~~~~~~~~~~~~ 
        \cBiggr{+ (X \lto Y)}^{1/2}
\~.
\label{eqn:sigma_ranapSD}
\eeqn
 Here a short--hand notation for the six quantities
 on which the estimate of $(\armn / \armp)_{\pm, n}^{\rm SD}$ depends
 has been introduced:
\beq
   c_{1, X}
 = I_{n, X}
\~,
   ~~~~~~~~~~~~ 
   c_{2, X}
 = I_{0, X}
\~,
   ~~~~~~~~~~~~ 
   c_{3, X}
 = r_X(\QminX)
\~;
\label{eqn:ciX}
\eeq
 and similarly for the $c_{i, Y}$.
 Estimators for ${\rm cov}(c_i, c_j)$ have been given
 in Eqs.~(\ref{eqn:cov_In}) and (\ref{eqn:cov_rmin_In}).
 Explicit expressions for the derivatives of $\calR_{n, X}$
 given in Eq.~(\ref{eqn:RnX_min})
 with respect to $c_{i, X}$ are:
\cheqnXa{A}
\beq
   \Pp{\calR_{n, X}}{\InX}
 = \frac{n + 1}{n}
   \bfrac{\FQminX}{2 \QminX^{(n + 1) / 2} r_X(\QminX) + (n + 1) \InX \FQminX}
   \calR_{n, X}
\~,
\label{eqn:dRnX_dInX}
\eeq
\cheqnXb{A}
\beq
   \Pp{\calR_{n, X}}{\IzX}
 =-\frac{1}{n}
   \bfrac{\FQminX}{2 \QminX^{1 / 2} r_X(\QminX) + \IzX \FQminX}
   \calR_{n, X}
\~,
\label{eqn:dRnX_dIzX}
\eeq
 and
\cheqnXc{A}
\beqn
        \Pp{\calR_{n, X}}{r_X(\QminX)}
 \=     \frac{2}{n}
        \bfrac{  \QminX^{(n + 1) / 2} \IzX        - (n + 1) \QminX^{1 / 2} \InX}
              {2 \QminX^{(n + 1) / 2} r_X(\QminX) + (n + 1) \InX \FQminX}
        \non\\
 \conti ~~~~~~~~~~~~~~~~ \times 
        \bfrac{\FQminX}{2 \QminX^{1 / 2} r_X(\QminX) + \IzX \FQminX}
        \calR_{n, X}
\~;
\label{eqn:dRnX_drminX}
\eeqn
\cheqnX{A}%
 explicit expressions for the derivatives of $\calR_{n, Y}$
 with respect to $c_{i, Y}$ can be given analogously.
 Note that,
 firstly,
 factors $\calR_{n, (X, Y)}$ appear in all these expressions,
 which can practically be cancelled by the prefactors
 in the bracket in Eq.~(\ref{eqn:sigma_ranapSD}).
 Secondly,
 all the $I_{0, (X, Y)}$ and $I_{n, (X, Y)}$ should be understood
 to be computed according to
 Eq.~(\ref{eqn:In_sum}) or (\ref{eqn:In_int})
 with integration limits $\Qmin$ and $\Qmax$
 specific for that target.

 Similarly,
 expressions for the derivatives of $\calR_{\sigma, X}$
 can be computed from Eq.~(\ref{eqn:RsigmaX_min}) as
\cheqnXa{A}
\beq
   \Pp{\calR_{\sigma, X}}{\IzX}
 = \bfrac{\FQminX}{2 \QminX^{1 / 2} r_X(\QminX) + \IzX \FQminX}
   \calR_{\sigma, X}
\~,
\label{eqn:dRsigmaX_dIzX}
\eeq
\cheqnXb{A}
\beq
   \Pp{\calR_{\sigma, X}}{r_X(\QminX)}
 = \bfrac{2 \QminX^{1 / 2}}{2 \QminX^{1 / 2} r_X(\QminX) + \IzX \FQminX}
   \calR_{\sigma, X}
\~;
\label{eqn:dRsigmaX_drminX}
\eeq
\cheqnX{A}%
 and similarly for the derivatives of $\calR_{\sigma, Y}$.
 Remind that
 factors $\calR_{\sigma, (X, Y)}$ appearing here
 can also be cancelled by the prefactors
 in the bracket in Eq.~(\ref{eqn:sigma_ranapSD}).
\subsection{Statistical uncertainty on
            \boldmath$(\armn / \armp)_{\pm}^{\rm SI + SD}$}
 From the expression (\ref{eqn:ranapSISD}),
 the statistical uncertainty
 on $(\armn / \armp)_{\pm}^{\rm SI + SD}$
 can be given by
\beqn
        \sigma\afrac{\armn}{\armp}_{\pm}^{\rm SI + SD}
 \=     \cleft{   \sum_{i = X, Y, Z}
                  \bleft{   \pp{\cpX}\afrac{\armn}{\armp}_{\pm}^{\rm SI + SD} \cdot
                            \Pp{\cpX}{r_i(Q_{{\rm min}, i})}}}
        \non\\
 \conti ~~~~~~~~~~~~~~~~~~ 
        \cright{  \bright{+ \pp{\cpY}\afrac{\armn}{\armp}_{\pm}^{\rm SI + SD} \cdot
                            \Pp{\cpY}{r_i(Q_{{\rm min}, i})}}^2
                  \sigma^2(r_i(Q_{{\rm min}, i}))}^{1/2}
\!\!.
\label{eqn:sigma_ranapSISD}
\eeqn
 Here,
 from the first and second lines of
 the expression (\ref{eqn:ranapSISD}),
 we have,
\cheqnXa{A}
\beqn
        \pp{\cpX}\afrac{\armn}{\armp}_{\pm}^{\rm SI + SD}
 \=    -\frac{1}{\abrac{\cpX \snpX^2 - \cpY \snpY^2}^2}
        \non\\
 \conti ~~~~~~ \times 
        \bBiggl{    \cpY \snpX \snpY \abrac{\snpX - \snpY}}
        \non\\
 \conti ~~~~~~~~~~~~~~~~ 
        \bBiggr{\pm \frac{1}{2} \sfrac{\cpY}{\cpX}
                    \abrac{\cpX \snpX^2 + \cpY \snpY^2}
                    \vbrac{\snpX - \snpY}}
        \non\\
 \= \cleft{\renewcommand{\arraystretch}{0.5}
           \begin{array}{l l l}
            \\
            \D \mp \frac{\sqrt{\cpX \cpY} \abrac{\snpX - \snpY}}
                        {2 \cpX \abrac{\sqrt{\cpX} \snpX \mp \sqrt{\cpY} \snpY}^2}\~, &
            ~~ & ({\rm for}~\snpX > \snpY), \\~\\~\\
            \D \pm \frac{\sqrt{\cpX \cpY} \abrac{\snpX - \snpY}}
                        {2 \cpX \abrac{\sqrt{\cpX} \snpX \pm \sqrt{\cpY} \snpY}^2}\~, &
               & ({\rm for}~\snpX < \snpY), \\~\\
           \end{array}}
\label{eqn:dranapSISD_dcpX}
\eeqn
 and
\cheqnXb{A}
\beqn
        \pp{\cpY}\afrac{\armn}{\armp}_{\pm}^{\rm SI + SD}
 \=     \frac{1}{\abrac{\cpX \snpX^2 - \cpY \snpY^2}^2}
        \non\\
 \conti ~~~~~~ \times 
        \bBiggl{    \cpX \snpX \snpY \abrac{\snpX - \snpY}}
        \non\\
 \conti ~~~~~~~~~~~~~~~~ 
        \bBiggr{\pm \frac{1}{2} \sfrac{\cpX}{\cpY}
                    \abrac{\cpX \snpX^2 + \cpY \snpY^2}
                    \vbrac{\snpX - \snpY}}
        \non\\
 \= \cleft{\renewcommand{\arraystretch}{0.5}
           \begin{array}{l l l}
            \\
            \D \pm \frac{\sqrt{\cpX \cpY} \abrac{\snpX - \snpY}}
                        {2 \cpY \abrac{\sqrt{\cpX} \snpX \mp \sqrt{\cpY} \snpY}^2}\~, &
            ~~ & ({\rm for}~\snpX > \snpY), \\~\\~\\
            \D \mp \frac{\sqrt{\cpX \cpY} \abrac{\snpX - \snpY}}
                        {2 \cpY \abrac{\sqrt{\cpX} \snpX \pm \sqrt{\cpY} \snpY}^2}\~, &
               & ({\rm for}~\snpX < \snpY). \\~\\
           \end{array}}
\label{eqn:dranapSISD_dcpY}
\eeqn
\cheqnX{A}
 Then,
 from the definition (\ref{eqn:cpX}) of $\cpX$,
 one can get directly
\cheqnXa{A}
\beq
   \Pp{\cpX}{r_X(\QminX)}
 = 0
\~,
\label{eqn:dcpX_drminX}
\eeq
\cheqnXb{A}
\beq
   \Pp{\cpX}{r_Y(\QminY)}
 = \FSIQminZ \FSDQminX \cdot \frac{4}{3} \Afrac{J_X + 1}{J_X} \bfrac{\SpX}{A_X}^2
   \cdot \frac{\calR_{m, YZ}}{r_Y(\QminY)}
\~,
\label{eqn:dcpX_drminY}
\eeq
 and
\cheqnXc{A}
\beq
   \Pp{\cpX}{r_Z(\QminZ)}
 =-\FSIQminZ \FSDQminX \cdot \frac{4}{3} \Afrac{J_X + 1}{J_X} \bfrac{\SpX}{A_X}^2
   \cdot \frac{\calR_{m, YZ}}{r_Z(\QminZ)}
\~.
\label{eqn:dcpX_drminZ}
\eeq
\cheqnX{A}
 Similarly, 
 from the definition (\ref{eqn:cpY}) of $\cpY$,
 we have
\cheqnXa{A}
\beq
   \Pp{\cpY}{r_X(\QminX)}
 = \FSIQminZ \FSDQminY \cdot \frac{4}{3} \Afrac{J_Y + 1}{J_Y} \bfrac{\SpY}{A_Y}^2
   \cdot \frac{\calR_{m, XZ}}{r_X(\QminX)}
\~,
\label{eqn:dcpY_drminX}
\eeq
\cheqnXb{A}
\beq
   \Pp{\cpY}{r_Y(\QminY)}
 = 0
\~,
\label{eqn:dcpY_drminY}
\eeq
 and
\cheqnXc{A}
\beq
   \Pp{\cpY}{r_Z(\QminZ)}
 =-\FSIQminZ \FSDQminY \cdot \frac{4}{3} \Afrac{J_Y + 1}{J_Y} \bfrac{\SpY}{A_Y}^2
   \cdot \frac{\calR_{m, XZ}}{r_Z(\QminZ)}
\~.
\label{eqn:dcpY_drminZ}
\eeq
\cheqnX{A}
\subsection{Statistical uncertainty on
            \boldmath$\sigmapSD / \sigmapSI$}
 Since $\calCp$ and $\calCn$
 defined in Eqs.~(\ref{eqn:Cp}) and (\ref{eqn:Cn})
 are functions of $\armn / \armp$,
 once the $\armn / \armp$ ratio has been estimated
 (from e.g., some other direct detection experiments
  by Eq.~(\ref{eqn:ranapSD})
  by assuming a dominant  SD WIMP--nucleus interaction),
 $\sigmapSD / \sigmapSI$ can then be estimated
 by Eq.~(\ref{eqn:rsigmaSDpSI})
 with the following statistical uncertainty:
\beqn
        \sigma\afrac{\sigmapSD}{\sigmapSI}
 \=     \cleft{    \bbrac{\pp{(\armn / \armp)} \afrac{\sigmapSD}{\sigmapSI}}^2
                   \sigma^2\abrac{\afrac{\armn}{\armp}_{\pm, n}^{\rm SD}}}
        \non\\
 \conti ~~~~~~~~~~~~ 
        \cright{+ \sum_{i = X, Y}
                  \bbrac{  \frac{1}{\calE_i m_i^2}~
                           \pp{\calR_{m, i}} \afrac{\sigmapSD}{\sigmapSI}}^2
                  \sigma^2(r_i(Q_{{\rm min}, i})) }^{1/2}
\~,
\label{eqn:sigma_rsigmaSDpSI_ranapSD}
\eeqn
 where
\beq
    \pp{(\armn / \armp)} \afrac{\sigmapSD}{\sigmapSI}
 =  \pp{\calCpX} \afrac{\sigmapSD}{\sigmapSI} \cdot
    \Pp{\calCpX}{(\armn / \armp)}
  + \pp{\calCpY} \afrac{\sigmapSD}{\sigmapSI} \cdot
    \Pp{\calCpY}{(\armn / \armp)}
\~.
\label{eqn:drsigmaSDpSI_dranapSISD}
\eeq
 Here,
 from the expression (\ref{eqn:rsigmaSDpSI})
 for estimating $\sigmapSD / \sigmapSI$,
 its derivatives with respect to ${\cal C}_{{\rm p}, (X, Y)}$
 can be given as
\cheqnXa{A}
\beq
   \pp{\calCpX} \afrac{\sigmapSD}{\sigmapSI}
 =-\frac{\FSDQminX \calR_{m, Y}}
        {\calCpX \FSDQminX \calR_{m, Y} - \calCpY  \FSDQminY \calR_{m, X}}
   \afrac{\sigmapSD}{\sigmapSI}
\~,
\label{eqn:drsigmaSDpSI_dCpX}
\eeq
 and
\cheqnXb{A}
\beq
   \pp{\calCpY} \afrac{\sigmapSD}{\sigmapSI}
 = \frac{\FSDQminY \calR_{m, X}}
        {\calCpX \FSDQminX \calR_{m, Y} - \calCpY  \FSDQminY \calR_{m, X}}
   \afrac{\sigmapSD}{\sigmapSI}
\~.
\label{eqn:drsigmaSDpSI_dCpY}
\eeq
\cheqnX{A}
 And
 the derivatives of $\sigmapSD / \sigmapSI$
 with respect to $\calR_{m, (X, Y)}$ are
\cheqnXa{A}
\beqn
 \conti \pp{\calR_{m, X}} \afrac{\sigmapSD}{\sigmapSI}
        \non\\
 \=    -\frac{\bBig{  \calCpX \FSDQminX \FSIQminY
                    - \calCpY \FSDQminY \FSIQminX}
              \calR_{m, Y} }
             {\bBig{  \calCpX \FSDQminX \calR_{m, Y}
                    - \calCpY \FSDQminY \calR_{m, X} }^2}
\~,
\label{eqn:drsigmaSDpSI_dRmX}
\eeqn
 and
\cheqnXb{A}
\beqn
 \conti \pp{\calR_{m, Y}} \afrac{\sigmapSD}{\sigmapSI}
        \non\\
 \=     \frac{\bBig{  \calCpY \FSDQminY \FSIQminX
                    - \calCpX \FSDQminX \FSIQminY}
              \calR_{m, X} }
             {\bBig{  \calCpX \FSDQminX \calR_{m, Y}
                    - \calCpY \FSDQminY \calR_{m, X} }^2}
\~.
\label{eqn:drsigmaSDpSI_dRmY}
\eeqn
\cheqnX{A}%
 Meanwhile,
 from expression (\ref{eqn:Cp}) for $\calCp$
 one can find that
\beq
   \Pp{\calCp}{(\armn / \armp)}
 = \frac{2 \calCp}{\Srmp / \Srmn + \armn / \armp}
\~,
\label{eqn:dCp_dranap}
\eeq
 and,
 since we estimate in fact always $\armn / \armp$,
 one needs practically
\beq
   \Pp{\calCn}{(\armn / \armp)}
 =-\frac{2 \calCn}{\armn / \armp + (\Srmn / \Srmp) (\armn / \armp)^2}
\~.
\label{eqn:dCp_dranap}
\eeq

 On the other hand,
 for estimating $\armn / \armp$ by Eq.~(\ref{eqn:ranapSISD}),
 the statistical uncertainty on $\sigmapSD / \sigmapSI$
 can be expressed as
\beqn
        \sigma\afrac{\sigmapSD}{\sigmapSI}
 \=     \cBiggl{  \sum_{i = X, Y, Z}
        \cleft{   \bbrac{\pp{(\armn / \armp)} \afrac{\sigmapSD}{\sigmapSI}}}}
                  \bbrac{\pp{r_i(Q_{{\rm min}, i})}
                         \afrac{\armn}{\armp}_{\pm}^{\rm SI + SD}}
        \non\\
 \conti ~~~~~~~~~~~~~~~~~~~~~~~~ 
        \cBiggr{
        \cright{+ \frac{1}{\calE_i m_i^2}
                  \bbrac{\pp{\calR_{m, i}} \afrac{\sigmapSD}{\sigmapSI}}}^2
                  \sigma^2(r_i(Q_{{\rm min}, i}))}^{1/2}
\~,
\label{eqn:sigma_rsigmaSDpSI_ranapSISD}
\eeqn
 with $\p (\sigmapSD / \sigmapSI) / \p (\armn / \armp)$
 given in Eq.~(\ref{eqn:drsigmaSDpSI_dranapSISD})
 and
\beq
    \pp{r_i(Q_{{\rm min}, i})} \afrac{\armn}{\armp}_{\pm}^{\rm SI + SD} \!\!\!\!
 =  \pp{\cpX} \afrac{\armn}{\armp}_{\pm}^{\rm SI + SD} \!\!\!\! \cdot
    \Pp{\cpX}{r_i(Q_{{\rm min}, i})}
  + \pp{\cpY} \afrac{\armn}{\armp}_{\pm}^{\rm SI + SD} \!\!\!\! \cdot
    \Pp{\cpY}{r_i(Q_{{\rm min}, i})}
\~,
\label{eqn:dranapSISD_drminX}
\eeq
 for $i = X,~Y,~Z$.
 Moreover,
 by using Eq.~(\ref{eqn:CpX_p})
 to eliminate the $\armn / \armp$ dependence of $\sigmapSD / \sigmapSI$,
 the statistical uncertainty
 given in Eq.~(\ref{eqn:sigma_rsigmaSDpSI_ranapSISD})
 can be reduced to
\beq
        \sigma\afrac{\sigmapSD}{\sigmapSI}
 \simeq \frac{\FSIQminY (\calR_{m, X} / \calR_{m, Y})}{\calCpX \FSDQminX}
        \bbrac{  \frac{\sigma^2(r_X(\QminX))}{r_X^2(\QminX)}
               + \frac{\sigma^2(r_Y(\QminY))}{r_Y^2(\QminY)}}^{1/2}
\~.
\label{eqn:sigma_rsigmaSDpSI_ranapSISD_even}
\eeq
\end{document}